\documentclass[twoside,12pt]{article}
\usepackage{epsfig}
\usepackage{cite}
\usepackage{axodraw}

\newcommand{\Alpgen}{{\sc Alpgen}}
\newcommand{\ggtoWW}{{\sc gg2WW}}
\newcommand{\ggtoZZ}{{\sc gg2ZZ}}
\newcommand{\Hdecay}{{\sc Hdecay}}
\newcommand{\Herwig}{{\sc Herwig}}
\newcommand{\Madgraph}{{\sc Madgraph}}
\newcommand{\MCatNLO}{{\sc MC@NLO}}
\newcommand{\Powheg}{{\sc Powheg}}
\newcommand{\Prophecy}{{\sc Prophecy4f}}
\newcommand{\Pythia}{{\sc Pythia}}
\newcommand{\Sherpa}{{\sc Sherpa}}
\newcommand{\Comphep}{{\sc Comphep}}

\newcommand{\TeV}{\unskip\,\mathrm{TeV}}
\newcommand{\GeV}{\unskip\,\mathrm{GeV}}
\newcommand{\MeV}{\unskip\,\mathrm{MeV}}
\newcommand{\pba}{\unskip\,\mathrm{pb}}

\newcommand{\ipba}{\unskip\,\mathrm{pb^{-1}}}
\newcommand{\ifba}{\unskip\,\mathrm{fb^{-1}}}

\def\mathswitchr#1{\relax\ifmmode{\mathrm{#1}}\else$\mathrm{#1}$\fi}
\newcommand{\Pf}{\mathswitch  f}
\newcommand{\Pfbar}{\mathswitch{\bar f}}
\newcommand{\Pq}{\mathswitch  q}
\newcommand{\Pqbar}{\mathswitch{\bar q}}
\newcommand{\PW}{\mathswitchr W}

\newcommand{\PZ}{\mathswitchr Z}
\newcommand{\PV}{\mathswitchr V}

\newcommand{\Pg}{\mathswitchr g}

\newcommand{\PH}{\mathswitchr H}

\newcommand{\Pl}{\ell}
\newcommand{\Plp}{\ell^+}
\newcommand{\Plm}{\ell^-}
\newcommand{\Plpm}{\ell^\pm}
\newcommand{\Pn}{\nu}
\newcommand{\Pnbar}{\mathswitch{\bar \nu}}
\newcommand{\Pe}{\mathswitchr e}

\newcommand{\Pd}{\mathswitchr d}

\newcommand{\Pu}{\mathswitchr u}

\newcommand{\Ps}{\mathswitchr s}

\newcommand{\Pc}{\mathswitchr c}
\newcommand{\Pcbar}{\bar{\mathswitchr c}}
\newcommand{\Pb}{\mathswitchr b}
\newcommand{\Pbbar}{\mathswitchr{\bar b}}
\newcommand{\Pp}{\mathswitchr p}
\newcommand{\Pt}{\mathswitchr t}
\newcommand{\Ptbar}{\mathswitchr{\bar t}}
\newcommand{\Pep}{\mathswitchr {e^+}}
\newcommand{\Pem}{\mathswitchr {e^-}}

\newcommand{\Pmup}{\mathswitchr {\mu^+}}
\newcommand{\Pmum}{\mathswitchr {\mu^-}}

\newcommand{\PWp}{\mathswitchr {W^+}}
\newcommand{\PWm}{\mathswitchr {W^-}}

\newcommand{\Htogg}{\mathswitchr {\PH \to \gamma \gamma}}
\newcommand{\Htofl}{\mathswitchr {\PH \to \PZ \PZ \to \ell^+\ell^- \ell^{\prime +} \ell^{\prime -}}}
\newcommand{\Htoln}{\mathswitchr {\PH \to \PZ \PZ \to \ell^+\ell^- \nu \nu}}
\newcommand{\Htolq}{{\PH \to \PZ \PZ \to \ell^+\ell^- \Pq\Pqbar}}
\newcommand{\Htolb}{\mathswitchr {\PH \to \PZ \PZ \to \ell^+\ell^- \Pb\Pbbar}}
\newcommand{\Htolt}{\mathswitchr {\PH \to \PZ \PZ \to \ell^+\ell^- \tau \tau}}
\newcommand{\Htowwll}{\mathswitchr {\PH \to \PWp \PWm \to \ell^+ \nu \ell^- {\bar \nu}}}
\newcommand{\Htowwlnqq}{{\PH\to\PWp \PWm\to \ell \nu \Pq\Pqbar}}
\newcommand{\Htott}{\mathswitchr {\PH\to \tau^+ \tau^-}}
\newcommand{\Htobb}{\mathswitchr {\PH\to \Pb \Pbbar}}

\newcommand{\Zhqqbb}{{\PZ\PH\to \Pq\Pqbar \Pb\Pbbar}}
\newcommand{\Zhqqqq}{{\PZ\PH\to \Pq\Pqbar \Pq\Pqbar}}

\newcommand{\Zheeqq}{{\PZ\PH\to \Pep \Pem \Pq\Pqbar}}
\newcommand{\Zhmmqq}{{\PZ\PH\to \mu^+\mu^-\Pq\Pqbar}}
\newcommand{\Zheebb}{\mathswitchr {\PZ\PH\to \Pep \Pem \Pb\Pbbar}}
\newcommand{\Zhmmbb}{\mathswitchr {\PZ\PH\to \mu^+\mu^-\Pb\Pbbar}}
\newcommand{\Zhqqtt}{{\PZ\PH\to \Pq\Pqbar \tau^+\tau^-}}

\newcommand{\Zhttqq}{{\PZ\PH\to \tau^+\tau^- \Pq\Pqbar }}
\newcommand{\Zhttbb}{\mathswitchr {\PZ\PH\to \tau^+\tau^- \Pb\Pbbar }}
\newcommand{\Zhnnqq}{{\PZ\PH\to \nu \bar{\nu} \Pq\Pqbar }}
\newcommand{\Zhnnbb}{\mathswitchr {\PZ\PH\to \nu \bar{\nu} \Pb\Pbbar }}

\def\mathswitch#1{\relax\ifmmode#1\else$#1$\fi}

\newcommand{\MV}{\mathswitch {M_\PV}}
\newcommand{\MW}{\mathswitch {M_\PW}}

\newcommand{\MZ}{\mathswitch {M_\PZ}}
\newcommand{\MH}{\mathswitch {M_\PH}}
\newcommand{\Me}{\mathswitch {m_\Pe}}
\newcommand{\Mmy}{\mathswitch {m_\mu}}
\newcommand{\Mka}{\mathswitch {m_\mathrm{K}}}
\newcommand{\Mpi}{\mathswitch {m_\pi}}

\newcommand{\Mt}{\mathswitch {m_\Pt}}

\newcommand{\GH}{\Gamma_{\PH}}

\newcommand{\rc}{{\mathrm{c}}}
\newcommand{\rI}{{\mathrm{I}}}
\newcommand{\ri}{{\mathrm{i}}}
\newcommand{\rd}{{\mathrm{d}}}

\newcommand{\rL}{{\mathrm{L}}}
\newcommand{\rR}{{\mathrm{R}}}
\newcommand{\rT}{{\mathrm{T}}}
\newcommand{\rY}{{\mathrm{Y}}}

\newcommand{\MSbar}{$\overline{\mathrm{MS}}$}
\newcommand{\BR}{{\mathrm{BR}}}
\newcommand{\LO}{{\mathrm{LO}}}
\newcommand{\NLO}{{\mathrm{NLO}}}

\newcommand{\SM}{{\mathrm{SM}}}
\newcommand{\YM}{{\mathrm{YM}}}
\newcommand{\Yuk}{{\mathrm{Yuk}}}
\newcommand{\ferm}{{\mathrm{ferm}}}

\newcommand{\gs}{g_{\mathrm{s}}}
\newcommand{\alphas}{\alpha_{\mathrm{s}}}
\newcommand{\muF}{\mu_{\mathrm{F}}}
\newcommand{\muR}{\mu_{\mathrm{R}}}
\newcommand{\rw}{\mathswitchr w}
\newcommand{\sw}{\mathswitch {s_\rw}}
\newcommand{\cw}{\mathswitch {c_\rw}}
\newcommand{\GF}{\mathswitch {G_\mu}}
\newcommand{\sweff}{\mathswitch {\sin^2\theta^{f}_{\mathrm{eff}}}}
\newcommand{\cweff}{\mathswitch {\cos^2\theta^{f}_{\mathrm{eff}}}}

\def\beq{\begin{equation}}
\def\eeq{\end{equation}}
\def\bit{\begin{itemize}}
\def\eit{\end{itemize}}
\def\beqar{\begin{eqnarray}}
\def\eeqar{\end{eqnarray}}
\def\barr#1{\begin{array}{#1}}
\def\earr{\end{array}}
\def\bfi{\begin{figure}}
\def\efi{\end{figure}}
\def\btab{\begin{table}}
\def\etab{\end{table}}
\def\bce{\begin{center}}
\def\ece{\end{center}}
\def\nn{\nonumber}

\def\text{\textstyle}

\def\refeq#1{\mbox{(\ref{#1})}}

\def\reffi#1{\mbox{Figure~\ref{#1}}}
\def\reffis#1{\mbox{Figures~\ref{#1}}}
\def\refta#1{\mbox{Table~\ref{#1}}}

\def\refse#1{\mbox{Section~\ref{#1}}}
\def\refses#1{\mbox{Sections~\ref{#1}}}

\def\citere#1{\mbox{Ref.~\cite{#1}}}
\def\citeres#1{\mbox{Refs.~\cite{#1}}}

\makeatletter

\newcommand{\lsim}
{\mathrel{\raisebox{-.3em}{$\stackrel{\displaystyle <}{\sim}$}}}
\newcommand{\gsim}
{\mathrel{\raisebox{-.3em}{$\stackrel{\displaystyle >}{\sim}$}}}
\def\asymp#1%
{\mathrel{\raisebox{-.4em}{$\widetilde{\scriptstyle #1}$}}}

\def\Nequal#1%
{\mathrel{\raisebox{-.5em}{$\stackrel{=}{\scriptstyle\rm#1}$}}}
\newcommand{\dsl}[1]{\not \hspace{-0.7mm}#1}
\def\dsl{\mathpalette\make@slash}
\def\make@slash#1#2{\setbox\z@\hbox{$#1#2$}%
  \hbox to 0pt{\hss$#1/$\hss\kern-\wd0}\box0}

\makeatother

\def\draftdate{\relax}
\def\lra{\mathop{\mathrm{\leftrightarrow}}\nolimits}
\def\mda{\relax}
\def\mua{\relax}
\def\mla{\relax}
\def\Mda{\relax}
\def\Mua{\relax}
\def\Mla{\relax}
\def\draft{
\def\thtystars{******************************}
\def\sixtystars{\thtystars\thtystars}
\typeout{}
\typeout{\sixtystars**}
\typeout{* Draft mode!
         For final version remove \protect\draft\space in source file *}
\typeout{\sixtystars**}
\typeout{}
\def\draftdate{\today}
\def\mua{\marginpar[\boldmath\hfil$\uparrow$]%
                   {\boldmath$\uparrow$\hfil}%
                    \typeout{marginpar: $\uparrow$}\ignorespaces}
\def\mda{\marginpar[\boldmath\hfil$\downarrow$]%
                   {\boldmath$\downarrow$\hfil}%
                    \typeout{marginpar: $\downarrow$}\ignorespaces}
\def\mla{\marginpar[\boldmath\hfil$\rightarrow$]%
                   {\boldmath$\leftarrow $\hfil}%
                    \typeout{marginpar: $\lra$}\ignorespaces}
\def\Mua{\marginpar[\boldmath\hfil$\Uparrow$]%
                   {\boldmath$\Uparrow$\hfil}%
                    \typeout{marginpar: $\uparrow$}\ignorespaces}
\def\Mda{\marginpar[\boldmath\hfil$\Downarrow$]%
                   {\boldmath$\Downarrow$\hfil}%
                    \typeout{marginpar: $\downarrow$}\ignorespaces}
\def\Mla{\marginpar[\boldmath\hfil$\Rightarrow$]%
                   {\boldmath$\Leftarrow $\hfil}%
                    \typeout{marginpar: $\lra$}\ignorespaces}
\overfullrule 5pt
\oddsidemargin -15mm
\evensidemargin-1cm
\marginparwidth 10mm
}

\begin{document}

\renewcommand{\O}{\mathswitch{{\cal{O}}}}
\renewcommand{\L}{{\cal{L}}}
\newcommand{\M}{{\cal{M}}}

\title{ \vspace{1cm} \bf The Higgs Boson in the Standard Model\\ 
                     --- \\
                     From LEP to LHC: Expectations, Searches, and Discovery of a Candidate}
\author{\bf S.\ Dittmaier and M.\ Schumacher \\[1em]
Physikalisches Institut, Albert-Ludwigs-Universit\"at Freiburg, \\
    Hermann-Herder-Strasse 3, 79104 Freiburg im Breisgau, Germany}
\maketitle

\thispagestyle{empty}

\vspace{2cm}

\begin{abstract}
The quest for the Higgs boson of the Standard Model, which was a cornerstone
in the physics programme at particle colliders operating at the energy frontier
for several decades, is the subject of this review. 
After reviewing the formulation of electroweak symmetry 
breaking via the Higgs mechanism within the Standard Model,
the phenomenology of the Higgs boson at colliders 
and the theoretical and phenomenological
constraints on the Standard Model Higgs sector are discussed.
General remarks on experimental searches and the methodology of statistical
interpretation are followed by a description of the phenomenology of Higgs-boson
production and the corresponding precise predictions. The strategies of the experimental searches and
their findings are discussed for
the Large Electron Positron Collider (LEP) at CERN, the proton--antiproton collider 
Tevatron at Fermilab, and the proton--proton Large Hadron Collider (LHC) at CERN. The article concludes 
with the description of the observation of a Higgs-like boson at the LHC.
\end{abstract}


\clearpage
\tableofcontents
\clearpage

\section{Introduction}

\paragraph{Electroweak history in a nutshell} 
\mbox{}\\*[.5em]
More than 40 years ago the standard theory of electroweak 
interaction~\cite{Glashow:1961tr}, 
the Glashow--Salam--Weinberg model, emerged from a longer 
theoretical development, starting from the old Fermi theory in the 1930s, 
being generalized to a phenomenological model of intermediate massive
vector bosons, and finally receiving the form of a spontaneously broken 
gauge theory in the 1960s. In those days the last bottleneck in the
mathematically consistent construction of the electroweak theory was 
the merge of the successful Yang--Mills structure for the interaction
of vector bosons with fermions with the experimental fact that the weak 
vector bosons possess mass. A historical milestone in this
development was laid with the realization that a spontaneous breakdown 
of the gauge symmetry, driven by a self-interacting
scalar field, can lend mass to gauge bosons---an idea nowadays known as 
the ``Englert--Brout--Higgs--Guralnik--Hagen--Kibble (EBHGHK) mechanism'' or simply
``Higgs mechanism''.

In the early 1960's it was realized by Nambu, Goldstone, Salam and 
Weinberg~\cite{Nambu:1960xd} that the
spontaneous breakdown of a {\it global} continuous symmetry necessarily leads to the 
existence of a massless scalar (``Goldstone'') particle---a statement
known as ``Goldstone's theorem''. This fact prevented
theorists from associating the postulated massive weak
vector bosons with a broken symmetry, because no corresponding
Goldstone bosons had been observed. 
Later it was realized by Brout, Englert~\cite{Englert:1964et}, 
Higgs~\cite{Higgs:1964ia}, and
Guralnik, Hagen, Kibble~\cite{Guralnik:1964eu}
that a spontaneous breakdown
of a {\it local} continuous symmetry does not require Goldstone bosons,
but rather their degrees of freedom deliver the longitudinal
polarization modes of the gauge bosons that become massive. The understanding of
this phenomenon, called the EBHGHK or Higgs mechanism, lead to the 
mathematical formulation of 
the class of spontaneously broken gauge field theories.
In his second paper \cite{Higgs:1964ia} P.W.~Higgs stated
``It is worth noting that an essential feature of
the type of theory which has been described in
this note is the prediction of incomplete multiplets
of scalar and vector bosons.'' It seems that it is this statement on the particle
structure of such models that finally gave the Higgs particle its name.

The qualitative picture of the Higgs mechanism is
that all massive elementary particles%
\footnote{Here we understand that ``elementary'' means ``point-like'' 
and thus non-composite particles, because the mass of the latter
typically receives large contributions from their binding energy.
In the case of nucleons, the binding energy even comprises the
major part of the mass.}
interact with the 
vacuum, which is characterized by a non-vanishing expectation value of the
{\it Higgs field}. This interaction with the vacuum prevents massive 
particles from acquiring the speed of light, in analogy to light rays 
proceeding through matter. At the same time, the Higgs field itself 
admits quantum-mechanical excitations which appear as new scalar particle 
types, the {\it Higgs bosons}. 

Based on the understanding of the Higgs mechanism to lend masses to
gauge bosons,
Weinberg and Salam
successfully completed Glashow's model for the unified
electroweak interaction to the spontaneously broken
SU(2)$\times$U(1) gauge theory that still represents the heart
of the electroweak part of our Standard Model (SM) of particle physics
today.
The remaining part of the SM is the gauge theory of strong 
interactions, known as quantum chromodynamics~\cite{Gross:1973id}, 
which is based on the unbroken SU(3) colour gauge symmetry.
The final breakthrough of gauge theories in particle physics came 
in the early 1970's with the proof of their renormalizability,
both for unbroken and broken symmetries by 't~Hooft,
Veltman~\cite{'tHooft:1972ue}, Lee and 
Zinn-Justin~\cite{Lee:1972fj}. 
This step raised gauge theories generally to the 
level of mathematically consistent quantum field theories,
a fact that serves as our basis to work out predictions
for collider experiments at a level of precision that is
essentially only limited by our technical capabilities to evaluate
higher orders in perturbation theory. Of course, some non-perturbative
input is needed, such as parton distribution functions for
hadronic collisions, but conceptually there is a solid 
field-theoretical basis.

\paragraph{From theory to Higgs phenomenology and searches} 
\mbox{}\\*[.5em]
The particle content of the SM comprises three charged leptons,
three corresponding neutrinos, and six quarks in the 
sector of fermions. These ``matter fermions'' interact via the
exchange of gauge bosons---the photon $\gamma$ for electromagnetic
interaction, the weak gauge bosons $\PZ$ and $\PW^\pm$,
and the eight gluons of the strong interaction.
Apart from this particle content, which is experimentally 
established after the discoveries of the top quark at the Fermilab 
Tevatron in 1995
and of the $\tau$ neutrino at the Fermilab DONUT experiment in 2000, 
there is just one 
postulated electrically neutral Higgs boson in the SM
which is the phenomenological footprint 
of the Higgs mechanism. Since masses play the role of coupling strengths
between particles and the vacuum, and 
Higgs bosons appear as
``vacuum excitations'', the coupling of the Higgs boson 
to any other particle is predicted 
to be proportional to the particle's mass term in the field equations. 
A phenomenological confirmation 
of the Higgs mechanism, thus, requires finding the Higgs boson(s),
measuring its/their quantum numbers such as spin and electrical charge, 
and finally verifying the proportionality of the Higgs coupling to the
mass of the attached particle.
Once a value for the hypothetical Higgs-boson mass is assumed, 
the profile of the SM Higgs boson is completely fixed, and precision 
predictions are possible for production and decay rates and signal process 
kinematics. Based on these precise predictions experimental searches have 
been performed for the signature of the SM Higgs boson at colliders operating at 
the energy frontier. It took 
25~years after the formulation of the 
Higgs mechanism until a significant mass range could be probed
with the start of the operation of the Large Electron Positron Collider (LEP)
at CERN in 1989. The search was continued
at the Tevatron proton--antiproton 
collider from 2002 to 2011 at Fermilab. In 2010 the Large Hadron Collider (LHC) 
started to take data in proton--proton collisions at unprecedented CM energies
with a primary goal of finally answering the question of whether a SM Higgs boson is 
realized in nature or not.

\section{The Standard Model Higgs boson}

Before we enter a thorough discussion of the phenomenology of
the Higgs boson in the SM of particle physics, we
briefly review the general structure of the model.
Here we focus on the features of the SM that are most relevant 
for Higgs-boson phenomenology and refer to
standard textbooks such as \citeres{Weinberg-Vol2,BDJ,Sterman,ESW,Collins}
for more details and theoretical background.

\subsection{The gauge structure of the Standard Model}

The Lagrangian $\L_\SM$ of the SM can be divided into four different 
parts,
\beq
\L_\SM = \L_\YM + \L_\ferm + \L_\PH + \L_\Yuk,
\eeq
which will be discussed one by one in the following. 
We start with the first two,
which contain the dynamics of the gauge bosons and matter fermions
as predicted by the unbroken gauge symmetry, before we turn to the
latter two which involve the Higgs field.

The Yang--Mills part $\L_\YM$ describes the genuine dynamics
of the gauge fields, i.e.\ their free propagation as well as
their self-interactions,
\beq
\L_\YM = -\frac{1}{4}W^i_{\mu\nu}W^{i,\mu\nu} 
-\frac{1}{4} B_{\mu\nu}B^{\mu\nu}
-\frac{1}{4}G^a_{\mu\nu}G^{a,\mu\nu},
\eeq
where
\beqar
W^i_{\mu\nu} &=& 
\partial_\mu W^i_\nu-\partial_\nu W^i_\mu -g\epsilon^{ijk} W^j_\mu W^k_\nu,
\qquad i,j,k=1,2,3,
\nn\\
B_{\mu\nu} &=& \partial_\mu B_\nu-\partial_\nu B_\mu,
\nn\\
G^a_{\mu\nu} &=& \partial_\mu G^a_\nu-\partial_\nu G^a_\mu 
-\gs f^{abc} G^b_\mu G^c_\nu,
\qquad a,b,c=1,\dots,8,
\eeqar
are the field-strength tensors of the gauge fields $W^i_\mu$
for the SU(2)$_\rI$ group of the weak isospin $I_\rw^i$, $B_\mu$ for the U(1)$_\rY$
of weak hypercharge $Y_\rw$, and $G^a_\mu$ for the SU(3)$_\rc$ of colour.
The respective gauge couplings of these groups are denoted
$g$, $g'$, and $\gs$, and the structure constants of the 
non-abelian groups SU(2) and SU(3) are 
$\epsilon^{ijk}$ and $f^{abc}$, following the usual notation.

The interaction of the gauge fields with the fermions
is encoded in
\beq
\L_\ferm = \ri\overline\Psi_L\dsl{D}\Psi_L 
+ \ri\overline\psi_{\Pl_\rR}\dsl{D}\psi_{\Pl_\rR}
+ \ri\overline\Psi_Q\dsl{D}\Psi_Q
+ \ri\overline\psi_{u_\rR}\dsl{D}\psi_{u_\rR}
+ \ri\overline\psi_{d_\rR}\dsl{D}\psi_{d_\rR},
\eeq
where $L=(\nu_{\Pl_\rL},\Pl_{\rL})^\rT$ are the left-handed SU(2)$_\rI$
doublets of charged leptons $\Pl=\Pe,\mu,\tau$ and neutrinos
$\nu_\Pl=\nu_\Pe,\nu_\mu,\nu_\tau$, 
$Q=(u_\rL,d_\rL)^\rT$ the left-handed SU(2)$_\rI$ doublets of up-type
quarks $u=\Pu,\Pc,\Pt$ and down-type quarks $d=\Pd,\Ps,\Pb$,
and $\Pl_\rR$, $u_\rR$, $d_\rR$ are the respective
right-handed SU(2)$_\rI$ singlets.%
\footnote{Right-handed neutrinos are omitted, since they do not
play any role at high-energy colliders.}
The actual interaction is contained in the covariant
derivative
\beq
D_\mu = \partial_\mu
+\ri g I^i_\rw W^i_\mu
+\ri g' \frac{Y_\rw}{2} B_\mu
+\ri\gs T^a_\rc G^a_\mu,
\label{eq:D}
\eeq
where $I^i_\rw$, $Y_\rw$, and $T^a_\rc$ are the generators
of the respective gauge groups in the representation of the
fermions they act on, i.e.\ 
$I^i_\rw=\sigma^i/2$ ($\sigma^i$ = Pauli matrices) for
the left-handed SU(2)$_\rI$ doublets and $I^i_\rw=0$ for the
right-handed singlets, the weak hypercharge $Y_\rw$ (=~numbers) is related to the
relative electric charge $Q$ by the Gell-Mann--Nishijima relation
$Q=I^3_\rw+Y_\rw/2$, and $T^a_\rc=\lambda^a/2$
($\lambda^a$ = Gell-Mann matrices) for SU(3)$_\rc$ quark triplets
and $T^a_\rc=0$ for the leptons.
The requirement that the coupling structure of the photon is parity blind
and proportional to $Q\overline\psi\dsl{A}\psi$, as in quantum 
electrodynamics, identifies the
photon field $A_\mu$ and the Z-boson field $Z_\mu$ by the
rotation
\beq
\pmatrix{Z_\mu \cr A_\mu} = \pmatrix{\cw & -\sw \cr \sw & \cw}
\pmatrix{W^3_\mu \cr B_\mu}
\eeq
with the weak mixing angle $\theta_\rw$ and electric unit charge $e$
fixed by
\beq
\cos\theta_\rw = \cw = \sqrt{1-\sw^2} = \frac{g}{\sqrt{g^2+(g')^2}},
\qquad
e = \frac{gg'}{\sqrt{g^2+(g')^2}}.
\eeq
Once the photon is identified among the gauge fields, it is
easy to see that the fields
\beq
W^\pm_\mu = (W^1_\mu\mp\ri W^2_\mu)/\sqrt{2}
\eeq
correspond to the charged weak gauge bosons $\PW^\pm$ of
charge $\pm e$.

Note that the part of the SM described so far, encoded in
$\L_\YM + \L_\ferm$, does not involve any mass terms.
Naive gauge-boson mass terms,
such as $W^i_\mu W^{i,\mu}$ for the $\PW$ bosons,
obviously violate gauge invariance.
Since left- and right-handed fermions transform differently
under SU(2)$_\rI\times$U(1)$_\rY$ gauge transformations,
naive fermion mass terms 
$\propto (\overline\psi_{f_\rL}\psi_{f_\rR}+\overline\psi_{f_\rR}\psi_{f_\rL})$
for a fermion $f$ are also ruled out by gauge invariance.
The introduction of particle masses, while still maintaining the
gauge invariance of the dynamics of the model,
requires extensions of the theory.

\subsection{Electroweak symmetry breaking and
the Higgs mechanism}

The Higgs part 
\beq
\L_\PH = (D_\mu\Phi)^\dagger(D^\mu\Phi)-V(\Phi)
\label{eq:LH}
\eeq
of the Lagrangian extends the
experimentally well established particle content of the SM
by the complex scalar SU(2)$_\rI$ doublet 
$\Phi = (\phi^+,\phi^0)^\rT$
of weak hypercharge $Y_{\rw,\Phi}=1$, so that $\phi^+$ carries
charge $+e$ and $\phi^0$ is neutral.
In total $\Phi$ involves four real degrees of freedom.
The self-interaction of $\Phi$ is described by the potential
\beq
V(\Phi) = -\mu^2(\Phi^\dagger\Phi) 
+ \frac{\lambda}{4}(\Phi^\dagger\Phi)^2,
\label{eq:V}
\eeq
whose form is constrained by gauge invariance and
renormalizability of the model. The latter is guaranteed
by the polynomial structure of degree four in the field
components, the former by the $\Phi$ dependence via the combination
$\Phi^\dagger\Phi$, leaving only the two real free parameters
$\mu^2$ and $\lambda$ in $V$. While $\lambda>0$ is required
by vacuum stability, the sign of $\mu^2$ is deliberately
taken positive in order to force a non-vanishing 
vacuum expectation value (vev) $\Phi_0$ of $\Phi$.
Minimizing $V$ yields the condition
\beq
\Phi_0^\dagger\Phi_0 = \frac{v^2}{2}, \qquad 
v = 2\sqrt{\frac{\mu^2}{\lambda}}.
\eeq
Requiring that the vev is electrically neutral, forces
the upper component of $\Phi_0$ to vanish, i.e.\
$\Phi_0$ is fixed up to a phase, with the usual
choice $\Phi_0=(0,v/\sqrt{2})^\rT$. This freedom in choosing
the vev $\Phi_0$ of $\Phi$ reflects the spontaneous breakdown
of the SU(2)$_\rI\times$U(1)$_\rY$ symmetry down to the
remaining electromagnetic U(1)$_{\mathrm{em}}$ invariance.
Splitting off the vev from $\Phi$, 
\beq
\Phi = \pmatrix{\phi^+ \cr \phi^0 = (v+H+\ri\chi)/\sqrt{2} },
\label{eq:Phiparam}
\eeq
we reparametrize $\Phi$ in terms of the real
physical Higgs field $H$ and the unphysical would-be
Goldstone boson fields $\phi^+$ and $\chi$, which are complex and real, respectively. 
The fact that $\phi^+$ and $\chi$
do not correspond to physical states can already be seen by the fact
that they are connected to the vev by gauge transformations; in
fact one can always find a gauge,
known as the ``unitary gauge'', in which $\phi^+$ and $\chi$ vanish.
Making use of this gauge and
inserting the parametrization \refeq{eq:Phiparam} of $\Phi$ and
the covariant derivative \refeq{eq:D} with
$I^i_{\rw,\Phi}=\sigma^i/2$, $Y_{\rw ,\Phi}=1$,
$T^a_{\rc,\Phi}=0$ into the Higgs Lagrangian \refeq{eq:LH},
we find
\beq
\L_{\PH,\mbox{\scriptsize{U-gauge}}} 
= \frac{1}{2}(\partial H)^2 
+\frac{g^2}{4}(v+H)^2 W^+_\mu W^{-,\mu}
+\frac{g^2}{8\cw^2}(v+H)^2 Z_\mu Z^\mu
+ \frac{\mu^2}{2} (v+H)^2
- \frac{\lambda}{16} (v+H)^2,
\eeq
which in particular contains bilinear terms in the
gauge fields $W^\pm$, $Z$ and in the Higgs field $H$, i.e.\ mass terms for
the corresponding weak gauge bosons $\PW^\pm$ and $\PZ$ as well as
for the Higgs boson $H$.
Identifying these masses according to
\beq
\MW = \frac{gv}{2}, \qquad \MZ = \frac{\MW}{\cw}, \qquad
\MH = \sqrt{2\mu^2},
\eeq
we can eliminate the parameters $\mu^2$, $\lambda$, $v$ completely
and get
\beqar
\L_{\PH,\mbox{\scriptsize{U-gauge}}} 
&=& \frac{1}{2}(\partial H)^2 -\frac{1}{2}\MH^2 H^2
+\MW^2 W^+_\mu W^{-,\mu}
+\frac{1}{2}\MZ^2 Z_\mu Z^\mu
\nn\\
&& {} + g\MW H W^+_\mu W^{-,\mu} + \frac{g^2}{4} H^2  W^+_\mu W^{-,\mu}
+ \frac{g\MZ}{2\cw} H Z_\mu Z^\mu + \frac{g^2}{4\cw^2} H^2  Z_\mu Z^\mu
\nn\\
&& {} -\frac{g\MH^2}{4\MW}H^3 -\frac{g^2\MH^2}{32\MW^2}H^4
+\mbox{const.},
\eeqar
where we have not spelled out an irrelevant constant.
In summary, the SM makes the following important phenomenological predictions 
in the Higgs sector which can be tested experimentally:
\begin{itemize}
\item
Associated to the Higgs field $H$, a physical neutral, spinless
particle of mass $\MH$ is postulated---the Higgs boson. 
Since it can be viewed as some kind of vacuum excitation, it carries
the quantum numbers of the vacuum and is thus even with respect to CP symmetry.
Note that $\MH$ is the only free SM parameter that 
is tied to a property of the Higgs boson,
while the other parameters are fixed by the
weak-gauge-boson masses and the gauge couplings.
Theoretical and phenomenological constraints of $\MH$ are discussed
below.
\item
The ratio $\rho=\MW^2/(\cw^2\MZ^2)$~\cite{Ross:1975fq}%
\footnote{The ratio $\rho$ is called $\beta$ in \citere{Ross:1975fq}.} 
is equal to one, which is a non-trivial
relation among the weak-gauge-boson masses and the gauge couplings
$g=e/\sw$ and $g'=e/\cw$. While $\rho=1$ in this form is used to
define $\cw=\MW/\MZ$ in the process of renormalization to all orders,
it nevertheless has important phenomenological consequences.
For instance, on top of the
Z-boson resonance, as measured at LEP and SLD via $\Pep\Pem\to\PZ\to f\bar f$,
an effective weak mixing angle can be defined for each fermion
species $f$, usually quantified via $\sweff$, which can be 
measured from various asymmetries. The ratios
$\rho_f=\MW^2/(\cweff\MZ^2)$ thus are predicted to be
equal to one up to radiative corrections, a fact that is experimentally
confirmed at a level of better than $10^{-3}$~\cite{:2005ema}.

The lowest-order property $\rho=1$ is not shared by all possible
scalar sectors that can be employed to lend masses to the $\PW^\pm$ and
$\PZ$ bosons. If not accidental or forced by fine-tuned parameters, 
it is rather a consequence of
a ``custodial symmetry''~\cite{Sikivie:1980hm} of the scalar sector.
In the SM this is an SO(4) symmetry of $\L_\PH$ 
with respect to the exchange of the four real components of $\Phi$, 
which holds up to U(1)$_{\mathrm{Y}}$ gauge interactions and differences in the fermion masses 
within SU(2)$_\rI$ doublets, both inducing custodial-symmetry-breaking effects, 
however, only in higher orders.
\item
The model predicts couplings of the Higgs boson to a massive
weak gauge boson $\PV=\PW,\PZ$ proportional to $\MV^2/v$,
because the $HV^\dagger V$ couplings originate from the factor
$(1+H/v)^2$ multiplying the gauge-boson mass term in $\L_\PH$.
Owing to the square in this factor, quartic couplings
of two Higgs bosons and two gauge bosons are also predicted, which are
proportional to $\MV^2/v^2$.
\item
Finally, triple and quartic Higgs-boson self-interactions
are predicted, both scaling with $\MH^2$. Since these couplings
are in one-to-one correspondence with the shape of the Higgs
potential that drives electroweak (EW) symmetry breaking, an
experimental reconstruction of these couplings from an analysis of
scattering processes would be part of an ultimate phenomenological
confirmation of the Higgs mechanism.
However, since the relevant processes involve multi-Higgs-boson final
states with very low cross sections, the LHC will at best be
able to give qualitative results here.
\end{itemize}

\subsection{Yukawa couplings and fermion masses}

Renormalizability and gauge invariance of the SM Lagrangian
allow for so-called Yukawa couplings of the Higgs doublet $\Phi$
to all fermions. The most general form of these interactions is
\beq
\L_\Yuk = -\overline\Psi_L G_\Pl \psi_{\Pl_\rR}\Phi
-\overline\Psi_Q G_u \psi_{u_\rR}\tilde\Phi
-\overline\Psi_Q G_d \psi_{d_\rR}\Phi
+\mbox{h.c.},
\eeq
where ``h.c.'' means hermitian conjugate and 
$\tilde\Phi=\ri\sigma^2\Phi^*=((\phi^0)^*,-\phi^-)^\rT$
denotes the charge-conjugate Higgs doublet with quantum numbers
opposite to $\Phi$.
The matrices $G_f$ ($f=\Pl,u,d$) represent arbitrary complex
$3\times3$ matrices, i.e.\ at first sight $\L_\Yuk$ involves
a large number of free parameters. However, most of them turn out 
to be not physically relevant and can be transformed to
canonical values or eliminated by appropriate field redefinitions.
We first note that each term in $\L_\Yuk$ involves terms that
are bilinear in the fermion fields because $\Phi$ and $\tilde\Phi$
contain a constant piece in the form of the vev $v$.
More precisely, the non-diagonal elements of 
$G_f$ mix the left- and right-handed
parts of the different generations of fermion type
$f=\Pl,u,d$, where $\Pl$ generically stands for charged leptons,
$u$ for up-type quarks, and $d$ for down-type quarks.
Owing to these mixing terms, a fermion of flavour $f_i$ ($i=1,2,3$)
of the $i$th generation oscillates into $f_j$ of the other
generations ($j\ne i$) even during a free propagation in space and time.
This oscillation can be removed upon transforming the
existing ``flavour basis'' 
$(\psi_{f_{\tau,1}},\psi_{f_{\tau,2}},\psi_{f_{\tau,3}})$ 
of left- and right-handed fields ($\tau=\rL,\rR$)
into a ``mass basis''
$(\hat\psi_{f_{\tau,1}},\hat\psi_{f_{\tau,2}},\hat\psi_{f_{\tau,3}})$ 
with a unitary matrix $U$,
\beq
\hat\psi_{f_{\tau,i}} = U^{f_{\tau}}_{ij} \,\psi_{f_{\tau,j}},
\qquad f=\Pl,u,d, \qquad \tau=\rL,\rR, 
\eeq
where in this process the matrices $G_f$ receive a diagonal form,
\beq
U^{f_{\rL}} G_f  (U^{f_{\rR}})^\dagger = 
\frac{\sqrt{2}}{v} \, \mathrm{diag}\{m_{f_1},m_{f_2},m_{f_3}\}.
\label{eq:Gf-diag}
\eeq
The diagonal value $m_{f_i}$, which can be chosen non-negative
by convention, is the mass of the fermion $f_i$.
For completeness we mention that the left-handed neutrino
fields are transformed with the same unitary matrix as their
charged counterparts; this is possible as long as we work in the
approximation of three massless (i.e.\ mass-degenerate)
neutrinos.
The effect of this field redefinition on the whole SM
Lagrangian can be summarized easily:
(i) The coupling matrices $G_f$ are replaced by their diagonal form
\refeq{eq:Gf-diag};
(ii) all fermion fields $\psi_{f_{\tau,i}}$ are replaced by their
counterparts $\hat\psi_{f_{\tau,i}}$ of the mass basis;
(iii) the only remnant of the $U$ matrices is the appearance
of the Cabibbo--Kobayashi--Maskawa matrix
$V=U^{u_\rL}(U^{d_\rL})^\dagger$
in fermion chains of type 
$\overline{\hat\psi}_{u_\rL}\dots V\hat\psi_{d_\rL}$
and of $V^\dagger$ in 
$\overline{\hat\psi}_{d_\rL}\dots V^\dagger\hat\psi_{u_\rL}$.
In other words, only charged-current interactions receive
modifications by $V$, while neutral currents remain unchanged.
In the following we adopt the common convention to omit
the clumsy hats on fermionic fields and assume the use of the
mass basis.

In the unitary gauge, where the would-be Goldstone fields are absent,
the Yukawa Lagrangian takes the simple final form
\beq
\L_{\Yuk,\mbox{\scriptsize{U-gauge}}} = -\sum_f m_f \,
(\overline\psi_{f_\rL}\psi_{f_\rR}+\overline\psi_{f_\rR}\psi_{f_\rL})
\, \biggl(1+\frac{H}{v}\biggr),
\eeq
where the sum over $f$ runs over all fermion flavours of all generations.
This form shows a distinctive footprint of the Higgs mechanism
in the fermionic sector:
The Higgs boson couples to each fermion $f$ of mass $m_f$ 
with the strength $y_f=m_f/v$. Moreover, the coupling is the one of
a pure scalar, i.e.\ the coupling to fermions does not have any 
pseudo-scalar admixture proportional to $\gamma_5$.
Testing these features offers a possibility to empirically tell 
a potential Higgs candidate from scalar particles predicted by
other models. Alternative models with non-minimal Higgs sectors 
often predict new pseudoscalars as well,
or even scalar particles without definite CP quantum numbers.
Moreover, the strict proportionality of the Yukawa coupling
strength to the fermion masses might be broken, as it is
for instance the case in (type-II) Higgs doublet models, where
the proportionality factor between $y_f$ and $m_f$ is different
for up- and down-type fermions.

\subsection{From input parameters to predictions for collider experiments}
\label{se:predictions}

\paragraph{Running parameters and renormalization scale} \mbox{}
\\*[.5em]
Like any gauge-field theory, the SM involves ultraviolet (UV) divergences that are
removed in the process of renormalization which ties the input parameters
of the theory to measurable quantities at some renormalization scale $\muR$.
The input parameters, which are taken from experiment, in this sense depend
on the arbitrary value of $\muR$ as well as on more details that fix the actual
procedure, thereby defining the {\it renormalization scheme}.
The fact that predicted physical observables,
on the other hand, cannot depend on $\muR$ is expressed in terms of 
renormalization group equations (RGE) for each scheme (if it allows for a flexible
renormalization scale).
Solving these RGE for any field-theoretical quantity naturally involves the
concept of running, i.e.\ $\muR$-dependent, parameters, such as the famous
strong coupling $\alphas(\muR)$. 
The running of each parameter from one scale 
to another can be predicted in perturbation theory. 
In actual predictions of observables, thus, there is an {\it explicit} and an 
{\it implicit} dependence on $\muR$, which compensate each other order by order
in perturbation theory---the former resulting from the renormalization
of the considered observable at the scale $\muR$, the latter from the input values 
matched to renormalized quantities at $\muR$.
A residual $\muR$ dependence arises from orders of perturbation theory that are
not completely taken into account.
If an observable involves a typical scale $Q$, such as a scattering cross section
at energy $Q$, it is advisable to fix $\muR$
in the vicinity of $Q$, since this minimizes potentially large missing corrections
that involve powers of logarithms $\ln(Q/\muR)$.
If a process involves many different scales, the benefit of adjusting $\muR$ is
limited.
For the precise mathematical formulation of this concept we refer to
standard literature on quantum field 
theory such as \citeres{Weinberg-Vol2,BDJ,Sterman,ESW,Collins}.

\paragraph{The input parameters of the SM} \mbox{}
\\*[.5em]
As is obvious from the construction of the model, 
the free input parameters
of the SM are the gauge couplings $g$, $g'$, $g_{\mathrm{s}}$, the parameters
$\mu^2$ and $\lambda$ of the Higgs sector, the fermion masses $m_f$, and
the CKM matrix $V$. For phenomenology it is much more convenient to take instead
the following parameters as input:
the electromagnetic coupling $\alpha=e^2/(4\pi)$,
the strong couplings constant $\alphas=g_{\mathrm{s}}^2/(4\pi)$,
the weak gauge-boson masses $\MW$ and $\MZ$,
the Higgs-boson mass $\MH$,
and finally $m_f$ and $V$.
The masses can all be defined as {\it pole masses}, defined from the locations
of the particle poles in the respective propagators, but for the heavy quarks
it is often useful to switch to a running mass at some appropriate scale.
For Higgs physics, the CKM matrix plays a minor role.
The couplings $\alpha$ and $\alphas$, however, have to be chosen 
thoughtfully. The strong coupling is usually defined as running coupling
$\alphas(\muR)$ in the so-called \MSbar\ scheme. Very often the
value of $\alphas$ at the Z~pole, $\alphas(\MZ)$, is used as numerical input
and transferred to some other value $\muR$ using the RGE.

For the electromagnetic coupling $\alpha$ basically the choice is between three
different values: the fine-structure constant $\alpha(0)\approx1/137$,
the effective value $\alpha(\MZ)\approx1/129$, where $\alpha(0)$ is evolved via RGE
from zero-momentum transfer to the Z~pole, and an effective value
derived from the Fermi constant $\GF$ leading to
$\alpha_{\GF}=\sqrt{2}\GF\MW^2(1-\MW^2/\MZ^2)/\pi\approx1/132$, 
defining the so-called ``$\GF$-scheme''. The various values of $\alpha$
differ by $2{-}6\%$.
Very often the actual value of $\alpha$ can be adjusted in such a way
that large universal EW corrections are already absorbed into the
lowest-order prediction. Even different values of $\alpha$ can be appropriate
in one calculation, but care has to be taken that the same choice is taken
within gauge-invariant subsets of diagrams. 
An external photon (virtuality $Q^2=0$) always effectively couples with $\alpha(0)$,
while internal photons with virtuality $Q^2$ at a high energy scale
effectively couple with $\alpha(Q)$, which is much closer to $\alpha(\MZ)$.
On the other hand, the couplings of W and Z bosons should be parametrized
with $\alpha_{\GF}$, which does not only take into account the running
from $Q=0$ to the EW scale, but also universal effects from the 
$\rho$-parameter.
Following these rules, in particular, avoids perturbative instabilities
due to the appearance of light-quark masses, as e.g.\ discussed in
\citere{Dittmaier:2001ay}.

\paragraph{Predictions for hadronic collisions and factorization scale} \mbox{}
\\*[.5em]
Hadronic collisions are treated within the {\it QCD-improved parton model},
whose full description is beyond the scope of this review
(see e.g.\ the standard textbooks~\cite{Sterman,ESW,Collins}).
In this framework hadronic cross sections are calculated in the factorized
form
\beq
\sigma_{h_1h_2} = \sum_{a,b}\int_0^1\rd x_1 \int_0^1\rd x_2\,
f_{a/h_1}(x_1,\muF)\, f_{b/h_2}(x_2,\muF)\, 
\int\rd\hat\sigma_{a_1a_2}(x_1 P_1,x_2 P_2,\muF),
\eeq
where the parton distribution functions (PDFs) $f_{a_i/h_i}(x_i,\muF)$
are generalized probability densities for finding a parton $a_i$ 
(quarks, antiquarks, gluons, photons) with momentum
$p_i=x_iP_i$ in the hadron $h_i$ which has momentum $P_i$.
The PDFs comprise the soft physics, which is dominated by small
energy scales and not accessible by perturbation theory,
and are process independent.
The partonic cross sections $\rd\hat\sigma_{a_1a_2}$, on the other hand,
contain the full information on the hard scattering reaction of the partons
into a specific final state and can be calculated within perturbation theory. 
The separation of the soft and hard domains
is performed at the {\it factorization scale} $\muF$, which effectively
is the upper bound on transverse momenta up to which outgoing hadronic 
particles are considered as parts of the struck hadron.
Similar to $\muR$, the scale 
$\muF$ is arbitrary to a large extent, and 
predictions---if calculated to arbitrary precision---cannot depend on it.
In $\rd\hat\sigma_{a_1a_2}(x_1 P_1,x_2 P_2,\muF)$ the value of $\muF$
can be trivially changed, in the PDFs the change of $\muF$ is governed
by the so-called Dokshitzer--Gribov--Lipatov--Altarelli--Parisi
(DGLAP) evolution, which ensures the $\muF$-independence of
observables.
Again a truncation of the perturbative series for $\rd\hat\sigma_{a_1a_2}$,
however, leads to some residual $\muF$ dependence of observables, which is
part of the intrinsic uncertainty in predictions.

The PDFs cannot be calculated from first principles, but are 
fitted to a large variety of data resulting from lepton--hadron and
hadron--hadron collisions. State-of-the-art PDF sets are provided by
several groups:
CT10~\cite{Lai:2010vv},
MSTW2008~\cite{Martin:2009iq},
NNPDF2.1~\cite{Ball:2011mu},
ABKM11~\cite{Alekhin:2012ig},
GJR08~\cite{Gluck:2008gs}, and
HERAPDF1.5~\cite{CooperSarkar:2010wm}.
Generally, there is good agreement between predictions, e.g.\ for the
LHC, based on those PDF sets, in particular between CT10, MSTW2008, and NNPDF2.1,
and differences between the PDF sets are continuously analyzed.
More details and references on PDFs and their application to Higgs physics
at the LHC can be found in \citeres{Botje:2011sn,Dittmaier:2011ti,Dittmaier:2012vm}.
In practice, each perturbative order requires its own set of PDFs for
consistency,
i.e.\ a LO cross section $\sigma_{\LO}$ has to be evaluated with LO PDFs
and the corresponding one-loop running $\alphas(\mu)$,
$\sigma_{\NLO}$ with NLO PDFs and two-loop $\alphas(\mu)$, etc.
In this context, QCD corrections are often quantified by the
{\it $K$-factor} 
\beq
K=\frac{\sigma_{\mathrm{(N)NLO}}}{\sigma_{\mathrm{LO}}},
\eeq
resulting from the cross sections consistently calculated in different orders.
While up-to-date PDF sets are available in LO, NLO, and NNLO QCD,
no current PDF set includes EW corrections, i.e.\ EW corrections are
both missing in the PDF fit to data as well as in the $\muF$-evolution of the PDFs.
The older PDF set MRST2004QED~\cite{Martin:2004dh} includes these effects,
but is outdated otherwise.
It is, thus, more advisable to use up-to-date PDFs, since
ignoring EW PDF corrections typically introduces uncertainties of the level of
$\lsim1\%$ only~\cite{Spiesberger:1995dm}.%
\footnote{Though outdated, the MRST2004QED PDF set is, however, the only one that provides
a photon PDF and, thus, offers the opportunity to at least approximately
calculate the effect from partonic channels involving initial-state photons.}

\paragraph{Parametric and theoretical uncertainties} \mbox{}
\\*[.5em]
{\it Parametric uncertainties} arise from the errors of the input parameters of the
SM model. For observables that involve strongly interacting objects
the largest errors usually originate from the uncertainties of
the strong coupling constant $\alphas$ and of the masses of the heavy quarks
that are possibly directly involved. 
In observables based on the EW interaction often the uncertainty of
the electromagnetic coupling $\alpha$ becomes significant if it is derived from the
value of the running coupling $\alpha(Q^2)$ at high energies, such as
$\alpha(\MZ^2)$. Parametric errors of different observables
are usually correlated, which has to be respected when their effect is combined.

For hadronic collisions the PDFs are
very often the largest source of parametric uncertainties, which are also
correlated with $\alphas$. The various groups
who deliver regular updates of PDFs fitted to data provide different 
``error sets'' of PDFs whose spread quantifies the PDF uncertainty. These
error PDFs also 
allow for a combined assessment of the PDF+$\alphas$ error
of observables. A proposal how
to combine errors from PDFs of different groups can be found in the report of
the PDF4LHC Working Group~\cite{Botje:2011sn}.

{\it Theoretical uncertainties} can be---to some extent---quantified by the
``scale uncertainties'' mentioned above.
The compensation between explicit and implicit $\muR$ and/or $\muF$
dependences remains necessarily
incomplete in perturbative calculations, because in practice not all (but 
only few) perturbative orders can be calculated completely. Thus, a
residual scale dependence remains in predictions, and, if the scales are varied
within a reasonable range, the corresponding variation in the result
quantifies theoretical uncertainties originating from missing higher-order effects.
One should, however, keep in mind that not all missing corrections are sensitive
to the residual scale dependences. For instance, missing finite corrections depend on
$\muR$ only very weakly, but might be large. Thus, overdoing a minimization
of the residual scale dependence does not remove the uncertainties, but
rather tends to hide them. 

Finally, parametric uncertainties and theoretical uncertainties have to be
combined. Since theoretical uncertainties have no true statistical meaning that
is safely tied to a probability distribution, the two errors should not be
added in quadrature, but linearly. A possible procedure to do that for
Higgs observables at the LHC was, e.g., proposed in \citere{Dittmaier:2011ti}.

\subsection{Theoretical and phenomenological constraints on the Higgs sector}
\label{se:Hconstraints}

Although the mass $\MH$ of the Higgs boson represents a free parameter
of the SM, it cannot assume arbitrary values. In the following we
briefly summarize the bounds on $\MH$ from above
and below that originate from theoretical considerations and 
phenomenological constraints. 

\paragraph{The heavy-Higgs limit of the SM} \mbox{}
\\*[.5em] 
Before looking more closely 
into the phenomenology of Higgs bosons, it is
interesting to discuss the behaviour of the model in the limit
$\MH\to\infty$, which means that $\MH$ is supposed to be much larger
than any other mass or energy scale involved in the considered 
physical system. In this limit the Higgs boson is effectively
removed from the physical particle spectrum, but its vev still 
delivers the masses to the weak gauge bosons and massive fermions.
Formally the Higgs potential tends to a delta
function proportional to $\delta(\Phi^\dagger\Phi-\Phi_0^\dagger\Phi_0)$,
i.e.\ the absolute value of the Higgs doublet field is constrained to its
vev. This constrained scalar sector is more conveniently
described in the framework of non-linear representations of the
doublet $\Phi$, where the physical Higgs boson plays the role of
the radial excitation in the space of its four real components
and gets fixed by the delta function. 
\newcommand{\tr}{\mathrm{tr}}
\newcommand{\bPhi}{\mathbf{\Phi}}

To illustrate this, we first rephrase the Higgs Lagrangian \refeq{eq:LH}
in matrix notation,
\beqar
\L_\PH = \frac{1}{2}\tr\{(D_\mu\bPhi)^\dagger(D^\mu\bPhi)\}-V(\Phi),
\qquad
D_\mu\bPhi = \partial_\mu\bPhi+\ri \frac{g}{2}\sigma^i W^i_\mu\bPhi
-\ri\frac{g'}{2}\bPhi\sigma^3 B_\mu,
\label{eq:LHnl}
\eeqar
where the hermitian matrix field $\bPhi$ is given by
\beq
\bPhi = \frac{1}{\sqrt{2}}\left[(v+H)\mathbf{1}+\ri\phi^i\sigma^i\right]
=  \pmatrix{(v+H - \ri\chi)/\sqrt{2} & \phi^+  \cr
                              -\phi^- & (v+H + \ri\chi)/\sqrt{2}}
= \left(\tilde\Phi,\Phi\right),
\eeq
with $\sigma^i$ again denoting the Pauli matrices and
the real scalar fields $\phi^i$, which are defined by
$\phi^\pm=(\phi^2\pm\ri\phi^1)/\sqrt{2}$ and $\chi=-\phi^3$.
In this linear representation the Lagrangian
and the meaning of the scalar field components are exactly the same as described 
before, only the bookkeeping is somewhat different. 
The non-linear representation of $\bPhi$ parametrizes the matrix field
according to 
\beq
\bPhi = \frac{1}{\sqrt{2}}(v+H)U, \qquad
U = \exp\left\{\frac{\ri\sigma^i\phi^i}{v}\right\},
\eeq
which results from the linear parametrization upon a non-trivial field
transformation. 
In this pa\-ra\-me\-tri\-za\-tion the field $H$ is gauge invariant.
Inserting the non-polynomial field $\bPhi$ into the 
Higgs Lagrangian \refeq{eq:LHnl}, leads to a theory that is physically equivalent
to the SM with the more common linear Higgs representation~\cite{Lee:1972yfa}, i.e.\ the
$S$-matrix is the same in the linear and non-linear representations, although
the Green functions do not coincide.
The new form of the Higgs Lagrangian is
\beq
\L_{\PH,\mathrm{nl}} = \frac{1}{4}(v+H)^2\tr\{(D_\mu U)^\dagger(D^\mu U)\}
+\frac{1}{2}(\partial_\mu H)(\partial^\mu H)
+\frac{\mu^2}{2}(v+H)^2-\frac{\lambda}{16}(v+H)^4.
\eeq
Note that here the fields $\phi^i$ of the unphysical Goldstone bosons,
which are contained in the matrix $U$, appear only
in their dynamical part $\propto\tr|DU|^2$, 
but not in the Higgs potential, where only $(v+H)$ survives owing to
the unitarity of $U$. For this reason this form is
very attractive for the discussion of the heavy-Higgs-boson limit, 
even in gauges different
from the unitary gauge, which corresponds to $U=\mathbf{1}$.

To lowest order of perturbation theory, 
the heavy-Higgs-boson limit of $\L_{\PH,\mathrm{nl}}$
is reached upon setting the Higgs field $H$ to zero, since all Higgs propagators are 
suppressed by factors $1/\MH^2$ in amplitudes---a suppression that cannot be 
compensated by the $\MH^2$ factors in the Higgs self-couplings, since there are always
more Higgs propagators than Higgs self-couplings in amplitudes without external 
Higgs bosons. In other words, a heavy Higgs boson decouples in tree-level
amplitudes.
The resulting theory without a Higgs field is called the
{\it gauged non-linear sigma model}~\cite{Appelquist:1980vg} 
and described by the scalar Lagrangian
\beq
\L_{\sigma,\mathrm{nl}} = \frac{1}{4}v^2\tr\{(D_\mu U)^\dagger(D^\mu U)\}
=\L_{\PH,\mathrm{nl}}(H=0).
\label{eq:gnlsm}
\eeq
This model is equivalent to the SM in the unitary gauge without Higgs boson,
but still represents an SU(2)$_\rI\times$U(1)$_\rY$ gauge theory. 
The model, though, is physically unsatisfactory, since it is neither
renormalizable, nor does its $S$-matrix respect unitarity. In fact these
two short-comings are closely related, as shown in \citere{Cornwall:1974km}.

Beyond leading order, a heavy Higgs boson does not decouple anymore
in amplitudes, a fact that is visible in the different UV behaviour
of the SM with and without Higgs boson---the latter being the
gauged non-linear sigma model. While the SM with Higgs boson is renormalizable,
i.e.\ closed in the UV sector, the gauged non-linear sigma model is
non-renormalizable, i.e.\ incomplete in the UV sector and thus at best
can serve as an effective field theory up to some finite energy.
Thus, removing the Higgs boson from the SM particle content upon
increasing its mass more and more, $\MH$ effectively works as
UV cutoff scale for the UV divergences of the gauged non-linear sigma model.
Specifically, using regularization in $D$ dimensions the 
uncancelled one-loop UV poles $1/(D-4)$ of the sigma model are in one-to-one
correspondence with one-loop corrections involving $\ln\MH$
in the heavy-Higgs SM. At one loop, these non-decoupling effects
of a heavy Higgs boson have been quantified in terms of effective
Lagrangians in the literature~\cite{Herrero:1993nc}, keeping also
finite terms that are not suppressed by powers of $\MH$.
At higher loop orders, non-decoupling effects can even grow with
powers of $\MH^2$, as e.g.\ seen in heavy-Higgs corrections
$\propto\MH^2$ and $\MH^4$ to the $\rho$-parameter
at the two-loop~\cite{vanderBij:1983bw} 
and three-loop level~\cite{Boughezal:2004ef}, 
respectively, suggesting a scaling law
of $\MH^{2(L-1)}$ at $L$-loop order, possibly modified by powers of $\ln\MH$.

The SM is of course not the only possible ``UV closure'' of the non-renormalizable
non-linear $\sigma$-model. Any renormalizable model containing it bears some
dimensionful quantity $\Lambda$, such as a particle mass, that acts effectively
as cut-off for UV
divergences, like the Higgs-boson mass in the SM. In the limit of large $\Lambda$,
the same structure of $\ln\Lambda$ terms will, thus, show up as in the 
heavy-Higgs SM with $\MH$ playing the role of $\Lambda$.

\paragraph{Constraints on $\mathbf{\MH}$ from unitarity arguments} \mbox{}
\\*[.5em]
The unitarity of the $S$-matrix
implies bounds on the high-energy behaviour of partial waves of 
$2\to2$ scattering processes, but a naive power counting of energy
factors in amplitudes with longitudinally polarized weak gauge bosons
$\PV_\rL=\PW^\pm_\rL,\PZ_\rL$
suggests that these bounds do not hold. The leading
high-energy term in the longitudinal polarization vector 
$\varepsilon_{\PV_\rL}(k)$ of an external $\PV_\rL$ with momentum $k^\mu$, which
behaves $\sim k^\mu/\MV$, introduces a dangerous energy factor which seems
to spoil the power counting. Unitarity is restored by cancellations
that are governed by gauge invariance in the form of 
Slavnov--Taylor identities, which relate the $k^\mu$ contraction of an
external $\PV_\mu$ leg of an amplitude to another amplitude
in which $\PV_\rL$ is replaced by the respective Goldstone-boson
field $\phi_\PV$. Schematically an amplitude $\M_{\PV_\rL\dots}$ behaves like
\beq
\M_{\PV_\rL\dots} = \varepsilon_{\PV_\rL,\mu}(k) \, T^\mu_{\PV_\rL\dots}
\;\asymp{k^0\gg\MV}\; \frac{k_\mu}{\MV}\, T^\mu_{\PV_\rL\dots}
=\M_{\phi_\PV\dots},
\label{eq:et}
\eeq
where $\M_{\phi_\PV\dots}$ is the ``scattering amplitude'' of
$\phi_\PV$ (defined with an appropriate phase)
at high energies with a decent scaling law in the
scattering energy, as demanded by unitarity.
Equation~\refeq{eq:et} expresses the so-called Goldstone-boson equivalence
theorem~\cite{Cornwall:1974km,Vayonakis:1976vz},
whose formulation to all orders requires some care~\cite{Yao:1988aj}.

It is interesting to note that the equivalence theorem can also
be formulated within the gauged non-linear sigma 
model~\cite{GrosseKnetter:1994mf}. 
Since the non-polynomial structure of the Goldstone-boson
Lagrangian \refeq{eq:gnlsm} involves arbitrarily high powers
of derivatives (momentum factors in amplitudes),
however, the power-counting of
energy factors in \refeq{eq:et} does not lead to a decent
high-energy behaviour of the amplitude $\M_{\PV_\rL\dots}$.
Amplitudes $\M_{\PV_\rL\dots}$ in fact potentially violate
unitarity, both in the gauged non-linear sigma
model and in the SM with a heavy Higgs boson, $\MH\to\infty$.
In the unitary gauge, where no Goldstone-boson fields are present,
the same conclusion is drawn from the ``bad
high-energy behaviour'' of the $\PW$ and $\PZ$ propagators.

The violation of unitarity can already be observed in the SM with finite
$\MH$ if the $\MH$ value is large enough.
The larger $\MH$, the more delayed 
is the necessary unitarity cancellation for large $k^0$.
If $\MH$ is too large, the amplitude starts to ``explode''
before the damping due to Higgs-boson exchange sets in.
In $2\to2$ particle scattering, the most delicate unitarity 
cancellations occur in longitudinal vector-boson scattering
$\PV_\rL\PV_\rL\to\PV_\rL\PV_\rL$ whose tree-level amplitudes respect
unitary only if $\MH\lsim1\TeV$~\cite{Lee:1977eg}.
Within pure perturbation theory it is not possible to give sharp
unitarity limits on $\MH$, because the range around $\MH\sim1\TeV$
represents the transition region where perturbation theory 
runs out of control in the Higgs sector due to the huge self-couplings.

\paragraph{Triviality and vacuum-stability bounds} \mbox{}
\\*[.5em]
It is interesting to apply the concept of a running coupling to the Higgs-boson
self-coupling $\lambda$. 
The running of $\lambda(\muR)$ is mainly driven by loop corrections to the
$H^4$-interaction involving the Higgs-boson self-couplings or
top quarks with the largest Yukawa coupling.
Present-day evaluations of $\lambda(\muR)$ involve (at least) two-loop
effects and RGE. The behaviour of the resulting running $\lambda(\muR)$ at
large values of $\muR$ contains information on the scale $\Lambda$
up to which the SM can exist as a self-consistent quantum field theory:
\begin{itemize}
\item
For a large Higgs-boson mass
the Higgs self-interactions drive $\lambda(\muR)$ to larger values with
increasing $\muR$, reflecting the lack of asymptotic freedom. 
Above some scale the coupling enters its non-perturbative domain, a fact that is 
most obvious at one-loop order where $\lambda(\muR)$ even reaches a so-called Landau pole
and diverges. At two-loop order the divergence is mitigated to a UV fixed-point, but
at three-loop order a Landau pole is expected to appear again.
More details on the running and matching issues can be found in \citere{Hambye:1996wb}.
Of course, the transition to the non-perturbative domain is a continuous process,
so that only a generic scale $\Lambda$ for the onset of this region can be given. 
The non-perturbative domain and the Landau pole 
could be avoided if the exchange of new particles (not part of the SM) in the 
loop corrections to the $H^4$ interaction counterbalances
the too rapid running induced by the graphs containing the
Higgs self-couplings.
Thus, $\Lambda$ can be interpreted as a lower
bound on the new-physics scale if the SM is supposed to stay in its
perturbative regime. Being perturbative or non-perturbative is certainly not
a physical criterion on a field theory. However, the appearance of a Landau pole
does not only signal non-perturbativity, but also a mathematical inconsistency.
Beyond the Landau pole the self-consistency
of the field theory seems to require a scalar self-coupling $\lambda$ that 
vanishes at all, leading to a {\it trivial} theory. The bound on $\Lambda$
to avoid the Landau pole is, thus, also called the
{\it triviality bound}~\cite{Maiani:1977cg}.
We are not aware of a field-theoretical proof of the triviality of scalar
field theories with $\phi^4$ interactions, such as the SM, but the triviality
hypothesis is supported by non-perturbative 
results from lattice field theory (see \citere{Fodor:2007fn} and earlier
references therein).
\item
For a small Higgs-boson mass the running of $\lambda(\muR)$ is dominated
by the top-quark loops which drive $\lambda(\muR)$ even to negative values 
for large $\muR$.
This fact has drastic consequences on the
Higgs potential that determines the non-vanishing vev of the Higgs field.
In lowest order the potential \refeq{eq:V} still shows
the asymptotics $\propto \lambda H^4$, but $\lambda(\muR)<0$ for large $\muR$
indicates
that higher-order effects bend the Higgs potential $V$ over to negative
values for large Higgs fields $H$, i.e.\ the potential is not bounded from
below anymore. Thus, a stable vacuum does no longer exist, and the whole model
becomes inconsistent. A thorough theoretical analysis of the vacuum structure
goes beyond a pure running-coupling analysis and takes into account 
the leading static (for Higgs fields at zero momentum) all-order-resummed 
corrections to the Higgs potential $V$, leading to the so-called
{\it effective Higgs potential} $V_{\mathrm{eff}}$~\cite{Coleman:1973jx}.
For small $\MH$ the SM can escape the
{\it vacuum-stability bound}~\cite{Lindner:1988ww} 
only if new-physics effects at some scale $\Lambda$
modify the vacuum structure determined by $V_{\mathrm{eff}}$ in such a way
that a stable vacuum exists.
This results in a lower bound on $\MH$ for a given value of $\Lambda$.
The bound can be relaxed to some extent
by assuming a metastable EW vacuum with 
a lifetime longer than the age of the Universe, where the vacuum decay may be
mediated either by zero-temperature quantum fluctuations or thermal fluctuations.
\end{itemize}
Figure~\ref{fig:trivstabMH}, which shows the results of the recent two-loop
RGE analysis of \citere{Ellis:2009tp}, illustrates the upper and lower bounds
on the SM Higgs-boson mass resulting from the hypothesis that the SM
Higgs sector should be perturbative (and thus escapes the triviality bound)
with a stable or at least metastable vacuum up to a potential new-physics scale
$\Lambda$.
\bfi
\centerline{
\includegraphics[width=.6\textwidth,height=.4\textwidth]%
{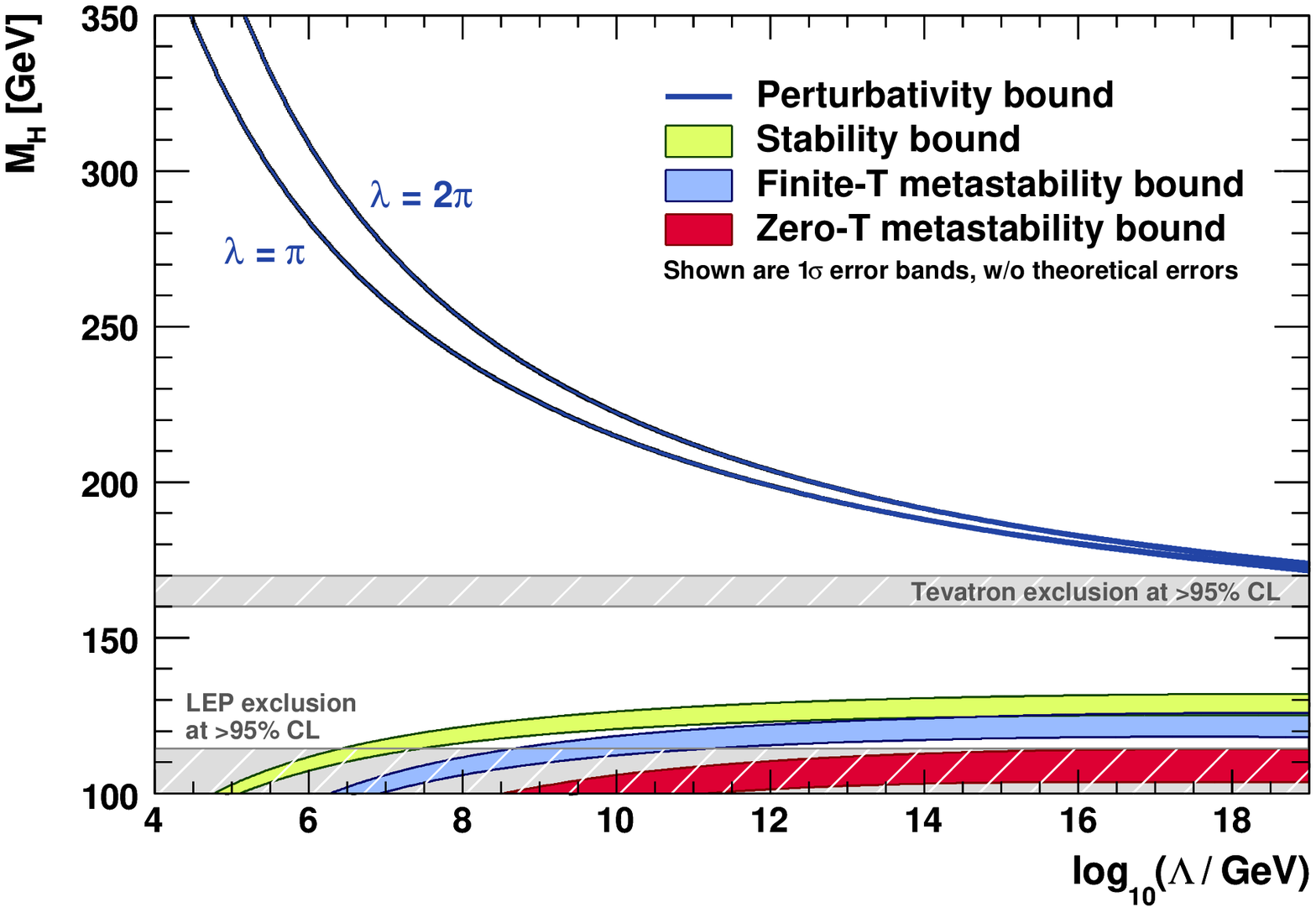}
}
\caption{Perturbativity (triviality) upper bound and vacuum-stability lower bound on $\MH$ as
function of the scale $\Lambda$ at which new physics has to appear (at the latest) 
in order to avoid the bounds (plot taken from \citere{Ellis:2009tp}).
The two different versions of the perturbativity bound reflects the intrinsic
uncertainty of the limit, and the bands of the (meta)stability bounds reflect
parametric uncertainties.}
\label{fig:trivstabMH}
\efi
A SM Higgs boson with $\MH\lsim200\GeV$ admits the SM to escape the 
non-perturbativity/triviality bound up to scales $\Lambda\lsim10^{12}\GeV$, 
far above any energies reachable in collider experiments.
On the other hand, the LEP exclusion limit $\MH>114.4\GeV$ pushes the vacuum
(meta)stability limit for the onset of new physics already to $\Lambda\sim10^6\GeV$, which is
out of reach for colliders as well.
For a Higgs boson within the mass window
$130\GeV\lsim\MH\lsim170\GeV$
the SM would be surely consistent up to scales as high as the Planck scale
$\Lambda_{\mathrm{Planck}}=2\times10^{18}\GeV$, where gravity effects should become as
large as the strong and EW interactions.
Finally, interpreting the recent LHC discovery as a SM Higgs boson with $\MH=126\GeV$
still allows the SM to be valid up to the Planck scale if our universe resides in a
metastable phase.

\paragraph{Constraints from precision data} \mbox{}
\\*[.5em]
\label{sec:blueband}
Experiments at LEP1, LEP2, SLC, and the Tevatron have performed a large variety
of high-precision measurements, such as the determination of the W, Z, and top-quark
masses and widths, measurements of cross sections and various asymmetries in the process
$\Pep\Pem\to\gamma^*/\PZ\to f\bar f$ at the Z pole, etc.
These measurements characterize the past two decades as the era of EW
precision physics.
Parametrizing SM predictions as described in
\refse{se:predictions}, thus, renders a fit of the complete SM to these data possible,
which highly constrains the SM input parameters (with the exception of the CKM matrix,
where dedicated observables from flavour physics are needed).
Figure~\ref{fig:BR-SMfit} shows, on the l.h.s., the status of March 2012
of such a fit projected on
the Higgs-boson mass, together with the exclusion limits from the direct searches
to be discussed later in detail.
\bfi
\raisebox{-1em}{\includegraphics[width=.49\textwidth,height=.45\textwidth]%
{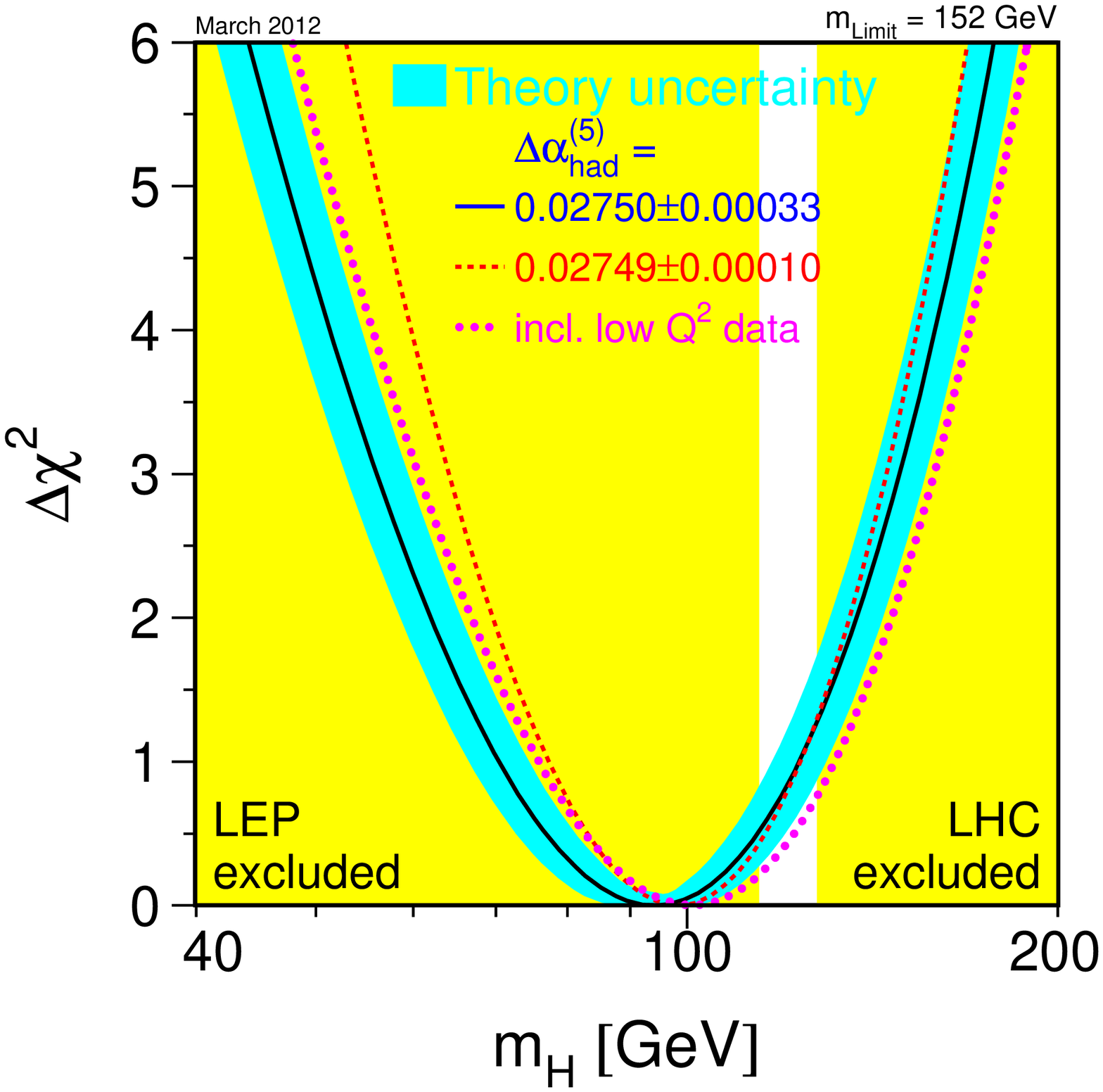}}
\includegraphics[width=.49\textwidth,height=.4\textwidth]%
{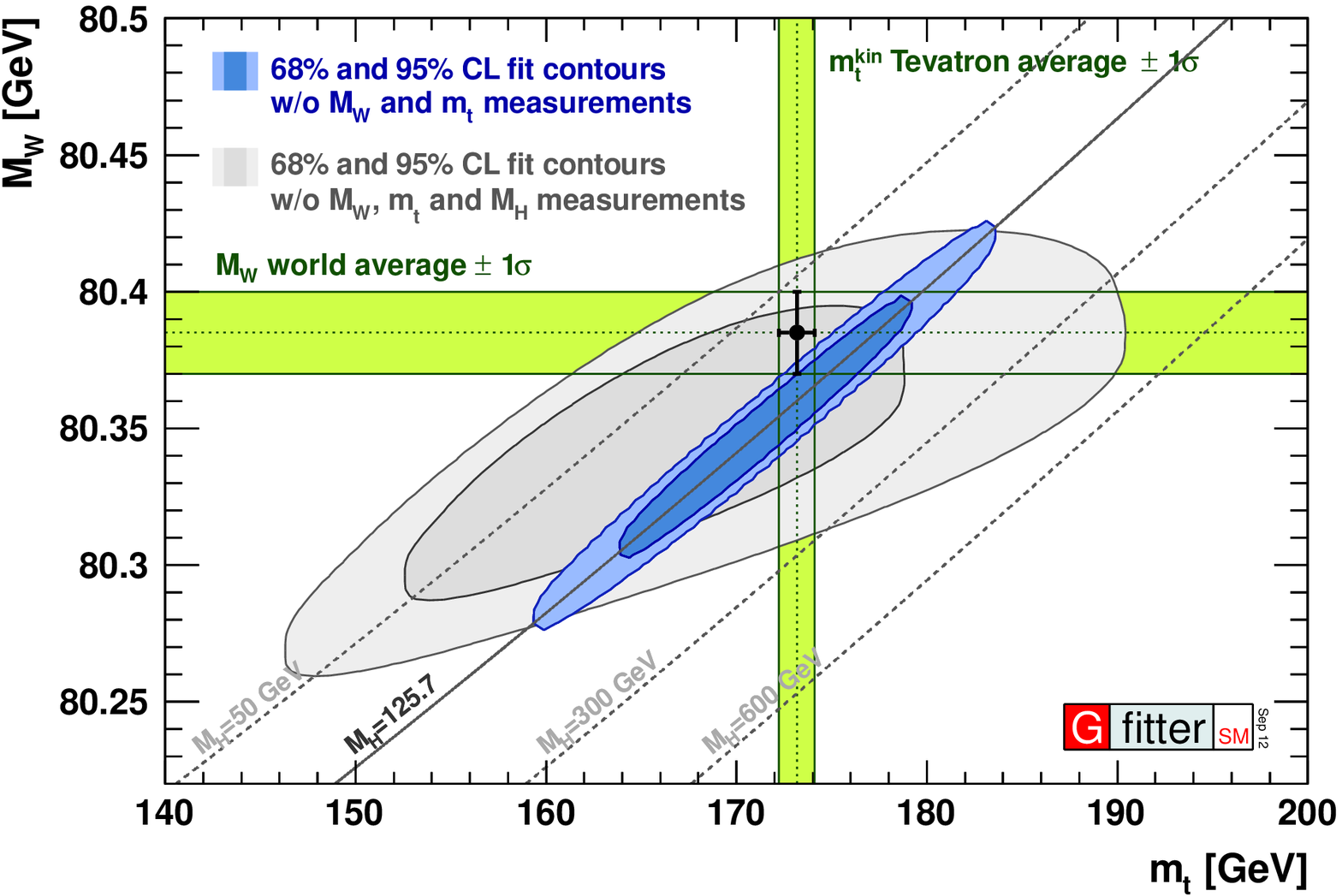}
\caption{The ``blueband plot'' (left, taken from 
\citere{lepewwg}) showing the result of a global
$\chi^2$ fit of the SM to precision data, projected onto the $\MH$ axis
and comparison of constraints on the top-quark and $\PW$-boson masses 
(right, taken from \citere{Baak:2012kk})
resulting from direct measurements and a global SM fit.}
\label{fig:BR-SMfit}
\efi
The blue band indicates theoretical uncertainties from missing higher-order
corrections in the (numerous) underlying precision calculations, which were 
condensed into the state-of-the-art codes 
{\sc Zfitter}~\cite{Arbuzov:2005ma}
and {\sc Topaz0}~\cite{Montagna:1998kp}
(see also \citere{BP} for details and references).
The best fit value together with its $68\%$ CL limits 
for the Higgs-boson mass is $\MH=94^{+29}_{-24}\GeV$, being compatible with the open mass
window $122\GeV<\MH<127\GeV$ (status July 2012, see \refse{se:lhcexclusion}) which is obtained from the 
95\% C.L.\ exclusion limits of the
LHC from below and above, respectively.
The r.h.s.\ of \reffi{fig:BR-SMfit} illustrates the SM overall fit,
as obtained by the {\sc Gfitter}~\cite{Baak:2012kk}%
\footnote{Similar results have been presented in \citere{Eberhardt:2012gv}.} 
collaboration, by its
projection into the $\Mt{-}\MW$ plane and compares the preferred fit region
with the directly measured values of the top-quark and W-boson masses.
The shown SM prediction for $\MW$ as function of $\Mt$ and $\MH$ (diagonal lines
for some fixed $\MH$ values) is obtained
from the measured muon lifetime (often translated into the Fermi constant $\GF$).
All these constraints on $\Mt$ and $\MW$ are perfectly compatible with
the recent observation of a Higgs-boson candidate of a mass around $\MH=126\GeV$,
to be discussed in detail below.

\subsection{Higgs-boson decays}
\label{se:Hdecays}

The Higgs boson of the SM is unstable and predominantly
decays into the heaviest particle--antiparticle pair
that is kinematically possible depending on the 
available energy $\MH$. The search for the Higgs boson,
which has to be reconstructed from its decay products
in detectors, thus depends on the Higgs-boson
mass, since the relevant final states strongly vary with $\MH$.
Since the branching ratio $\BR(\PH\to X)=\Gamma_{\PH\to X}/\GH$,
which is the probability for the Higgs boson to decay into the
final state $X$, weights the relevant Higgs-boson production cross section
for the exclusive channel $\PH\to X$, theory has to provide precise
predictions for all significant Higgs-boson decay widths---irrespective
of whether 
the channel could be detected or not---since they all enter the
total decay width $\GH$, which is the sum of all partial widths
$\Gamma_{\PH\to X}$.
To get the global picture we show the SM Higgs-boson branching ratios and
the total decay width in \reffis{fig:BR-lowMH} and \ref{fig:BR-LHC},
where we separate the low-$\MH$ range that was accessible at the time
of the $\Pep\Pem$ collider LEP (\reffi{fig:BR-lowMH}) from the
$\MH$ range that is challenged by the hadron colliders
Tevatron and LHC (\reffi{fig:BR-LHC}).
\bfi
\includegraphics[width=.49\textwidth,height=.4\textwidth]%
{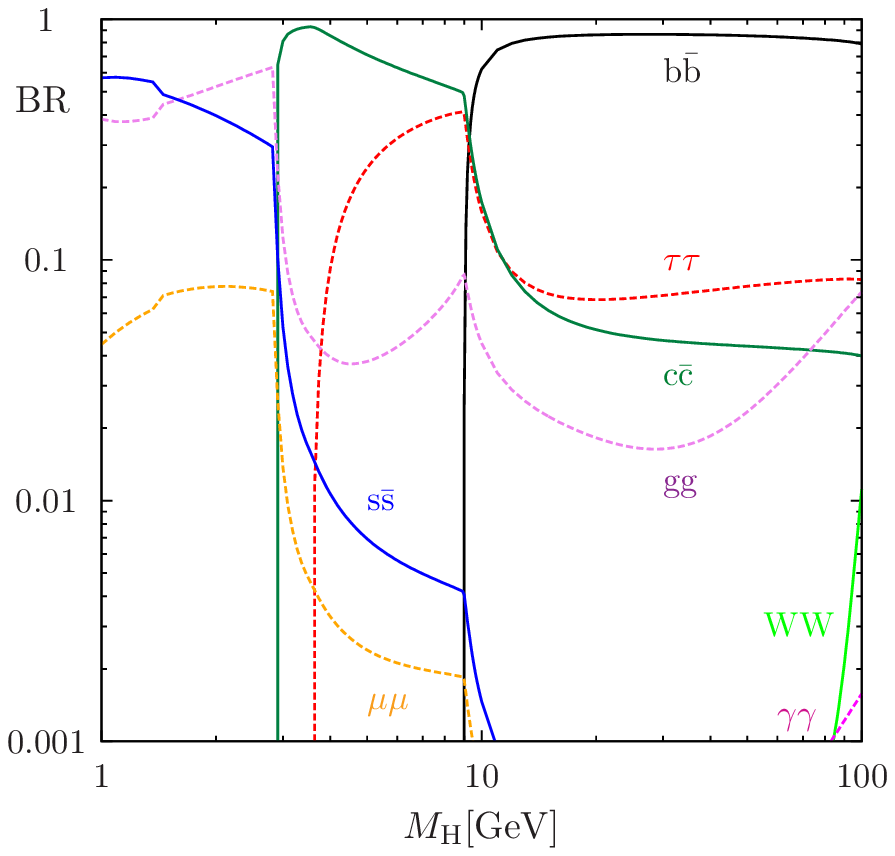}
\includegraphics[width=.49\textwidth,height=.4\textwidth]%
{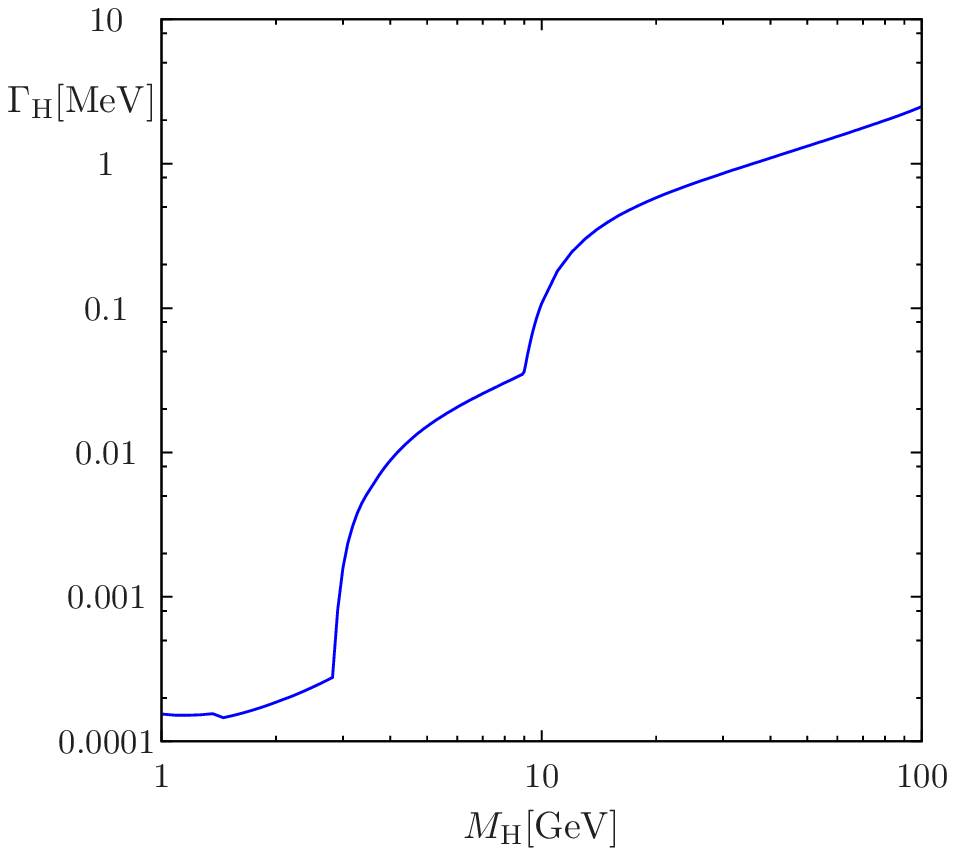}
\caption{Branching ratios of the SM Higgs boson (left)
and total decay width (right) for Higgs-boson masses
accessible at LEP and before, calculated with the program 
{\sc Hdecay}~\cite{Djouadi:1997yw}.}
\label{fig:BR-lowMH}
\vspace*{3em}
\includegraphics[width=.49\textwidth,height=.4\textwidth]%
{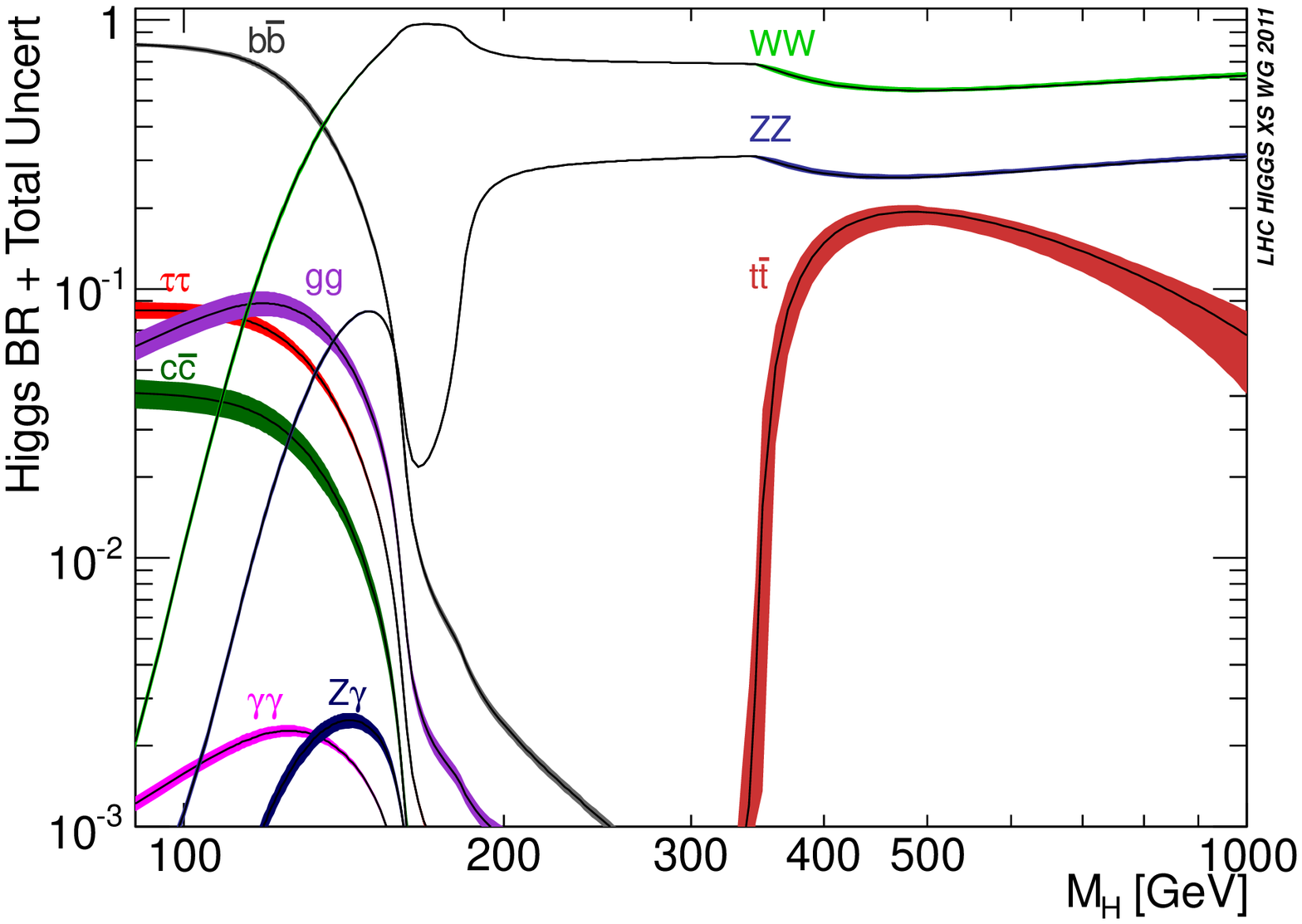}
\includegraphics[width=.49\textwidth,height=.4\textwidth]%
{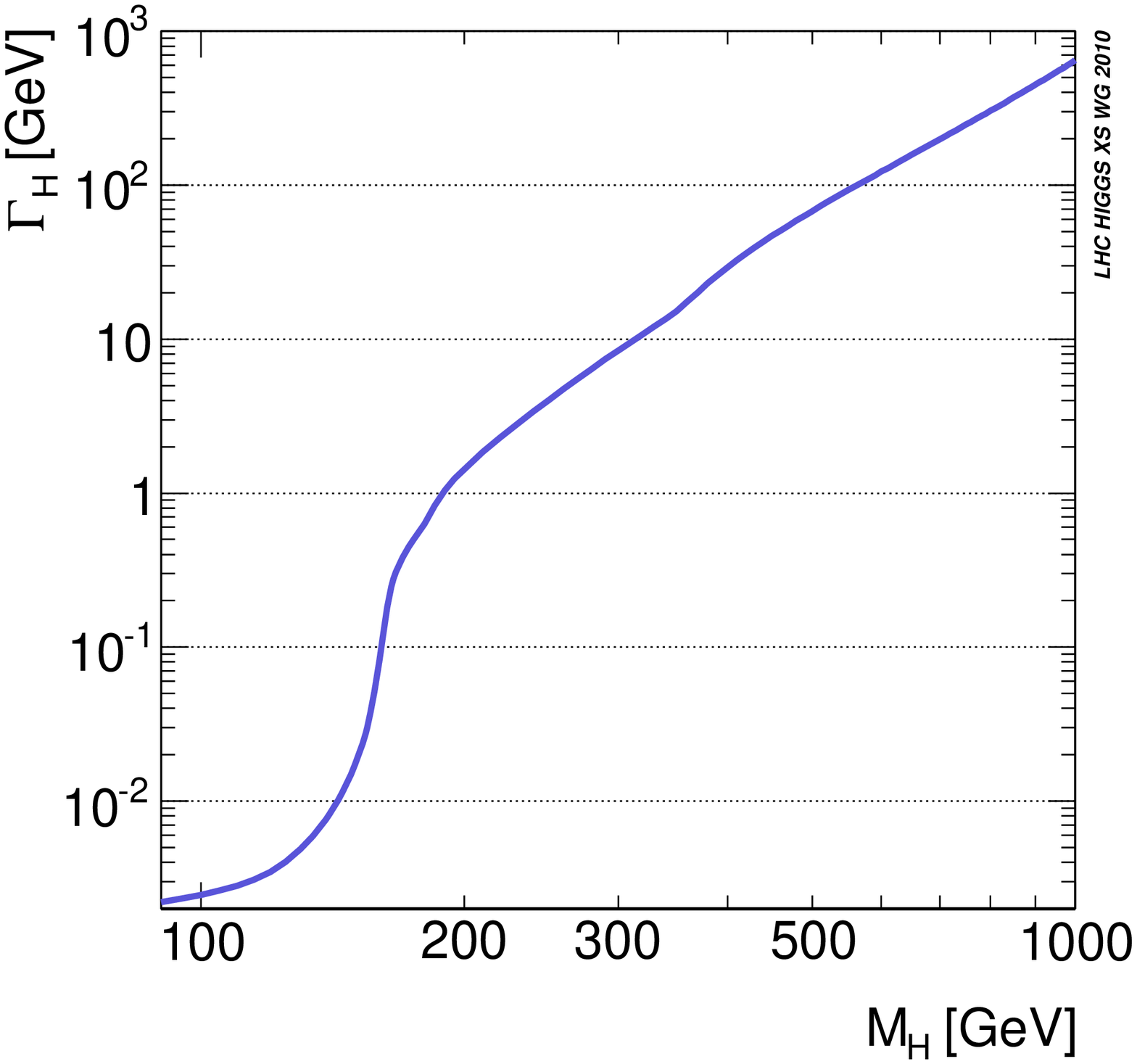}
\caption{Branching ratios of the SM Higgs boson
(left, taken from \citeres{Denner:2011mq,Dittmaier:2012vm}),
with the band widths illustrating the parametric and theoretical uncertainties, 
and total decay width
(right, taken from \citere{Dittmaier:2011ti})
in the Higgs-boson mass range accessible by the LHC.} 
\label{fig:BR-LHC}
\efi
The relevant leading-order Feynman diagrams for the various decay
channels are depicted in \reffi{fig:Hdecaydiags}.
\bfi
\centerline{
{
\begin{picture}(90,80)(10,0)
\ArrowLine( 90,10)(60,40)
\ArrowLine(60,40)( 90,70)
\DashLine(30,40)(60,40){5}
\Vertex(60,40){2}
\put( 95,66){$f$}
\put( 15,36){$\PH$}
\put( 95, 5){$\bar f$}
\end{picture}
\SetScale{1}
}
\hspace{.5em}
{
\begin{picture}(115,80)(10,0)
\ArrowLine( 90,20)(60,40)
\ArrowLine(60,40)( 90,60)
\ArrowLine( 90,60)(90,20)
\DashLine(30,40)(60,40){5}
\Gluon(90,20)(120,10){2}{4}
\Gluon(90,60)(120,70){2}{4}
\Vertex(60,40){2}
\Vertex(90,20){2}
\Vertex(90,60){2}
\put(125,66){$\Pg$}
\put( 15,36){${\PH}$}
\put( 95,36){$Q$}
\put(125, 5){$\Pg$}
\end{picture}
\SetScale{1}
}
\hspace{.5em}
{
\begin{picture}(125,80)(10,0)
\DashLine(30,40)(60,40){5}
\DashLine( 90,20)(60,40){1.5}
\DashLine(60,40)( 90,60){1.5}
\DashLine( 90,60)(90,20){1.5}
\Photon(90,20)(120,10){2}{4}
\Photon(90,60)(120,70){2}{4}
\Vertex(60,40){2}
\Vertex(90,20){2}
\Vertex(90,60){2}
\put(125,66){$\gamma$}
\put( 15,36){${\PH}$}
\put( 95,36){$Q,{\PW}$}
\put(125, 5){$\gamma,\PZ$}
\end{picture}
\SetScale{1}
}
\hspace{.5em}
{
\begin{picture}(115,80)(0,0)
\Photon( 90,10)(60,40){2}{5}
\Photon(60,40)( 90,70){2}{5}
\DashLine(20,40)(60,40){5}
\ArrowLine(120, 0)( 90,10)
\ArrowLine(90,10)(120,25)
\ArrowLine( 90,70)(120,80)
\ArrowLine(120,55)(90,70)
\Vertex(60,40){2}
\Vertex(90,10){2}
\Vertex(90,70){2}
\put( 43,56){${\PW,\PZ}$}
\put(  5,36){${\PH}$}
\put( 43,15){${\PW,\PZ}$}
\end{picture}
\SetScale{1}
}
}
\vspace*{-.5em}
\caption{Leading-order diagrams for the various SM Higgs-boson
decay channels, where $Q$ denotes any heavy quark.}
\label{fig:Hdecaydiags}
\efi
As anticipated above, the highest branching ratios typically
correspond to decays into the heaviest particle--antiparticle pair
that are kinematically allowed for a given value of $\MH$. 
However, as far as predictions are concerned,
the regions where new channels become significant are theoretically
delicate. For heavy-quark pairs the threshold regions are
particularly sensitive to the heavy-quark masses whose proper
perturbative treatment within QCD is highly non-trivial.
For the Higgs-boson decays into the weak-gauge-boson pairs $\PW\PW$ and $\PZ\PZ$,
even the regions near and below the $\PW\PW/\PZ\PZ$
thresholds become important which implies
that the decays of the $\PW/\PZ$ bosons have to be included in the
predictions.

Owing to the proportionality of the Higgs couplings to the
masses of the decay products, the total Higgs-boson width $\Gamma_\PH$
stretches over many orders of magnitude, viz.\ from
$10^{-4}\GeV$ for $\MH\lsim10\GeV$ to $10^3\GeV$ for $\MH\sim 1\TeV$. 
Because of the finite energy resolution of detectors, the width of a
light Higgs boson is too small to be resolved in invariant-mass 
distributions of its decay products. 
On the theoretical side this
fact justifies the approach of dealing with stable, on-shell Higgs
bosons in production processes and factorizing the subsequent decay
from the production. For large invariant masses of the Higgs-boson
decay products resulting from $\PH\to VV$ ($V=\PW,\PZ$), i.e.\ for
$M_{VV}\gsim2M_V>\MH$, off-shell effects can be quite large,
${\cal O}(10\%)$, but this region can be excluded by dedicated 
cuts~\cite{Kauer:2012hd}.
On the other hand, a heavy Higgs
boson would show up as an extremely broad resonance, whose width
becomes of the same size as its mass for $\MH\lsim1\TeV$.
Theoretically this situation is very challenging, because production and
decay processes do not factorize anymore from each other.
Instead, a proper treatment of the broad resonance has to deal with
the signal, consisting of Higgs-boson production, propagation, and decay,
background, comprising non-resonant diagrams with the same final state
as the Higgs-boson decay, and of interference effects between signal and
background at the same time. Even the proper field-theoretical
definition of mass and width of a heavy Higgs boson to parametrize
the resonance becomes subtle. For more details about these issues,
which are still under investigation, we have to refer to the
literature (see e.g.\ \citeres{Passarino:2010qk,Dittmaier:2011ti,Dittmaier:2012vm,Anastasiou:2012hx} 
and earlier references therein).
In the following we put the emphasis on the Higgs-boson mass range
$100\GeV<\MH<200\GeV$, which is favoured by the overall fit of the
SM to precision data and the results of the direct searches.

To deliver precise predictions for the Higgs-boson decay widths and branching ratios,
a huge effort was made by many theorists. The
decay channels that are most important for Tevatron and the LHC are:
\begin{itemize}
\item
$\PH\to f\bar f$ (mainly $f=\tau,\Pb,\Pt$)
\cite{Resnick:1973vg,Ellis:1975ap,Braaten:1980yq,Drees:1989du,Gorishnii:1983cu,Anastasiou:2011qx,Djouadi:1994gf,Melnikov:1995yp,Bernreuther:1997af,Fleischer:1980ub,Kniehl:1991ze,Kniehl:1994ph,Ghinculov:1994se}
\\
In describing a Higgs boson decaying into bottom (or even lighter) quarks it is essential
to base the Yukawa coupling on the running quark mass at the relevant scale,
which is set by the Higgs-boson mass. For instance, for $\MH\gsim100\GeV$ the transition
from the pole to the running bottom mass $\overline{m}_\Pb(\MH)$
reduces the $\PH\to\Pb\bar\Pb$ partial decay width by $\sim60\%$ or more. 
Starting from this improved LO prediction, the perturbative QCD series,
which is known in NLO~\cite{Braaten:1980yq} and beyond that even up to
NNNNLO~\cite{Gorishnii:1983cu}, shows nice convergence with a small residual
scale uncertainty of $\sim0.1\%$. 
Recently, the NNLO QCD corrections to $\PH\to\Pb\bar\Pb$ became available
for fully differential observables~\cite{Anastasiou:2011qx} as well.
Generally, a proper treatment of the $q\bar q$ threshold region 
deserves particular attention~\cite{Drees:1989du}.

For the decay into top quarks, the full mass dependence of the $\Pt\bar\Pt$
final state has to be included, and the
issue of a running mass is not as
pronounced as for the lighter quarks.
The QCD corrections, which are available at NLO~\cite{Braaten:1980yq,Djouadi:1994gf}
and NNLO~\cite{Melnikov:1995yp}, turn out to be moderate, but still $\sim5\%$ at NNLO.
The decay of the top quarks should be taken into account
in predictions as well, not only for precision, but also since it would offer the
possibility to analyze the CP properties of a possible heavy-Higgs-boson candidate
(see \citeres{Bernreuther:1997af,Djouadi:2005gi} and references therein).

The theoretical structure of the NLO EW corrections~\cite{Fleischer:1980ub,Kniehl:1991ze}
to $\PH\to f\bar f$ is very similar for all fermions, and their size is typical of
the order of a few percent. Beyond NLO, some QCD and EW corrections that are enhanced by
factors of $\GF\Mt^2$ are 
known~\cite{Kniehl:1994ph} as well as the leading two-loop heavy-Higgs-boson effects 
$\propto\GF^2\MH^4$~\cite{Ghinculov:1994se} that
reflect the failure of perturbation theory in the TeV range for $\MH$, where
they are as large as their NLO counterparts.
\item
$\PH\to\Pg\Pg$
\cite{Rizzo:1979mf,Inami:1982xt,Spira:1995rr,Chetyrkin:1997iv,Actis:2008uh,Actis:2008ts,Steinhauser:1998cm}
\\
The NLO QCD (two-loop) corrections to the $\PH\to\Pg\Pg$ decay were first
calculated as an asymptotic expansion in $\MH/\Mt$~\cite{Inami:1982xt}, which
is applicable only for Higgs-boson masses below the $\Pt\bar\Pt$ threshold
($\MH<2\Mt$), and later generalized to the full mass
dependence~\cite{Spira:1995rr}. These corrections are very large, ranging
from $70\%$ to $40\%$ for $\MH=100\GeV$ to $1\TeV$.
QCD effects beyond NLO~\cite{Chetyrkin:1997iv},
which are known 
in the form of an expansion in small values of $\MH$ valid up to $\MH\lsim\Mt$,
add roughly another $20\%$ to the partial decay width of a light Higgs boson to gluons.
Electroweak NLO corrections are known with the full mass dependence and amount
to $\sim5\%$~\cite{Actis:2008uh,Actis:2008ts}. Mixed QCD--EW beyond NLO corrections were calculated 
in the heavy-top limit, i.e.\ for light Higgs bosons, but found to be very
small~\cite{Steinhauser:1998cm}.
\item
$\PH\to\gamma\gamma/\gamma\PZ$
\cite{Resnick:1973vg,Ellis:1975ap,Shifman:1979eb,Zheng:1990qa,Melnikov:1993tj,Spira:1995rr,Melnikov:1994jb,Aglietti:2004nj,Actis:2008uh,Actis:2008ts,Steinhauser:1996wy,Spira:1991tj}
\\
Similar to the gluonic Higgs-boson decay,
the NLO QCD corrections to the $\PH\to\gamma\gamma$ decay were first
calculated in the large-top-quark-mass limit~\cite{Zheng:1990qa}
and later with the full mass 
dependence~\cite{Melnikov:1993tj,Spira:1995rr}. For small $\MH$ these corrections
amount to only $\lsim5\%$, but for masses above the $\Pt\bar\Pt$ threshold they
can be as large as $50{-}100\%$ around $\MH\sim600\GeV$, where the LO prediction
receives some suppression due to a destructive interference between quark and
W-boson loops.
QCD effects beyond NLO are also known in the form of an expansion in small values of 
$\MH$~\cite{Steinhauser:1996wy}.
NLO EW corrections, which were presented in \citeres{Aglietti:2004nj,Actis:2008uh,Actis:2008ts},
turn out to be of the order of a few percent.
In the actual calculations, particular care has to be taken in the vicinity
of the $\Pt\bar\Pt$ and $\PW\PW$ thresholds, where the finite decay widths
of the top-quarks and W-bosons in the loop have to be taken into 
account~\cite{Melnikov:1994jb,Actis:2008uh,Actis:2008ts}.

Because of its lower phenomenological importance,
the predictions to the decay $\PH\to\gamma\PZ$ are much less
advanced than in the $\gamma\gamma$ case. The NLO QCD corrections, which turn 
out to be below $1\%$, can be found in \citere{Spira:1991tj}.
Many leading higher-order corrections are very similar or even identical in the
$\gamma\gamma/\gamma\PZ$ cases.
\item
$\PH\to\PW\PW/\PZ\PZ\to4f$
\cite{Nelson:1986ki,Fleischer:1980ub,Kniehl:1990mq,Bredenstein:2006rh,Ghinculov:1995bz}
\\
Since the Higgs-boson decay channels into weak-gauge-boson pairs
are also phenomenologically relevant 
in the WW/ZZ threshold region and below, i.e.\ for $\MH\lsim2\MV$
($\PV=\PW,\PZ$), the W/Z decays have to be taken into account in
precision calculations. Moreover, it was extensively discussed in the
literature~\cite{Nelson:1986ki} that the kinematical details (energy,
angular, and invariant-mass distributions) of the
four-fermion final state offer the possibility to extract the spin and
CP properties of a Higgs-boson candidate.
Radiative corrections have first been calculated for on-shell
W/Z bosons~\cite{Fleischer:1980ub,Kniehl:1990mq}, the results of which are only
applicable for $\MH>(2\MV+\mbox{some GeV})$, and later 
including the full off-shell effects of the W/Z bosons~\cite{Bredenstein:2006rh}
in the framework of the complex-mass scheme~\cite{Denner:1999gp}
for treating the W/Z resonances.
For $\PH\to\PW\PW/\PZ\PZ$ the heavy-Higgs-boson limit is of particular interest,
because the dominating decay into longitudinal W/Z bosons leads to the
above-mentioned rise $\propto\MH^3$ of the total decay width $\Gamma_\PH$.
The leading heavy-Higgs-boson corrections to $\Gamma_{\PH\to\PW\PW/\PZ\PZ}$
are proportional to $\GF\MH^2$ and $\GF^2\MH^4$ at the one- and two-loop
level~\cite{Ghinculov:1995bz}. The fact that this two-loop term
becomes as large as the one-loop correction for
$\MH\sim1\TeV$ signals again the onset of the non-perturbative regime
for a heavy Higgs boson.
The leading two-loop effect, which is negligible for $\MH\lsim400\GeV$
and amounts to $\sim4\%$ for $\MH\sim700\GeV$,
can serve as an estimate for the theoretical uncertainty for large $\MH$.
For $100\GeV<\MH<200\GeV$ the radiative corrections to the partial widths
range from some percent 
to $\lsim12\%$, depending on the four-fermion final state
and can be even much higher in distributions.
\end{itemize}
More details can be found in reviews or 
reports~\cite{Spira:1997dg,Djouadi:2005gi,Dittmaier:2011ti,Dittmaier:2012vm}.

State-of-the-art predictions can be obtained upon
combining results from the two public programmes
\Hdecay~\cite{Djouadi:1997yw} and
\Prophecy~\cite{Bredenstein:2006rh}:
\begin{itemize}
\item
\Hdecay\ is designed to calculate all relevant Higgs-boson decay
widths and branching ratios in the SM and its minimal supersymmetric
extension. On the QCD side, all relevant known corrections are
included, while the EW corrections, which are smaller
for most of the channels, are approximated. Specifically,
the EW (two-loop) corrections to $\PH\to\Pg\Pg/\gamma\gamma$
are reconstructed via a numerical grid obtained from the full results
of \citere{Actis:2008uh,Actis:2008ts},
and the corrections to $\PW\PW/\PZ\PZ$ (considered as two-body decays)
are fitted to the corrections provided by \Prophecy.
\item
\Prophecy\ is a Monte Carlo event generator for all four-particle
decays $\PH\to\PW\PW/\PZ\PZ\to4f$ that takes into account 
all interferences and
all off-shell effects of the intermediate $\PW/\PZ$ bosons, so that
the obtained results are valid
above, near, and below the gauge-boson pair thresholds. 
As an event generator, \Prophecy\ can simulate the four-body 
decays with its full kinematics with NLO QCD and EW corrections; 
for leptonic final states
unweighted events can be produced as well.
For large $\MH$ the leading two-loop EW
corrections from \citere{Ghinculov:1995bz} are included as well.
\end{itemize}

All results shown in \reffi{fig:BR-lowMH} 
and all but the decay widths for $\PH\to\PW\PW,\PZ\PZ$ 
in \reffi{fig:BR-LHC} are based on \Hdecay,
while the results on $\PH\to\PW\PW,\PZ\PZ$ 
are obtained upon summing over the relevant four-fermion final
states in $\PH\to\PW\PW/\PZ\PZ\to4f$ whose decay widths are
calculated with \Prophecy. More technical details about the
calculations for \reffi{fig:BR-LHC} can be found in
\citeres{Dittmaier:2011ti,Denner:2011mq,Dittmaier:2012vm} from 
which the plots are taken.

\btab
\caption{Estimated relative uncertainties of Higgs-boson branching ratios 
due to theoretical plus parametric errors
for low and intermediate Higgs-boson
masses, as given by the LHC Higgs Cross Section
Working Group~\mbox{\cite{Denner:2011mq,Dittmaier:2012vm}}.}
\label{tab:BR}
\begin{center}
\begin{tabular}{crcccccc}
\hline
$\MH[\mathrm{GeV}]$ & $\PH\to\;\Pb\bar\Pb$ & $\tau^+\tau^-$ &
$\Pc\bar\Pc$ & $\Pg\Pg$ & $\gamma\gamma$ & $\PW\PW$ & $\PZ\PZ$ \\
\hline
$120$ & $3\%$ & $6\%$ & $12\%$ & $10\%$ & $5\%$ & $5\%$ & $5\%$ \\
$150$ & $4\%$ & $3\%$ & $10\%$ & $8\%$  & $2\%$ & $1\%$ & $1\%$ \\
$200$ & $5\%$ & $3\%$ & $10\%$ & $8\%$  & $2\%$ & $<0.1\%$ & $<0.1\%$
\\
\hline
\end{tabular}
\end{center}
\etab
The bands in the prediction of the branching ratios 
shown in \reffi{fig:BR-LHC} reflect
the combined uncertainties due to missing higher-order effects
in the calculations and parametric uncertainties in the input
parameters, as assessed by the LHC Higgs Cross Section 
Working Group~\cite{Denner:2011mq,Dittmaier:2012vm}.
Table~\ref{tab:BR} briefly summarizes the estimated total
uncertainties for $\MH$ values $120\GeV$, $150\GeV$, and
$200\GeV$, which are most interesting for SM Higgs physics
at Tevatron and the LHC.
It is important to note that the uncertainties of the branching ratios
result from an interplay of the uncertainties of the corresponding
partial decay width and the normalization to the total width $\GH$.
The decay widths into hadronic final states ($\Pb\bar\Pb$,
$\Pc\bar\Pc$, $\Pg\Pg$) possess relatively large uncertainties
via the parametric uncertainties from the heavy-quark masses and 
the strong coupling. As long as $\GH$ is dominated by 
$\Gamma_{\PH\to\Pb\bar\Pb}$ this uncertainty is mitigated in
$\BR(\PH\to\Pb\bar\Pb)$, but feeds into the branching ratios
of the other channels,
as, e.g., clearly visible for $\PH\to\PW\PW/\PZ\PZ$ at $\MH=120\GeV$.
The opposite is observed at $\MH=200\GeV$, where $\GH$ is dominated by
$\Gamma_{\PH\to\PW\PW/\PZ\PZ}$, 
which are very precisely known, i.e.\ in this case
the uncertainties of the other branching ratios entirely reflect the
one of the respective partial widths.
A similar estimate of the uncertainties of the branching ratios
has been presented in \citere{Baglio:2010ae} in the Higgs-boson mass
range of $\MH=100{-}200\GeV$, where somewhat larger uncertainties
are found, mainly due to larger assumed errors on the b- and c-quark
masses.

For very low and very high Higgs-boson masses the theoretical
uncertainties of the branching ratios become larger and larger and
hard to quantify. 
Figure~\ref{fig:BR-lowMH} shows the branching ratios without
error bands, mainly because the predictions 
in the vicinity of the $q\bar q$ thresholds, i.e.\ for $\MH\lsim15\GeV$,
become more of qualitative nature with uncertainties of $100\%$.
This theoretical shortcoming was balanced experimentally by
inclusive search strategies in the low-$\MH$ range, where the 
individual decay channels of the Higgs boson did not play a role.
For large Higgs-boson masses the uncertainty of predictions increases
more and more
when the non-perturbative regime of $\MH\sim1\TeV$ is approached.

\section{General considerations for Higgs-boson searches}

\subsection{Challenges}

\begin{figure}
  \centering
  \begin{minipage}[b]{.47\textwidth}
    \includegraphics[width=0.9\textwidth]{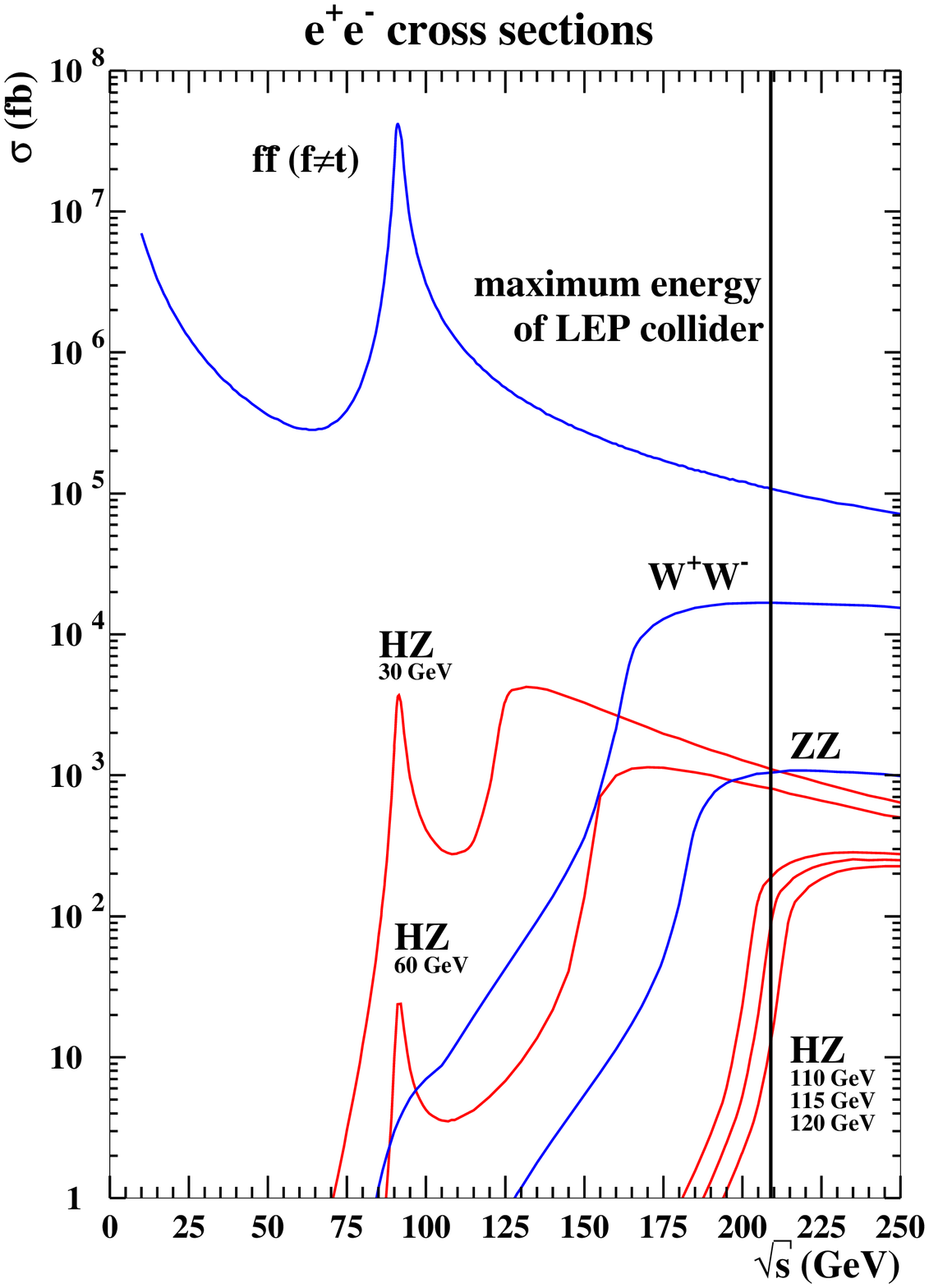}
  \end{minipage}
  \begin{minipage}[b]{.47\textwidth}
    \includegraphics[width=0.9\textwidth]{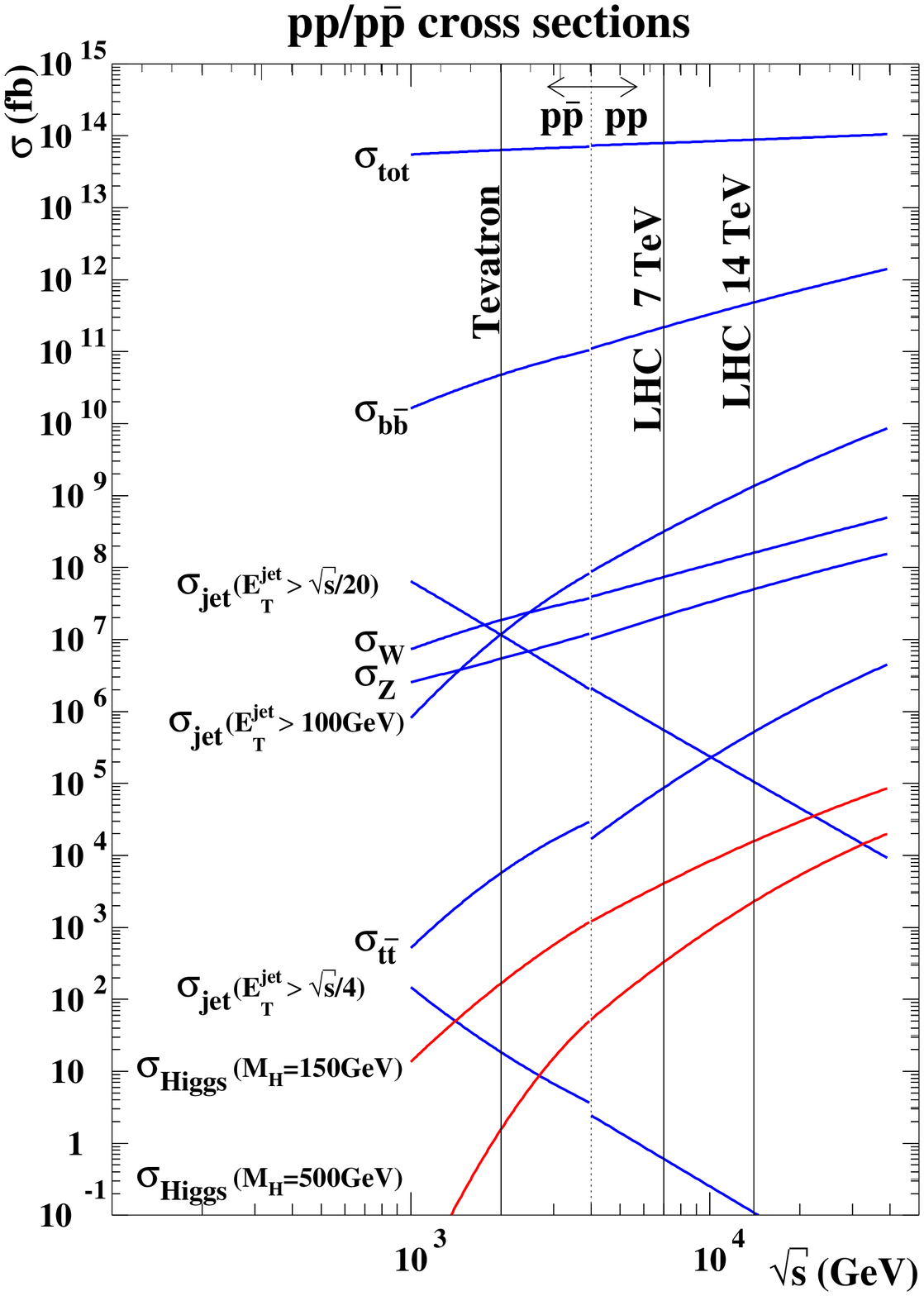}
  \end{minipage}
  \caption{Signal and background cross sections at leading order 
(produced with \Pythia~\cite{Sjostrand:2006za,Sjostrand:2007gs})
at $\Pe^+\Pe^-$ colliders (left) 
   and at hadron colliders (proton--antiproton and proton--proton colliders). }
  \label{fig:sigbgxs}
\end{figure}

The challenge in searches for the Higgs boson is to develop a selection that allows to increase the a priori very small 
signal-to-background ratio (see \reffi{fig:sigbgxs}) to a level of the order  1/1 to 1/10. In order to reach this goal,
sophisticated selection strategies have been developed over the last decades at the experiments at LEP, Tevatron, and the LHC. 
In the beginning only simple rectangular cuts were used to separate the signal process from background processes. By now many 
search strategies make use of multivariate techniques such as likelihood ratios, artificial neural networks, or boosted decision trees.

The combination of production mechanisms and decay modes of the Higgs boson and the particles that are eventually
produced in association with the Higgs particle yields a plenitude of final-state topologies which can be used to 
search for Higgs-boson production. The following considerations have to be taken into account when choosing the most promising 
search channels: sufficient a priori signal rate determined by cross section and branching ratios, means to trigger on signal 
events with high efficiency, means to suppress the background to an acceptable level, means to estimate the uncertainties
in the background prediction, and in the case of exclusion also the uncertainties 
in the signal prediction. It is desirable to
be able to construct a final discriminating variable to distinguish the two alternative hypotheses of background-only
and signal+background in addition to simple event counting. An obvious choice is the reconstructed mass of the 
Higgs-boson candidate events. Depending on the final-state particles this is not always possible. Alternatively or in addition
other powerful observables can be used, e.g.\ the response of a multivariate event classifier.

For the discovery of a new resonance, which means that the probability to observe the actual number of events is less than $2.87\times10^{-7}$ 
under the background-only hypothesis, only the background prediction and its uncertainty need to be known. Wrong assumptions on the 
shape of discriminating observables for the signal process will lead to a reduced sensitivity. Hypothetically one observed event can yield 
the claim of discovery given the search is basically background free ($\approx 3\times10^{-7}$ expected background events) with vanishing
systematic uncertainty, which unfortunately is never the case in the environments at colliders. For the exclusion of a signal hypothesis, 
neglecting systematic uncertainties and performing a counting-only experiment with negligible background, 
a minimal signal yield of 3 expected events after application of 
the full selection chain is needed in order to exclude the signal+background hypothesis
at the 95\% confidence level, when using the Bayesian or ``CL$_{\mathrm{S}}$'' 
technique (see below). In this case the uncertainties on the signal
rate and shape of the discriminating distributions will also 
influence the derived exclusion limits; 
for instance, a too optimistic assumption 
for the mass resolution for Higgs-boson candidates would yield a too stringent exclusion.

Although 
the Higgs mechanism had been incorporated in the electroweak theory in the late 1960s, it lasted until 1989,
when LEP started its operation, before 
significant parts of the mass range of the SM Higgs boson could be investigated.%
\footnote{A search carried out by the CDF experiment at Tevatron with up to $4.4\ipba$ at CM energy of 1.8 $\TeV$
excluded the mass ranges $2\Mmy{-}818\MeV$ and $846\MeV{-}2\Mka$ at 90\% CL~\cite{Abe:1989xa}. 
However, in this mass range the prediction of the branching ratios have large theoretical uncertainties, as discussed in 
\refse{se:Hdecays}.} 
The particle data group in 1988 \cite{Yost:1988ke} concluded on Higgs-boson 
searches before the start of LEP ``In summary, the only cast-iron constraint on the Higgs 
mass is $\MH > 14\MeV$. A combination of theoretical arguments and bounds from $B$, $\Upsilon$, and $K$ decays probably excludes the range below
$4\GeV$.'' 
Hence the mission of LEP was 
to probe the mass range of the SM Higgs boson starting from
a zero mass to the highest possible mass values.
For a discussion of Higgs-boson searches before the start of the LEP accelerator we, e.g., refer to 
\citeres{Schwarz:1985ca,Gunion:1989we,Cahn:1989tv,Gross:1992vq,Kado:2002er}.
In this article we review the most important results of the searches for the SM Higgs boson at the
colliders LEP, Tevatron, and LHC. For details of the individual analyses, in particular
identification criteria and systematic uncertainties, the original literature should be consulted.

\subsection{Statistical interpretation of search results}

The result
of a search for the Higgs boson after applying the whole selection algorithms is, in the simplest case,
given by the observed event yield, and the expected event yield for the background processes and the signal process, 
which depends on the hypothetical Higgs-boson mass. Such analyses are often called ``counting experiments''.  Most of
the searches performed during the LEP1 era were of this type.  Very often the sensitivity of the search can be increased 
by using a final discriminating observable, which can be one-dimensional or, in rare cases, is chosen to be multi-dimensional. 
Here not only expected and observed event yields, but also the observed and expected shapes of the distributions in the final
observable are used, when deriving the statistical interpretation of the results. This final discriminant 
is often given by the mass distribution of the Higgs-boson candidates, reconstructed from the decay products,
or by the response of a multivariate event classifier (e.g.\ likelihood, artificial neural network, boosted decision tree).
All searches during the LEP2 era, at Tevatron, and at the LHC use the distribution of a discriminating observable. 
Usually the expected event yields from signal and background processes as well as the shape of the distribution of 
the discriminating observable are affected by several systematic uncertainties, comprising, e.g., 
predictions of cross sections and acceptances due to kinematical requirements by theoretical calculations, knowledge of 
the detector performance in reconstruction and identification efficiencies, energy resolutions, and energy scales. 
Their influence on the signal and background expectations can be described by nuisance parameters $\theta$, which can 
be common in signal and background. The expected distribution for a Higgs-boson signal with SM strength
$s(\MH,\theta)$ depends on the Higgs-boson mass $\MH$ and can either be given by a simple event count, a distribution
of the final discriminant provided as a histogram, or an analytic function. In the following we will suppress the $\MH$ 
dependence in the equations, but it should be kept in mind.
The signal-strength parameter $\mu = \sigma_\PH / \sigma_{\PH_{\SM}}$  allows to describe a signal strength different from 
the SM expectation, where $\mu =1$ holds.  In the same way the background expectation is given by $b(\theta)$ and the 
data by $n$. The statistical interpretation uses the frequentist concept of hypothesis testing for claiming evidence or discovery.
When setting exclusion limits either the concept of hypothesis testing is used as well, or the closely related frequentist derivation
of an upper limit on the signal yield or signal-strength parameter $\mu$. For setting limits the Bayesian concept
of providing a one-sided credibility interval or Bayesian upper limit is used as well. 
Below a partial overview of the main methods 
used at LEP, Tevatron, and the LHC is given. For more details we refer to the excellent review~\cite{Beringer:1900zz}
and to the experiments' publications. A general introduction into the topics of frequentist hypothesis tests, and 
Bayesian and frequentist limit setting and their interpretation can be found, e.g., in \citere{sda}. 

In a hypothesis test the consistency of the observed data with a given hypothesis, which one tries to falsify (null hypothesis H$_0$),
is evaluated via the so-called $p$-value. The null hypothesis is the opposite of the statement one is aiming at, because hypotheses can only be rejected,
but not be approved: When looking for evidence or discovery, it is given by the existence of only background processes called 
``background-only'' hypothesis (``$b$''-only), whereas when excluding a signal, it is given by the  ``signal+background'' hypothesis (``$s+b$'').
A test statistic $t$ is constructed from the 
observed results (event yields and shapes of final discriminating observables), which is used to evaluate the consistency with the null hypothesis. 
Given that the probability density function 
$f(t(\mu)|{\mathrm H_0})$ 
for the test statistic $t$ under the null hypothesis  
is known, and knowing
that small (large)  values of $t$ indicate consistency (inconsistency) with the null hypothesis the $p$-value for an actual observed value $t_{\mathrm{obs}}$ is defined by
\beq
p = \int_{t_{\mathrm{obs}}}^\infty f(t(\mu)|{\mathrm H_0})\, {\mathrm d}t.
\eeq
The $p$-value is the probability to observe a value of the test statistic $t$
at least as large as observed in data assuming the
null hypothesis is realized in nature. Equivalently it quantifies the probability to wrongly reject the null hypothesis, if it is actually true.   
The $p$-value can be translated into a so-called Gaussian significance $Z$ via $Z=\Phi^{-1}(1-p)$, where $\Phi^{-1}$ is the inverse
of the cumulative distribution for the standard Gaussian probability density function. For claiming discovery one requires 
the $p$-value to be less than or equal to
$2.87\times10^{-7}$ or equivalently the significance $Z$ to be at least 5.
For exclusion, when one wants to reject the signal+background hypothesis the $p$-value is usually 
required to be less than 0.05. 
The hypothesis is then called excluded with a confidence level CL${}=1-p$. Alternatively one can determine the signal strength for which
the $p$-value is exactly 0.05. This is equivalent to constructing a one-sided frequentist confidence interval for $\mu$ or an upper limit
on $\mu$ called $\mu_{95}$ to a confidence level CL${}=95\%$, which fulfills the condition
\beq
0.05=1-\mathrm{CL}= \int_{t_{\mathrm{obs}}}^\infty f(t(\mu_{95})|{\mathrm H_0})\, {\mathrm d}t, \quad
\mbox{i.e.} \quad
0.95 = \mathrm{CL}= \int^{t_{\mathrm{obs}}}_{-\infty} f(t(\mu_{95})|{\mathrm H_0})\, {\mathrm d}t.
\eeq
All values of $\mu$ larger than $\mu_{95}$ are then excluded, i.e.\ the SM Higgs-boson hypothesis is excluded if $\mu_{95}\le 1$.
The Bayesian technique to derive a one-sided credibility interval or a Bayesian upper limit on $\mu$  at credibility level CL
relies on the likelihood function ${\cal{L}} (\mathrm{data}|\mu)$, which gives the conditional probability to observe the data for a
given value of $\mu$. Using Bayes' theorem the posterior probability for the degree of belief in a particular value of 
$\mu$ in the light of the data is given by 
\beq
p(\mu|\mathrm{data}) = \mathrm{const.} \times {\cal{L}} (\mathrm{data}|\mu) \,\pi(\mu),
\eeq
where $\pi(\mu)$ is the a priori belief about the knowledge of $\mu$ and $const.$ is an irrelevant constant to normalize $p(\mu|\mathrm{data})$.
Very often and in all results discussed in this review, $\pi(\mu)$ is chosen to be uniform for non-negative and hence physically
meaningful values of $\mu$ and zero otherwise. The Bayesian upper limit $\mu_{95}$ to a credibility level CL${}=95\%$ is then obtained via
\beq
0.95 = \frac{\int_{-\infty}^{\mu_{95}} p(\mu|\mathrm{data})\,{\mathrm d}\mu }{ \int_{-\infty}^\infty p(\mu|\mathrm{data})\, {\mathrm d}\mu}
     = \frac{\int_0^{\mu_{95}} {\cal{L}} (\mathrm{data}|\mu)\,{\mathrm d}\mu }{ \int_0^\infty {\cal{L}} (\mathrm{data}|\mu)\, {\mathrm d}\mu}.
\eeq
The procedure described above is repeated for each hypothetical value of $\MH$ in all approaches 
(frequentist hypothesis test, frequentist upper limit, and Bayesian upper limit).

Let us first consider a counting experiment as was done during the LEP1 era and partially at the beginning of the LEP2 era. 
The discussion will be restricted to rejecting the signal+background hypothesis and hence to the derivation of exclusion limits,
as during that era no significant excess was observed.
The background $b$ is assumed to be known without uncertainties and the expected signal yield is given by $s$. 
The upper limits on the signal yield is denoted by $s_{95}=\mu_{95} s$.
In the frequentist approach one has to specify the test statistic, which is chosen to be the number of observed events 
$n$, which in this case does not depend on the signal hypothesis.
The probability density function for $n$ under the signal+background hypothesis is simply the Poisson probability  
$f(n|\mu s+b) =  e^{-(\mu s+b)}(\mu s+b)^n/n!$. The likelihood function needed in the Bayesian approach is also given by the Poisson 
probability ${\cal{L}} (\mathrm{data}|\mu s+b) = e^{-(\mu s+b)}(\mu s+b)^n/n!$, where $n$ plays the role of ``data''.
A low number of observed events indicates 
an inconsistency with the signal+background hypothesis. Hence the frequentist 
upper $s_{95}$ limit is derived from
\beq
   0.05 = \sum^{n_{\mathrm{obs}}}_{n=0} e^{-(s_{95}+b)}(s_{95}+b)^n / n! \, .
\eeq
The analytic solution of this equation for  $s_{95}$ gives
\beq
s_{95} = \frac{1}{2}  F^{-1}_{\chi^2}(0.95;2(n_{\mathrm{obs}}+1)) - b,
\eeq
where $F^{-1}_{\chi^2}$ denotes the inverse of 
the cumulative $\chi^2$ probability density distribution 
evaluated for \mbox{$2(n_{\mathrm{obs}}+1)$} degrees of freedom.
The Bayesian upper limit has to be derived from \cite{Helene:1982pb}
\beq
0.05 = \frac{\sum^{n_{\mathrm{obs}}}_{n=0} e^{-(s_{95}+b)} (s_{95}+b)^n /n!}
     {\sum^{n_{\mathrm{obs}}}_{n=0} e^{-b} b^n /n!} \,.
\eeq
For zero background ($b = 0$) the two approaches give identical results.
For $n _{\mathrm{obs}} = 0$ ($1$) observed events the $s_{95}$ is then given by $3.0$ ($4.7$).
These are the numbers that are used in the derivation of the final
LEP1 exclusion limits, where all observed events were considered 
as candidate events or equivalently using $b = 0$ in the above equations
for $s_{95}$. A Higgs-boson mass hypothesis is excluded 
if the expected signal event yield for the value of $\MH$ tested
is larger than $3.0$ ($4.7$) given $0$ ($1$) events are observed in data.
The expected signal event yields are decreased by their systematic uncertainty
before the determination of $s_{95}$ and its comparison with $\mu s$.

For non-vanishing background the frequentist approach due to the $-b$ 
term can give arbitrarily small values $s_{95}$, e.g.\ $s_{95}=0$ for $b=3$
and $n=0$, if the observed event yield is significantly lower than 
the one expected from background processes. In the Bayesian approach $s_{95}$
can never get smaller than 3. Even if completely correct from a 
purely frequentist point of view, people felt uneasy with quoting a very
small $s_{95}$ in a situation where the average experiment has no 
sensitivity to exclude such a small signal yield, but excludes 
it due to a downward fluctuation of the observed event yield with 
respect to the background-only hypothesis. Such situations are avoided
in the Bayesian approach. An alternative derivation of the Bayesian 
formula in a pseudo-frequentist ansatz and its interpretation are 
given in \citere{Zech:1988un}.
Formally the right-hand side of the Bayesian formula can be 
rewritten in the language of frequentist hypothesis tests as 
$p_{\mathrm{s+b}}/(1-p_\Pb)$, where $p_{\mathrm{s+b}}$
and $p_\Pb$ denote the $p$-values under the signal+background and background-only
hypotheses, respectively, keeping in mind that small (large) observed event yields 
mean inconsistency with the signal+background (background-only) hypothesis.

At LEP2 and later, multiple channels with different signal-to-background ratios 
were combined, and the distributions of the final discriminants were also used.
At the beginning of the LEP2 era different techniques \cite{statother} were used 
for the construction of 
a test statistic that takes into account the different
signal-to-background ratios in the various search channels and bins of
the histograms of the final discriminants. 
The likelihood functions ${\cal L}(\mathrm{data} | \mu s+b)$
and ${\cal L}(\mathrm{data}|b)$ are given by the product of
Poisson probabilities running over all channels and bins. 
Neglecting systematic uncertainties, the optimal test statistic 
is given, according to the Neyman--Pearson lemma \cite{neymanpearson},
by the ratio of the two likelihoods to observe the data under the signal+background hypothesis 
and background-only hypothesis, its reciprocal, or any monotonic function of it. 
This is the statistic 
\beq
Q = {\cal{L}}(\mathrm{data} | \mu s + b)/{\cal{L}}(\mathrm{data} | b),
\eeq
or its transform $-2\ln Q$ used at LEP.
For a single-channel/single-bin counting experiment the Neyman--Pearson test statistic
derived from the ratio of two Poisson probabilities is (setting $\mu = 1$)
\beq
Q = e^{-s}(1+s/b)^n, \quad \mbox{i.e.} \quad  -2\ln Q = 2s - 2 n \ln(1+s/b).
\eeq
Hence as $Q$ is monotonic in $n$, $n$ is the optimal test statistic for
a simple counting experiment in one channel. Each observed event enters
the calculation of $-2\ln Q$ with a weight of $\ln(1+s/b)$ which 
takes into account the local $s/b$ ratio.
Large values of $-2\ln Q$ are obtained for ``b-only''-like data and 
small values for ``s+b''-like data. The median of the $-2\ln Q$
distribution is positive under the background-only hypothesis, whereas the 
median is negative under the signal+background hypothesis. 
The probability density functions for $-2\ln Q$ under the signal+background hypothesis
and background-only hypothesis are obtained from simulation of toy MC experiments.

Translating the Bayesian formula in the interpretation of \citere{Zech:1988un}
to the test statistic  $-2\ln Q$ yields the so-called 
``CL$_{\mathrm{S}}$'' technique \cite{Read:2002hq,Junk:1999kv}, which avoids exclusion without sensitivity.
The $p$-values for the LEP test statistic $-2\ln Q$ are calculated according to:
\beq
p_{\mathrm{s+b}}
 = \int_{-2\ln Q_{\mathrm{obs}}}^{\infty} f(-2\ln Q|{s+b})\, {\mathrm d}(-2\ln Q)
\eeq
and
\beq
p_\Pb = \int^{-2\ln Q_{\mathrm{obs}}}_{-\infty} f(-2\ln Q|{b}) \,{\mathrm d}(-2\ln Q).
\eeq
The ``CL$_{\mathrm{S}}$'' value is then calculated as
\beq
\mathrm{CL}_{\mathrm{S}} = p_{\mathrm{s+b}}/ (1-p_\Pb)   .
\eeq
A Higgs-boson mass hypothesis is called excluded at least to 95\% CL if CL$_{\mathrm{S}} \le 0.05$.
The probability to reject the signal+background hypothesis, if it is true, is less than 5\%, as
the ``ad hoc'' correction in the denominator destroys the strict frequentist interpretation.
Even if not proven in a mathematically rigorous way, good agreement has been found between the 
Bayesian approach and ``CL$_{\mathrm{S}}$'' technique in practice at LEP, Tevatron, and the LHC. 
Systematic uncertainties (denoted by $\theta$) were incorporated at LEP in the hybrid 
frequentist--Bayesian approach described in \citere{Cousins:1991qz}, which results in modified probability 
density functions for the unchanged test statistic $-2\ln Q$,
\beq
f(-2\ln Q | s+b ) = \int f(-2\ln Q | s+b, \theta) \pi(\theta)\, {\mathrm d}\theta
\quad \mbox{and} \quad
f(-2\ln Q | b ) = \int f(-2\ln Q | b, \theta) \pi(\theta) \,{\mathrm d}\theta,
\eeq
which is a so-called marginalization with respect to the parameters $\theta$ with a priori probabilities $\pi(\theta)$.
The explicit choice of $\pi(\theta)$ is specific to the sources of the systematic uncertainties and thus
depends on the analysis.

At Tevatron the Neyman--Pearson test statistic was modified to include the parameters $\theta$, describing the systematic uncertainties,
in the likelihood, which yields a better discrimination between signal+background and background-only hypotheses in the presence of systematic uncertainties
compared to LEP. The likelihood function now reads ${\cal{L}}(\mathrm{data}|\mu s(\theta) + b(\theta)) \pi(\theta)$,
with a priori probabilities $\pi(\theta)$ for the nuisance parameters $\theta$, and the test statistics $\mathrm{LLR}$, in the 
Tevatron convention, 
is given by ratio of two profiled likelihoods
\newcommand{\hhtheta}{\hspace{.2em}\hat{\hspace{-.2em}\hat{\theta}}}
\beq 
\mathrm{LLR} = -2\ln\left(\frac{{\cal{L}}(\mathrm{data} | \mu s(\hhtheta) + b(\hhtheta)) \pi(\hhtheta)}
           { {\cal{L}}({\mathrm{data}} | b(\hat{\theta})) \pi(\hat{\theta})}\right).
\eeq
The $\hhtheta$ and $\hat{\theta}$ are, respectively, the conditional maximum-likelihood estimators of two fits performed
(i) under the signal+background hypothesis in the numerator and (ii) under the background-only hypothesis in the denominator.
The best fit values $\hhtheta$ and $\hat{\theta}$ for $\theta$ are usually different under the two hypotheses.
The determination of the probability density functions for $\mathrm{LLR}$ and of 
the ``CL$_{\mathrm{S}}$'' value proceeds as was done at LEP.
The primary choice for deriving upper limits on $\mu$, which is called $R$ in the Tevatron analyses,
in the latest Tevatron combinations is the use of the Bayesian method with the posterior probabilities for $\mu$ given by 
\beq
p(\mu|\mathrm{data}) = \mathrm{const.} \times \int {\cal{L}}({\mathrm{data}}| \mu s(\theta)+b(\theta)) \pi(\theta) \pi(\mu) \,{\mathrm d}\theta.
\eeq
Again a uniform prior for positive $\mu$ is used and a marginalization with respect to the $\theta$ is performed.
For deriving $p$-values under the background-only hypothesis and calculating significances, a fit of 
${\cal{L}}(\mathrm{data}| \mu s (\theta) + b(\theta)) \pi(\theta)$ to the data is performed, and the best 
fit value $\hat{\mu}$ for the signal-strength parameter $\mu$ is used as the test statistic.
The probability density functions for $\hat{\mu}$ are derived again from toy MC experiments.

At the LHC two different test statistics are used for the two cases, setting limits and claiming 
evidence for discovery~\cite{Cowan:2010js}. The likelihood function is the same as at Tevatron,
${\cal{L}}(\mathrm{data} | \mu s(\theta) +b(\theta)) \rho(\tilde{\theta} | \theta)$, but the  
a priori probabilities for the nuisance parameters $\pi(\theta)$ are reinterpreted or replaced
by conditional probabilities $\rho(\tilde{\theta}|\theta)$ to observe a 
value $\tilde{\theta}$ in an auxilary measurement given the value $\theta$.
For setting limits, i.e.\ testing a signal strength $\mu$, the test statistic is given by
\beq
\tilde{q}_\mu = -2\ln\left(\frac{{\cal{L}}(\mathrm{data} |       \mu s(\hhtheta) + b(\hhtheta)) \rho(\tilde{\theta} | \hhtheta)}
                      {{\cal{L}}(\mathrm{data} | \hat{\mu} s(\hat{\theta}) + b(\hat{\theta})) \rho(\tilde{\theta} | \hat{\theta})}\right).
\eeq
In the numerator the likelihood under the signal+background hypothesis with fixed signal strength $\mu$, to be tested, 
is evaluated, while in the denominator (in contrast to Tevatron) the likelihood is evaluated after fitting also for the best 
signal strength consistent with the data $\hat{\mu}$.  The range for $\hat{\mu}$ is restricted to $0\le\hat{\mu}\le \mu$,
as $\hat\mu < 0$ is unphysical and as a one-sided hypothesis test is performed, where $\hat{\mu} > \mu$ is 
considered as being consistent with the signal+background hypothesis (see \citere{Cowan:2010js} for details).
The value of the test statistic $\tilde{q}_\mu$ is restricted to the range between zero and infinity. 
In order to be able to define a one-sided hypothesis test, the value of $\tilde q_\mu$ is set to zero for $\hat{\mu}$ 
values larger than $\mu$. Background-like data yield large values of $\tilde q_\mu$, whereas 
``$s+b$''-like data accumulate at low values of $\tilde q_\mu$. The advantage of not fixing $\mu$ to $0$
in the denominator, as done at LEP and Tevatron, is that the probability density function for $\tilde{q}_\mu$
in the limit of large event yields are known analytically due to Wilks' \cite{wilks} and Wald's  \cite{wald} theorems
(see \citere{Cowan:2010js} for the approximative analytic expressions). The determination of frequentist upper limits
proceeds as at LEP and Tevatron via the ``CL$_{\mathrm{S}}$'' technique.

For testing the background-only hypothesis, when looking for discovery or evidence at the LHC, the test statistic is calculated 
as a different ratio of profiled likelihoods: 
\beq
q_0 = -2\ln\left(\frac{{\cal{L}}( \mathrm{data} | b(\hhtheta) ) \rho(\tilde{\theta}|\hhtheta)}
            {{\cal{L}}( \mathrm{data} | \hat{\mu} s(\hat{\theta}) + b(\hat{\theta})) \rho(\tilde{\theta}|\hat{\theta})}\right).
\eeq
Now $\mu$ is fixed in the numerator under the null hypothesis (background-only) and is determined via a fit to the data 
$\hat{\mu}$ in the denominator. The range for $\hat{\mu}$ is restricted to $\ge 0$, as $\mu < 0$ is unphysical and as a one-sided 
hypothesis test is performed, where $\hat{\mu} < 0$ is considered as being consistent with the background-only hypothesis 
(see \citere{Cowan:2010js} for details).
The values of $q_0$ are restricted to the range between zero and infinity. In order to be able to define a one-sided hypothesis test, 
the value of $q_0$ 
is set to zero for $\hat{\mu}<0$. Now ``$b$-only''-like data accumulate at low values of 
$q_0$, whereas ``$s+b$''-like data yield large values of $q_0$. Again the test statistic $q_0$ has the advantage that its probability 
density function is know analytically due to Wilks' and Wald's theorems, i.e.\
\beq
f(q_0 | b ) =  \frac{1}{2} \delta(q_0) + \frac{1}{2} f_{\chi^2}(q_0;1),
\eeq
by a combination of $\frac{1}{2}\delta$-function and $\frac{1}{2}\chi^2$-probability density function with 1 degree of freedom 
(see \citere{Cowan:2010js} for details). The $\frac{1}{2}\delta$-function is due to the fact that $q_0$ is set to $0$ 
for $\hat{\mu} < 0$. Hence any observed $q_0\ne0$ yields a maximal $p$-value of 0.5.  

So far in the likelihood functions that enter the construction of test statistics used for testing the background-only hypothesis 
($-2\ln Q$ at LEP, $\hat{\mu}$ at Tevatron, $q_0$ at the LHC) the Higgs-boson mass hypothesis $\MH$ is 
fixed under the alternative signal+background hypothesis. The corresponding $p$-values are hence called
{\it local} $p$-values. It gives the probability to observe such a fluctuation for a fixed $\MH$ value.
The so-called {\it global} $p$-value denotes the probability to find such an excess anywhere in the mass spectrum or 
more precisely for arbitrary Higgs-boson mass hypothesis under the alternative hypothesis within a certain $\MH$ range. 
The relation between the two $p$-values is given by $p_{\mathrm{global}} = f_{\mathrm{trial}}$ $p_{\mathrm{local}}$. The trial factor 
$f_{\mathrm{trial}}$ can be interpreted as the  effective number of independent searches performed. This enhancement in the $p$-value is commonly 
called the ``look-elsewhere effect''. A technique to derive approximate global $p$-values and trial factors 
used at the LHC can be found in \citere{Gross:2010qma},
and in \citere{Dunn:1961my} for Tevatron.

Often expected limits on the signal strength, expected $p$-values, and significances are quoted. These are the median values for those
quantities derived under the alternative hypothesis: {\it Expected limit} is the median limit  under the background-only hypothesis, 
{\it expected significance} is the median significance under the  signal+background hypothesis. 

A subtle difference should be noted when generating toy MC experiments at Tevatron and the LHC.
The central values of the nuisance parameters, which are used as input to the MC experiments, are fixed to the a priori
best values at Tevatron and are extracted from a fit to the data at the LHC (see \citere{comproc} for details).

In the following, all limits are given to 95\% CL, if not stated otherwise.

\section
{\boldmath{Higgs-boson production at $\mathrm{e^+e^-}$ colliders}}

\label{se:lepprod}
At $\Pep\Pem$ colliders the importance of the various Higgs production
mechanisms strongly depends on the centre-of-mass (CM) energy of
the collider and to some extent also on the Higgs-boson mass.
Figure~\ref{fig:epemcolldiags} illustrates the production channels
by some representative lowest-order diagrams.
\bfi[b]
\centerline{
{\unitlength .8pt
\SetScale{0.8}
\begin{picture}(150,100)(-5,-10)
\ArrowLine(20, 5)(50,40)
\ArrowLine(50,40)(20,75)
\Photon(50,40)(90,40){2}{5}
\Photon(120,20)(90,40){2}{5}
\DashLine(90,40)(120,60){5}
\Vertex(50,40){2}
\Vertex(90,40){2}
\put(125,56){${\PH}$}
\put( 65,22){${\PZ}$}
\put(125,12){${\PZ}$}
\put( 5,70){${\Pep}$}
\put( 5, 2){${\Pem}$}
\put(15,-15){\footnotesize (a)}
\end{picture}
\SetScale{1}
}
{\unitlength .8pt
\SetScale{0.8}
\begin{picture}(150,100)(-5,-10)
\ArrowLine(20, 7)(50,10)
\ArrowLine(50,70)(20,73)
\ArrowLine(50,10)(110, 5)
\ArrowLine(110,75)(50,70)
\Photon(50,70)(80,40){2}{5}
\Photon(50,10)(80,40){2}{5}
\DashLine(80,40)(115,40){5}
\Vertex(80,40){2}
\Vertex(50,70){2}
\Vertex(50,10){2}
\put(120,35){${\PH}$}
\put(117,75){${\bar\nu_\Pe/\Pep}$}
\put(117, 0){${\nu_\Pe/\Pem}$}
\put(75,52){${\PW/\PZ}$}
\put(75,16){${\PW/\PZ}$}
\put( 5,70){${\Pep}$}
\put( 5, 2){${\Pem}$}
\put(15,-15){\footnotesize (b)}
\end{picture}
\SetScale{1}
}
{\unitlength .8pt
\SetScale{0.8}
\begin{picture}(150,100)(-5,-10)
\ArrowLine(20, 5)(50,40)
\ArrowLine(50,40)(20,75)
\Photon(50,40)(90,40){2}{5}
\ArrowLine(120,10)(90,40)
\ArrowLine(90,40)(110,60)
\ArrowLine(110,60)(130,80)
\DashLine(110,60)(130,40){5}
\Vertex(50,40){2}
\Vertex(90,40){2}
\Vertex(110,60){2}
\put(133,76){${\Pt}$}
\put(133,35){${\PH}$}
\put( 55,22){${\gamma,\PZ}$}
\put(125, 5){${\bar\Pt}$}
\put( 5,70){${\Pep}$}
\put( 5, 2){${\Pem}$}
\put(15,-15){\footnotesize (c)}
\end{picture}
\SetScale{1}
}
{\unitlength .8pt
\SetScale{0.8}
\begin{picture}(150,100)(-5,-10)
\ArrowLine(20, 5)(50,40)
\ArrowLine(50,40)(20,75)
\Photon(50,40)(90,40){2}{5}
\DashLine(90,40)(110,60){1.5}
\DashLine(90,40)(110,20){1.5}
\DashLine(110,60)(110,20){1.5}
\DashLine(110,60)(130,80){5}
\Photon(130,0)(110,20){2}{4}
\Vertex(50,40){2}
\Vertex(90,40){2}
\Vertex(110,60){2}
\Vertex(110,20){2}
\put(133,76){${\PH}$}
\put( 55,22){${\gamma,\PZ}$}
\put(133,-3){${\gamma}$}
\put(113,35){${f,\PW}$}
\put( 5,70){${\Pep}$}
\put( 5, 2){${\Pem}$}
\put(15,-15){\footnotesize (d)}
\end{picture}
\SetScale{1}
}
}
\vspace*{.2em}
\caption{Representative lowest-order diagrams for the main SM Higgs-boson production channels
at $\Pep\Pem$ colliders:
(a)~Higgs-strahlung, (b) vector-boson fusion, (c) top-quark
associated production, (d) $\PH\gamma$ production.}
\label{fig:epemcolldiags}
\efi
Up to LEP energies, $\sqrt{s}\lsim209\GeV$, $\PZ\PH$ production
via ``Higgs-strahlung'' with a variety of Higgs and $\PZ$-boson decays
was the main search channel. 
Some events from $\PH\gamma$ production were expected in spite of
a very small cross section; owing to the extremely suppressed
electron Yukawa coupling the leading order of the channel is 
loop-induced by massive-fermion and W-boson loops.
At future $\Pep\Pem$ linear colliders, such as the ILC or CLIC,
two additional types of production channels become relevant: 
the fusion of W or Z bosons and the associated
production with $\Pt\bar\Pt$
pairs. For instance, the cross section of W-boson fusion at an ILC with a CM energy
$\sqrt{s}=500\GeV$ exceeds the one of Higgs-strahlung off Z~bosons for Higgs-boson
masses larger than $\sim160\GeV$.
While the $\PH\PZ$ production cross section falls off $\propto 1/s$,
like any typical $s$-channel process, the $t$-channel-like W/Z-fusion
processes receive contributions that grow like $\ln(s/\MH^2)/\MW^2$ for
large energies.
Above the $\Pt\bar\Pt\PH$ production threshold, $\sqrt{s}>2\Mt+\MH$,
Higgs production in association with $\Pt\bar\Pt$ pairs offers the possibility
to directly measure the top-quark Yukawa coupling with good precision.
Figure~\ref{fig:HXS_ee} (r.h.s.) presents an overview over the various cross sections
as function of the collider energy $\sqrt{s}$, covering the range from LEP
energies to the TeV region of potential future colliders.
\bfi
\centerline{
\includegraphics[bb = 125 130 460 335,clip,width=.5\textwidth,height=.3\textwidth,angle=0]%
{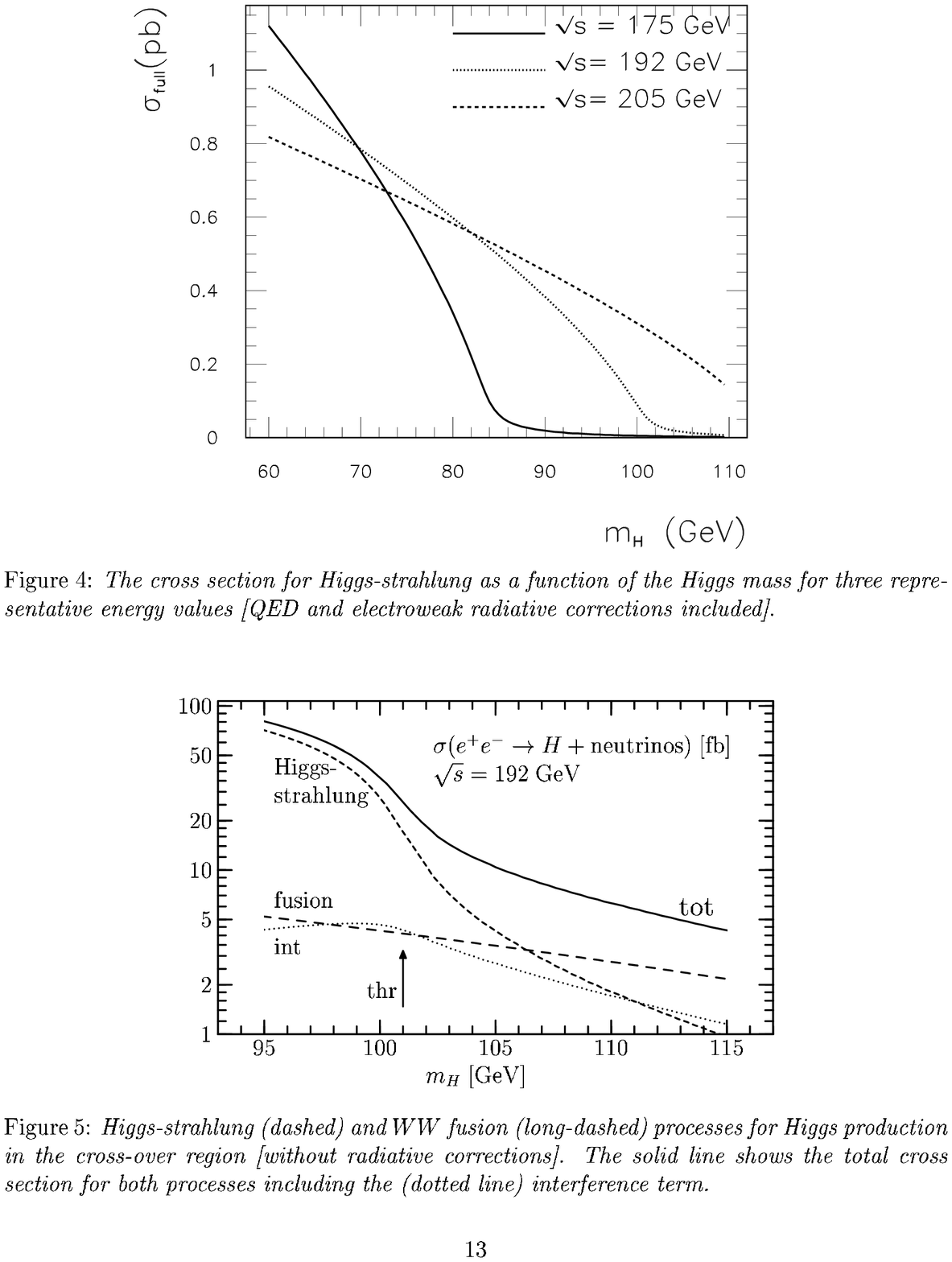}
\includegraphics[width=.5\textwidth,height=.3\textwidth,angle=0]%
{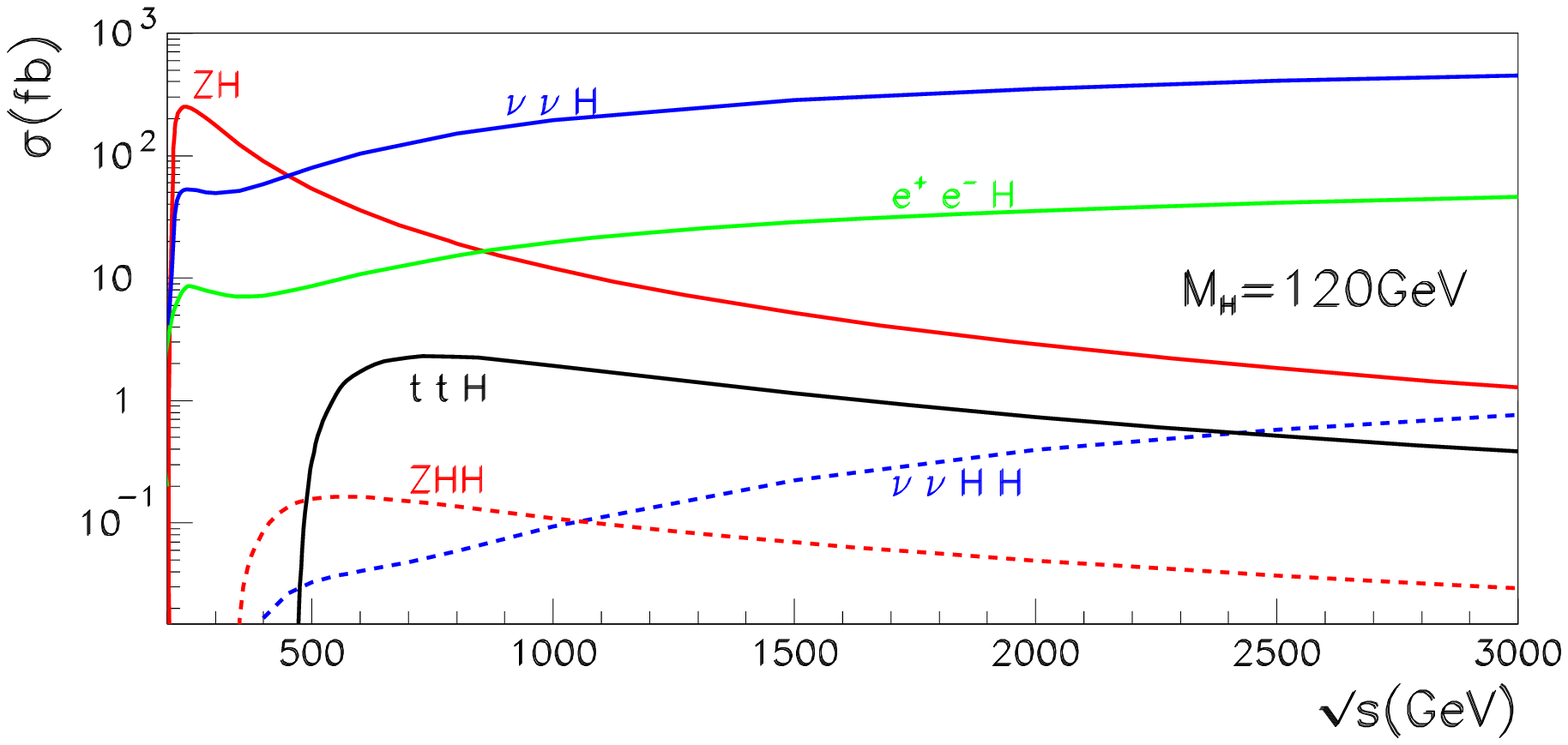}
}
\caption{Left: cross section for $\Pep\Pem\to\PH\nu\bar\nu$ for a typical LEP2 energy
and its breakup into Higgs-strahlung, W-fusion, and interference contribution
(taken from \citeres{Carena:1996bj,Kilian:1995tr});
right: cross sections for the various SM single-Higgs-boson and Higgs-boson pair 
production channels for a future $\Pep\Pem$ linear collider
(taken from \citere{Boudjema:2005rk}).
}
\label{fig:HXS_ee}
\efi
The theoretical preparation of the LEP analysis is described in detail in
\citeres{Bernreuther:1989rt,Carena:1996bj}, and a more detailed
survey of both the physics potential of linear colliders
and features of the predictions can be found in 
\citeres{Accomando:1997wt,Djouadi:2005gi}. 

Figure~\ref{fig:HXS_ee} is based on lowest-order predictions, but state-of-the-art
calculations include higher-order improvements as well. Collinear photonic
initial-state radiation (ISR) generically delivers corrections at the level of
$10\%$ or more, which are particularly pronounced in region where cross sections 
or distributions show strong variations, as it is, e.g., the case near kinematical 
thresholds. The by far dominating part of ISR is of universal origin and
consists of logarithmically enhanced terms $\alpha^n\ln^m(\Me/Q)$ with $(m\le n)$, where 
$Q$ is a typical scale of the process. This important effect can be taken into
account by structure functions, which are known to quite high order
(see, e.g., \citere{Beenakker:1996kt} and references therein). 
The full inclusion of NLO corrections in predictions,
however, require dedicated calculations:
\begin{itemize}
\item
{\it Higgs-boson--photon production $\Pep\Pem\to\PH\gamma$}
\cite{Cahn:1978nz,Abbasabadi:1995rc}
\\
The production of a Higgs boson in association with a photon mainly proceeds
via top-quark and W-boson loops, where a significant destructive interference occurs.
At LEP energies, $\PH\gamma$ production only contributed 
in the search for a 
light Higgs boson with a mass well below the Z-boson mass, i.e.\ 
$\PH\gamma$ production was mainly treated as potential decay mode of the 
Z~boson~\cite{Cahn:1978nz}.
At a high-luminosity $\Pep\Pem$ linear collider $\PH\gamma$ should,
however, also be observable at energies far above the 
Z~resonance~\cite{Abbasabadi:1995rc}, where $\Pep\Pem\to\PH\gamma$ is a true
$2\to2$ scattering process.
Since this rare process is already loop induced, full NLO predictions 
would require a two-loop calculation with multiple scales
that is technically still out of reach
with present calculational techniques. On the other hand, taking into account
the universal ISR effects most likely will also be sufficient to match the
expected experimental precision at a future collider.
\item
{\it Higgs-strahlung $\Pep\Pem\to\PH\PZ$}
\cite{Ellis:1975ap,Jones:1979bq,Fleischer:1982af,Jegerlehner:2005gf}
\\
The actual Higgs-strahlung signal that was searched for at LEP was mainly based
on the $\PH\to\Pb\bar\Pb$ decay and various Z-boson decays
$\PZ\to f\bar f$.
Higher-order effects from ISR lead to relevant corrections of $10{-}20\%$
at LEP energies and have been taken into account in the LEP analyses.
The remaining genuine weak corrections, which comprise weak-boson
exchange etc., have been first calculated in the
approximation of on-shell Z and Higgs bosons~\cite{Fleischer:1982af}
and amount to $1{-}2\%$ only at LEP energies.
Later the EW corrections have also been evaluated
including the leading off-shell effects of $\PZ\to\mu^+\mu^-$ and 
$\PH\to\Pb\bar\Pb$~\cite{Jegerlehner:2005gf}, which will be relevant at possible
precision studies at a future linear collider. The EW corrections
become more and more relevant with increasing collider energy.
\item
{\it W fusion $\Pep\Pem(\PW\PW)\to\PH\nu\bar\nu$}
\cite{Kilian:1995tr,Jones:1979bq,Dawson:1984ta,Hikasa:1985ee,Dicus:1985zg,Altarelli:1987ue,Belanger:2002ik,Denner:2003yg}
\\
Actually the physically observable process involves a sum over all three
neutrino species in the final state. While W fusion is only possible for
$\nu_\Pe\bar\nu_\Pe$ pairs in the final state, all three $\nu\bar\nu$
channels receive contributions from $s$-channel $\PH\PZ$ production
with $\PZ\to\nu\bar\nu$ decays; for $\PH\nu_\Pe\bar\nu_\Pe$ $t$- and
$s$-channel interfere. 
The l.h.s.\ of \reffi{fig:HXS_ee} illustrates the interplay of
$\PH\PZ(\to\nu\bar\nu)$ production and W fusion for a typical LEP2
energy near the $\PH\PZ$ threshold region where W~fusion becomes
significant. For $\sqrt{s}\lsim\MH+\MZ$ the $\PH\PZ$ production
process can only proceed via off-shell Z~bosons (the Higgs boson
is too narrow to get off shell for the $\MH$ values relevant at LEP).
Sufficiently above the $\PH\PZ$ threshold
($\sqrt{s}\gsim\MH+\MZ+100\GeV$ for $\MH\sim100{-}200\GeV$), which
is an interesting region for future linear colliders,
W fusion dominates (see r.h.s.\ of \reffi{fig:HXS_ee}). 
The NLO corrections are of pure EW origin and
known~\cite{Belanger:2002ik,Denner:2003yg}.
Owing to the dominating $t$-channel kinematics,
which prefers forward/backward-produced (anti)neutrinos and 
low-virtuality intermediate W bosons, the corrections to W fusion 
can be described within few percent by an approximation based on
universal ISR (with appropriate scale $Q$) and a simple constant correction
factor from heavy top loops~\cite{Denner:2003yg}.
\item
{\it Z fusion $\Pep\Pem(\PZ\PZ)\to\PH\Pep\Pem$}
\cite{Hikasa:1985ee,Dicus:1985zg,Altarelli:1987ue,Boudjema:2004eb}
\\
The Z-fusion cross section is suppressed with respect to the one of
W fusion by roughly a factor of 10 as a mere consequence of the different size
of the charged- and neutral-current couplings of the electron.
However, Z fusion bears the advantage that it
also would be observable at the $\Pem\Pem$ variant of a linear collider.
Similar to $\PH\nu\bar\nu$ production, $\PH\Pep\Pem$ production
receives contributions from $s$-channel $\PH\PZ$
production diagrams and $t$-channel Z fusion graphs, where the
logarithmic rise of the latter dominates at high energies.
The EW corrections~\cite{Boudjema:2004eb} again receive only small 
contributions $\lsim5\%$ beyond ISR.
\item
{\it $t\bar t$ associated 
production $\Pep\Pem\to\Pt\bar\Pt\PH$}
\cite{Djouadi:1991tk,Gunion:1996vv,Baer:1999ge,Battaglia:2000jb,Dittmaier:1998dz,You:2003zq,Farrell:2005fk,Moretti:1999kx}
\\
A precise measurement of
the top-quark Yukawa coupling represents the main motivation
to analyze $\Pt\bar\Pt\PH$ production in the clean environment of
an $\Pep\Pem$ collider~\cite{Gunion:1996vv}. 
Assuming, for instance, $\MH=120\GeV$
the Yukawa coupling could be measured with an accuracy of $\sim5\%$
at a linear collider operating at the energy of $800\GeV$ after
collecting the integrated luminosity of 
$1\,\mathrm{ab}^{-1}$~\cite{Baer:1999ge};
in combination with other measurements even a higher accuracy is
possible~\cite{Battaglia:2000jb}.
Owing to the $\Pt\bar\Pt$ pair in the final state, the $\Pt\bar\Pt\PH$
cross section receives both QCD~\cite{Dittmaier:1998dz}
and EW corrections~\cite{You:2003zq}
at the NLO level, which are completely known. Sufficiently above the
$\Pt\bar\Pt$ threshold the generic size of QCD, photonic, and remaining
electroweak corrections
is $5{-}10\%$ on the cross section, but the effect on kinematical
distributions can be much larger.  In particular, the QCD correction
in the $\Pt\bar\Pt$ threshold region deserves particular attention,
since it is dominated by the Coulomb singularity. A careful treatment
of these long-distance effects within non-relativistic QCD~\cite{Farrell:2005fk}  
reveals corrections of $\sim70\%$ relative to the NLO cross 
section for slowly moving $\Pt\bar\Pt$ pairs. For a Higgs boson with
$\MH=120\GeV$ at a $500\GeV$ collider this pushes the NLO $K$~factor 
of $1.7$ further up to $2.4$.

On the other hand, a realistic, precise description of the process 
has to include the top-quark and Higgs-boson decays, leading to a 10-particle
final state already at LO, assuming the Higgs-boson decay into
$\Pb\bar\Pb$ pairs. Corresponding multi-particle Monte Carlo
simulations are described in \citere{Moretti:1999kx}.
\end{itemize}

A future $\Pep\Pem$ linear collider could
also allow for a determination of
the triple Higgs-boson self-coupling via Higgs-boson pair 
production~\cite{Gounaris:1979px,Boudjema:1995cb,Belanger:2003ya,Boudjema:2005rk} 
if the collider energy is high enough.
As illustrated in \reffi{fig:epemcolldiagsHH}, the promising production channels
are Higgs-strahlung
and W fusion, 
where diagrams with triple Higgs couplings
compete with other pair production diagrams.
\bfi
\centerline{
{\unitlength .8pt
\SetScale{0.8}
\begin{picture}(180,100)(-5,-10)
\ArrowLine(20, 5)(50,40)
\ArrowLine(50,40)(20,75)
\Photon(50,40)(90,40){2}{5}
\Photon(120,20)(90,40){2}{5}
\DashLine(90,40)(120,60){5}
\DashLine(150,40)(120,60){5}
\DashLine(150,80)(120,60){5}
\Vertex(50,40){2}
\Vertex(90,40){2}
\Vertex(120,60){2}
\put(155,76){${\PH}$}
\put(155,36){${\PH}$}
\put( 65,22){${\PZ}$}
\put(125,12){${\PZ}$}
\put( 5,70){${\Pep}$}
\put( 5, 2){${\Pem}$}
\put(15,-15){\footnotesize (a)}
\end{picture}
\SetScale{1}
}
{\unitlength .8pt
\SetScale{0.8}
\begin{picture}(180,100)(-5,-10)
\ArrowLine(20, 7)(50,10)
\ArrowLine(50,70)(20,73)
\ArrowLine(50,10)(110, 5)
\ArrowLine(110,75)(50,70)
\Photon(50,70)(80,40){2}{5}
\Photon(50,10)(80,40){2}{5}
\DashLine(80,40)(115,40){5}
\DashLine(145,20)(115,40){5}
\DashLine(145,60)(115,40){5}
\Vertex(80,40){2}
\Vertex(50,70){2}
\Vertex(50,10){2}
\Vertex(115,40){2}
\put(150,55){${\PH}$}
\put(150,15){${\PH}$}
\put(117,75){${\bar\nu_\Pe/\Pep}$}
\put(117, 0){${\nu_\Pe/\Pem}$}
\put(75,52){${\PW/\PZ}$}
\put(75,16){${\PW/\PZ}$}
\put( 5,70){${\Pep}$}
\put( 5, 2){${\Pem}$}
\put(15,-15){\footnotesize (b)}
\end{picture}
\SetScale{1}
}
}
\vspace*{.2em}
\caption{Lowest-order diagrams for Higgs-boson pair production
at $\Pep\Pem$ colliders:
(a)~Higgs-strahlung, (b) W/Z-boson fusion.}
\label{fig:epemcolldiagsHH}
\efi
Figure~\ref{fig:HXS_ee} includes the corresponding Higgs pair 
production cross sections for a linear collider up to energies
in the few-TeV range.
Similar to single-Higgs-boson production, the Higgs-strahlung mechanism
delivers the largest production cross section for Higgs-boson pairs
for energies directly above the $\PH\PH\PZ$ threshold, but this
cross section decreases $\propto 1/s$.
W~fusion again shows a logarithmic rise with increasing energy and
takes over the leading role for energies $\sqrt{s}\gsim1\TeV$
for $\MH=120\GeV$.
It should be noted, however, that for a low Higgs-boson mass the cross 
sections are $\lsim 0.2\,\mathrm{fb}^{-1}$ for $\sqrt{s}\lsim1\TeV$ and get even more
suppressed for larger $\MH$. 
Thus, a determination of the triple-Higgs coupling to
$\sim20\%$ for $\MH=120{-}180\GeV$ requires a high-luminosity linear collider
$(L\sim 1\,\mathrm{ab}^{-1})$ operating at TeV energies~\cite{Boudjema:1995cb}.
The EW corrections both to $\PH\PH\PZ$ production~\cite{Belanger:2003ya} 
and to W fusion to Higgs-boson pairs~\cite{Boudjema:2005rk} 
are known.
The results resemble the qualitative features of their single-Higgs
counterparts, but the size of the effects get somewhat larger.
In addition to the obligatory large ISR effects, the genuine weak corrections
are at the generic level of $5{-}10\%$. For $s$-channel $\PH\PH\PZ$ production the 
weak corrections show the tendency to grow negative with increasing energy, while
the weak corrections to $t$-channel W~fusion hardly depend on the energy sufficiently
above the $\PH\PH$ threshold.

A direct experimental investigation of the quartic Higgs self-coupling via
triple Higgs production processes seems out of reach, since the cross
sections are too much suppressed and are not very sensitive to
the $H^4$ coupling.

\section{Searches at LEP}

\subsection{The LEP1 era}

From autumn 1989 to summer 1995 the LEP accelerator was operated at CM  energies in the vicinity 
of the mass of the $\PZ$ boson, and in total 15.5 million hadronic $\PZ$-boson decays $\PZ \to \Pq \Pqbar$ and 1.7 million leptonic 
decays $\PZ \to \Plp \Plm$ were collected by the four experiments ALEPH, DELPHI, L3, and OPAL. 
At the end of LEP1 data taking a peak luminosity of 2 $\times$ 10$^{31}$ cm$^{-2}$s$^{-1}$ was reached, corresponding to approximately 
1000 $\PZ$ bosons being recorded per hour by each experiment. Even for very low Higgs-boson masses very efficient triggering of the 
final states of interest was possible and did not impose a severe problem for the analysis.

\begin{figure}
%
  \centering
  \begin{minipage}[b]{.34\textwidth}
    {\includegraphics[bb=  160 530 452 755,clip,width=0.97\textwidth]{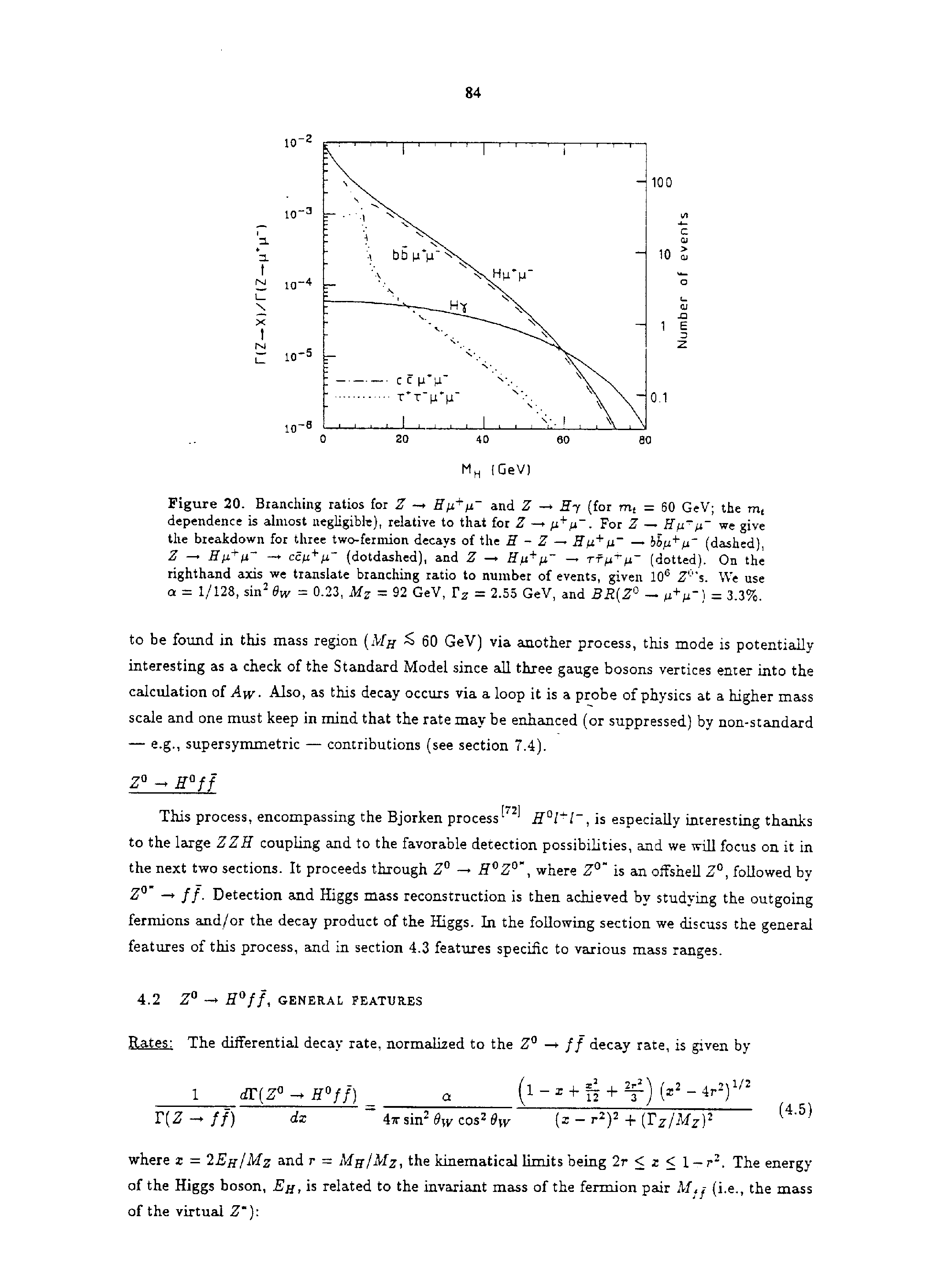}}
  \end{minipage}
  \begin{minipage}[b]{.64\textwidth}
    {{\includegraphics[bb=85 547 530 730, clip,width=0.97\textwidth]{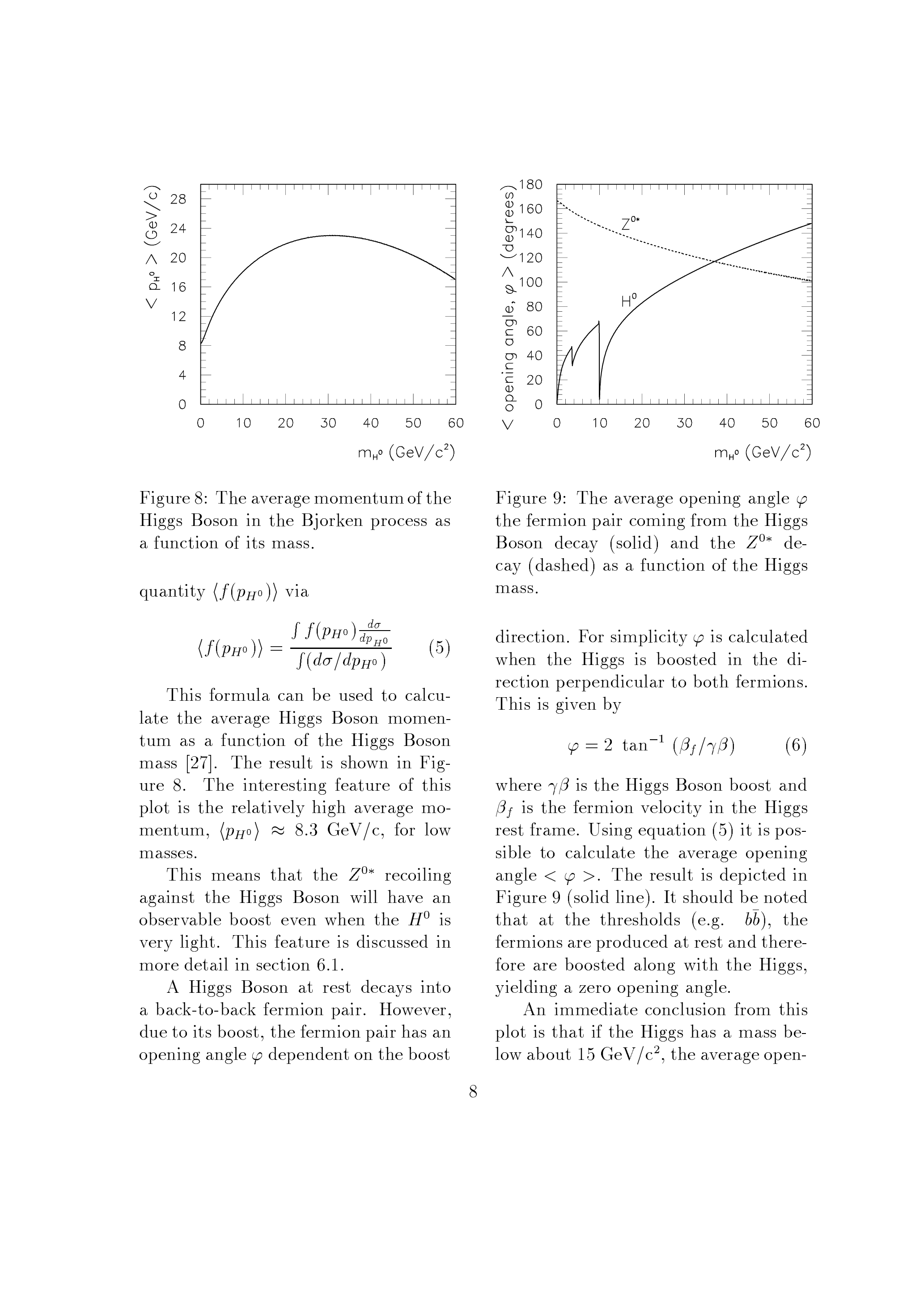}}}
  \end{minipage}
  \caption{Relative decay rates for $\PZ \to \PZ^* \PH \to \mu \mu \PH$ and $\PZ \to \gamma \PH$ 
  and number of expected events per $10^6$ hadronic $\PZ$ decays~\cite{Altarelli:1989wt} (left),
  average momentum of the Higgs boson (middle), and average opening angle of the $\PZ^*$ and $\PH$-boson
  decay products~\cite{Gross:1992vq} (right) as a function of the hypothetical Higgs-boson mass.}
  \label{fig:lep1xs}
\end{figure}

As detailed in \refse{se:lepprod}, the dominant production mode for the Higgs boson is the rare 
decay $\PZ \to \PZ^* \PH$. The total expected event rate for the decay $\PZ \to \PZ^* \PH\to \mu \mu \PH$
for one million of hadronic Z decays is approximately $75\,(12,0.4)$ for a Higgs-boson mass of $10\,(30,60)\GeV$
(see \reffi{fig:lep1xs}, left). For a Higgs-boson mass of roughly $60\GeV$ the event rates in $\PZ \to \PZ^* \PH\to \mu \mu \PH$ and 
$\PZ \to \gamma \PH$ are of comparable size (see \reffi{fig:lep1xs}, left).  However, considering also the decay of the $\PZ^*$  
into electrons and neutrinos increases the rate in accessible final states arising from $\PZ \to \PZ^* \PH$ by a factor of eight. 
In addition to the smaller rate for $\PZ \to \gamma \PH$ the background from $\Pf  \Pfbar \gamma$ was also 
more severe for this final state.
Hence no dedicated search for the decay $\PZ \to \gamma \PH$ was performed, as the contribution to 
the sensitivity is negligible given the integrated luminosity collected at LEP1 of up to approximately 160$\ipba$.
At LEP1 the Higgs-boson mass range from 0 to  approximately $65\GeV$ could be probed.%
\footnote{In parallel to the start of LEP a search for light Higgs bosons, 
performed in an electron beam-dump experiment, was published, which unambiguously excludes the mass range $1.2{-}52\MeV$ \cite{Davier:1989wz}.}
The a priori signal-to-background ratio corresponds to 10$^{-2}$ for low masses and 10$^{-6}$ at the edge of the mass sensitivity (see also \reffi{fig:sigbgxs}). 
The kinematics in Higgs-boson production and decay is characterized by the average values of the
Higgs-boson momentum and the opening angle of the $\PH$- and $\PZ$-boson decay products, which are shown in \reffi{fig:lep1xs} (middle and right)
as function of the Higgs-boson mass.

Three Higgs-boson mass ranges can be roughly distinguished: (a) $0\le\MH\le2\Mmy$,
(b) $2\Mmy\le\MH\le3\GeV$, and (c) $\MH\ge3\GeV$. These mass ranges 
differ in the dominant Higgs-boson branching ratios and the robustness of their theoretical predictions,
the kinematics and topology, the composition of the dominant backgrounds, and the search strategies applied.

\begin{itemize}
\item For $2\Mmy\le\MH\le3\GeV$ a perturbative description of the Higgs-boson decay, the prediction 
of the branching ratios, and hence the estimation of the detection efficiencies have large uncertainties. 
The Higgs boson decays dominantly into several hadrons (e.g.\ a pair of pions, kaons, etc.)  and for 
$2\Mmy\le\MH\le 2\Mpi$ 
into a pair of muons. Different approaches were used by the four experiments to cover this mass range.
ALEPH \cite{Decamp:1989nt} and DELPHI \cite{Abreu:1990zc} extended their standard searches in the lepton $\PZ^*\to\Plp\Plm$ and missing energy $\PZ^*\to\Pn\Pnbar$ 
final states (see below) to the mass range below $3\GeV$, allowing only two tracks being assigned to the Higgs-boson candidate. They 
relied on some modelling of Higgs-boson decays in this non-perturbative regime. In addition, 
for mass values below $2\GeV$ 
the decay $\PZ^*\to\Pq\Pqbar$ was considered by asking for three- or four-jet events and assigning the jet with the lowest track multiplicity 
to the Higgs-boson candidate, which is only allowed to have two tracks. L3 \cite{Adeva:1990vp} only utilized the decay $\PZ^*\to\Plp\Plm$ 
with two energetic acoplanar leptons  and considered Higgs-boson decays to a least one track, covering the decay into muons and hadrons. 
OPAL \cite{Acton:1991pd} performed a decay-mode-independent search by two complementary analyses: (a) acoplanar leptons with vetoing on $\PH\to$ ``electromagnetic''
$\Pep\Pem/\gamma\gamma/\pi^0\pi^0$ decays, which is also sensitive to a long-lived Higgs boson decaying outside the detector, and (b) a 
dedicated search  $\PH \to$ ``electromagnetic'' decays accompanied by missing energy from $\PZ^*\to \Pn\Pnbar$. The OPAL approach does not rely 
on the prediction of branching ratios and details of modelling the hadronic Higgs-boson decay in the non-perturbative regime, and hence 
is almost model independent.

\item  For $\MH\le 2\Mmy$ the dominant decay is to a pair of electrons, and for $\MH\le 2\Me$ only the loop-induced decay to a pair 
of photons is possible.  In this mass range the Higgs boson gets long-lived (average decay length of 1 (100) meter(s) for 
$\MH=100\,(10)\MeV$),
and a significant fraction of Higgs bosons decay outside the detector, leaving an invisible Higgs-boson decay as signature. 
ALEPH \cite{Decamp:1989nt,Decamp:1990ch} and DELPHI \cite{Abreu:1990bq} extended the above searches in the $\PZ^*\to\Pq\Pqbar$ and $\PZ^*\to \Pn\Pnbar$ final states and performed a dedicated 
analysis, which looked for a $V^0$ signature  (i.e.\ a displaced vertex with two tracks) accompanied by any decay mode of the $\PZ^*$ boson.  The di-track invariant 
mass of the $V^0$ candidate should not be consistent with that of a long-lived hadron ($K^0_s$, $\Lambda$, etc.), and the tracks should be not identified as hadrons. 
L3 \cite{Adeva:1990jw} used the acoplanar lepton search from the 
$2\Mmy\le\MH\le3\GeV$ range allowing also for $\PH \to \Pep\Pem$ decays.
OPAL \cite{Akrawy:1990bj} looked for $\PZ^*\to \Pn\Pnbar$ and $\PH\to\Pep\Pem/\gamma\gamma$ decays with at least one energy deposit in the electromagnetic 
calorimeter, to which tracks can be assigned, and not much additional detector activity.  All experiments complemented this 
analysis  by a search for invisible Higgs-boson decays in the acoplanar lepton topology with no other significant activity in the detector. 
A combination of the ``invisible Higgs-boson'' search and the other ones discussed above exclude all mass values down to 0. 

\item  For $\MH\ge3\GeV$ the perturbative description of the Higgs-boson decay is more reliable
(apart from the mass values corresponding to $\Pb\Pbbar$ and $\Pc\Pcbar$ bound states). For low Higgs-boson 
masses ($\MH\le 10{-}15\GeV$) the Higgs-boson decay products build a mono-jet topology due to the large Lorentz boost. 
For larger Higgs-boson masses the topology tends to two separated jets, which have an opening angle less than 180 degree
in space (``acollinearity'') and also in the projection into the plane perpendicular to the beam axes (``acoplanarity'').
The final state with largest sensitivity is the decay $\PZ \to \PZ^* \PH \to \nu \bar{\nu} \PH$, where for $\MH$ larger 
than $15\GeV$ the decay into a pair of $\Pb$-quarks is dominant. 
A production rate that is smaller by a factor three is provided by the decays
$\PZ \to \PZ^* \PH \to \Pmup\Pmum\PH/\Pep\Pem\PH$. This ratio of three is roughly retained
at the end of the event selections. Together the two final states cover approximately 25\% of the Higgs-boson production rate.
Final states involving tau leptons, 
$\PZ^* \PH \to \tau^+\tau^- \Pq \Pqbar$, $\PZ^* \PH \to \Pq \Pqbar \tau^+\tau^-$,
$\PZ^* \PH \to \ell^+ \ell^- \tau^+\tau^-$, and $\PZ^* \PH \to \nu \bar{\nu} \tau^+\tau^-$, 
have either a small production rate 
or a quite low selection efficiency. The expected signal event yields after the full selection are approximately one
order of magnitude smaller than that expected in the $\PZ^* \PH \to \nu \bar{\nu} \PH$ search. Hence final states with tau leptons
only contribute marginally to the combined sensitivity. All collaborations investigated final states with tau leptons 
\cite{Decamp:1990ch,Abreu:1991dg,Adeva:1992gt,Akrawy:1990bt} at intermediate stages of the Higgs-boson search at LEP1, 
but not included them in the analysis providing the final mass limit.
Fully hadronic final states from $\PZ \to \PZ^* \PH$ are not considered due to the overwhelming background.

The $\PZ \to \PZ^* \PH \to \nu \bar{\nu} \PH$ topology is characterized by missing energy, missing (transverse) 
momentum, and the detector activity from the Higgs-boson decay. With increasing Higgs-boson mass the particle
multiplicity from the Higgs-boson decay is increased. Due to the smaller Lorentz boost the topology changes 
from a mono-jet to an acollinear and acoplanar di-jet final state. At the same time the missing energy and missing 
mass decrease, as the $\PZ^*$ boson has to be more off shell. Almost no irreducible backgrounds exist, and the 
background contributions are mostly of instrumental nature due to mismeasurements, wrong identification of final-state 
particles, and incomplete detector coverage.
The dominant background for large Higgs-boson masses is due to $\PZ \to \Pq \Pqbar$ decays, where missing energy arises
from neutrinos produced in heavy-flavour decays and mismeasurements, and from four-fermion production via 
$\PZ^* \to \PZ^* \gamma^* \to \nu \bar{\nu} \Pq \Pqbar$ and  $\Pep\Pem \to \Pep\Pem \gamma^* \gamma^* \to \Pep\Pem \Pq \Pqbar$. 
The four-fermion processes are characterized by a low mass of the hadronic system and hence
are mostly relevant for small Higgs-boson masses, and
events of the $\Pep\Pem \Pq \Pqbar$ channel typically show
small (missing) transverse momentum.
The mass resolution for the hadronic system is of the order of $10\%$. 
DELPHI also applied b-tagging for the final analysis exploiting the large branching ratio $\PH \to \Pb\Pbbar$.

The $\PZ \to \PZ^* \PH \to \ell^+ \ell^- \PH$ topology is characterized by two isolated leptons, whose acollinearity 
and acoplanarity increase with increasing Higgs-boson mass, and a recoiling hadronic system. The mass of the Higgs-boson 
candidate can be reconstructed from the four-momenta of the leptons using energy--momentum conservation (``recoil mass technique'')
with a precision of $1{-}2\%$. 
The dominant background from four-fermion production has either a small mass of the leptonic or of the hadronic 
system. The $\PZ \to \Pq \Pqbar$ background with leptonic heavy-flavour decays possesses a large mass of the hadronic 
system, and the leptons do not originate from the primary collision vertex. For the final analysis performed at LEP1 most 
experiments exploited b-tagging to further discriminate the signal especially from four-fermion background processes.

The event selections
were mostly designed in such a way that the expected background was at the level of one event or less 
at the edge of the Higgs-boson mass sensitivity for a given data set. With increased integrated luminosities 
this required more sophisticated and refined analyses and tighter identification and isolation requirements
to exploit the above-mentioned differences between signal and background topologies and their 
dependence on the Higgs-boson mass. Hence the selection efficiency for signal events at the sensitivity edge dropped
from roughly 60\% or higher in the first searches to a level of 30\% in the final LEP1 searches. Only a few events 
survived all selection requirements in agreement with expectations from background processes.
\end{itemize}

\begin{figure}
  \centering
  \begin{minipage}[b]{.47\textwidth}
    \includegraphics[width=7.5cm,height=7.5cm]{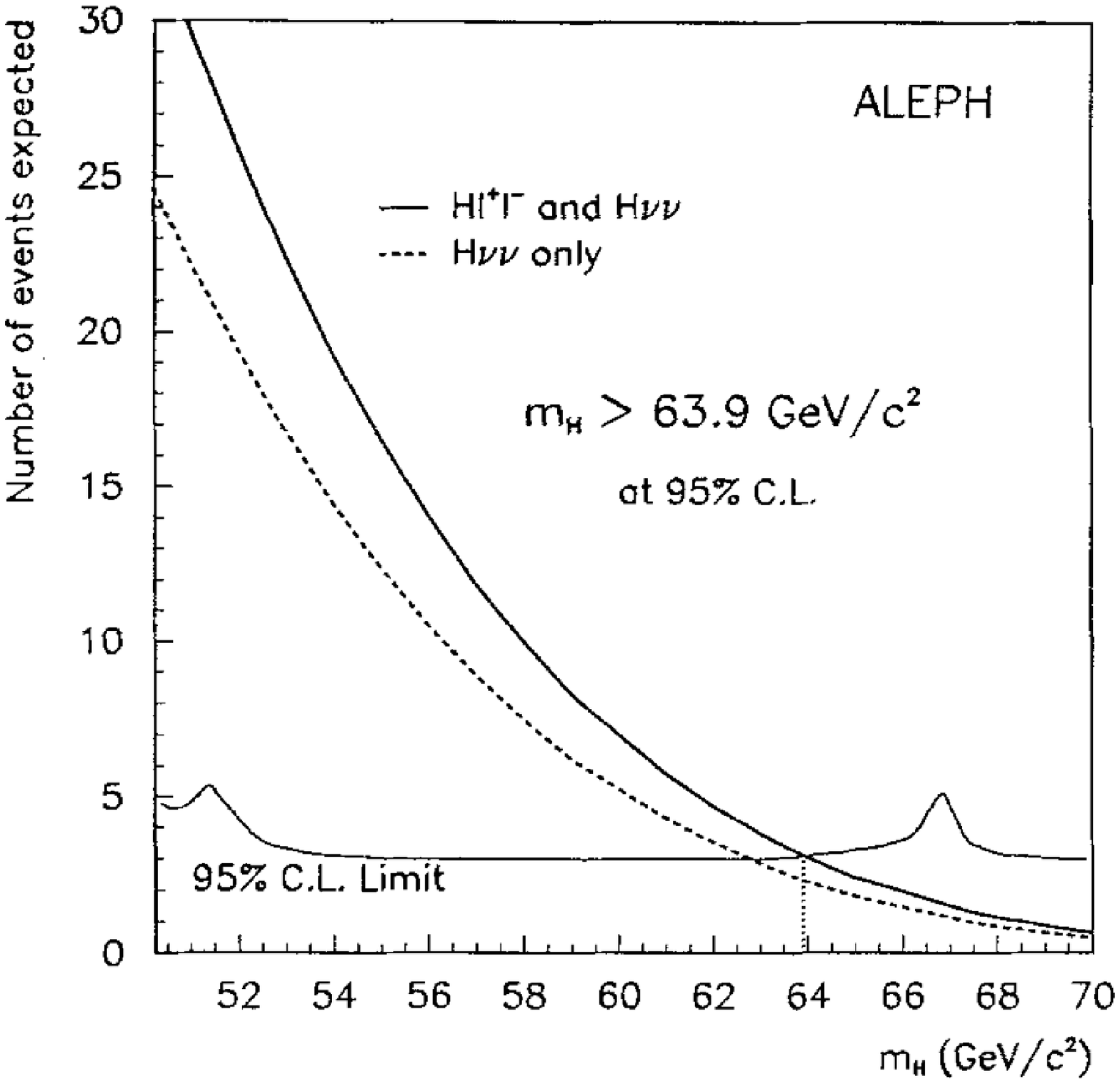}
  \end{minipage}
  \begin{minipage}[b]{.47\textwidth}
    \includegraphics[width=7.5cm,height=7.5cm]{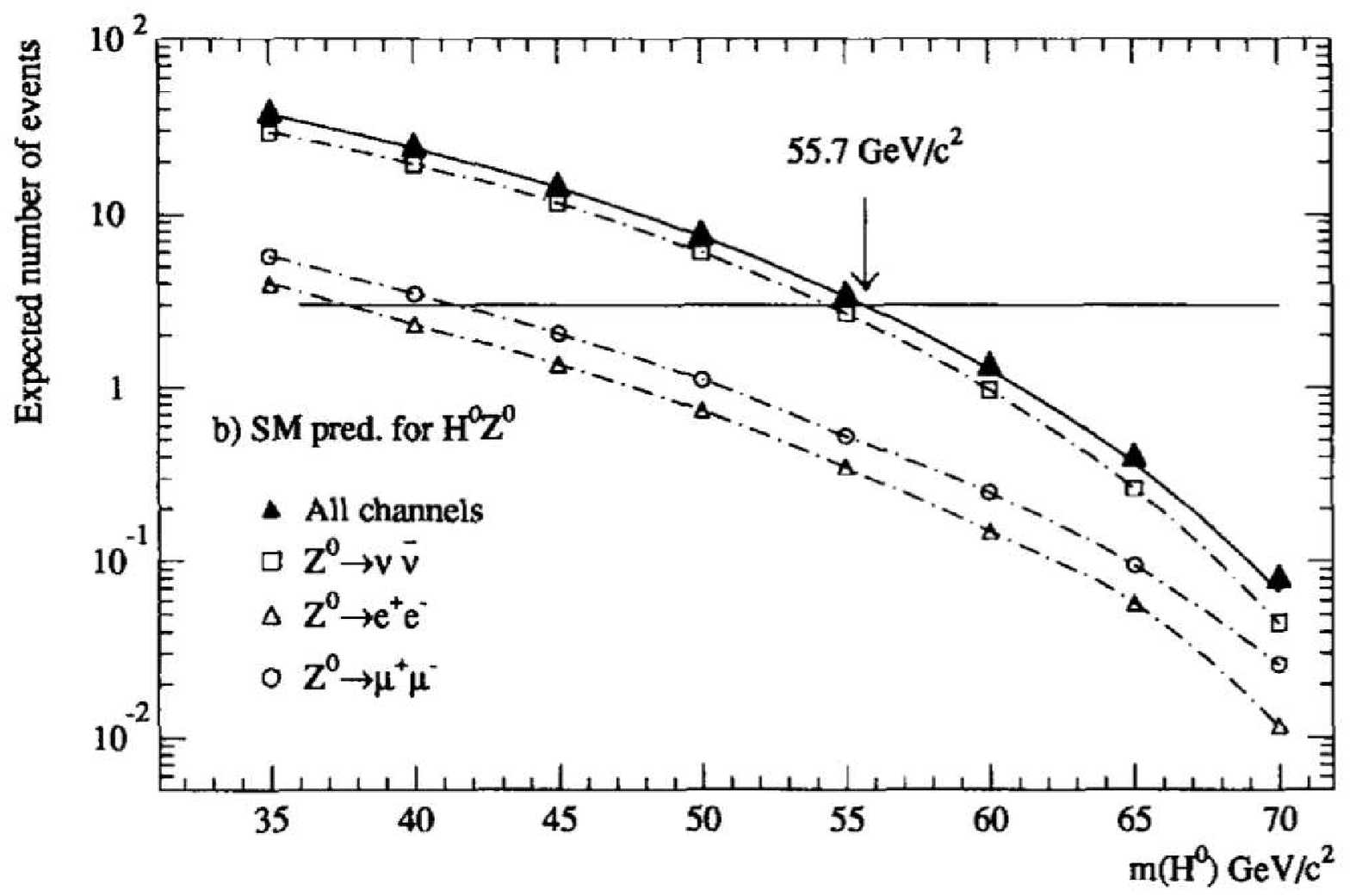}
  \end{minipage}
  \begin{minipage}[b]{.47\textwidth}
    \includegraphics[width=7.5cm,height=7.5cm]{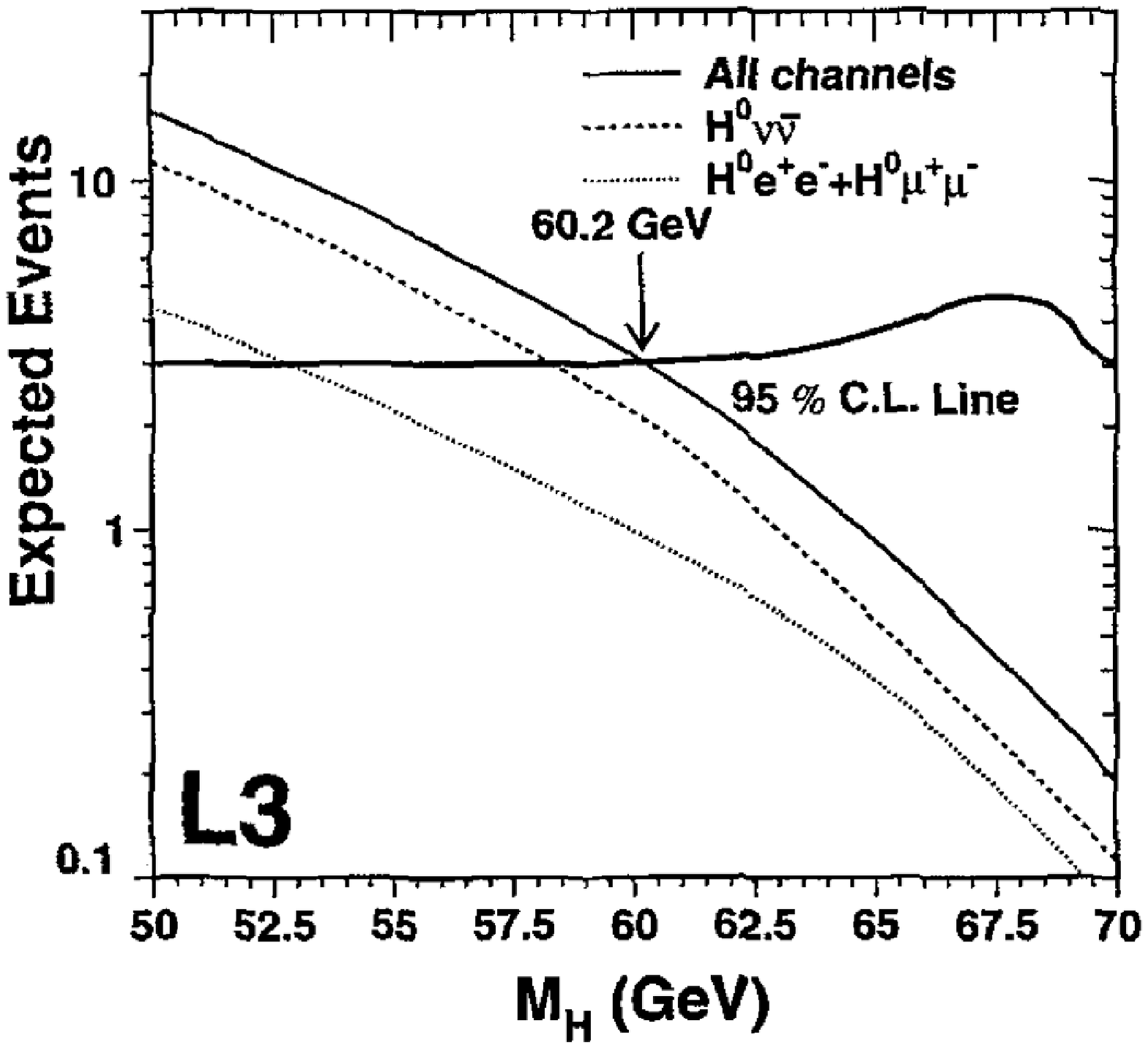}
  \end{minipage}
  \begin{minipage}[b]{.47\textwidth}
    \includegraphics[width=7.5cm,height=7.5cm]{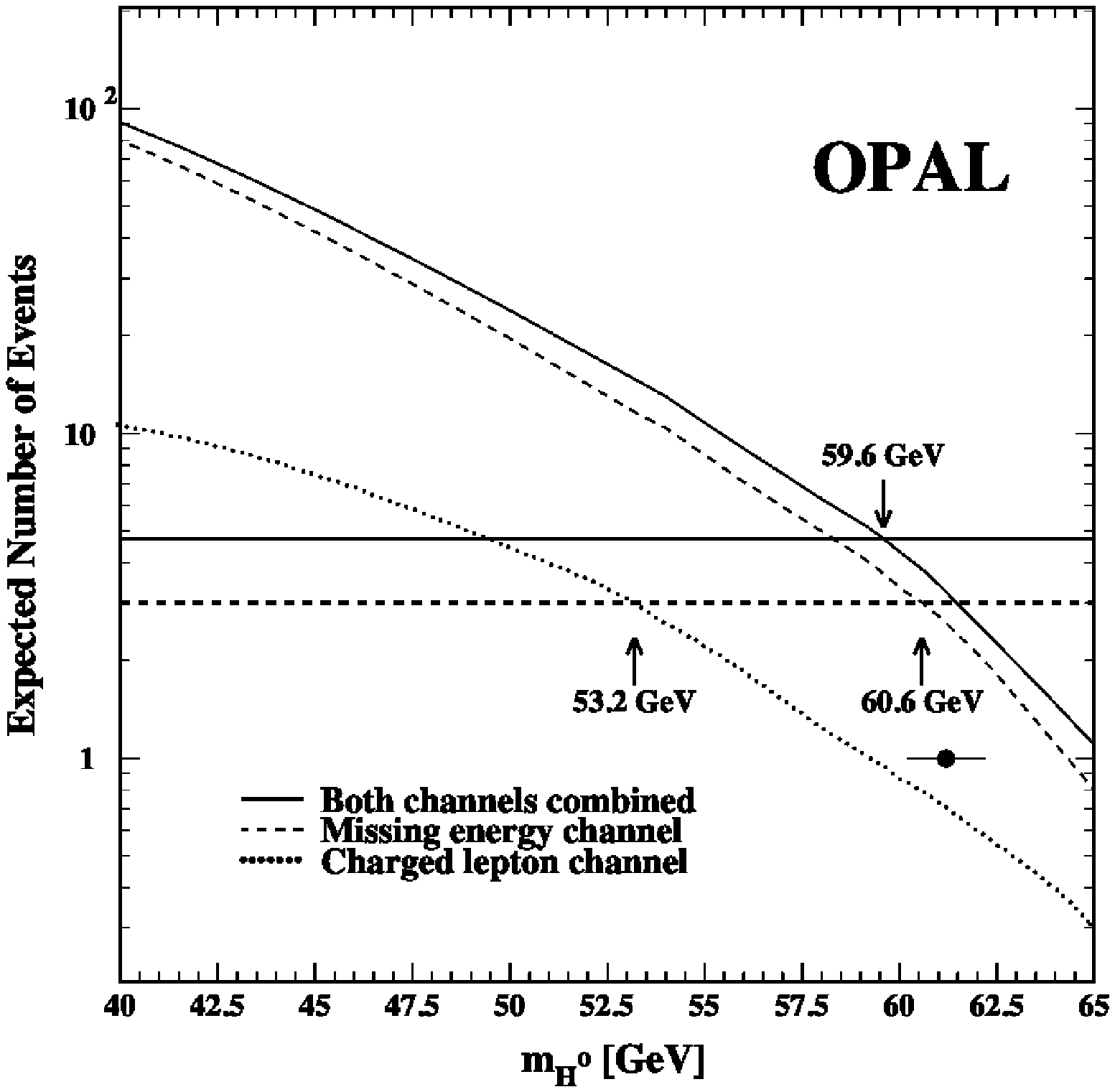}
  \end{minipage}
  \caption{Excluded  and expected event yields for Higgs-boson production at LEP1 
   by ALEPH (top-left) \cite{Buskulic:1996hz}, DELPHI (top-right) \cite{Abreu:1994rp},
   L3 (bottom-left) \cite{Acciarri:1996um}, and OPAL (bottom-right) \cite{Alexander:1996ai}.}
  \label{fig:lep1results}
\end{figure}

All observed events are considered as Higgs-boson candidates and the expected signal yields are reduced by their 
systematic uncertainty. In summary, the searches performed by the four experiments allowed to exclude all Higgs-boson mass hypotheses 
below $63.9\GeV$ (ALEPH \cite{Buskulic:1996hz}), $55.7\GeV$ (DELPHI \cite{Abreu:1994rp}), $60.2\GeV$ (L3 \cite{Acciarri:1996um}),
and $59.6\GeV$  (OPAL \cite{Alexander:1996ai}), as shown in \reffi{fig:lep1results} and in \refta{tab:lep1results}.  
\begin{table}
\caption{Excluded Higgs-boson mass ranges  from LEP1 searches: experiment, excluded mass range, number (in thousands) of recorded hadronic Z decays,
integrated luminosity, years of data taking and references. The entry ``--'' indicates that the information is not available.}
\label{tab:lep1results}
\begin{center}
\begin{tabular}{c|c|c|c|c|c}
\hline
Experiment & excluded $\MH$ range  & $10^3$ $\PZ\to\Pq\Pqbar$ & int.~lumi.$\,[\ipba]$ &  data from year & Ref.\\ \hline 
ALEPH      & $0.032{-}15\GeV$      & $   12$ & $0.5 $  & $89        $   & \cite{Decamp:1989nt} \\ 
           & $11{-}24   \GeV$      & $   25$ & $1.2 $  & $89        $   & \cite{Decamp:1990mt} \\ 
           & $ 0{-}57   \MeV$      & $   23$ & $1.2 $  & $89        $   & \cite{Decamp:1990zp} \\ 
           & $< 41.6 \GeV$         & $  100$ & $-         $ & $89{-}90   $   & \cite{Decamp:1990ch} \\ 
           & $< 48   \GeV$         & $  185$ & $-         $ & $89{-}90   $   & \cite{Decamp:1991uy}  \\ 
           & $< 58.4 \GeV$         & $ 1233$ & $-         $ & $89{-}92   $   & \cite{Buskulic:1993gf} \\ 
           & $< 63.9 \GeV$         & $ 4500$ & $-         $ & $89{-}95   $   & \cite{Buskulic:1996hz} \\ \hline 
DELPHI     & $2\Mmy{-}14 \GeV$     & $   11$ & $0.5 $  & $89        $   & \cite{Abreu:1990zc}  \\ 
           & $0{-}2 \Mmy$          & $  119$ & $-         $ & $90        $   & \cite{Abreu:1990bq} \\ 
           & $< 38 \GeV$           & $  119$ & $-         $ & $90        $   & \cite{Abreu:1991dg} \\  
           & $< 55.7 \GeV$         & $ 1000$ & $34.6 $ & $91{-}92   $   & \cite{Abreu:1994rp}   \\ \hline 
L3         & $2{-}32 \GeV$         & $   50$ & $-         $ & $90        $   & \cite{Adeva:1990vp}\\ 
           & $0{-} 2 \GeV$         & $   70$ & $-         $ & $90        $   & \cite{Adeva:1990jw}\\ 
           & $< 41.8 \GeV$         & $  111$ & $5.3  $ & $90        $   & \cite{Adeva:1991vx}\\ 
           & $< 52 \GeV$           & $  408$ & $17.5 $ & $90{-}91   $   & \cite{Adeva:1992gt}   \\ 
           & $< 57.7 \GeV$         & $ 1062$ & $39.0 $ & $90{-}92   $   & \cite{Adriani:1993gp} \\ 
           & $< 60.2 \GeV$         & $ 3000$ & $114  $ & $91{-}94   $   & \cite{Acciarri:1996um}\\ \hline 
OPAL       & $3{-} 19.3 \GeV$      & $ -   $ & $0.8 $  & $89        $   & \cite{Akrawy:1989jw}\\ 
           & $0{-}2 \Mmy$          & $ 25  $ & $1.2 $  & $89        $   & \cite{Akrawy:1990bj}\\ 
           & $3{-}25.3\GeV$        & $ 25  $ & $1.2 $  & $89        $   & \cite{Akrawy:1990gm}\\ 
           & $3{-}44 \GeV$         & $ 170 $ & $8   $  & $89{-}90   $   & \cite{Akrawy:1990bt}\\ 
           & $0{-}11.3\GeV$        & $ -   $ & $6.8 $  & $90        $   & \cite{Acton:1991pd} \\ 
           & $< 56.9 \GeV$         & $1900 $ & $78 $   & $90{-}93   $   & \cite{Akers:1994wi}   \\ 
           & $< 59.6 \GeV$         & $5000 $ & $160$   & $89{-}95   $   & \cite{Alexander:1996ai} \\ \hline 
\end{tabular}
\begin{minipage}[t]{16.5 cm}
\vskip 0.5cm
\noindent
\end{minipage}
\end{center}
\end{table}     
The relative contribution to the sensitivity of the missing energy and leptonic final states can be deduced 
from \reffi{fig:lep1results}. In ALEPH and L3 the mass resolution of candidates close to the edge of the excluded mass range is taken into account 
following the technique described in \citere{Grivaz:1992ng}, whereas OPAL did not apply such a technique. 
No official combination of the findings of the four LEP experiments was performed. Owing to the exponential decrease of 
the Higgs-strahlung cross section with increasing Higgs-boson mass, the excluded mass range only extended slowly with 
collected integrated luminosity (see \refta{tab:lep1results}). Hence an extension of data taking at LEP1 would not have 
significantly increased the mass reach of the searches, which for the first time excluded a significant fraction 
of the allowed Higgs-boson mass range in an unambiguous way.

\subsection{The LEP2 era}

\begin{figure}
  \centering
  \begin{minipage}[b]{.4\textwidth}
    \includegraphics[width=0.97\textwidth]{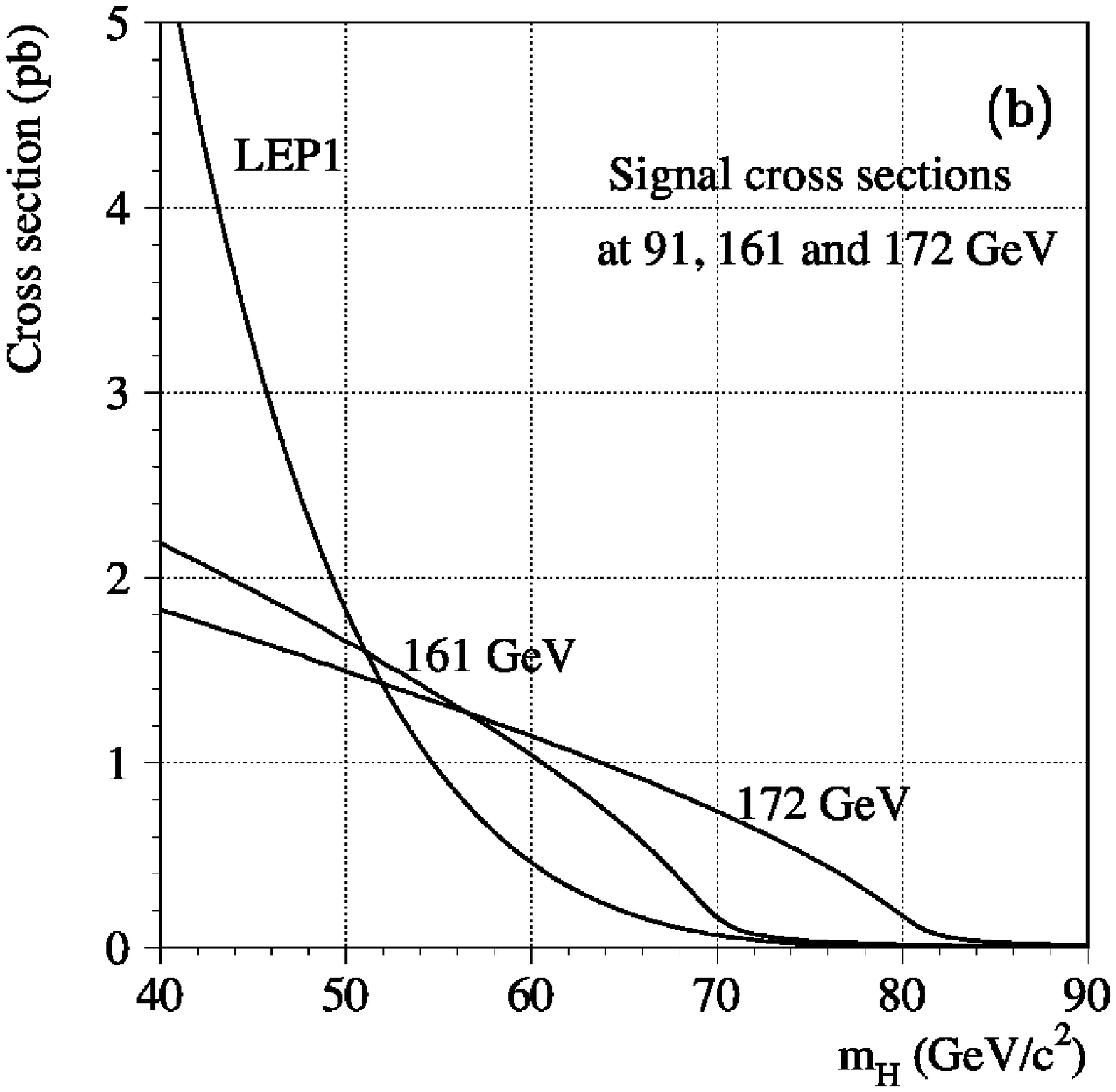}
  \end{minipage}
  \begin{minipage}[b]{.4\textwidth}
    \includegraphics[width=0.97\textwidth]{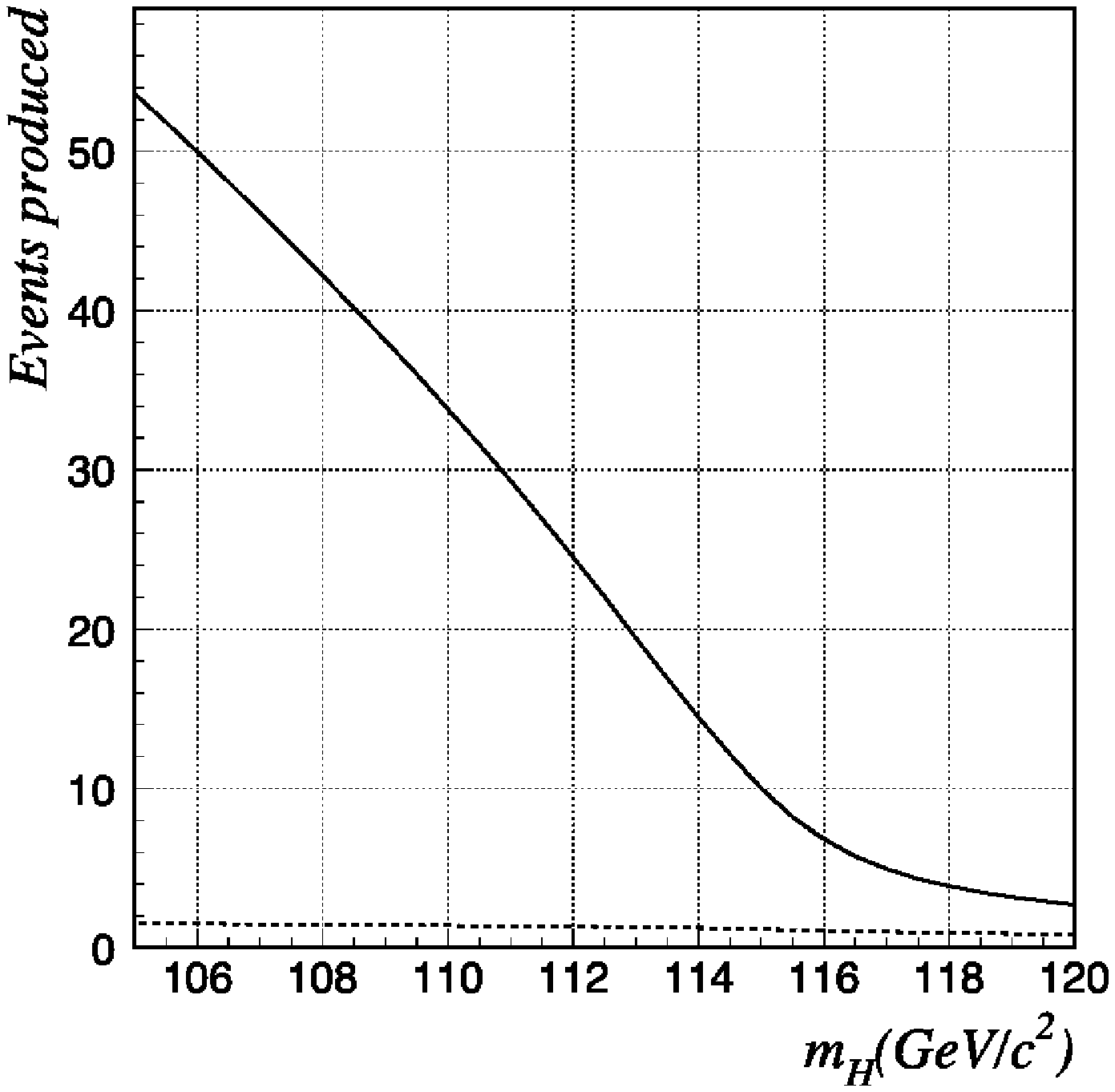}
  \end{minipage}
  \caption{Comparison of Higgs-strahlung production cross sections at LEP1 and at the beginning of LEP2 (left from \citere{Barate:1997mb}).
           Number of expected Higgs-boson events per experiment produced in 2000 data taking from Higgs-strahlung (full line) 
           and from vector-boson fusion and interference (dashed line) (right from \citere{Barate:2000ts}).} 
  \label{fig:lep2xs}
\end{figure}

Starting in summer 1995 at $130/136\GeV$, the CM energy of the LEP collider was continuously increased over the next years up to $209\GeV$ in 2000.
The data taken at $130/136\GeV$ were not used for SM Higgs-boson searches. A summary of the data sets recorded at different CM energies 
used for SM Higgs-boson searches during the LEP2 programme is given in \refta{tab:lep2data}. Approximately 0.5 (2.5)\,fb$^{-1}$ at CM energies 
in excess of $206\,(189)\GeV$ were collected in total by the four experiments ALEPH, DELPHI, 
L3, and OPAL. 
\begin{table}
\caption{CM energies and integrated luminosities collected in each experiment at LEP2. Numbers in brackets give 
         the luminosity-weighted CM energy of the corresponding year. The luminosity ranges are caused by the different
         data-taking efficiencies in the four experiments.}
\label{tab:lep2data}
\begin{center}
\begin{tabular}{c|c|c|c|c|c}
\hline
year                         &  $1996        $  & $1997 $ & $1998   $ & $1999         $ & $2000   $  \\ \hline
CM energy       [GeV]     &  $161/170/172 $  & $183  $ & $189    $ & $192{-}202 (198)$ & $202{-}209  (206)$  \\
int.~luminosity [$\ipba$] &  $10{-}11/1/9{-}10$  & $54{-}57$ & $158{-}176$ & $228{-}237      $ & $217{-}224$     \\
\hline 
\end{tabular}
\begin{minipage}[t]{16.5 cm}
\vskip 0.5cm
\noindent
\end{minipage}
\end{center}
\end{table}     
At LEP2 CM energies the dominant production process is again Higgs-strahlung. 
But in contrast to LEP1 the Z boson in the final state is dominantly produced on its mass shell. 
The difference in the $\MH$ dependence of $\Pep\Pem \to \PZ \to \PZ^* \PH$ at LEP1 and 
$\Pep\Pem \to \PZ^* \to \PZ \PH$ at LEP2 is shown in \reffi{fig:lep2xs} (left).
Additional contributions from the vector-boson-fusion processes
$\Pep\Pem \to \Pep\Pem \PZ^* \PZ^* \to \Pep\Pem  \PH$ and 
$\Pep\Pem \to \Pn\Pnbar\PW^* \PW^* \to \Pn\Pnbar \PH$ and their interference 
with the respective final states from Higgs-strahlung are at the level of maximally 
$10{-}20\%$ (see \reffi{fig:HXS_ee}) at the sensitivity edge, which is approximately given by $\sqrt{s}-\MZ-\mbox{few GeV}$.
The contribution to the 
total event yield (per experiment)
for the production of a hypothetical Higgs boson in 2000 data is shown 
in \reffi{fig:lep2xs} (right). No dedicated searches for production in vector-boson fusion were performed. 
Signal cross sections were calculated and signal event were generated with {\sc HZHA}~\cite{Altarelli:1996ww}.

In contrast to LEP1, now the hadronic decays of the Z boson can be utilized in the search, as the a priori signal-to-background 
ratio is larger by two to three orders of magnitude compared to LEP1 (see l.h.s.\ of \reffi{fig:lep1xs}) at the sensitivity edge. 
Furthermore 
the mass constraint of two jets to stem from the decay $\PZ\to\Pq\Pqbar$ helps to discriminate the signal process 
from background processes. This new channel delivers the largest contribution to the sensitvity at LEP2.
At LEP1 the background was mostly reducible, at least in principle, and caused by mismeasurements and hence
difficult to model in all details. 
After crossing the thresholds for $\PW\PW$ and $\PZ\PZ$ production 
at the LEP2 
CM energies of $161\GeV$ and $183\GeV$, respectively, the dominant backgrounds are due to di-boson production 
and to a large extent irreducible. These could reliably be predicted via simulation.

Searches were performed in the ``four-jet channel'' $\Zhqqbb$ ($\Zhqqqq$), ``missing-energy channel'' $\Zhnnbb$ ($\Zhnnqq$), 
``tau channels'' $\Zhqqtt$ and $\Zhttbb$ ($\Zhttqq$), and ``lepton channels'' $\Zheebb$  and $\Zhmmbb$ ($\Zheeqq$  and $\Zhmmqq$). 
Those final states together cover approximately 80\% of the total production rate (see \refta{tab:lep2rates}) for a Higgs boson with a mass 
of $115\GeV$, when only taking into account the decay $\PH\to\Pb\Pbbar$.  Given the long time intervals between bunch crossings of 22 $\mu\mathrm{s}$, 
the instantaneous luminosity of $10^{32}$ cm$^{-2}$s$^{-1}$, and the size of the total cross section, triggering on all signal topologies 
was not a problem. 
\begin{table}
\caption{Fraction of production rates in the different search channels at LEP2. 
The number outside (inside) brackets correspond to $\PH\to\Pq\Pqbar$ ($\PH\to\Pb\Pbbar$) decays.
}
\label{tab:lep2rates}
\begin{center}
\begin{tabular}{c|c|c }
\hline
Channel         & $\MH = 60 \GeV$ & $\MH = 115 \GeV$ \\ \hline
four jets       &  $64 (60) \%$    &  $ 57 (50) \% $ \\
missing energy  &  $18 (17) \%$    &  $ 16 (14) \% $ \\
taus            &  $ 9  (9) \%$    &  $ 6 (5)   \% $ \\
leptons         &  $ 6  (6) \%$    &  $ 8 (8)   \% $ \\ \hline
sum             &  $97  (91)\%$    &  $ 87 (77) \% $ \\ \hline
\end{tabular}
\begin{minipage}[t]{16.5 cm}
\vskip 0.5cm
\noindent
\end{minipage}
\end{center}
\end{table}     
Due to the presence of irreducible backgrounds from pair production of weak gauge bosons, in particular for CM energies
of $183\GeV$ and larger, it was not possible to design analyses that yield background expectations at the level of few events
as at LEP1, while retaining a large fraction of the signal process. Extensive use was made of largely improved capabilities
to identify b-flavoured jets in order to suppress backgrounds, in particular from $\PW\PW$ production. 
The reconstruction of the mass of Higgs-boson candidates was greatly improved in many channels by kinematic fits which required
energy and momentum conservation and often the consistency of a di-fermion mass with $\MZ$. Many analyses exploited the advantages 
of multivariate techniques, which were trained for various different Higgs-boson mass hypotheses. In several cases the 
mass-sensitive observables together with the response of the multivariate technique were combined in a final 
discriminant, which was 
used in the hypothesis tests.  

Below the main characteristics of the four search channels, which are ordered by their sensitivity, and the most important 
means to discriminate them from the dominant backgrounds are discussed: 
\begin{itemize}
\item ``Four-jet channel'': The topology is characterized by four jets, two consistent with the decay $\PZ\to\Pq\Pqbar$ and 
two consistent with the decay $\PH\to\Pb\Pbbar$. Requiring two jets to be identified as b-flavoured suppresses strongly backgrounds
from $\PW\PW$ production and partially from $\PZ\PZ$ production, as the branching ratios for $\PZ\to\Pb\Pbbar$ of 15\% is much smaller than 
the corresponding one for the Higgs boson. The sensitivity is enhanced by a precise reconstruction of the 
di-jet invariant mass of the
jets assigned to the Higgs-boson decay. Exploiting energy and momentum conservation and constraining the other di-jet mass to $\MZ$,
significantly improves the mass resolution. Contributions from $\Pep\Pem \to \PZ/\gamma^*\to4\,$jets are 
small for CM energies beyond $183\GeV$.
\item 
``Missing-energy channel'': The topology is characterized by large missing energy, missing (transverse) momentum,
and missing mass (derived from energy and momentum conservation)
due to the decay $\PZ\to\Pn\Pnbar$, two acoplanar b-flavoured jets from the decay of the Higgs boson, and no leptons being present. 
Requiring b-tagged jets greatly reduces backgrounds from $\PW\PW\to\Pq\Pqbar\Pl\Pn$, where the charged lepton escapes detection, and 
partially helps to reduce the contribution from $\PZ\PZ\to\Pq\Pqbar\Pn\Pnbar$. The reconstruction of the mass of the Higgs boson candidate 
from the di-jet system is often significantly improved by exploiting energy and momentum conservation and constraining the missing mass to $\MZ$. 
The remaining background is dominated by di-boson production.
\item ``Lepton channels'': The topology is characterized by two acoplanar leptons consistent with the decay $\PZ\to\Plp\Plm$ and two acoplanar 
b-flavoured jets from the decay of the Higgs boson. Requiring b-tagged jets reduces background from $\PZ\PZ\to\Pq\Pqbar\Plp\Plm$. 
The mass of the Higgs-boson candidate is reconstructed as the recoil to  the di-lepton system. In some cases kinematic fits exploiting
energy and momentum resolution improve the sensitivity. For CM energies larger than $183\GeV$ the final background is, 
to a very large extent, due to $\PZ\PZ$ production.
\item ``Tau channels'': The topology is characterized by two hadronic jets and two tau-lepton candidates, where one
pair is assigned to the $\PZ$ decay and the other
to the $\PH$ decay. The assignment to the $\Zhqqtt$ or $\Zhttbb$ channels 
is based on the invariant masses of the tau and jet systems, and the quality of kinematic fits requiring energy and momentum 
conservation and applying the $\MZ$ constraint. B-tagging improves the sensitivity in the $\Zhttbb$ final state.  
Selected background events are largely due to di-boson production.
\end{itemize}

After all selection requirement a typical signal-to-background ratio of the order between  
2-to-1 and 1-to-1 could be achieved 
at the edge  of the LEP sensitivity for masses of $115\GeV$  still retaining $30{-}60\%$ of the eventually produced signal events depending 
on the final state. No significant
deviations from the background-only hypothesis were observed in the data collected 
up to 1999 (see first four rows in \refta{tab:lep2indfinal}). 
\begin{table}
\caption{Excluded mass ranges in the individual experiments at various stages during the LEP2 programme:
data set used, observed (expected) mass limit, and reference. The entry ``--'' indicates that this information 
is not published.}
\label{tab:lep2indfinal}
\begin{center}
\begin{tabular}{c|c|c|c|c }
\hline
CM energy          & ALEPH                             & DELPHI                              & L3                                  & OPAL                           \\ \hline 
$\le 161 \GeV$         &    --                             & --                                  & --                                  & 65.0 (--) \cite{Ackerstaff:1996db}     \\ 
$\le 172 \GeV$         &  70.7  (--)  \cite{Barate:1997mb} & 66.2  (--)    \cite{Abreu:1997km}   & 69.5 (--)     \cite{Acciarri:1997nf} & 69.4 (65)  \cite{Ackerstaff:1997cza} \\ 
$\le 183 \GeV$         &  87.9  (85.3)\cite{Barate:1998gw} & 85.7  (86.5)  \cite{Abreu:1999uh}   & 87.6 (86.8)   \cite{Acciarri:1998wd} & 88.3 (86.1) \cite{Abbiendi:1998rd}   \\
$\le 189 \GeV$         &  92.9  (95.9)\cite{Barate:2000na} & 94.6  (94.4)  \cite{Abreu:2000kg}   & 95.3 (94.8)   \cite{Acciarri:1999by} & 91.0 (94.9) \cite{Abbiendi:1999sy}   \\ 
$\le 202 \GeV$         & 107.7 (107.8)\cite{Barate:2000zr} & 107.3 (106.4) \cite{Abdallah:2001ux}& 107.0 (105.2) \cite{Acciarri:2000hv} &       --                             \\ 
$\le 209 \GeV$  & 111.1 (114.2)\cite{Barate:2000ts} &114.3  (113.5) \cite{Abreu:2000fw}   & --   & 109.7 (112.5 )\cite{Abbiendi:2000ac} \\ \hline 
final             & 111.5 (114.2)\cite{Heister:2001kr}& 114.1 (113.3)  \cite{Abdallah:2003ip} & 112.0 (112.4) \cite{Achard:2001pj}   & 112.7 (112.7) \cite{Abbiendi:2002yk}
\\   \hline 
\end{tabular}
\begin{minipage}[t]{16.5 cm}
\vskip 0.5cm
\noindent
\end{minipage}
\end{center}
\end{table}     
When comparing the sensitivity of the four experiments one should keep in
mind that slightly different techniques for deriving the limits are used: 
for example ALEPH uses the technique discussed in \citere{jinmac}
instead of the CL$_{\mathrm{S}}$ technique yielding a better expected limit by up to $0.5\GeV$.  Shortly after the end of data taking in 2000, ALEPH \cite{Barate:2000ts} reported 
an excess of events, dominated by the observation in the ``four-jet channel'', with a minimal local $p$-value of $1.5\times 10^{-3}$ for $\MH= 116\GeV$ and stated 
that the observed rate would be compatible with Higgs-boson production for $\MH = 114\GeV$. The other three experiments also published their initial
findings based 
on 2000 data with focus on the mass range around $115\GeV$. L3 \cite{Acciarri:2000ke} reported an excess with a $p$-value at $\MH= 114.5\GeV$ for 
the background-only hypothesis 
of 0.09 (for signal+background of 0.62). DELPHI and OPAL did not observe any excess.
DELPHI \cite{Abreu:2000fw} preferred the background-only hypothesis ($p$-value = 0.23) over the signal+background hypothesis 
($p$-value = 0.03) for $\MH = 115 \GeV$, and OPAL \cite{Abbiendi:2000ac}
reported a similar $p$-value of 0.2 (0.4) for the background-only (signal+background) hypothesis. The corresponding observed and expected mass limits are shown in the fifth row of 
\refta{tab:lep2indfinal}. The final LEP analyses 
(see last row of \refta{tab:lep2indfinal}) used the latest and more accurate 
determinations of the LEP beam energies, better detector calibrations, a higher-statistics sample
for simulated events and in some cases performed an optimization of the 
selection strategies for high $\MH$, whereas non-optimized analyses were used in the 
first analysis of the 2000 data.

\begin{figure}
  \centering
  \begin{minipage}[b]{.4\textwidth}
    \includegraphics[width=0.97\textwidth]{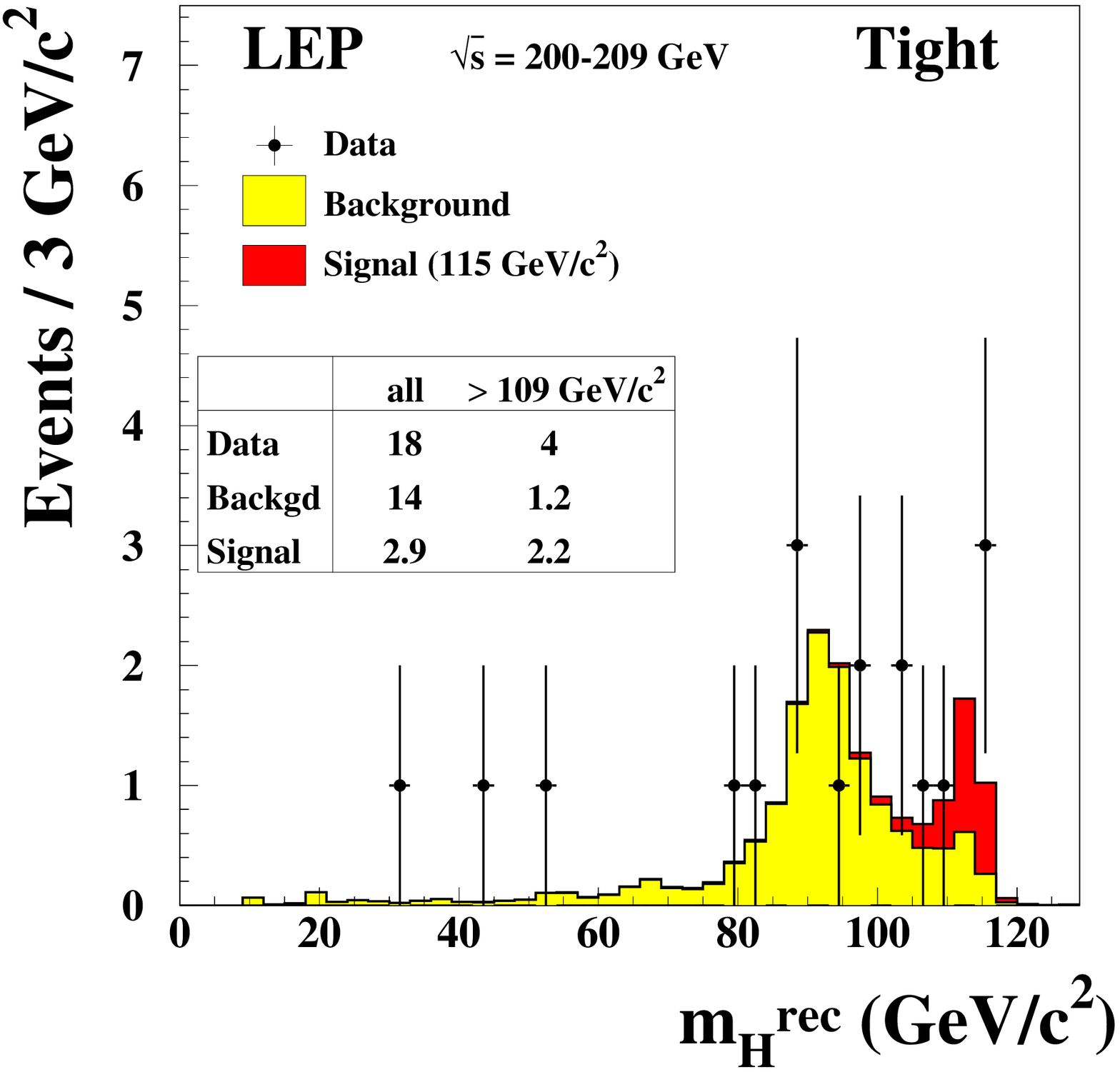}
  \end{minipage}
  \begin{minipage}[b]{.4\textwidth}
    \includegraphics[width=0.97\textwidth]{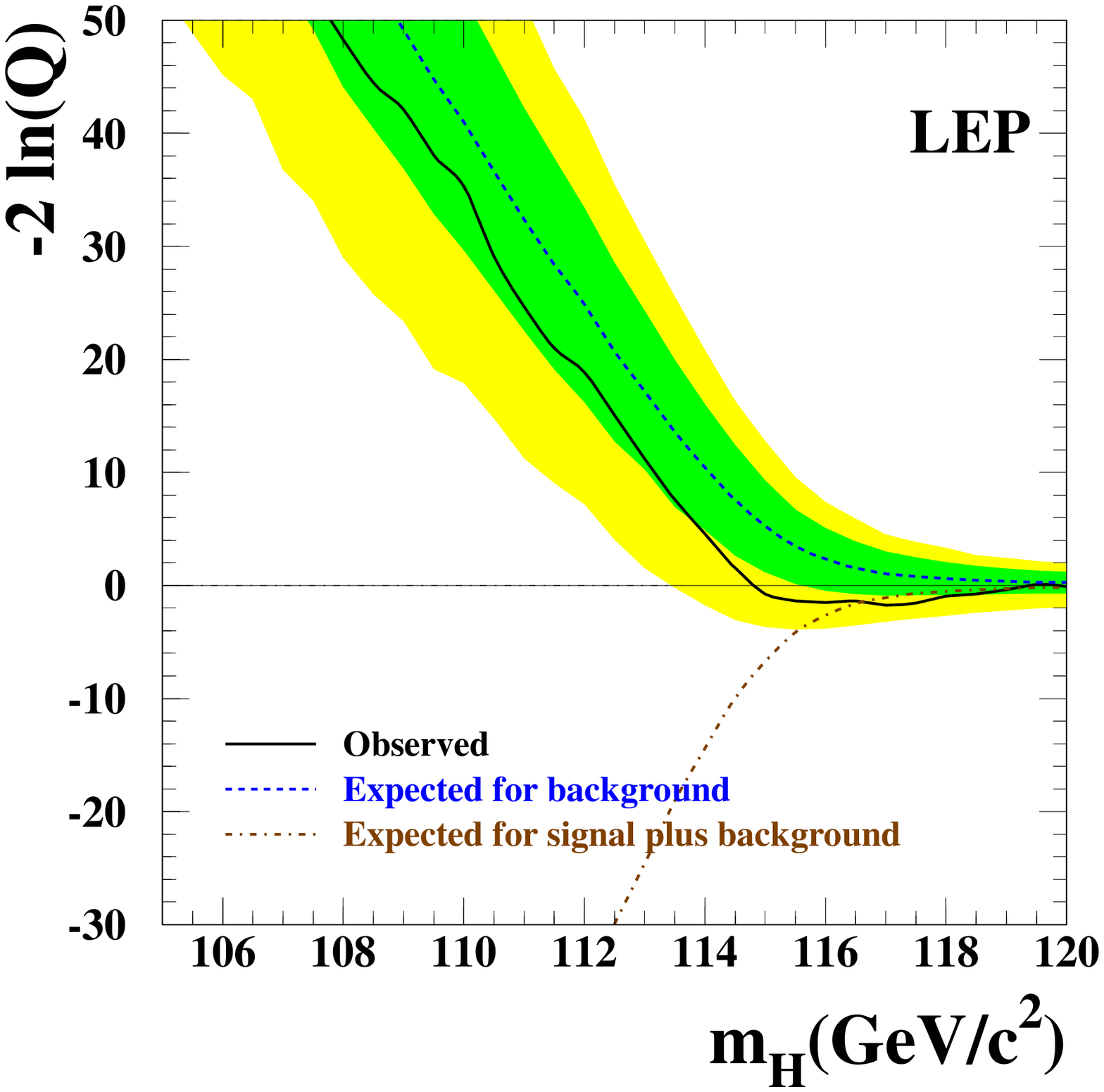}
  \end{minipage}
  \begin{minipage}[b]{.4\textwidth}
    \includegraphics[width=0.97\textwidth]{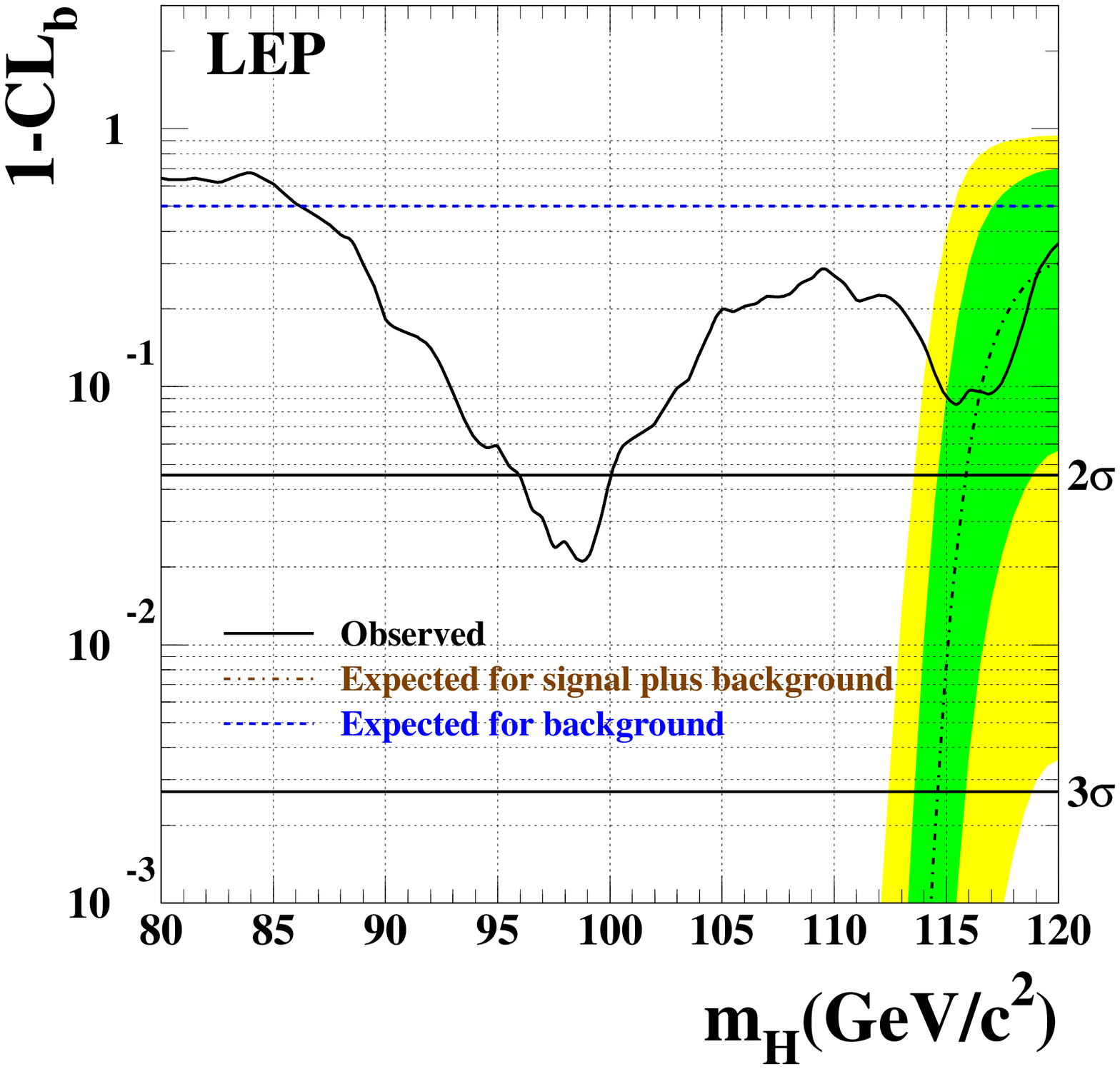}
  \end{minipage}
  \begin{minipage}[b]{.4\textwidth}
    \includegraphics[width=0.97\textwidth]{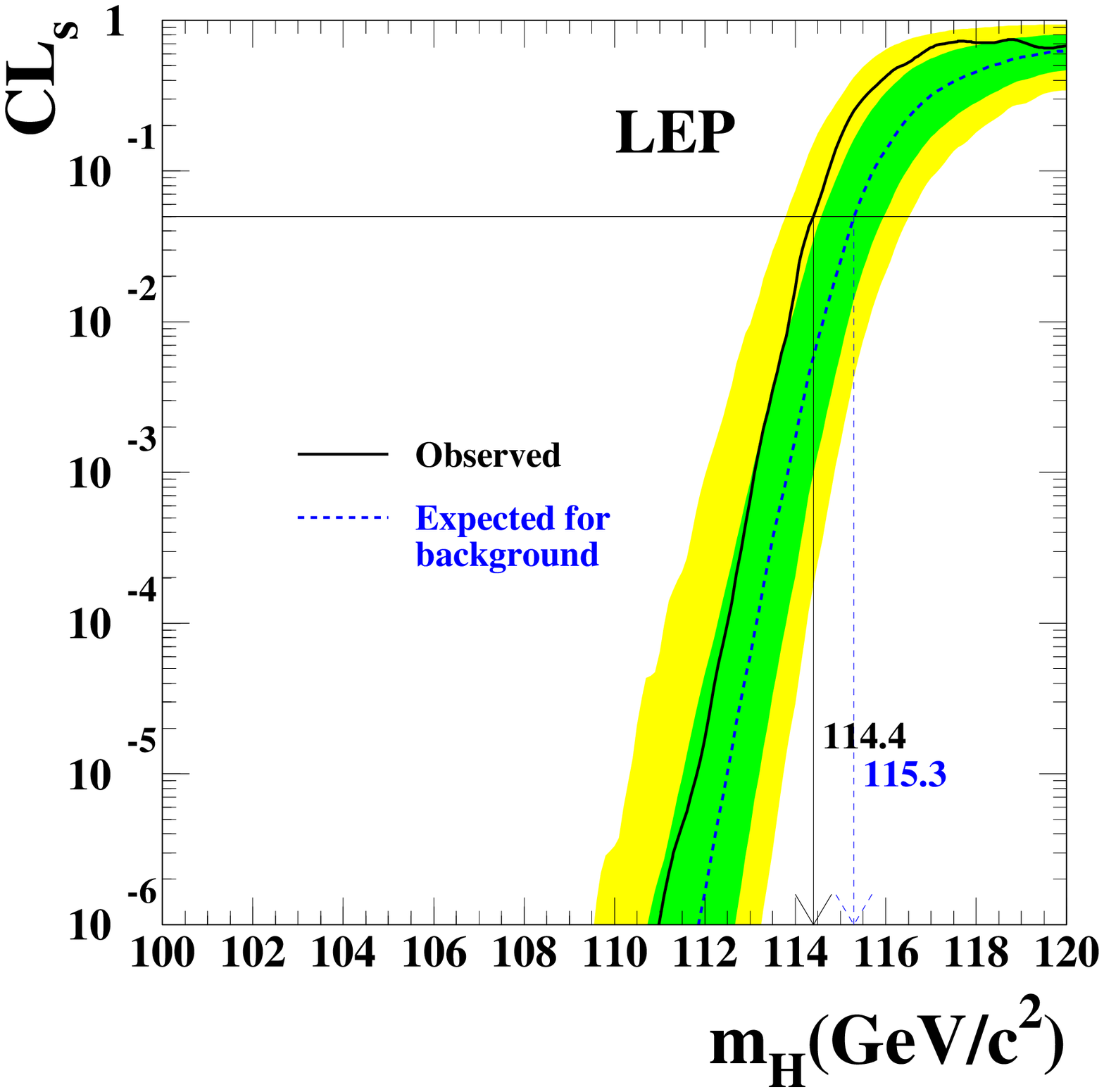}
  \end{minipage}
  \caption{Final results of the Higgs-boson searches at LEP \cite{Barate:2003sz}: 
           mass distribution of tightly selected Higgs-boson candidates compared
           to signal and background expectations (top-left), mean value of the 
           test statistic $-2\ln Q$ as function of $\MH$ for data and
           expectations from the signal+background hypothesis and the
           background-only hypothesis with 68\% and 95\% CL error bands (top-right),
           local $p$-values as function of $\MH$ for data and expectations from the
           background-only hypothesis and
          signal+background hypothesis with 68\% and 95\% CL error bands (bottom-left),
           CL$_{\mathrm{S}}$-values as function of $\MH$ for data and expectations from the 
           signal+background hypothesis and 
          background-only hypothesis with 68\% and 95\% CL error bands (bottom-right).}
  \label{fig:lep2results}
\end{figure}

The final mass distribution obtained from the combination of the four LEP experiments after tight selection requirements is shown in \reffi{fig:lep2results} (top-left). 
The slight excess at large values is dominated by the ``four-jet channel'' in ALEPH. 
The expected and observed values of the test statistic $-2 \ln Q$ depending on $\MH$ (\reffi{fig:lep2results}, top-right),
show the limited capabilities to distinguish the two hypotheses for large Higgs-boson masses. 
The two bands show the 68\% and 95\% probability bands for the background-only hypothesis. 
The final local $p$-values for the combination of all experiments are shown in \reffi{fig:lep2results} (bottom,left) and detailed in \refta{tab:lep2final} for the combination and the individual
experiments.  
\begin{table}
\caption{Final results of the searches at LEP2: local $p$-values for the consistency with the background-only and signal+background
hypotheses assuming $\MH = 115\GeV$,
expected and observed mass limits as derived in the LEP Higgs Working Group \cite{Barate:2003sz}.}
\label{tab:lep2final}
\begin{center}
\begin{tabular}{c|c|c|c|c }
\hline
                  & P$_{\mbox{\scriptsize b-only}}$    &P$_{\mathrm{s+b}}$ & exp.~limit   & obs.~limit \\ \hline 
LEP               & $0.09$              & $0.15$    & $115.3 \GeV$ & $114.4 \GeV$  \\ 
ALEPH             & $3.3\times10^{-3}$  & $0.87$    & $113.5 \GeV$ & $111.5 \GeV$  \\
DELPHI            & $0.79$              & $0.03$    & $113.3 \GeV$ & $114.3 \GeV$  \\
L3                & $0.33$              & $0.30$    & $112.4 \GeV$ & $112.0 \GeV$  \\
OPAL              & $0.50$              & $0.14$    & $112.7 \GeV$ & $112.8 \GeV$ \\  \hline 
four jets         & $0.05$              & $0.44$    & $114.5 \GeV$ & $113.3 \GeV$ \\  
all but four jets & $0.37$              & $0.10$    & $114.2 \GeV$ & $114.2 \GeV$ \\  \hline 
\end{tabular}
\begin{minipage}[t]{16.5 cm}
\vskip 0.5cm
\noindent
\end{minipage}
\end{center}
\end{table}     
No significant excess is observed in the combination. The $p$-values for the two alternative hypotheses are very similar, indicating again the difficulty to discriminate 
the background-only and signal+background hypotheses for $\MH = 115 \GeV$.  
All SM Higgs-boson mass hypotheses 
below $114.4\GeV$ were excluded with at least 95\% CL from the combination of the four experiments (see \reffi{fig:lep2results}, bottom-right)~\cite{Barate:2003sz}. 
The expected combined limit was $115.3\GeV$.  The very small differences in the limits from the LEP Higgs Working Group for each individual experiment in \refta{tab:lep2final} 
and the final ones published by each experiment, as shown in \refta{tab:lep2indfinal}, are due to differences in the statistical methods applied.

At the beginning of the LEP programme no solid limit existed on the mass of the Higgs boson. The searches for the SM Higgs boson carried out by the four LEP experiments 
extended the sensitive range well beyond that anticipated at the beginning of the LEP programme. This is due to the higher energy achieved and to more sophisticated detectors and analysis techniques.
The range below $114.4\GeV$ was and is difficult to probe 
at past and current hadron colliders. 

\section{Higgs-boson production at hadron colliders}
\label{sec:hadroncollcrosssec}

\subsection{Higgs-boson production mechanisms and cross-section overview}

The four main production mechanisms for SM Higgs bosons at hadron colliders
are illustrated by some representative LO diagrams in \reffi{fig:hadroncolldiags}.
\bfi[b]
\centerline{
{\unitlength .85pt
\SetScale{0.85}
\begin{picture}(120,100)(0,-10)
\Line(  0, 2)(20, 2)
\Line(  0, 5)(20, 5)
\Line(  0, 8)(20, 8)
\Line(  0,72)(20,72)
\Line(  0,75)(20,75)
\Line(  0,78)(20,78)
\Line(20, 7)(60,-2)
\Line(20, 4)(60,-5)
\Line(20, 1)(60,-8)
\Line(20,80)(60,89)
\Line(20,77)(60,86)
\Line(20,74)(60,83)
\Gluon(20, 7)(50,25){2}{5}
\Gluon(20,73)(50,55){2}{5}
\ArrowLine(50,25)(50,55)
\ArrowLine(50,55)(80,40)
\ArrowLine(80,40)(50,25)
\DashLine(80,40)(115,40){5}
\SetColor{Black}
\Vertex(80,40){2}
\Vertex(50,55){2}
\Vertex(50,25){2}
\GOval(20, 5)(10,5)(0){.5}
\GOval(20,75)(10,5)(0){.5}
\put(105,20){${\PH}$}
\put(35,35){${Q}$}
\put(5,-25){\footnotesize (a)}
\end{picture}
\SetScale{1}
}
{\unitlength .85pt
\SetScale{0.85}
\begin{picture}(140,100)(0,-10)
\Line(  0, 2)(20, 2)
\Line(  0, 5)(20, 5)
\Line(  0, 8)(20, 8)
\Line(  0,72)(20,72)
\Line(  0,75)(20,75)
\Line(  0,78)(20,78)
\Line(20, 7)(60, 3)
\Line(20, 4)(60, 0)
\Line(20,77)(60,81)
\Line(20,74)(60,78)
\ArrowLine(20, 5)(50,40)
\ArrowLine(50,40)(20,75)
\Photon(50,40)(90,40){2}{5}
\Photon(120,20)(90,40){2}{5}
\DashLine(90,40)(120,60){5}
\Vertex(50,40){2}
\Vertex(90,40){2}
\GOval(20, 5)(10,5)(0){.5}
\GOval(20,75)(10,5)(0){.5}
\put(125,56){${\PH}$}
\put( 52,20){${\PW/\PZ}$}
\put(105, 2){${\PW/\PZ}$}
\put(5,-25){\footnotesize (b)}
\end{picture}
\SetScale{1}
}
{\unitlength .85pt
\SetScale{0.85}
\begin{picture}(140,100)(0,-10)
\Line(  0, 2)(20, 2)
\Line(  0, 5)(20, 5)
\Line(  0, 8)(20, 8)
\Line(  0,72)(20,72)
\Line(  0,75)(20,75)
\Line(  0,78)(20,78)
\Line(20, 7)(60,-2)
\Line(20, 4)(60,-5)
\Line(20,77)(60,86)
\Line(20,74)(60,83)
\ArrowLine(20, 7)(50,10)
\ArrowLine(20,73)(50,70)
\ArrowLine(50,10)(110, 5)
\ArrowLine(50,70)(110,75)
\Photon(50,70)(80,40){2}{5}
\Photon(50,10)(80,40){2}{5}
\DashLine(80,40)(115,40){5}
\Vertex(80,40){2}
\Vertex(50,70){2}
\Vertex(50,10){2}
\GOval(20, 5)(10,5)(0){.5}
\GOval(20,75)(10,5)(0){.5}
\put(125,35){${\PH}$}
\put(122,75){${q}$}
\put(122, 0){${q}$}
\put(75,52){${\PW/\PZ}$}
\put(75,16){${\PW/\PZ}$}
\put(5,-25){\footnotesize (c)}
\end{picture}
\SetScale{1}
}
{\unitlength .85pt
\SetScale{0.85}
\begin{picture}(130,100)(0,-10)
\Line(  0, 2)(20, 2)
\Line(  0, 5)(20, 5)
\Line(  0, 8)(20, 8)
\Line(  0,72)(20,72)
\Line(  0,75)(20,75)
\Line(  0,78)(20,78)
\Line(20, 7)(60,-2)
\Line(20, 4)(60,-5)
\Line(20, 1)(60,-8)
\Line(20,80)(60,89)
\Line(20,77)(60,86)
\Line(20,74)(60,83)
\Gluon(20, 7)(50,10){2}{5}
\Gluon(20,73)(50,70){2}{5}
\ArrowLine(110, 5)(50,10)
\ArrowLine(50,70)(110,75)
\ArrowLine(80,40)(50,70)
\ArrowLine(50,10)(80,40)
\DashLine(80,40)(115,40){5}
\Vertex(80,40){2}
\Vertex(50,70){2}
\Vertex(50,10){2}
\GOval(20, 5)(10,5)(0){.5}
\GOval(20,75)(10,5)(0){.5}
\put(125,35){${\PH}$}
\put(122,75){${Q}$}
\put(122, 0){${\bar Q}$}
\put(75,55){${Q}$}
\put(75,16){${Q}$}
\put(5,-25){\footnotesize (d)}
\end{picture}
\SetScale{1}
}
\vspace*{1em}
}
\caption{Representative leading-order diagrams for the main SM Higgs-boson production channels
at hadron colliders, where $q$ and $Q$ denote light and heavy quarks, respectively:
(a) gluon fusion, (b)~Higgs-strahlung, (c)~vector-boson fusion, (d) heavy-quark
associated production.}
\label{fig:hadroncolldiags}
\efi
The size of the respective cross sections depends both on the type of colliding
hadrons and on the collision energy. Figures~\ref{fig:HXS_Tev} and
\ref{fig:HXS_LHC} show the total cross sections of the various channels
for the $\Pp\bar\Pp$ collider Tevatron at its CM energy of $\sqrt{s}=1.96\TeV$
and for the $\Pp\Pp$ collider LHC at the two energies $\sqrt{s}=7\TeV$ and
$14\TeV$.
\bfi
\centerline{
\includegraphics[width=.5\textwidth,height=.6\textwidth,angle=270]%
{plots/theory/tev.epsi}
}
\caption{Cross sections for the various SM Higgs-boson production channels at
Tevatron with a CM energy of $1.96\TeV$, as predicted by the 
TeV4LHC Section Working Group~\cite{Aglietti:2006ne}.
}
\label{fig:HXS_Tev}
\efi
\bfi
\includegraphics[width=.49\textwidth,height=.4\textwidth]%
{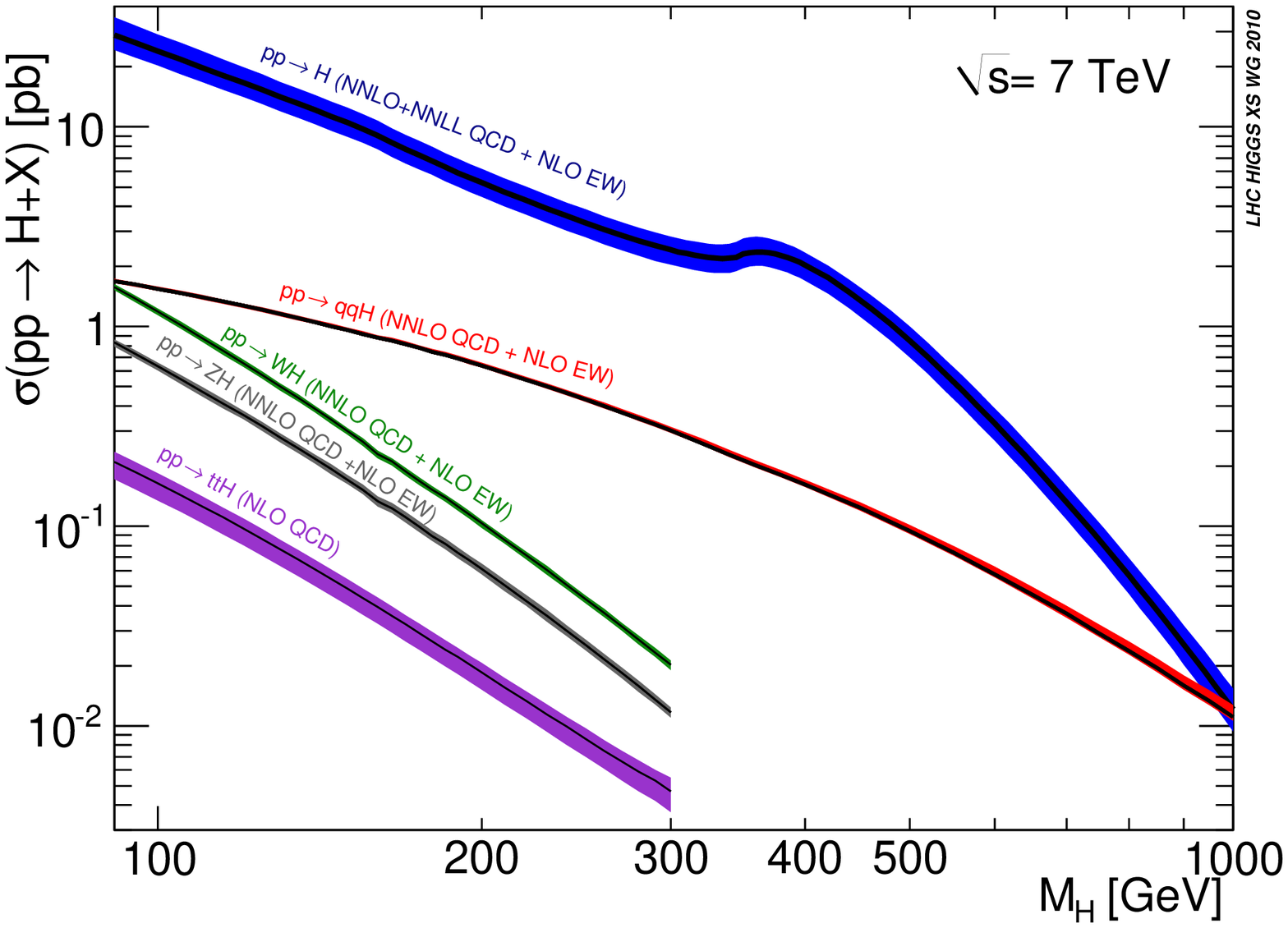}
\includegraphics[width=.49\textwidth,height=.4\textwidth]%
{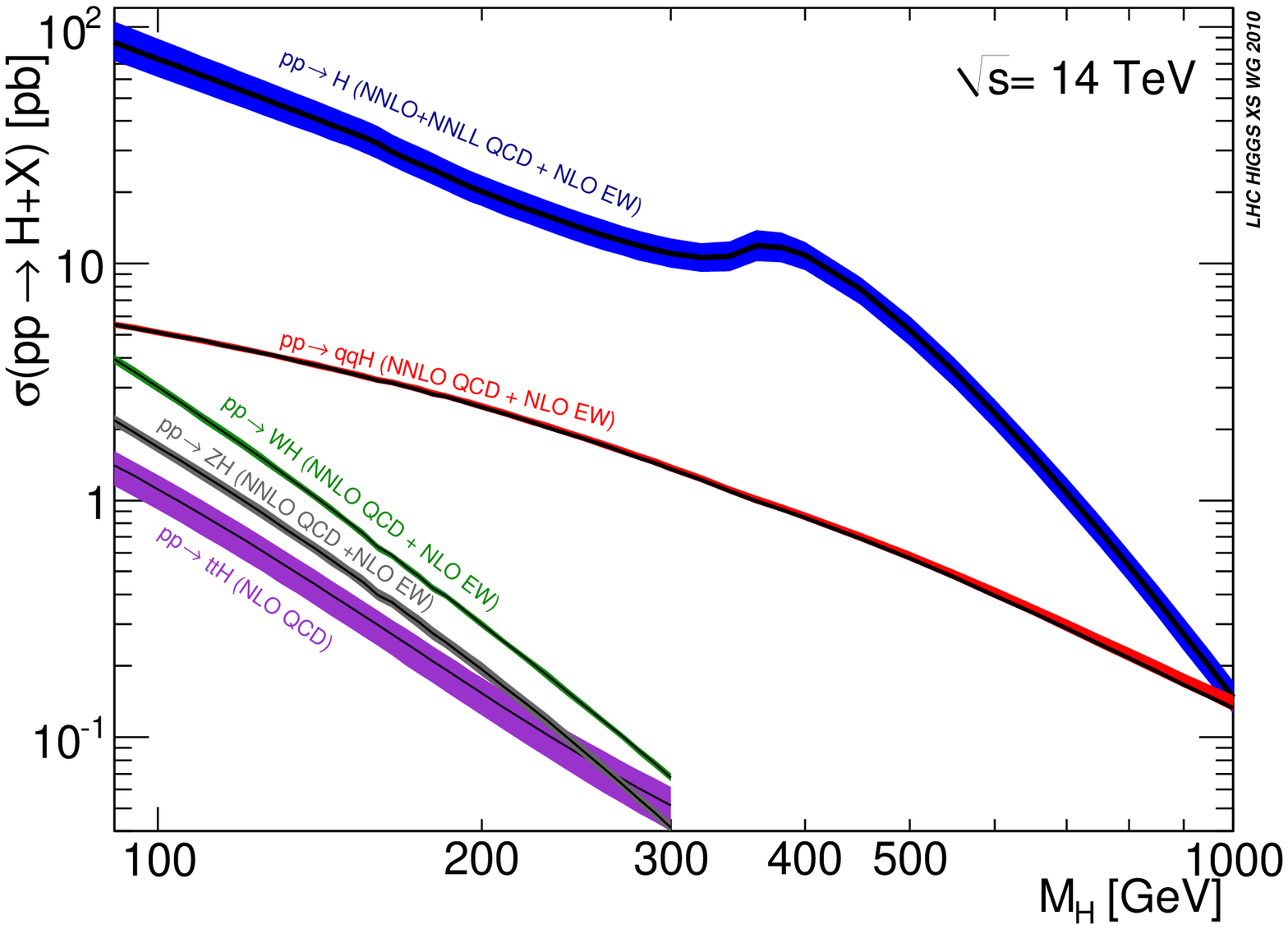}
\caption{Cross sections and respective uncertainties, indicated by the band widths,
for the various SM Higgs-boson production channels at
the LHC with a CM energy of $7\TeV$ and $14\TeV$, as predicted by the 
LHC Higgs Cross Section Working Group~\cite{Dittmaier:2011ti}.
}
\label{fig:HXS_LHC}
\efi
At the LHC, the energy increase from $7\TeV$ to $8\TeV$ leads to an increase
of $20{-}30\%$ in the Higgs-boson production cross sections for $\MH\sim100{-}200\GeV$. 
The transition from $7\TeV$ to $14\TeV$ in energy increases
the cross sections even
by a factor of about $3{-}4$ for these Higgs-boson masses,
with the exception of $\Pt\bar\Pt\PH$ production, where the factor is roughly 8.
Globally, loop-induced Higgs-boson production via gluon fusion delivers the largest cross section
owing to the large gluon flux in high-energy 
proton--(anti)proton collisions.
The respective cross section is typically 
an order of magnitude or more larger
than the remaining production cross sections. 
At the LHC, the vector-boson fusion (VBF) cross section competes in size with gluon fusion 
only for Higgs bosons as heavy as nearly $1\TeV$. 
VBF delivers the second
largest cross section, showing a much slower decrease with increasing Higgs-boson
mass owing to its $t$-channel dominance, which leads to a logarithmic rise
of its partonic cross section with increasing partonic CM energy $\sqrt{\hat s}$.
At Tevatron the Higgs-strahlung channels of $\PH\PW/\PH\PZ$ production
compete with VBF in size for Higgs masses $\MH\lsim100{-}200\GeV$ mainly
due to the different combinations of PDFs. For a $\Pp\bar\Pp$ initial state
high-energy $q\bar q$ collision, which is needed for Higgs-strahlung in LO,
is preferred over $qq$ scattering, since in this case the $q\bar q$ channel can proceed
via two valence quarks, which carry much more momentum than sea quarks on
average. The total Higgs-strahlung cross section for $\PH\PW$
production, where the sum over $\PW^\pm$ is taken, is larger than the one for
$\PH\PZ$ production by roughly a factor of two.
The smallest relevant cross section in all cases is provided by Higgs
production in association with $\Pt\bar\Pt$ pairs, whose suppression is mainly
due to the large invariant mass required to produce the three-particle final
state of heavy objects. Similar to pure $\Pt\bar\Pt$ pair production,
$\Pt\bar\Pt\PH$ production is largely dominated by 
$q\bar q$ annihilation at Tevatron, but by $\Pg\Pg$ fusion at the LHC. 
In the transition from Tevatron to the LHC and with increasing 
LHC energy, the suppression of the $\Pt\bar\Pt\PH$ cross section 
steadily decreases with respect to the other channels,
since the phase-space suppression of the heavy $\Pt\bar\Pt\PH$
final state is fading for larger collider energy.
The main motivation to measure the
$\Pt\bar\Pt\PH$ cross section clearly rests in the direct access to the
top Yukawa coupling, without contamination from other couplings;
even a qualitative
measurement at the LHC would be a great success.

Predictions for hadronic collisions in general involve
several serious sources of 
uncertainties that are tied to the hadronic
environment, as explained in more detail in \refse{se:predictions}.
Perturbative corrections, especially of the strong interactions, 
have to be taken into account as much as possible, in order to minimize 
the uncertainties, and the residual uncertainties, which are due
to missing higher-order effects and parametric errors have to be
quantified carefully. 
Figure~\ref{fig:HXS_LHC} illustrates the error estimate as assessed by the LHC Higgs
Cross Section Working Group for the total cross sections at the 
LHC~\cite{Dittmaier:2011ti}, comprising both theoretical and parametric
errors (PDF+$\alphas$). 
Table~\ref{tab:errors_HXS_LHC} gives a brief overview of theoretical
and parametric PDF+$\alphas$ uncertainties, for the latter 
following the recipe of the PDF4LHC Working Group~\cite{Botje:2011sn};
the explicit numbers are based on the results given in \citere{Dittmaier:2011ti}.
\btab
\caption{Theoretical (THU) and parametric PDF+$\alphas$ uncertainties (PU) for the
total Higgs-boson production cross sections at the LHC with CM energies
$7\TeV$ and $14\TeV$, as assessed by the LHC Higgs Cross Section
Working Group~\cite{Dittmaier:2011ti} employing the PDF4LHC~\cite{Botje:2011sn} recipe for PU,
as well as the typical size
of radiative corrections of the strong (QCD) and electroweak (EW)
interactions. The colour coding 
NLO/\Blue{NNLO}/\Green{NNLO+} refers to the respective perturbative order
included in the predictions, where NNLO+ means that resummations beyond the fixed-order correction
are included.}
\label{tab:errors_HXS_LHC}
\begin{center}
\begin{tabular}{lc||cc|cc||cc|cc}
\hline
&& \multicolumn{4}{c||}{LHC @ $\sqrt{s}=7\TeV$} &
\multicolumn{4}{c}{LHC @ $\sqrt{s}=14\TeV$}
\\
&& \multicolumn{2}{c|}{uncertainties} &
\multicolumn{2}{c||}{corrections}
& \multicolumn{2}{c|}{uncertainties} &
\multicolumn{2}{c}{corrections}
\\[-.3em]
& $\MH$[GeV] & THU & PU & QCD & EW & THU & PU & QCD & EW
\\
\hline
ggF &$<500$ & $6{-}10\%$ & $8{-}10\%$ & \Green{${\gsim}100\%$} & $5\%$ 
                 & $6{-}14\%$ & $7\%$ & \Green{${\gsim}100\%$} & $5\%$
\\
VBF &$<500$ & $1\%$ & $2{-}7\%$ & \Blue{$5\%$} & $5\%$ 
                 & $1\%$ & $3{-}4\%$ & \Blue{$5\%$} & $5\%$
\\
HW  &$<200$ & $1\%$ & $3{-}4\%$ & \Blue{$30\%$} & $5{-}10\%$ 
                 & $1\%$ & $3{-}4\%$ & \Blue{$30\%$} & $5{-}10\%$
\\
HZ  &$<200$ & $1{-}2\%$ & $3{-}4\%$ & \Blue{$40\%$} & $5\%$ 
                 & $2{-}4\%$ & $3{-}4\%$ & \Blue{$45\%$} & $5\%$
\\
ttH &$<200$ & $10\%$ & $9\%$ & $5\%$ & ? 
                 & $10\%$ & $9\%$ & $15{-}20\%$ & ?
\\
\hline
\end{tabular}
\end{center}
\etab
For later reference, the table also illustrates the typical size of the
known QCD and EW corrections.
A similar, but rather conservative estimate
of cross-section uncertainties for Tevatron can be found in
\citere{Baglio:2010um}.

The overview over the total cross sections shown in \reffis{fig:HXS_Tev}
and \ref{fig:HXS_LHC}
can only give a rough idea about
the importance the respective production channels for Higgs-boson discovery
or for later precision studies. The total cross sections provide the total
production rates for Higgs bosons, but in practice the rates of potentially
observable events is relevant. 
Moreover, the signatures left in the detectors are rather different for
the various production channels:
\begin{itemize}
\item
Single-Higgs production in $\Pg\Pg$ fusion strongly prefers
Higgs bosons at low transverse momenta plus some hadronic activity recoiling
against the Higgs boson.
Owing to the overwhelming hadronic background
it is impossible to access the $\PH\to\Pb\bar\Pb$
decay channel here.
Instead the $\PH\to\gamma\gamma$ decay yields a promising signal at small
Higgs-boson masses in spite of its small branching ratios.
At higher values of $\MH$ the decays $\PH\to\PW\PW/\PZ\PZ\to 4\,$fermions via
weak gauge bosons lead to rather clean signatures, in particular in the ``gold-plated''
four-muon channel of the $\PZ\PZ$ decay. 
\item
VBF offers the possibility to tag on the two accompanying jets which
strongly prefer the forward--backward directions owing to the $t$-channel
topology. Colour exchange between the two colliding partons is largely
suppressed also in higher orders, further pronouncing this signature.
Special VBF selection cuts on the two jets exploit this feature to
suppress background.
Apart from the $\PW\PW/\PZ\PZ$ decays, VBF offers the best possibility
to search for Higgs-boson decays into $\tau^+\tau^-$ pairs.
\item
Higgs-strahlung represents the main search channel for SM Higgs bosons
near the LEP exclusion limit at $\MH=114\GeV$ at Tevatron, but at the
LHC the higher background rates pose serious problems. In order to 
make use of this channel, a sophisticated novel search technique is promising
that first focuses on ``fat jets'' (with large cones) at large transverse momenta
which show an internal structure with B~mesons~\cite{Butterworth:2008iy}.
Heavy objects decaying into a
$\Pb\bar\Pb$ quark pair, such as a light SM Higgs boson, are very likely to
fit into this selection pattern.
\item
$\Pt\bar\Pt\PH$ production seems at best to be accessible at very low
Higgs-boson masses owing to the cross-section suppression, so that one is
forced to search for the $\PH\to\Pb\bar\Pb$ decay.
This leads to the partonic final state
$\Pt\bar\Pt\PH\to\PW\PW\Pb\bar\Pb\Pb\bar\Pb\to4f+4\Pb$, where $4f$ stands for a
four-fermion final state of the type $4\ell$, $2\ell2q$, or $4q$.
It turns out that it will be extremely difficult to establish a signal
over the serious background created by $\Pt\bar\Pt\Pb\bar\Pb$ and
$\Pt\bar\Pt+2\,$jet production. At present, here the hope again rests on
a selection focusing on fat jets with a $\Pb\bar\Pb$ substructure~\cite{Plehn:2009rk}.
However, owing to these difficulties, $\Pt\bar\Pt\PH$ production cannot
contribute to the Higgs-boson discovery, but would be very interesting in later
precision analyses to assess Higgs-boson couplings.
\item
Finally, according to the SM, 
Higgs-boson production with $\Pb\bar\Pb$ pairs is most likely not
accessible at hadron colliders 
owing to the overwhelming hadronic background at hadron colliders.
However, it plays an important role in
extensions of the SM where the $\PH\Pb\bar\Pb$ Yukawa coupling is enhanced,
such as in supersymmetric or non-supersymmetric two-Higgs doublet models.
\end{itemize}

\subsection{Survey of precision calculations}

The features reviewed above make it clear that theoretical predictions
have to account for kinematical details of the various production
mechanisms and of the Higgs-boson decay channels, because many of the specific
event selections heavily make use of those properties.

For the individual Higgs-boson production channels in the SM, theory
predictions are accurate within $5{-}20\%$, depending in detail on the 
production mechanism:
\begin{itemize}
\item
{\it Gluon fusion} $\Pp\Pp/\Pp\bar\Pp\to\PH+X$
\cite{Georgi:1977gs,Graudenz:1992pv,Spira:1995rr,Harlander:2000mg,Marzani:2008az,Catani:2003zt,Ahrens:2008qu,Aglietti:2004nj,Degrassi:2004mx,Actis:2008ug,Actis:2008uh,Anastasiou:2008tj,Keung:2009bs,Accomando:2007xc,deFlorian:2009hc,Baglio:2010um,Baglio:2010ae,deFlorian:1999zd,Bozzi:2005wk,deFlorian:2011xf,Berger:2010xi,Anastasiou:2004xq,Catani:2007vq,Anastasiou:2011pi,deFlorian:2012mx,mcfm,Berger:2006sh,Campbell:2006xx,Frixione:2002ik,Bagnaschi:2011tu}
\\
Higgs-boson production via gluon fusion is strongly dominated by top-quark loops in the
SM and receives very large QCD corrections, which are mainly due to soft-gluon
exchange and emission. The NLO corrections, which are known with their
full quark-mass dependence~\cite{Graudenz:1992pv,Spira:1995rr}, increase the cross section by
$80{-}100\%$. The NNLO QCD corrections, which were first
evaluated in the large-top-mass limit~\cite{Harlander:2000mg} and later on
refined by top-mass corrections~\cite{Marzani:2008az}, add another $25\%$ to
the cross section.
The still significant scale uncertainty of the NNLO QCD cross section can be reduced
by soft-gluon resummation. The resummation effects, known up to the
next-to-next-to-next-to-leading logarithmic (N${}^3$LL) 
level~\cite{Catani:2003zt},
add another $6{-}9\%$ to the cross section at the LHC, leading to the 
$6{-}10\%(6{-}14\%)$ of residual scale uncertainties at $7(14)\TeV$
quoted in \refta{tab:errors_HXS_LHC}.
The seemingly bad convergence in the sequence LO$\to$NLO$\to$NNLO of fixed-order predictions
can be cured by an analytical continuation of the cross-section result to an imaginary
renormalization scale, which results in a resummation of large corrections
$\propto(\alphas\pi^2)^n$~\cite{Ahrens:2008qu}. 
However, the optimistic estimate of uncertainties in
the effective-field theory prediction of \citere{Ahrens:2008qu}, which is also
accurate to the N${}^3$LL level, is still
under debate~\cite{Dittmaier:2011ti}.

EW corrections, which are known at 
NLO~\cite{Aglietti:2004nj,Degrassi:2004mx,Actis:2008ug,Actis:2008uh},
turn out to be of the size of $\sim5\%$ (relative to LO) and strongly depend on the Higgs-boson mass.
The extreme size $\sim100\%$ of the QCD corrections poses the question how to combine the NLO
${\cal O}(\alphas)$ QCD and ${\cal O}(\alpha)$ EW corrections: additively or in factorized form?
An investigation of mixed QCD--EW corrections for small $\MH$ clearly favours the
factorized approach~\cite{Anastasiou:2008tj}.
Electroweak corrections to Higgs+jet production~\cite{Keung:2009bs} 
influence the total single-Higgs production cross section at the negligible $\lsim1\%$ level,
but become relevant at large Higgs-boson transverse momenta.

Finally, at the level of some percent, interference effects between Higgs-boson
signal and irreducible background, and Higgs-boson off-shell effects should be taken into account,
as studied for $\Pg\Pg\to\PH\to\PW\PW/\PZ\PZ$ in 
\citeres{Kauer:2012hd,Passarino:2010qk,Accomando:2007xc}.

Various predictions of 
total Higgs-boson production cross 
sections~\cite{Baglio:2010um,Baglio:2010ae,Ahrens:2008qu,Anastasiou:2008tj,deFlorian:2009hc}
were described and partially updated in the first review~\cite{Dittmaier:2011ti} 
of the LHC Higgs Cross Section Working Group, where many details, in particular, about the
various error estimates can be found.
The second review~\cite{Dittmaier:2012vm} summarizes differential cross sections, such
as the Higgs-boson-transverse-momentum ($p_{\mathrm{T,H}}$)
distribution, and related cut-based observables, such as 
cross sections with a jet veto or definite jet multiplicities---a broad subject where
we can only make brief statements.

Predictions for the $p_{\mathrm{T,H}}$ distribution were first worked out in
fixed-order NLO QCD~\cite{deFlorian:1999zd} and were refined later by
soft-gluon resummations~\cite{Bozzi:2005wk,deFlorian:2011xf}.
For high $p_{\mathrm{T,H}}$, the application of the large-top-quark-mass limit in predictions
becomes subtle, and calculations should include the full quark-mass dependence as
far as possible; errors from missing corrections due to finite quark masses
do not necessarily show up in the scale uncertainty.
Moreover, differential $K$-factors~\cite{Anastasiou:2004xq,Catani:2007vq,deFlorian:2012mx},
i.e.\ corrections to distributions, in general
show significant variations over phase space or depend on cuts, clearly disfavouring
the use of a simple uniform rescaling of distributions by global $K$-factors.
Even for integrated quantities, in particular, if vetoes or cuts are involved,
a naive scale variation can underestimate theoretical uncertainties significantly.
For the jet multiplicities and jet vetoes the discussion of appropriate
error estimates and resummations started in
\citere{Dittmaier:2012vm} and went on in the literature since then, 
with proposals for solutions based on
traditional resummations~\cite{Banfi:2012yh} 
and effective-theory approaches~\cite{Berger:2010xi}.

Several public programmes that include higher-order corrections to Higgs-boson production via gluon fusion
are used in the experimental anaylses:
\begin{itemize}
\item
{\sc Higlu}~\cite{Spira:1995rr} for calculating the total NLO QCD cross section with the
full mass dependence; 
\item
{\sc MCFM}~\cite{mcfm,Campbell:2006xx} for Higgs+2jet production at NLO QCD;
\item
{\sc FEHiP}~\cite{Anastasiou:2004xq} and
{\sc HNNLO}~\cite{Catani:2007vq} for 
differential quantities, based on the NNLO QCD calculation in the heavy-top limit,
but supporting the full kinematical information on the Higgs-boson decays 
$\PH\to\gamma\gamma$ and $\PH\to\PW\PW/\PZ\PZ\to4\,$leptons;
\item
{\sc HqT}~\cite{deFlorian:2011xf} and {\sc HRes}~\cite{deFlorian:2012mx} for the
NNLL resummation matched to NLO QCD for Higgs-transverse-momentum spectrum,
where {\sc HRes} supports the decays $\PH\to\gamma\gamma$ and $\PH\to\PW\PW/\PZ\PZ\to4\,$leptons;
\item
{\sc iHixs}~\cite{Anastasiou:2012hx,Anastasiou:2011pi}
for the inclusive Higgs-boson production cross section using state-of-the-art fixed-order results, i.e.\
QCD corrections to NNLO, 
EW, and mixed QCD--EW corrections, as well as quark-mass 
and finite width effects;
\item
{\sc MC@NLO}~\cite{Frixione:2002ik} and
{\Powheg}~\cite{Bagnaschi:2011tu} for differential predictions in
NLO QCD properly matched to QCD parton showers.
\end{itemize}
\item
{\it Vector-boson fusion} $\Pp\Pp/\Pp\bar\Pp\to\PH+2\mbox{jets}+X$
\cite{Spira:1997dg,Dicus:1985zg,Cahn:1983ip,Altarelli:1987ue,Barger:1994zq,DelDuca:2001fn,Nikitenko:2007it,Berger:2006sh,Campbell:2006xx,Han:1992hr,Figy:2003nv,Ciccolini:2007jr,Harlander:2008xn,Bolzoni:2010xr,Bolzoni:2011cu,Andersen:2007mp,Nason:2009ai,Figy:2010ct}
\\
Higgs-boson production via VBF experimentally means 
Higgs-boson production in association with two hard jets which are forward--backward
pointing with some rapidity gap and an appropriate central-jet veto, in order to
drastically reduce background~\cite{Barger:1994zq}. 
The VBF cuts suppress $\PH+2\,$jet production via top-quark-loop-induced $\Pg\Pg$
fusion~\cite{DelDuca:2001fn} as well, which for $\MH=120\GeV$ contributes
about $4{-}5\%$~\cite{Nikitenko:2007it} to the cross section with a non-negligible residual
NLO scale uncertainty of about $35\%$~\cite{Berger:2006sh,Campbell:2006xx}.
While $s$-channel contributions from $\PH\PW/\PH\PZ$ with hadronically decaying
$\PW/\PZ$ bosons deliver up to $\sim30\%$ to the total cross section at 
low $\MH\sim120\GeV$, VBF cuts suppress this Higgs-strahlung contamination to
$\lsim0.6\%$.
Thus, $\PH+2\,$jet production with VBF cuts is nearly a pure EW process proceeding
via W- or Z-boson fusion to a Higgs boson with almost no colour exchange
between two forward-scattered (anti)quark lines. 

The suppressed colour exchange
and the forward kinematics suggest that QCD corrections are small after
an appropriate scale setting (to small scales of the order of $\MW$)
in the PDF redefinition of QCD factorization.
In fact this expectation is confirmed by explicit 
NLO QCD~\cite{Spira:1997dg,Han:1992hr,Figy:2003nv,Ciccolini:2007jr} 
and NNLO QCD~\cite{Harlander:2008xn,Bolzoni:2010xr,Bolzoni:2011cu} 
calculations, which are available for
differential cross sections in the former case and for the total cross section in the latter. 
The NLO QCD corrections are of the order of $5{-}10\%$, depending in detail on the
VBF cuts, with interferences between $t$- and $u$-channels---including gluon
exchange between the (anti)quark lines at NLO---at the negligible level of
$\lsim0.1\%$~\cite{Ciccolini:2007jr}.
This suppression of colour exchange is also confirmed at NNLO, where large
parts of the corrections with non-trivial colour flow were 
calculated or estimated~\cite{Harlander:2008xn,Bolzoni:2011cu}, so that
the NNLO QCD corrections can be treated in the so-called structure-function approach
as used in \citeres{Bolzoni:2010xr,Bolzoni:2011cu} for the total cross section.
Likewise, interference effects between amplitudes with top-quark-loop-induced 
$\Pg\Pg\PH$ couplings and VBF amplitudes with one-loop gluon exchange
turn out to be negligible as well~\cite{Andersen:2007mp}.
The remaining scale uncertainty of the total cross section at NNLO QCD, 
which is at the percent level, turns out to be small as compared to the 
parametric PDF+$\alphas$ uncertainty of $2{-}7\%(3{-}4\%)$ at the
LHC at $7(14)\TeV$.
For differential quantities, i.e.\ 
for cross sections with VBF cuts or for distributions, QCD
corrections at fixed order 
are only known to NLO, but including the
matching to QCD parton showers at this order~\cite{Nason:2009ai}.

The NLO EW corrections have been calculated for all $t/u$- and $s$-channel
contributions, including all interferences, in \citere{Ciccolini:2007jr}
and for the $t/u$-channels in \citere{Figy:2010ct}, where 
both calculations are valid for integrated and differential quantities. 
In the presence of
VBF cuts, the approximation by $t/u$-channels only is sufficient.
The EW corrections to integrated cross sections
amount to $5{-}10\%$ and are, thus, of the same generic 
size as the QCD corrections, but in distributions at high transverse momenta
of the Higgs boson or the jets
the EW corrections can even be larger.
Similar to the situation in the related decays
$\PH\to\PW\PW/\PZ\PZ\to4f$, discussed in \refse{se:Hdecays},
the leading two-loop corrections of the order 
$\GF^2\MH^4$~\cite{Ghinculov:1995bz} become of the
same size as 
their one-loop counterpart of order $\GF\MH^2$
at $\MH\sim700\GeV$, signalling the breakdown of perturbation theory.
For $\MH\lsim500\GeV$, on the other hand, perturbation theory is
safely applicable.

The described state-of-the-art is widely available by public programmes.
The fixed-order NLO QCD+EW predictions can be obtained with the Monte Carlo programmes
{\sc VBF@NLO}~\cite{Figy:2003nv,Figy:2010ct,vbfnlo} and
{\sc Hawk}~\cite{Ciccolini:2007jr,hawk}, the former in the $t/u$-channel
approximation, the latter including $s$-channels and all interferences as well.
The (approximated) NNLO QCD corrections to the total cross section can be obtained with the programme
{\sc VBF@NNLO}~\cite{Bolzoni:2010xr,Bolzoni:2011cu,vbfnnlo}.
Presently, the best prediction for differential distributions should be obtained
upon reweighting the parton-shower-improved NLO predictions of
{\Powheg}~\cite{Nason:2009ai} with the relative EW correction
from {\sc Hawk} or {\sc VBF@NLO}.

\item
{\it Higgs-strahlung} $\Pp\Pp/\Pp\bar\Pp\to\PH\PW/\PH\PZ+X$
\cite{Glashow:1978ab,Han:1991ia,Brein:2003wg,Ferrera:2011bk,%
Altenkamp:2012sx,Brein:2011vx,Ciccolini:2003jy,Denner:2011id,Banfi:2012jh,Brein:2012ne}
\\
As far as higher-order QCD corrections are concerned,
Higgs-strahlung is very similar to the well-known Drell--Yan process.
In LO there is just quark--antiquark annihilation to an off-shell
weak gauge boson, $q\bar q\to\PW^*/\PZ^*\to\PH\PW/\PH\PZ$,
and the structure and size ($\sim30\%$) of NLO QCD corrections~\cite{Han:1991ia}
are identical to the Drell--Yan case.
At NNLO QCD, the Drell--Yan-like corrections, which are
known for the total $\PH\PW/\PH\PZ$ cross sections~\cite{Brein:2003wg}
and the differential $\PH\PW$ cross section~\cite{Ferrera:2011bk},
dominate, but there are additional corrections to Higgs-strahlung
that have no counterparts in Drell--Yan production. 
Firstly, loop-induced $\Pg\Pg$ fusion contributes to $\PH\PZ$
production about $2{-}6\%(4{-}12\%)$ for CM energies of $7\TeV(14\TeV)$
at the LHC~\cite{Brein:2003wg}%
\footnote{These numbers, which were used also in \reffi{fig:HXS_LHC},
are based on the LO one-loop diagrams of the $\Pg\Pg$ channel.
A recent evaluation~\cite{Altenkamp:2012sx} 
of the two-loop NLO corrections in the heavy-top limit,
however, roughly doubles this contribution, i.e.\ in order to reach
the precision tag of \refta{tab:errors_HXS_LHC} these corrections should be
taken into account.}, and secondly 
diagrams involving
Higgs couplings to top-quark loops correct the $\PH\PW/\PH\PZ$
cross sections by another $1{-}3\%$~\cite{Brein:2011vx}.
As far as EW corrections are concerned, Higgs-strahlung is more 
involved than the Drell--Yan process. The NLO EW corrections to the
total cross sections have been calculated in \citere{Ciccolini:2003jy}
and are of the order of $-(5{-}10)\%$, only weakly depending on the
collider type and energy. The NLO EW corrections to differential
cross sections, calculated in \citere{Denner:2011id},
can be significantly larger, in particular for the 
kinematics relevant in the ``boosted-Higgs analysis'' which requires
large transverse Higgs momenta. For instance, the EW corrections
to $\PH\PW$ production with $\MH=120\GeV$ are
$\sim-14\%$ for $p_{\rT,\PH}\gsim200\GeV$.

From \refta{tab:errors_HXS_LHC} one can read off that theoretical
uncertainties at the LHC are of the order of $1{-}2\%(2{-}4\%)$ and the PDF+$\alphas$ 
uncertainties of the order of $3{-}4\%$ for a CM energy of $7(14)\TeV$
for the total HW/HZ cross sections (see \citere{Dittmaier:2011ti} for details). 
Note, however, that the uncertainties
for the scenario of a boosted Higgs boson $(p_{\rT,\PH}\gsim200\GeV)$
are somewhat larger~\cite{Dittmaier:2012vm}. 
Specifically, the issue of final-state radiation off the Higgs-boson decay products
deserves particular attention in the boosted-Higgs scenario~\cite{Banfi:2012jh}.
The PDF uncertainties are estimated to $\sim5\%$, and
for $\PH\PZ$ production, where differential
QCD corrections are only known to NLO, the theoretical uncertainty
is estimated to $\sim7\%$; 
a reduction of these theoretical uncertainties
to the few-percent level requires the inclusion of
NNLO QCD corrections. 

To a large extent,
the described precision calculations are available via
public programmes. The NLO QCD corrections for total and differential
cross sections can be calculated with
{\sc V2HV}~\cite{v2hv} and {\sc MCFM}~\cite{mcfm}, respectively, and 
the total cross section, including NNLO QCD and NLO EW corrections,
with the programme {\sc VH@NNLO}~\cite{Brein:2012ne}.
Finally, the NLO QCD+EW corrections to fully differential observables
(with decaying W/Z bosons) can be computed with the Monte Carlo programme
{\sc Hawk}~\cite{hawk}.
\item
{\it Associated Higgs-boson production with} $\Pt\bar\Pt$ pairs
$\Pp\Pp/\Pp\bar\Pp\to\Pt\bar\Pt\PH+X$
\cite{Raitio:1978pt,Beenakker:2001rj,Frederix:2011zi,Garzelli:2011vp,Plehn:2009rk}
\\
Both the total and differential cross sections to $\Pt\bar\Pt\PH$ production
are known to NLO QCD~\cite{Beenakker:2001rj}. The size of the QCD corrections
is moderate, i.e.\ typically $10{-}20\%$ (positive at the LHC and negative 
at Tevatron), which holds also 
true for differential
distributions if appropriate dynamical scales are used. 
Theoretical uncertainties from missing higher-order effects as well as
PDF+$\alphas$ uncertainties are estimated to be of the order of $\sim10\%$
for total cross sections at the LHC (see \refta{tab:errors_HXS_LHC} 
and \citere{Dittmaier:2011ti}). 
For differential distributions the theoretical uncertainty is somewhat
larger ($10{-}20\%$) and together with the parametric PDF+$\alphas$ uncertainties
can add up to $20{-}50\%$~\cite{Dittmaier:2012vm}.
The NLO QCD predictions were matched to parton showers,
employing the MC@NLO~\cite{Frederix:2011zi} and \Powheg~\cite{Garzelli:2011vp} concepts.
The results obtained with the two shower variants typically agree to better than $10\%$.
Electroweak corrections to $\Pt\bar\Pt\PH$ production are yet unknown, but their
effect should be covered by the size of the quoted QCD uncertainties.

As already mentioned above, suppressing the background to $\Pt\bar\Pt\PH$
production is a real challenge, since the final state involves four b~quarks,
where three or four bottom quarks have to be tagged, and two out of the four have to be
identified as $\PH\to\Pb\bar\Pb$ decay products. To properly simulate this
theoretically, precise predictions are required that take into account the
decays of both the top quarks and the Higgs boson. At present, such predictions
exist only at LO. Similar to the HW/HZ analysis, searching first for
fat jets at high transverse momentum and a $\Pb\bar\Pb$ substructure
can lift the signal-to-background ratio to $S/B\sim1/2.4$, assuming
three b~tags, $\MH=120\GeV$, and an integrated luminosity of 
$100\,\mathrm{fb}^{-1}$, as shown in the parton-level study of \citere{Plehn:2009rk}.
The main background results from 
$\Pt\bar\Pt\Pb\bar\Pb$ and $\Pt\bar\Pt{+}2$jet production within pure
QCD, whose cross sections have recently been calculated at NLO in
\citeres{Bredenstein:2008zb} and \cite{Bevilacqua:2010ve}, respectively.
However, the residual scale dependence of the background cross-section normalization
is about $\sim20\%$, and a careful study of the uncertainties of possible
extrapolation procedure from signal-free control regions to signal regions
has not yet been carried out. To this end, for the background the
differential information on the top-quark decays has to be taken into account as well.
The feasibility of the $\Pt\bar\Pt\PH$ measurement is still not established.
\end{itemize}

\subsection{Towards couplings analyses after discovery}

Although the difficult hadronic environment of the LHC limits the precision
of Higgs-boson studies, the global pattern of Higgs couplings
could be tested at a qualitative level in a global fit to various channels 
with only mild theory assumptions.
Assuming, for instance,
that the Higgs-boson decay widths into weak gauge bosons are at most as large as their
SM values, $\Gamma_{\PH\to\PW\PW/\PZ\PZ}\le\Gamma_{\PH\to\PW\PW/\PZ\PZ}^\SM$,
which is the case in many models with extended Higgs sectors,
the Higgs couplings of a not too heavy Higgs boson ($110\GeV<\MH<190\GeV$)
to top quarks, $\tau$-leptons, and W and
Z bosons could be determined to $10{-}40\%$ in an LHC run at $\sqrt{s}=14\TeV$
with an integrated 
luminosity of $300\,\mathrm{fb}^{-1}$~\cite{Duhrssen:2004cv,ATLAS-collaboration:2012iza,cms-note-2012-006}.
An increase of the integrated luminosity to $3000\,\mathrm{fb}^{-1}$
reduces the experimental uncertainties in the signal strengths 
$\sigma/\sigma_{\SM}$ by roughly a factor of 
$2$~\cite{ATLAS-collaboration:2012iza}, so that
the uncertainties of the Higgs-boson couplings should be reduced
by a factor of about $1.5$, since the couplings enter the signal
strengths quadratically.
Constraints on Higgs couplings might be even possible with lower luminosity
or tighter constraints with the same luminosity if more and more
sophisticated analysis techniques are exploited (see, e.g., \citere{Lafaye:2009vr}).
Currently, the ATLAS~\cite{ATLAS:2012wma} 
and CMS~\cite{cms-pas-hig-12-045} collaborations follow the strategy of a
first model-independent analysis of Higgs couplings,
which has been put forward by the LHC Higgs Cross Section Working Group
\cite{LHCHiggsCrossSectionWorkingGroup:2012nn} and is
based on a simple rescaling of SM cross sections.
Ultimate precision Higgs coupling analyses, however, should
go beyond such simple phenomenological studies, since their
informative value is rather limited in the interesting case of
significant discrepancies between SM and data.
Proper model-independent non-standard coupling analyses should
be thoroughly based on effective-field theories, including higher-order
effects in a consistent way as much as possible.
First steps into this direction are proposed in
\citere{Passarino:2012cb}.

Among the various Higgs couplings, the self-interactions are of particular
interest, since they determine the Higgs potential that drives EW
symmetry breaking. As already explained for $\Pep\Pem$ colliders,
such studies require multiple-Higgs-boson production, e.g.\ 
Higgs-boson pair production to access the triple-Higgs coupling.
The various channels for Higgs-boson pairs at hadron colliders are illustrated by some
LO diagrams in \reffi{fig:hadroncollHHdiags}.
\bfi
\centerline{
{\unitlength .85pt
\SetScale{0.85}
\begin{picture}(155,100)(0,-10)
\Line(  0, 2)(20, 2)
\Line(  0, 5)(20, 5)
\Line(  0, 8)(20, 8)
\Line(  0,72)(20,72)
\Line(  0,75)(20,75)
\Line(  0,78)(20,78)
\Line(20, 7)(60,-2)
\Line(20, 4)(60,-5)
\Line(20, 1)(60,-8)
\Line(20,80)(60,89)
\Line(20,77)(60,86)
\Line(20,74)(60,83)
\Gluon(20, 7)(50,25){2}{5}
\Gluon(20,73)(50,55){2}{5}
\ArrowLine(50,25)(50,55)
\ArrowLine(50,55)(80,40)
\ArrowLine(80,40)(50,25)
\DashLine(80,40)(115,40){5}
\DashLine(140,60)(115,40){5}
\DashLine(140,20)(115,40){5}
\SetColor{Black}
\Vertex(80,40){2}
\Vertex(115,40){2}
\Vertex(50,55){2}
\Vertex(50,25){2}
\GOval(20, 5)(10,5)(0){.5}
\GOval(20,75)(10,5)(0){.5}
\put( 95,20){${\PH}$}
\put(145,10){${\PH}$}
\put(145,60){${\PH}$}
\put(35,35){${Q}$}
\put(5,-25){\footnotesize (a)}
\end{picture}
\SetScale{1}
}
{\unitlength .85pt
\SetScale{0.85}
\begin{picture}(160,100)(0,-10)
\Line(  0, 2)(20, 2)
\Line(  0, 5)(20, 5)
\Line(  0, 8)(20, 8)
\Line(  0,72)(20,72)
\Line(  0,75)(20,75)
\Line(  0,78)(20,78)
\Line(20, 7)(60, 3)
\Line(20, 4)(60, 0)
\Line(20,77)(60,81)
\Line(20,74)(60,78)
\ArrowLine(20, 5)(50,40)
\ArrowLine(50,40)(20,75)
\Photon(50,40)(90,40){2}{5}
\Photon(120,20)(90,40){2}{5}
\DashLine(90,40)(120,60){5}
\DashLine(145,40)(120,60){5}
\DashLine(145,80)(120,60){5}
\Vertex(50,40){2}
\Vertex(90,40){2}
\Vertex(120,60){2}
\GOval(20, 5)(10,5)(0){.5}
\GOval(20,75)(10,5)(0){.5}
\put( 95,56){${\PH}$}
\put(150,35){${\PH}$}
\put(150,75){${\PH}$}
\put( 52,20){${\PW/\PZ}$}
\put(105, 2){${\PW/\PZ}$}
\put(5,-25){\footnotesize (b)}
\end{picture}
\SetScale{1}
}
{\unitlength .85pt
\SetScale{0.85}
\begin{picture}(160,100)(0,-10)
\Line(  0, 2)(20, 2)
\Line(  0, 5)(20, 5)
\Line(  0, 8)(20, 8)
\Line(  0,72)(20,72)
\Line(  0,75)(20,75)
\Line(  0,78)(20,78)
\Line(20, 7)(60,-2)
\Line(20, 4)(60,-5)
\Line(20,77)(60,86)
\Line(20,74)(60,83)
\ArrowLine(20, 7)(50,10)
\ArrowLine(20,73)(50,70)
\ArrowLine(50,10)(110, 5)
\ArrowLine(50,70)(110,75)
\Photon(50,70)(80,40){2}{5}
\Photon(50,10)(80,40){2}{5}
\DashLine(80,40)(115,40){5}
\DashLine(140,60)(115,40){5}
\DashLine(140,20)(115,40){5}
\Vertex(115,40){2}
\Vertex(80,40){2}
\Vertex(50,70){2}
\Vertex(50,10){2}
\GOval(20, 5)(10,5)(0){.5}
\GOval(20,75)(10,5)(0){.5}
\put( 90,28){${\PH}$}
\put(145,60){${\PH}$}
\put(145,10){${\PH}$}
\put(122,75){${q}$}
\put(122, 0){${q}$}
\put(75,54){${\PW/\PZ}$}
\put(75,14){${\PW/\PZ}$}
\put(5,-25){\footnotesize (c)}
\end{picture}
\SetScale{1}
}
\vspace*{1em}
}
\caption{Representative LO diagrams for the main SM Higgs-boson pair production channels
at hadron colliders, where $q$ and $Q$ denote light and heavy quarks, respectively:
(a) gluon fusion, (b)~Higgs-strahlung, (c)~vector-boson fusion.}
\label{fig:hadroncollHHdiags}
\efi
The cross sections for 
(a)~gluon fusion~\cite{Glover:1987nx,Dawson:1998py},
(b)~Higgs-strahlung~\cite{Barger:1988jk,Djouadi:1999rca},
(c)~vector-boson fusion~\cite{Keung:1987nw,Figy:2008zd}
are, however, very small and further suppressed by branching ratios of the
many-particle final states.
At present, it seems that only gluon fusion may be able to establish
a signal for Higgs-boson pairs and to allow for a qualitative measurement
of the triple-Higgs coupling. 
Higher-order QCD corrections to $\Pg\Pg\to\PH\PH$ are very similar to
single-Higgs production $\Pg\Pg\to\PH$, though kinematically more complicated;
the NLO QCD corrections to gluon fusion to Higgs-boson pairs, which are known
in the heavy-top limit~\cite{Dawson:1998py}, confirm this expectation and 
yield a $K$-factor of the order of $2$.
Based on the $\PH\PW\PW$ decays,
a non-vanishing $H^3$ coupling could be established for $\MH=150{-}200\GeV$
after collecting a high integrated 
luminosity of $300\mathrm{fb}^{-1}$ at the LHC
with design the energy of $14\TeV$~\cite{Baur:2002qd}.
Although the theoretical simulation \cite{Baur:2003gp} suggests that
for a low mass of $\MH\sim120\GeV$ the rare decay 
$\PH\PH\to\Pb\bar\Pb\gamma\gamma$ could establish the coupling for
an integrated luminosity of $600\mathrm{fb}^{-1}$,
an experimental sensitivity study~\cite{ATLAS-collaboration:2012iza}
for an integrated luminosity of $3000\mathrm{fb}^{-1}$
at a CM energy of $14\TeV$ indicates that more than one channel will be
necessary to empirically establish a non-vanishing triple-Higgs-boson coupling. 

Since triple-Higgs-boson production cross sections are way too 
small~\cite{Plehn:2005nk} at hadron colliders as well,
an experimental test of quartic Higgs couplings at colliders is out of reach.

\section{Searches at Tevatron}

The Tevatron Run II started in $2001$ with a pilot run at a new CM energy of $1.96\TeV$. 
From spring $2002$ until its shutdown in autumn $2011$ it provided proton--antiproton collisions 
corresponding to a data set of up to $10\ifba$ to the two experiments CDF and D0. In contrast to 
the searches at LEP, where the a~priori signal-to-background ratio was at the level of $10^{-4}$ to $10^{-6}$ 
as both signal and background processes are mediated by the electroweak interaction, the huge total 
inelastic cross section due to processes mediated by the strong force is more than ten orders of magnitude 
higher than the cross section for Higgs-boson production (see \reffi{fig:sigbgxs}). Owing to the large cross 
section, the instantaneous luminosity of up to $4\times 10^{32}\,\mathrm{cm^{-2}s^{-1}}$ and the short time interval 
between bunch crossings of $396\,\mathrm{ns}$ not all interactions can be recorded.
In order to limit the recorded data volume to a manageable size severe requirements on the 
presence and quality of final-state objects have to be imposed already at the trigger level. This online selection for 
Higgs-boson searches relies mostly on the presence of a colour-neutral object in the final state, 
i.e.\ photons, electrons, muons, or significant missing transverse momentum arising from neutrinos. 
Hence, although Higgs-boson topologies with 
fully hadronic final states, which comprise production in gluon
fusion with subsequent decays via  $\Htobb$ or $\PH \to \PW \PW (\PZ \PZ) \to4\,$quarks,
lead to the highest production rates, these 
cannot be exploited at Tevatron. Exceptions to this rule of thumb are 
the searches in the 
associated production $\PW\PH (\PZ\PH) \to \Pq\Pqbar \Pb\Pbbar$ and in vector-boson fusion 
with the decay $\PH\to\Pb\Pbbar$ performed by CDF~\cite{Aaltonen:2012ji}, where events were recorded by requiring 
multi-jets and a large scalar sum of transverse energies at the trigger level. 

\subsection{Overview of search channels}

Generically, the Higgs-boson mass range $100{-}200\GeV$ was investigated by both experiments CDF and D0. An overview of the 
production rates for most of the search channels considered at Tevatron is shown in \reffi{fig:tevatronrates}. 
\begin{figure}
  \centering
  \begin{minipage}[b]{.75\textwidth}
    \includegraphics[width=0.97\textwidth]{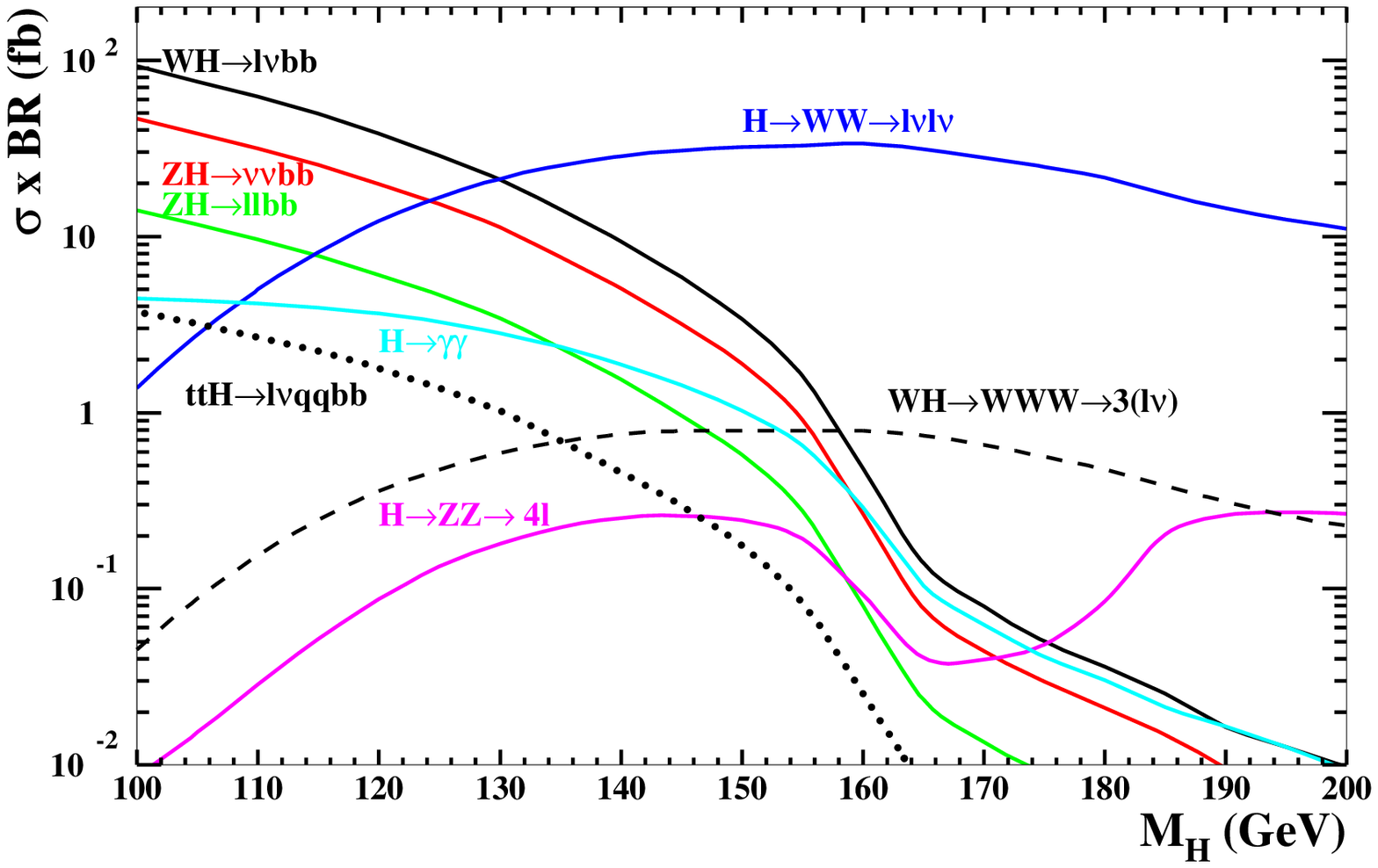}
  \end{minipage}
\vspace*{-1em}
  \caption{Production cross sections in the most sensitive search channels at the Tevatron. The numbers are extracted from~\citere{:2012zzl}.}
  \label{fig:tevatronrates}
\end{figure}
For $\MH \le 130\GeV$ the most abundant and most sensitive search channels are $\Htobb$ produced in association with 
a weak gauge boson $\PW$ or $\PZ$, which decay leptonically via $\PW \to \Pl \Pn$, $\PZ \to \Plp\Plm$, or $\PZ \to \Pn\Pnbar$. 
Here and in the following $\Pl$ denotes an electron or muon. The relative magnitude of the rates is due to different couplings 
of the Higgs bosons to $\PW$ and $\PZ$ and the branching ratios BR$(\PW \to \Pl \Pn)=22\%$,  BR$(\PZ \to \Plp \Plm)=6.7\%$,
and BR$(\PZ \to \Pn \Pnbar)=20\%$. For $\MH$ in the range $130{-}200\GeV$ the sensitivity is dominated by the search 
for $\Htowwll$ produced in gluon fusion. The discussion below focuses on these four final states. 
In addition to those high-sensitivity channels a plenitude of other channels was analyzed which yield  
cross-section limits for Higgs-boson production with a value of at best 7 times the SM prediction, and a 
sensitivity of a single channel (expressed in expected limits) of at least a factor 5 worse compared to a combination 
of the four channels mentioned above.  Those channels include 
$\Htogg$ (CDF~\cite{Collaboration:2012pa} and D0~\cite{Abazov:2011ix}),
$\Htofl$ (CDF~\cite{Aaltonen:2012ya}), 
$\Pq\Pqbar\Pb\Pbbar$ from $\PW\PH (\PZ\PH)$ and in VBF with the decay $\PH\to\Pb\Pbbar$ (CDF~\cite{Aaltonen:2012ji}),
$\Pt\Ptbar\PH$ with $\PH\to\Pb\Pbbar$ (CDF~\cite{Collaboration:2012bk}), 
$\Htott$ (CDF \cite{Aaltonen:2012jh}, D0~\cite{Abazov:2012zj}),
$\Htowwlnqq$ (D0~\cite{Abazov:2011bc}) and $\PW(\PZ)\PH\to\PW(\PZ)\PW\PW\to\Plpm\Plpm+X$ (CDF~\cite{Aaltonen:2010cm}, and
D0~\cite{Abazov:2011ed}). Additional channels are included in the preliminary combinations in each experiment and 
the Tevatron combination.
All those channels contribute to the overall sensitivity at the level of $10{-}20$\%. In most analyses the final discriminant
is given by an event classifier based on the combination of multivariate techniques, which combines information from kinematical 
properties, such as observables sensitive to the Higgs-boson mass and topological information  from flavour tagging. 
Usually the pre-selected event samples, where the presence of the basic physics objects of the final state under 
consideration are required, are divided in categories based on jet multiplicity, lepton flavour, multiplicity of 
b-tagged jets and the tightness of their selection requirements, and on additional kinematical properties, such as low or large 
missing transverse momentum. Often the selection strategy using multivariate techniques is a multi-step procedure. 
In the first step either dominant reducible backgrounds are greatly reduced, or the events are divided in categories,
in which a specific  background is enriched or depleted. In the next selection step the signal can then be separated 
from the remaining backgrounds in a more efficient way. Sometimes a sequence of multivariate event classifiers 
is also applied, 
which one after the other try to suppress the dominant background classes. The final discriminants are then constructed 
from the responses of various multivariate techniques. This sophisticated selection procedure improves significantly  
the signal-to-background ratios in particular categories with respect to a more inclusive analysis and hence 
the overall sensitivity of the combined search. Higgs-boson production is simulated with 
\Pythia~\cite{Sjostrand:2006za,Sjostrand:2007gs}, to which a reweighting of the transverse-momentum spectrum of the 
Higgs boson is applied. QCD multi-jet production is estimated entirely using daten-driven techniques. 
Other backgrounds are estimated from simulated events, which are generated with a variety of event generators such as 
\Alpgen~\cite{Mangano:2002ea},  \Comphep~\cite{Boos:2004kh}, \Herwig~\cite{Corcella:2000bw,Bahr:2008pv}, 
\MCatNLO~\cite{Frixione:2002ik,Frixione:2010wd}, \Pythia~\cite{Sjostrand:2006za,Sjostrand:2007gs}, and 
\Sherpa~\cite{Gleisberg:2008ta}. The misidentification probabilities for leptons 
and b-flavoured jets, as measured in data, are applied to the simulation.

\subsection[Searches for $\mathrm{H\to b \bar b}$]{\boldmath{Searches for $\mathrm{H\to b \bar b}$}}

The searches for $\PH\to\Pb\Pbbar$ in the mass range $100{-}150\GeV$ are performed in three non-overlapping 
final-state topologies: (a) $1$ lepton and large missing transverse momentum designed for 
$\PW\PH\to \Pl\Pn\Pb\Pbbar$, but also selecting contributions from $\PZ\PH\to \Plp\Plm\Pb\Pbbar$, where one 
lepton is not reconstructed and identified, (b) $2$ leptons with an invariant mass consistent with $\MZ$ 
optimized for $\PZ\PH\to \Plp\Plm\Pb\Pbbar$ production, and (c) $0$ leptons and large missing transverse 
momentum designed for $\PZ\PH\to \Pn\Pnbar\Pb\Pbbar$, but also receiving contributions from 
$\PW\PH\to \Pl\Pn\Pb\Pbbar$ and $\PZ\PH\to \Plp\Plm\Pb\Pbbar$, where one or two leptons are not 
reconstructed and identified. The suppression of background processes, in particular from 
$\PW(\PZ)+\mbox{light-flavour production}$, depends crucially on the performance 
of the algorithm to identify the $\Pb$-flavoured jets from the Higgs-boson decay. Significant improvements 
were achieved over the last years, which greatly enhanced the sensitivity. The mass of the Higgs-boson 
candidate can be reconstructed from the invariant mass of the two jets assigned to the Higgs-boson decay with 
a mass resolution of $10{-}15$\%. This di-jet mass is an important observable to discriminate the signal 
process from background processes, in particular from di-boson production VZ with $\PZ \to \Pb\Pbbar$. 
In addition to the multiplicity of leptons and the presence of missing transverse momentum exactly two or three jets are 
required with at least one being identified as b-flavoured. The events are further divided into categories depending on lepton 
flavour, lepton quality, b-tag multiplicity, and b-tag quality, in order to improve the signal-to-background ratio. 
For each category and Higgs-boson-mass hypothesis multivariate event classifiers are trained, which use the 
di-jet invariant mass and other kinematical quantities as input observables.  The response of these multivariate techniques is used as the 
final discriminant. For the purpose of visualization of the sensitivity and results, 
especially when several channels and categories are combined, the bins of each final discriminating observable 
are ordered with respect to
the expected signal-to-background ($\mathrm{s/b}$) ratio. The expected and observed event yields in each 
$\ln(\mathrm{s/b})$ 
bin are then added over the discriminating variables in all non-overlapping final states and event categories. 

The expected and observed limits on the signal-strength parameter $\mu=\sigma/\sigma_{\SM}$
(called $R$ in the Tevatron analyses), which relates the excluded cross section $\sigma$ to the SM
expectation $\sigma_{\SM}$ for Higgs-boson production, from the analysis of full data set collected during Run~II at Tevatron 
($9.45\ifba$ in CDF and $9.5{-}9.7\ifba$ in D0)
are summarized in \refta{tab:tevhbblim}. 
\begin{table}
\caption{Expected and observed limits on the signal-strength parameter $\mu=\sigma/\sigma_{\SM}$ in the search for $\Htobb$ in D0 and CDF.
The entry ``--'' indicates that this information is not available.}
\label{tab:tevhbblim}
\begin{center}
\begin{tabular}{c|c|c}\hline
                             &CDF                                                     & D0                                \\  
channel                      & obs.~(exp.) limit  & obs.~(exp.) limit  \\[-.2em] 
                             & at $\MH=115/125\GeV$ & at $\MH=115/125\GeV$  \\ \hline
\PW\PH\ (\Htobb, $\PW\to \Pl\Pn)$ & 3.1 (2.0)    /  4.9 (2.8)  \cite{Aaltonen:2012ic}  & 3.7 (3.2) / 5.2 (4.7)  \cite{:2012meb}             \\
\PZ\PH\ (\Htobb, $\PZ\to \Pn\Pn)$ & 2.7 (2.7)    /  6.7 (3.6)  \cite{Aaltonen:2012ii}  & 3.0 (2.7) / 4.3 (3.9)  \cite{Collaboration:2012hv}  \\
\PZ\PH\ (\Htobb, $\PZ\to \Pl\Pl)$ & 4.7 (2.7)    /  7.1 (3.9)  \cite{Aaltonen:2012id}  & 4.3 (3.7) / 7.1 (5.1)  \cite{:2012kg}               \\  \hline
combined                     & --            /   --                                   & 1.9 (1.6) / 3.2 (2.3)  \cite{:2012tf}               \\  \hline
\end{tabular}
\end{center}
\end{table}     
In both experiments and in all three search topologies the observed limit is 
weaker than the expected one over a broad mass range. The deviation is less than $1$ standard deviation over the full 
mass range in the D0 $\Pl\Pn\Pb\Pbbar$~\cite{:2012meb} and $\Pn\Pn\Pb\Pbbar$ searches~\cite{Collaboration:2012hv}, 
and within $2$ standard deviations in the 
D0 $\Plp\Plm\Pb\Pbbar$~\cite{:2012kg} search and the CDF $\Pl\Pn\Pb\Pbbar$~\cite{Aaltonen:2012ic} 
and $\Pn\Pnbar\Pb\Pbbar$~\cite{Aaltonen:2012id} searches. In the CDF $\Plp\Plm\Pb\Pbbar$~\cite{Aaltonen:2012id} search 
a maximum excess corresponding to a significance of 2.4 is observed at $\MH$= $135\GeV$, reduced to 2.1 when taking into 
account the ``look-elsewhere effect'' (LEE).
When combining the three channels, CDF~\cite{Aaltonen:2012if} observes a limit weaker than the expected 
by 2 standard deviations for $\MH = 110{-}150\GeV$  and a  maximal deviation from the background-only
hypothesis for $\MH = 135\GeV$ with a significance of $2.7$ ($2.5$ after including the LEE in the mass range $115{-}150$). In D0~\cite{:2012tf} 
the combination yields a limit weaker than the expected by $1{-}1.7$ standard deviations in the mass range $120{-}145\GeV$  
and a  maximal deviation from the background-only hypothesis for $\MH=135\GeV$ with a significance of 1.7 (1.5 with LEE in the 
mass range $115{-}150\GeV$). D0 excludes the mass range $\MH = 100{-}102\GeV$. Both experiments conclude 
that the excess is compatible with the production of the SM Higgs boson with a mass of $125 \GeV$ within large uncertainties.
The best fit values for the signal strength at $\MH=125\GeV$ is $\mu=1.2^{+1.2}_{-1.1}$ in D0, and in CDF $\mu$ is approximately 
2.5 consistent with the SM expectation of $1$ within 2 standard deviations.

The combination of the observed event yields from CDF and D0~\cite{Aaltonen:2012qt} in the $\ln(\mathrm{s/b})$ distribution of 
$\Htobb$ searches after subtraction of the expected background in \reffi{fig:tevbbres} (top-left) shows a positive 
deviation from the expectation of zero for the background-only hypothesis in the bins of large signal-to-background ratio.
\begin{figure}
  \centering
  \begin{minipage}[b]{.47\textwidth}
    \includegraphics[width=0.97\textwidth]{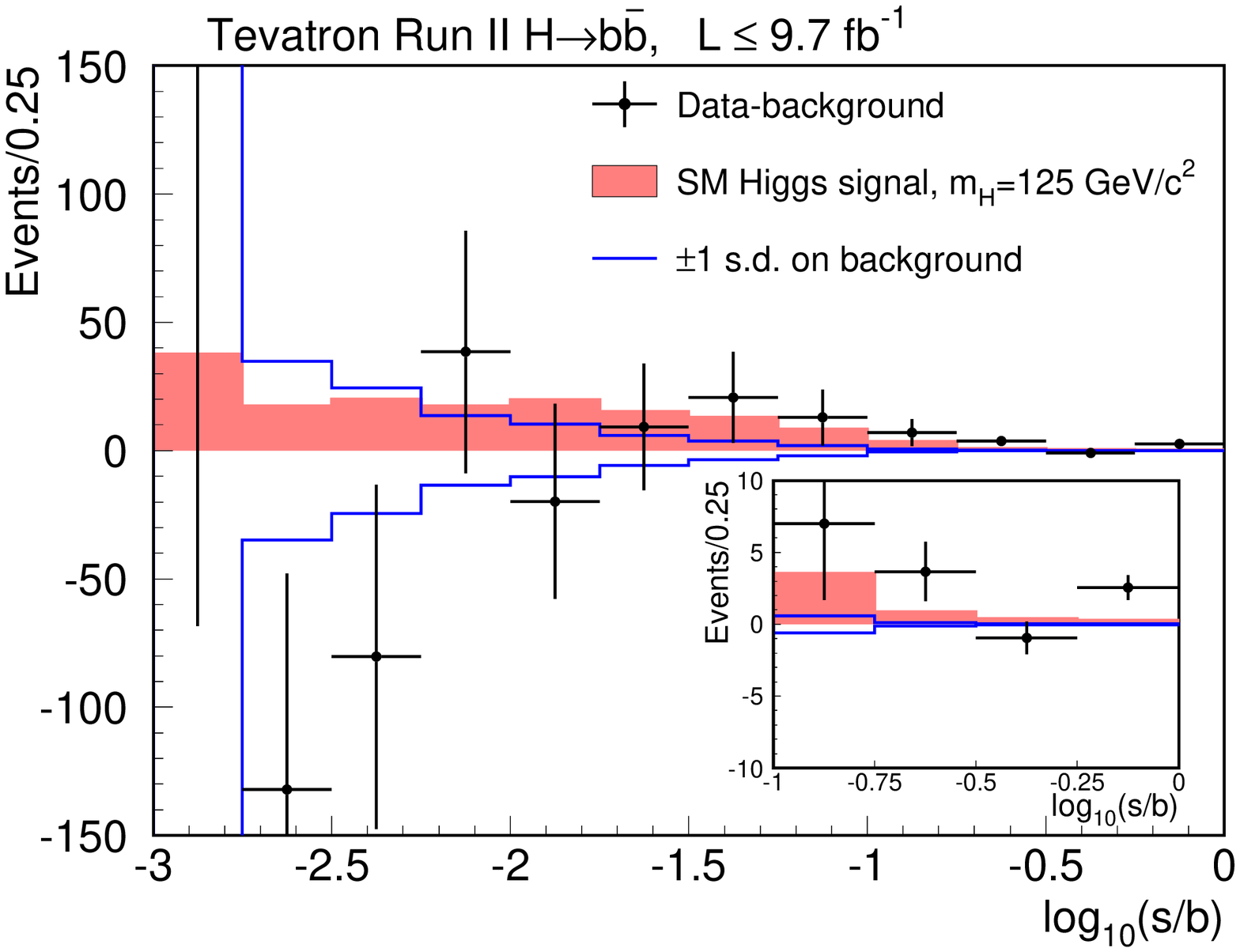}
  \end{minipage}
  \begin{minipage}[b]{.47\textwidth}
    \includegraphics[width=0.97\textwidth]{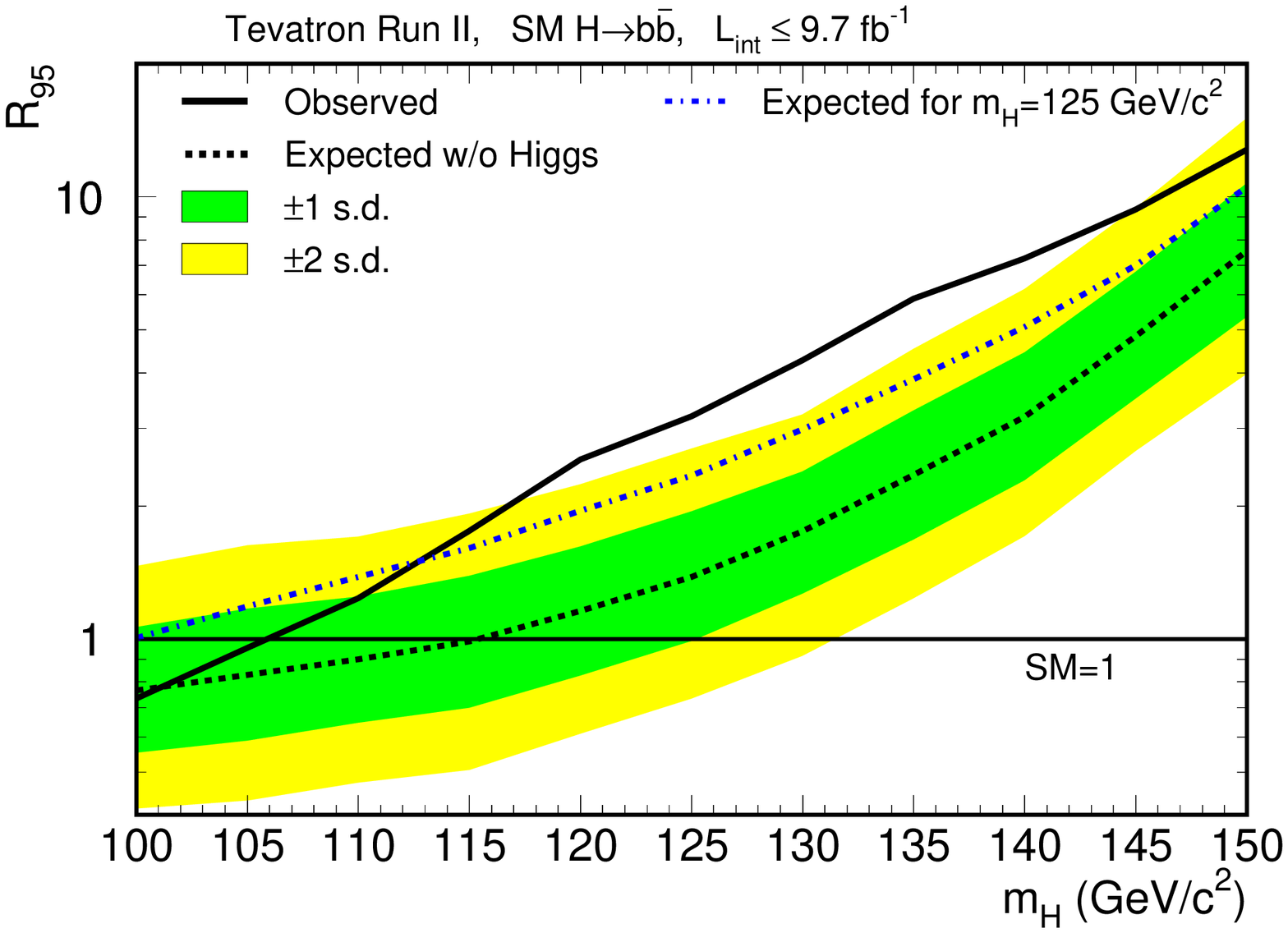}
  \end{minipage}
  \begin{minipage}[b]{.47\textwidth}
    \includegraphics[width=0.97\textwidth]{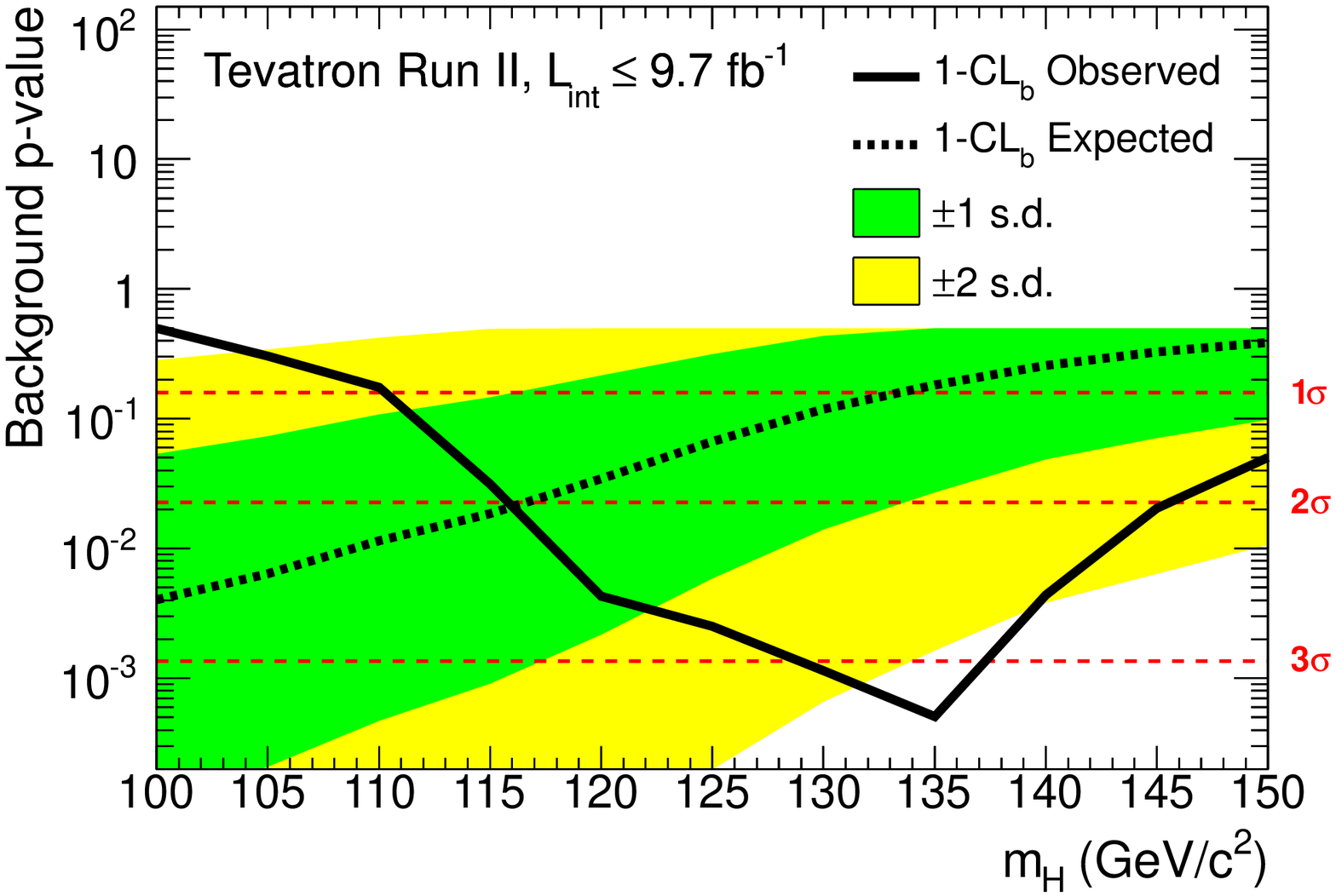}
  \end{minipage}
  \begin{minipage}[b]{.47\textwidth}
    \includegraphics[width=0.97\textwidth]{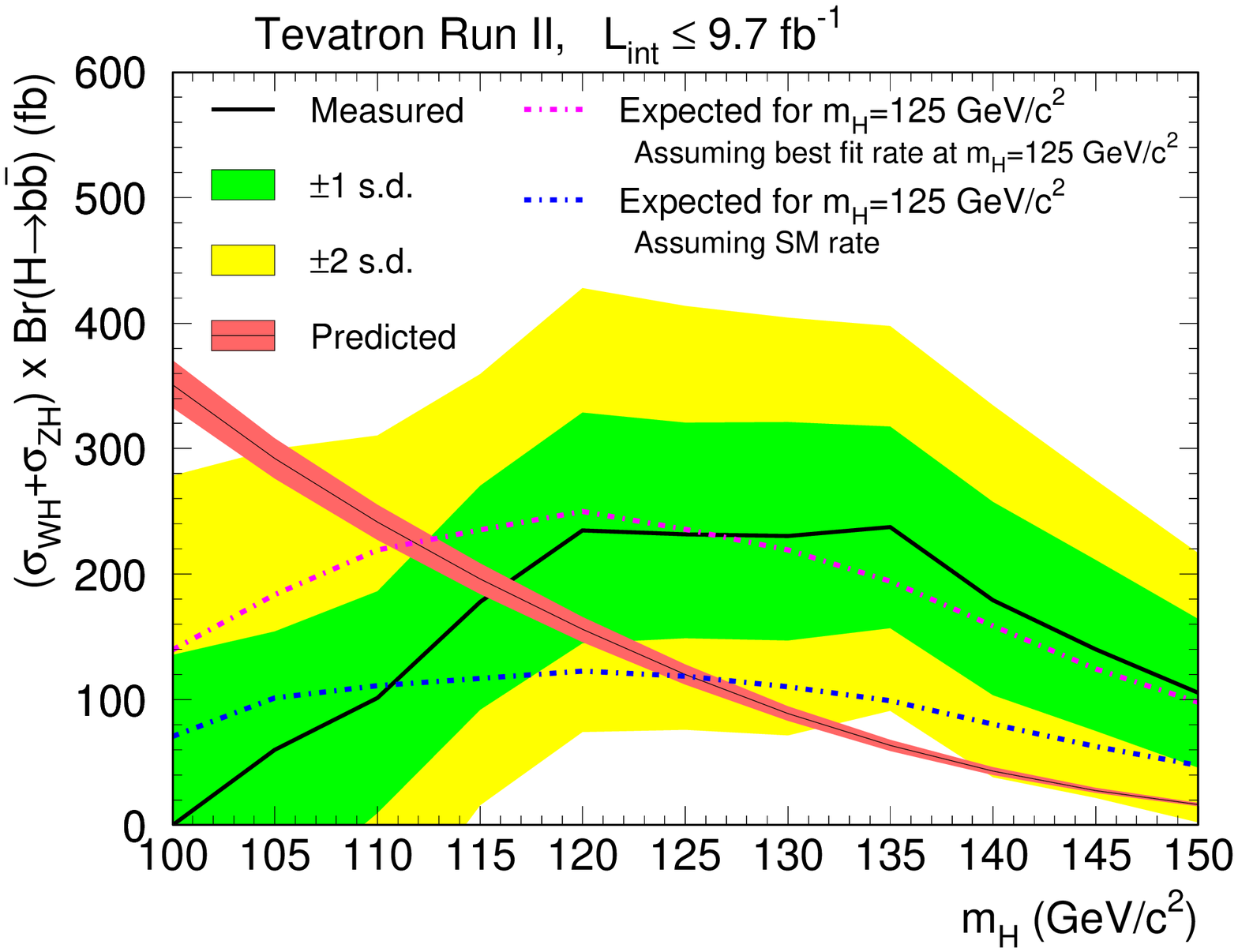}
  \end{minipage}
  \caption{Results of the search for $\Htobb$ at Tevatron from the combination of CDF and D0 full Run~II data~\cite{Aaltonen:2012qt}.
   Expected and observed event yields classified by signal-to-background ratio 
   after background subtraction (top-left), exclusion limits on the signal strength $R\equiv\mu=\sigma/\sigma_{\SM}$
   as function of $\MH$ (top-right), local $p$-value for the background-only
   hypothesis as function of $\MH$ (bottom-left), and best fit values for the signal cross section
   as a function of $\MH$, compared to the expectation for a SM Higgs boson as a function of $\MH$
   and the expectations for a SM Higgs boson with mass of $125\GeV$ evaluated at different $\MH$ (bottom-right).}
  \label{fig:tevbbres}
\end{figure}
The observed limit on the signal strength $\mu$ 
is weaker than the expected by more than two standard deviations in the range 
$115{-}145\GeV$, as shown in \reffi{fig:tevbbres} (top-right). Under the background-only hypothesis 
one expects to exclude the Higgs-boson-mass hypothesis in the range $100{-}116\GeV$. Assuming a Higgs boson with mass of 
$125\GeV$ no excluded range is expected. The data allow the mass hypothesis below $106\GeV$ to be excluded.   
The local $p$-values in testing the background-only hypothesis as a function of $\MH$ (see \reffi{fig:tevbbres} top-left) 
show deviations from the background-only hypothesis with a significance of larger than 2   in the mass range $115{-}145\GeV$ 
with the smallest $p$-value at $135\GeV$ corresponding to a local significance of $3.3$ ($2.8$ at $\MH=125\GeV$). 
After taking into account the 
LEE in the mass range $115{-}150\GeV$ the significance is decreased to $3.1$. The best fit value for  the signal cross section 
obtained in data is compared to various types of expectations for signal+background hypothesis in \reffi{fig:tevbbres} 
(bottom-right). The fit values agree with the expectation for the SM Higgs boson at the $68$ ($95$)\% CL
in the approximate mass ranges $112{-}120\GeV$ ($105{-}132\GeV$ and $140{-}150\GeV$). As $\MH=135\GeV$ was already excluded by 
both LHC experiments and guided by the excess observed in each experiment at the LHC with a local significance 
of approximately $3$ at $\MH\approx125\GeV$
already in the analysis of the $2011$ data, the Tevatron experiments quote a best fit value for the signal 
cross section  $(\sigma_{\PW\PH}+\sigma_{\PZ\PH})\times\mathrm{BR}(\Htobb)$
at $\MH =125\GeV$ of  $0.23^{+0.09}_{-0.08}\pba$ to be compared to
the SM Higgs boson of $0.12\pm 0.01\pba$, consistent within $1.5$ standard deviations.
Despite having no sensitivity to exclude the SM Higgs boson 
hypothesis for masses larger than $116\GeV$, the Tevatron observes a broad excess in the $\Htobb$ final state, 
consistent with a wide range of SM Higgs-boson-mass hypothesis. Based on these findings the Tevatron experiments reported
evidence for the production of a new particle produced in association with weak bosons and decaying to a b-quark 
pair~\cite{Aaltonen:2012qt}.

\subsection[Searches for $\mathrm{H \to W^+ W^- \to \ell^+ \nu \ell^- \bar\nu}$]{\boldmath{Searches for $\mathrm{H \to W^+ W^- \to \ell^+ \nu \ell^- \bar\nu}$}}

The final-state topology is characterized by two isolated leptons of opposite charge and large missing transverse momentum.
Due to the presence of two neutrinos in the final state the mass of the Higgs-boson candidates cannot be 
reconstructed and only the transverse mass of the di-lepton and missing-transverse-momentum system can be reconstructed. 
Observables that allow to suppress background processes comprise, among others, 
the di-lepton invariant mass, the transverse mass, 
and the azimuthal opening angle between the two leptons~\cite{Dittmar:1996ss}. The latter observables exploit the scalar 
nature of the Higgs boson and the parity violating $V-A$ structure of the $\PW\to\Pl\Pn$ decay. Due to the spin correlations 
the signal accumulates at small values of the opening angle and di-lepton invariant mass. The search is optimized in different 
categories, distinguished, e.g., by lepton flavour and jet multiplicity (0, 1, $\ge$2). 
The latter aims to enhance the 
sensitivity for associated $\PW\PH(\PZ\PH)$ and VBF production with two quarks in the final state. The composition of the
remaining background is very different in the different jet categories: In the 0 (2) jet category the dominating 
background is given by $\PW\PW$ ($\Pt\Ptbar$) production. The final discriminating observables based on multivariate 
techniques in each event topology, which are separately optimized for all Higgs-boson-mass hypotheses tested, are used as 
input for the statistical interpretation of the results. In addition searches for final states with two leptons with same 
electric charge or with three leptons were performed to enhance the sensitivity to $\PW\PH(\PZ\PH)$ production,
where the associated $\PW (\PZ)$ boson decays to one (two) charged leptons. 
Due to the comparably large production rate, 
$\Htowwll$ was the first channel to reach the sensitivity to exclude SM Higgs-boson
cross sections in the vicinity of $\MH=160\GeV$. The only published combined Tevatron 
result~\cite{Aaltonen:2010yv} including $\Htowwll$ search channels from spring 2010, 
which is based on a data sets of up to $4.8\ifba$ in CDF \cite{Aaltonen:2010cm} 
and $5.4\ifba$ in D0 \cite{Abazov:2010ct}, excludes $\MH=162{-}166\GeV$ with an expected exclusion of $\MH=159{-}169\GeV$
(see \reffi{fig:tevwwpub}, right). 
\begin{figure}
  \centering
  \begin{minipage}[b]{.3\textwidth}
   \raisebox{0em}{\includegraphics[width=0.97\textwidth,height=5.5cm]{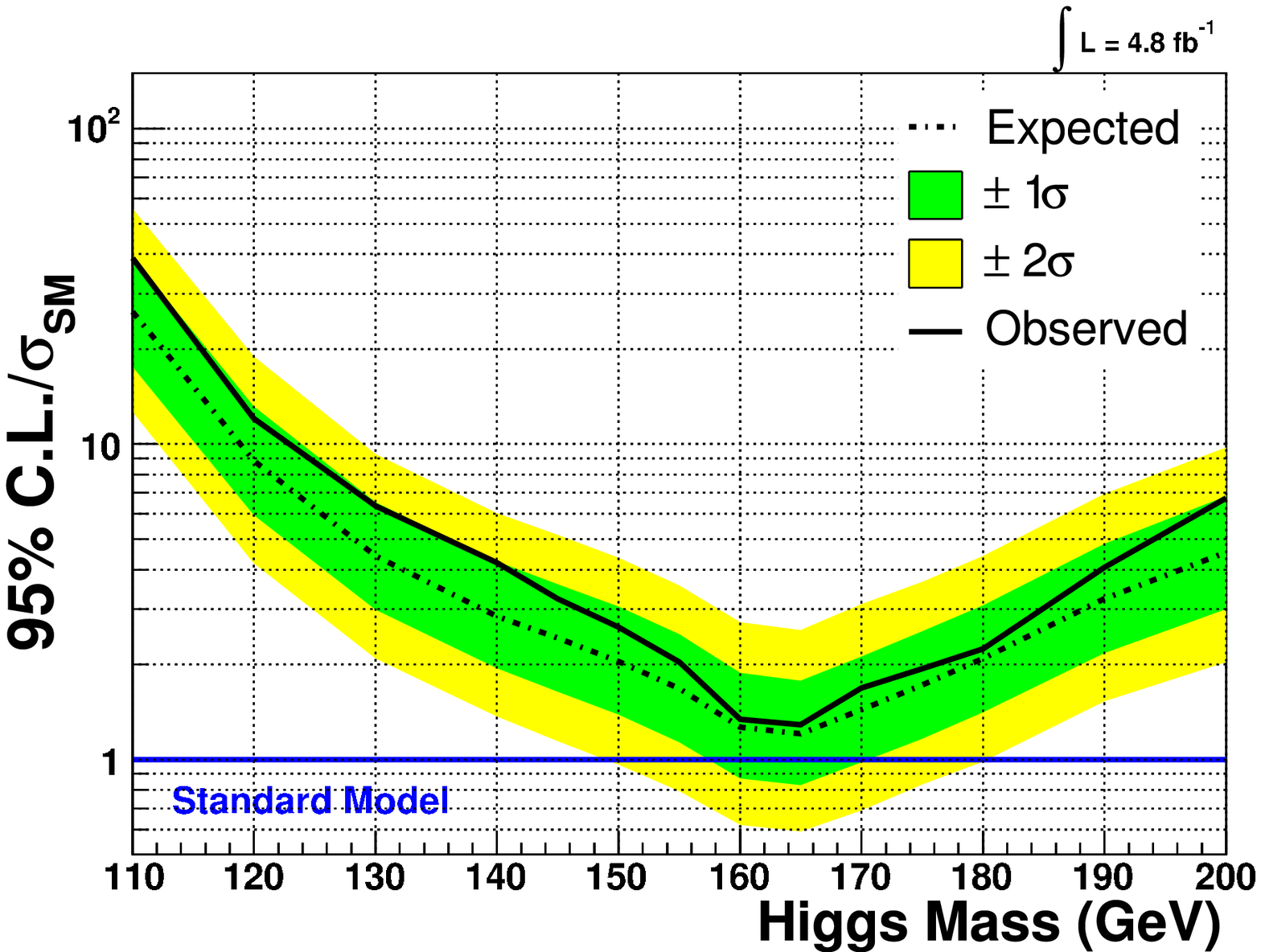}}
  \end{minipage}
  \begin{minipage}[b]{.3\textwidth}
    \raisebox{-.7em}{\includegraphics[width=0.97\textwidth,height=5.5cm]{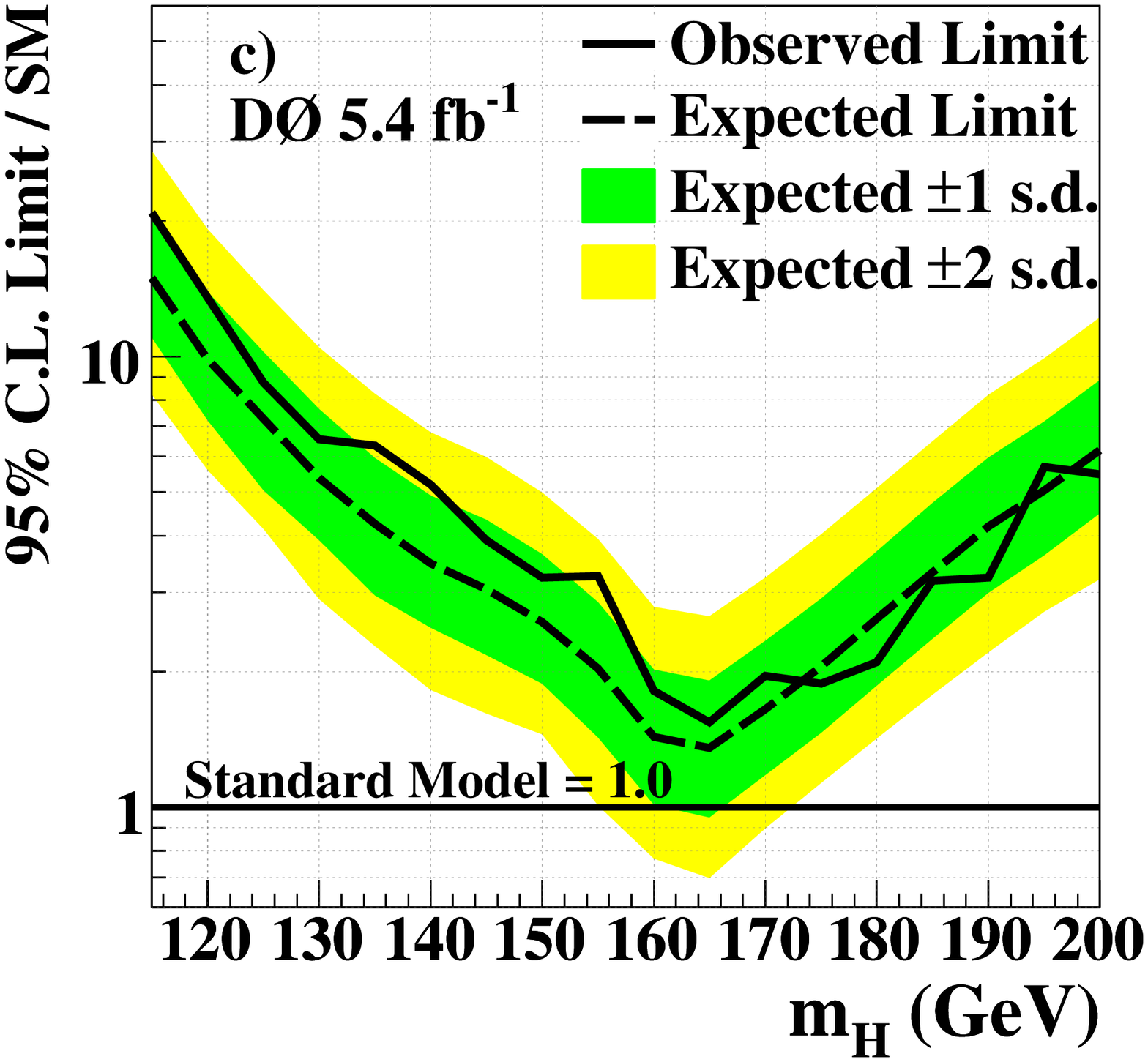}}
  \end{minipage}
  \begin{minipage}[b]{.3\textwidth}
    \raisebox{-.9em}{\includegraphics[width=0.97\textwidth,height=5.5cm]{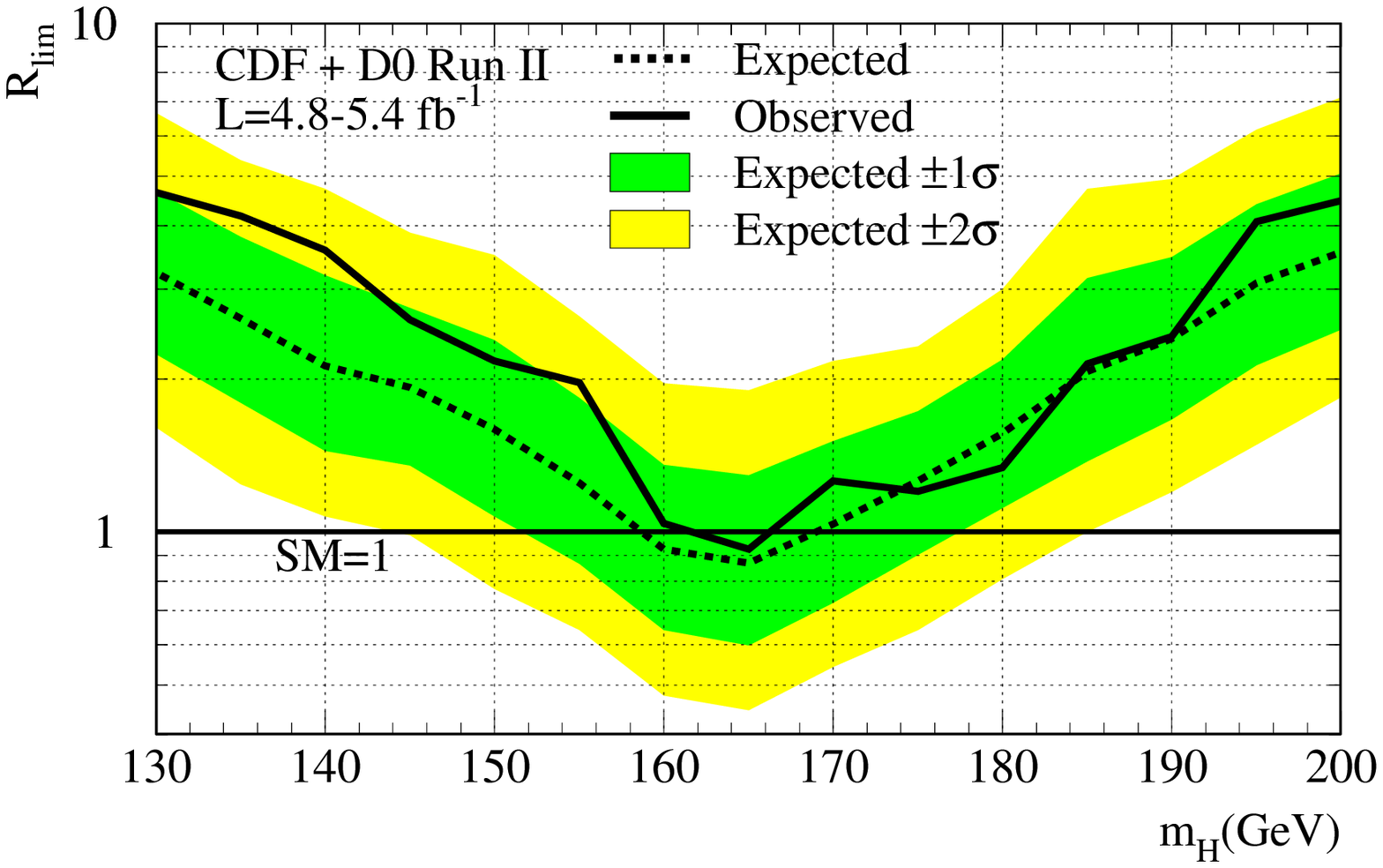}}
  \end{minipage}
  \caption{Cross-section limits with respect to the SM expectation 
   for $\Htowwll$ production in CDF~\cite{Aaltonen:2010cm} (left), 
   in D0 \cite{Abazov:2010ct}~(middle), and in the first Tevatron combination~\cite{Aaltonen:2010yv} (right).}
  \label{fig:tevwwpub}
\end{figure}
This was the first time that Tevatron expected 
to exclude a Higgs-boson mass range. The individual experiments were not yet able to exclude a mass range, but gave observed 
(expected) limits on the signal-strength parameter at $\MH=165\GeV$ of 1.29 (1.20) in CDF (see \reffi{fig:tevwwpub}, left) 
and  1.55 (1.36) in D0 (see \reffi{fig:tevwwpub}, middle), to be compared with the Tevatron combination 0.93 (0.87). 
Since then, many preliminary updates of searches for $\Htowwll$ were performed extending the sensitivity for SM
Higgs-boson production to a mass window of size of roughly $20\GeV$ around $2\MW$. Only D0 published a new result 
based on a data set of $8.6\ifba$~\cite{:2012qq} yielding an expected excluded mass range of $\MH=159{-}169\GeV$.
However, no mass range could be excluded due to an upwards fluctuation of the observed event yields by one standard deviation
with respect to the background expectation.

\subsection{Preliminary combined results}
Over the years the findings in the plenitude of the decay channels and final-state topologies were
combined regularly by CDF and D0 separately, and also in a common effort as the Tevatron combination. 
The correlations in the systematic uncertainties across channels and also across experiments were
carefully taken into account. In the following the latest preliminary Tevatron combination (July 2012)~\cite{:2012zzl}
for the mass range $\MH=100{-}200\GeV$ is briefly discussed. A comparison of the expected and observed event yields classified 
by signal-to-background ratios after background subtraction for $\MH=125\GeV$ and $\MH=165\GeV$ is shown
in \reffi{fig:tevprel2012} (top row). 
\begin{figure}
  \centering
  \begin{minipage}[b]{.47\textwidth}
    \includegraphics[width=0.97\textwidth]{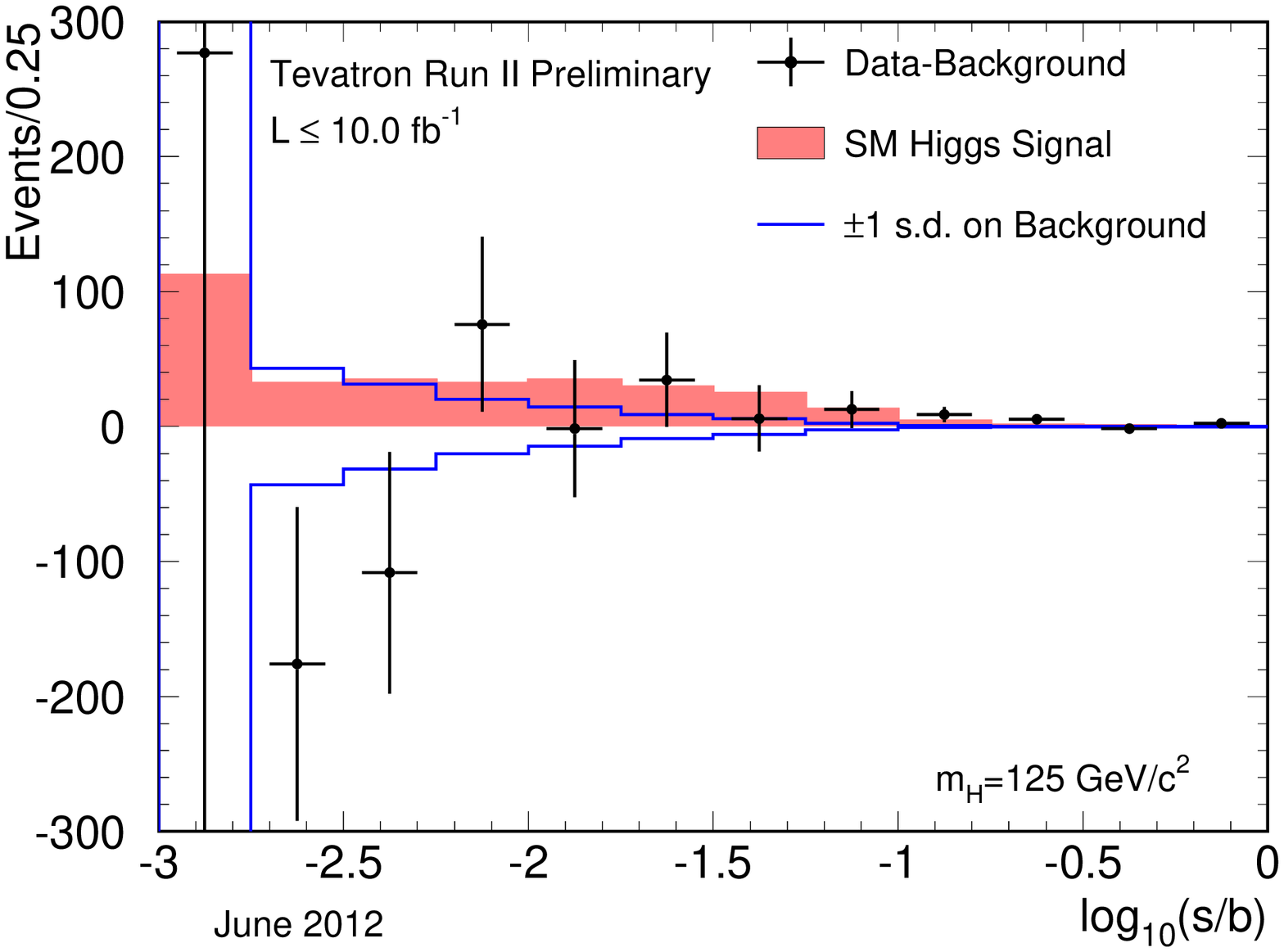}
  \end{minipage}
  \begin{minipage}[b]{.47\textwidth}
    \includegraphics[width=0.97\textwidth]{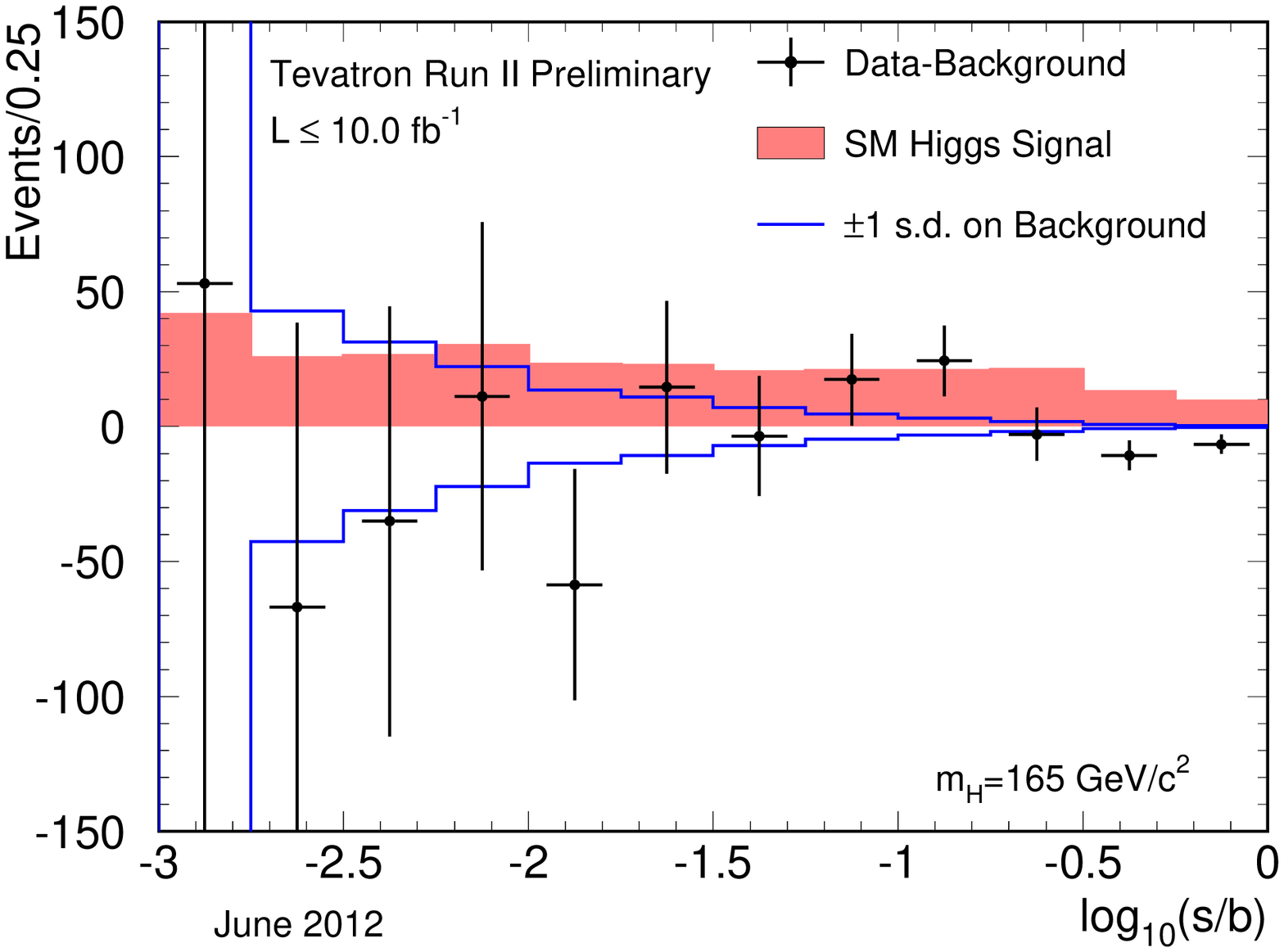}
  \end{minipage}
  \begin{minipage}[b]{.47\textwidth}
    \includegraphics[width=0.97\textwidth]{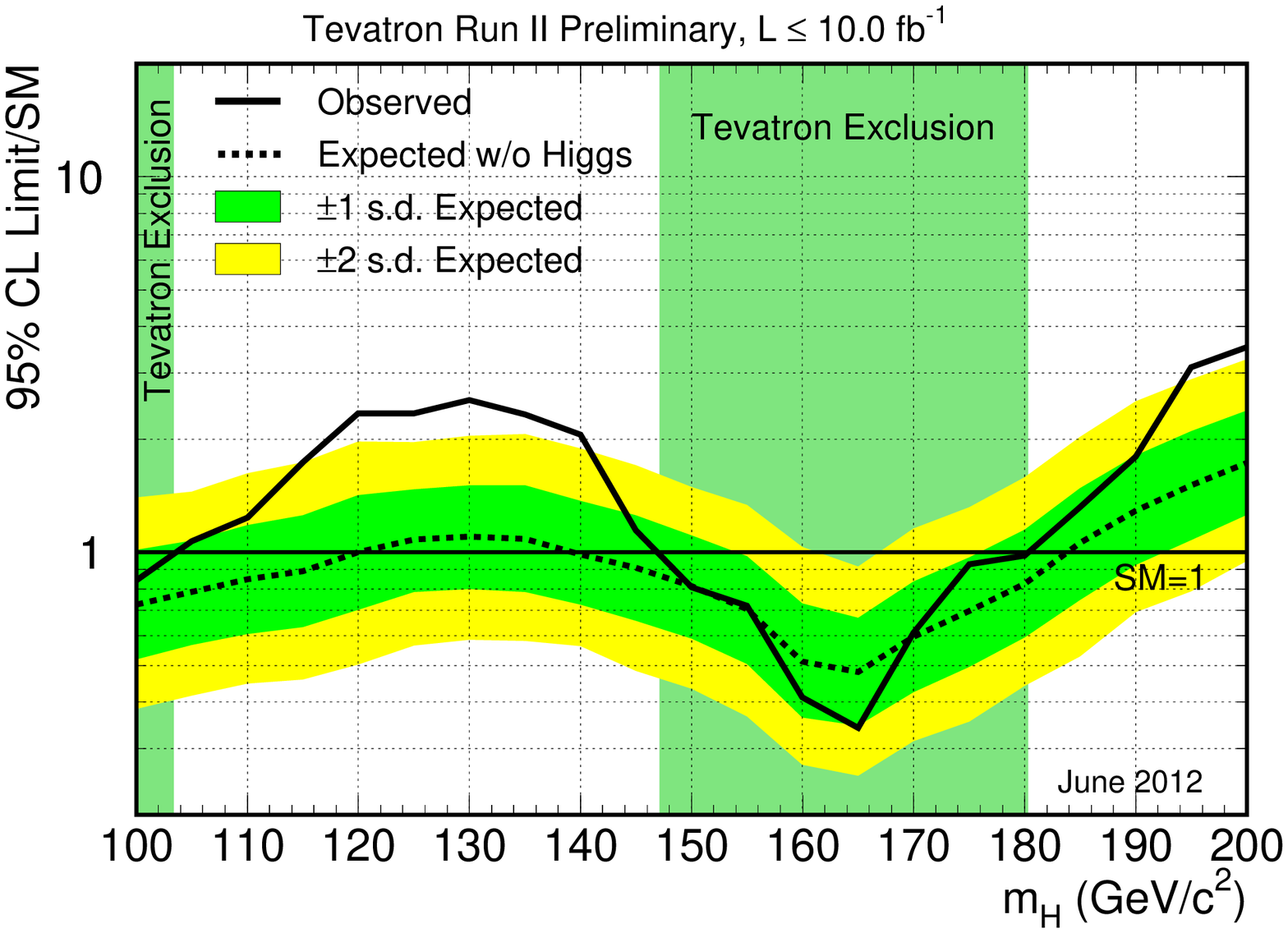}
  \end{minipage}
  \begin{minipage}[b]{.47\textwidth}
    \includegraphics[width=0.97\textwidth]{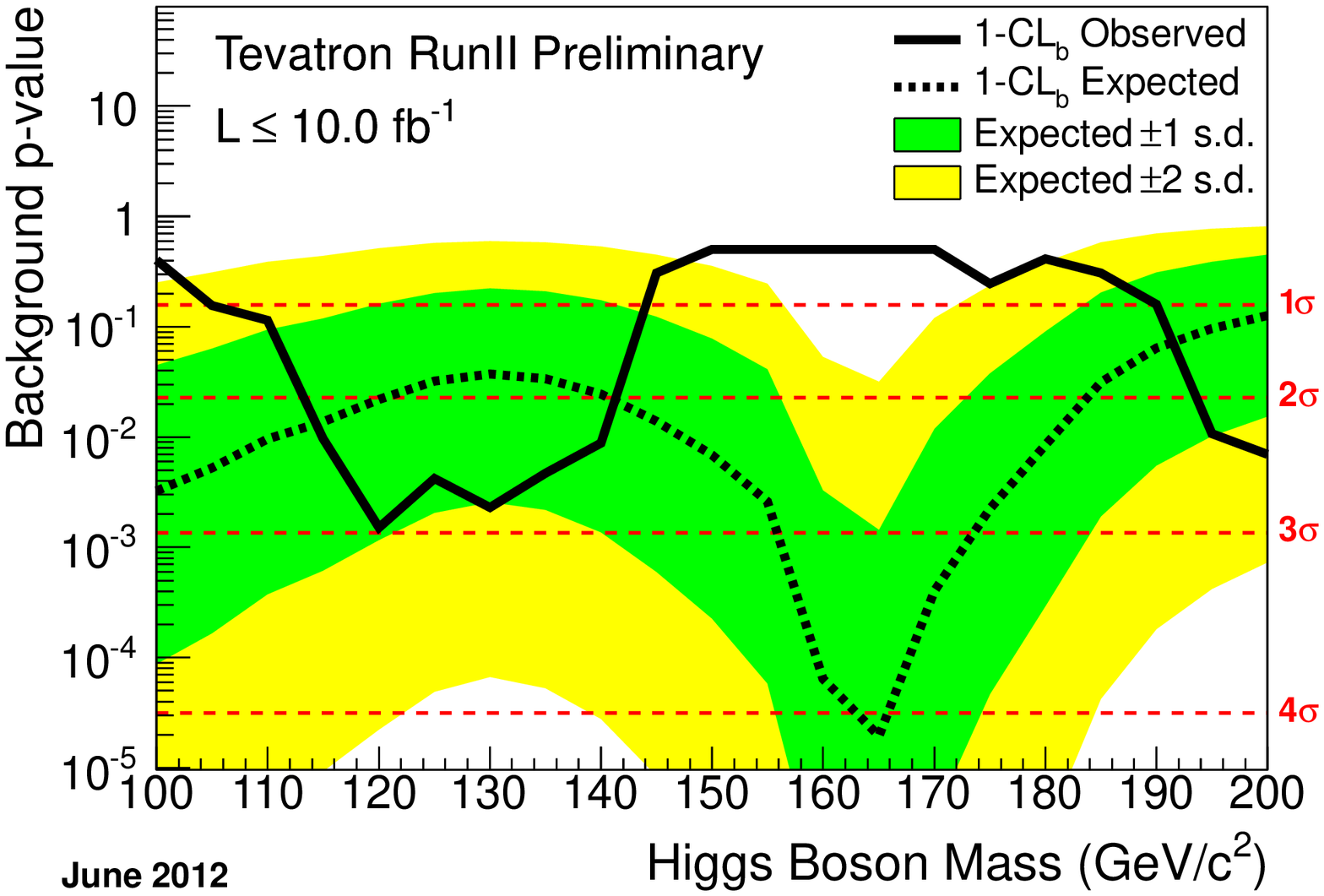}
  \end{minipage}
  \begin{minipage}[b]{.47\textwidth}
    \includegraphics[width=0.97\textwidth]{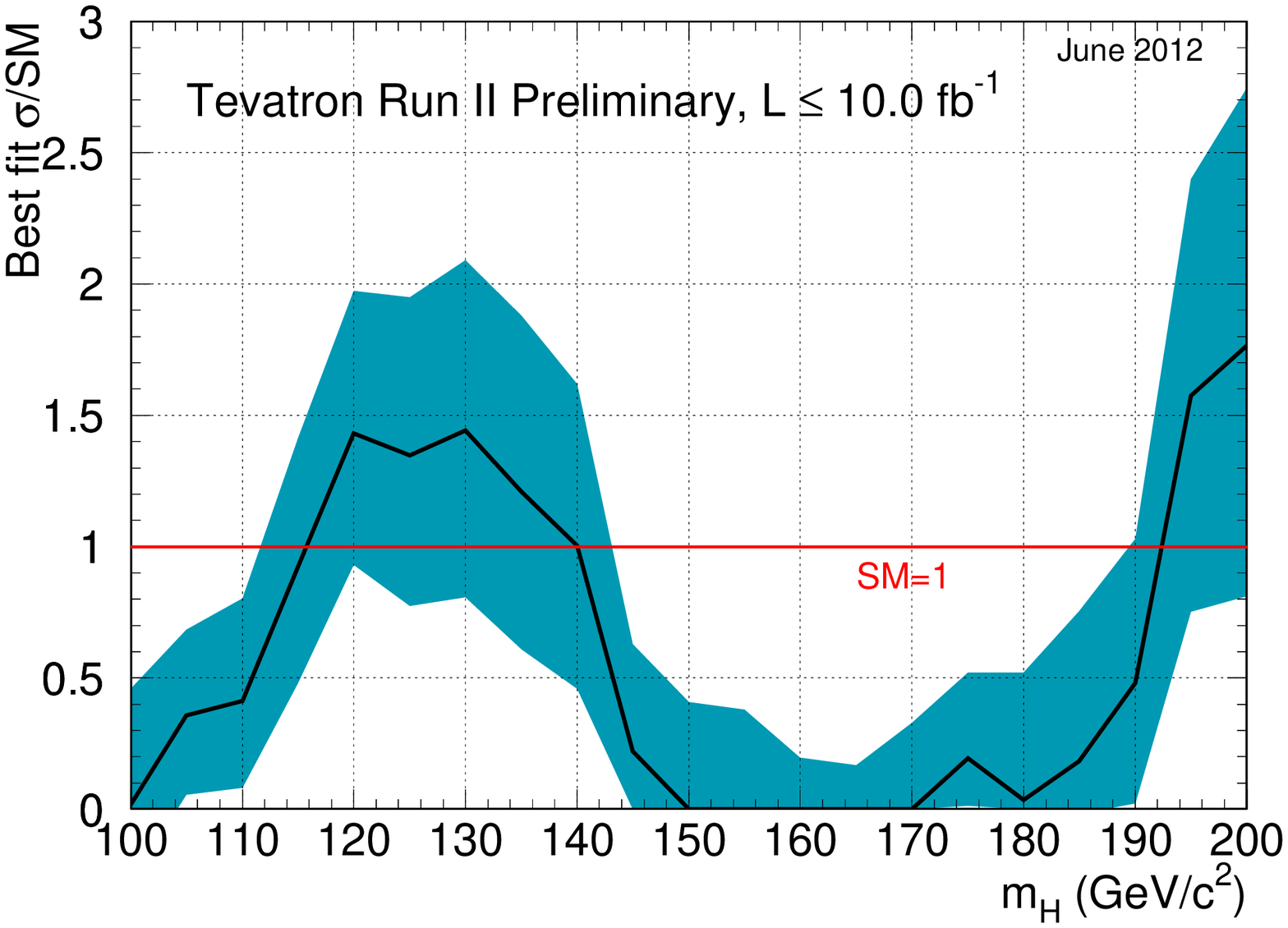}
  \end{minipage}
  \begin{minipage}[b]{.47\textwidth}
    \includegraphics[width=0.97\textwidth,height=6cm]{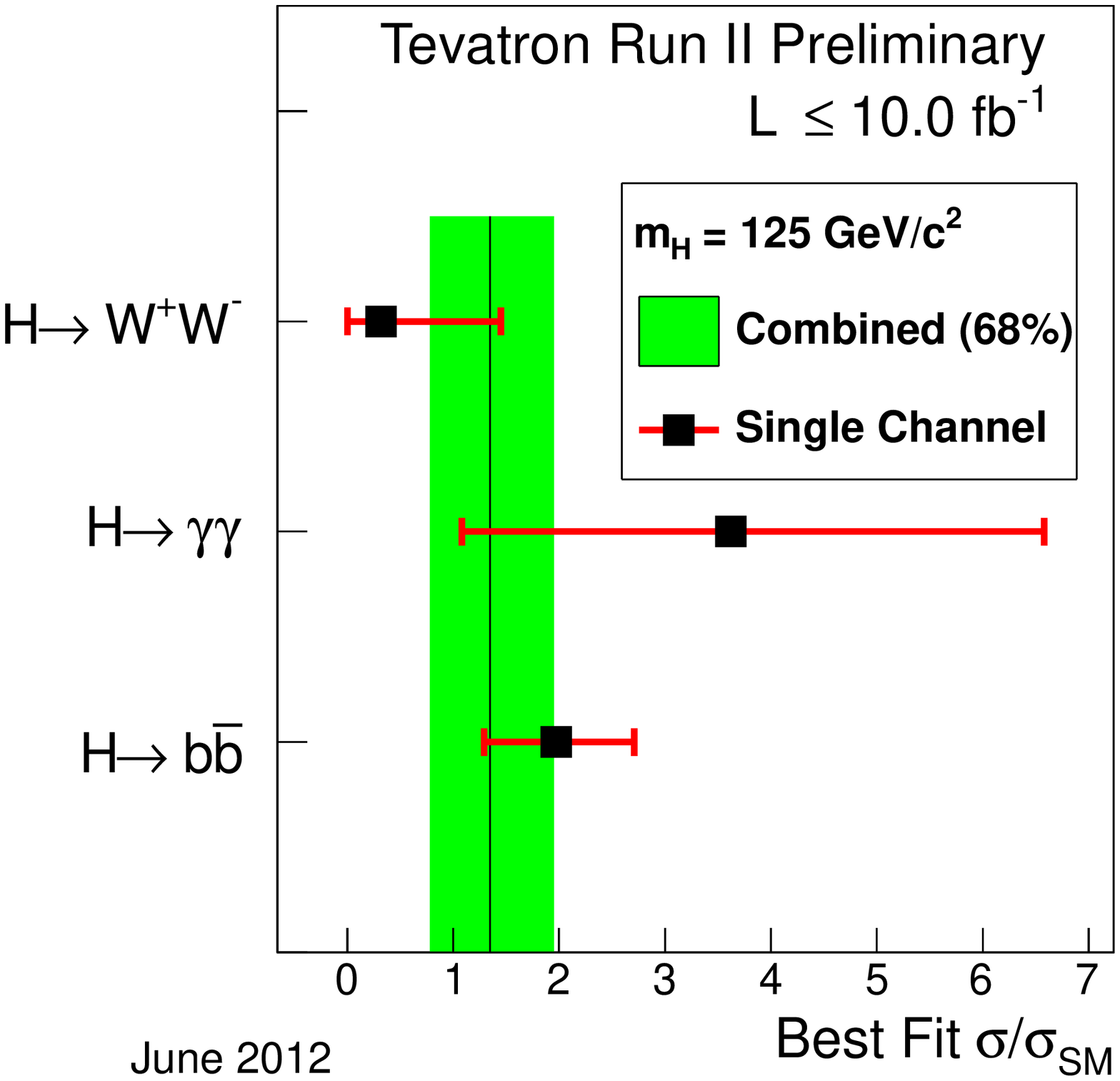}
  \end{minipage}
  \caption{Preliminary results from the Tevatron combination in July 2012~\cite{:2012zzl}.
   Expected and observed event yields classified by the signal-to-background ratio 
   after background subtraction for $\MH=125\GeV$ (top-left) and $\MH=165\GeV$ (top-right);
   exclusion limit on the signal strength $\sigma/\sigma_{\SM}$ as a function of $\MH$ (middle-left),
   local $p$-value for the background-only hypothesis as a function of $\MH$ (middle-right),
   best fit values of the signal strength as a function $\MH$ (bottom-left),
   and best fit values for the signal strength in different decay modes
   for $\MH=125\GeV$ (bottom-right).}
  \label{fig:tevprel2012}
\end{figure}
The Tevatron combination extends the exclusion sensitivity significantly 
with respect to the one in the individual experiments. The mass regions  $\MH=100{-}120\GeV$ and $\MH=139{-}184\GeV$ 
are expected to be excluded as shown in \reffi{fig:tevprel2012} (middle-left). The observed excluded mass ranges are significantly 
weaker and extend only over the mass ranges $\MH= 100{-}103\GeV$ and $\MH=147{-}180\GeV$. 
The largest deviation from the background-only hypothesis is observed for $\MH=120\GeV$ with a local $p$-value 
of $1.5 \times 10^{-3}$, corresponding to a local significance of 3.0 as shown in \reffi{fig:tevprel2012} (middle-right). 
The excess is dominated by the search
for $\Htobb$ discussed above. Due to the limited mass resolution the significance is larger than 2
in the approximate mass range $\MH=115{-}145\GeV$. Hence the influence of the LEE is also rather 
modest  with a trial factor of approximately 2 yielding a global significance of $2.5$. Another deviation from the 
background-only hypothesis with a significance of larger than $2$ is observed for a Higgs-boson mass larger than $195\GeV$, 
which was already excluded by the LHC experiments from 2011 data. The best fit value for the signal strength 
as a function of $\MH$, shown in \reffi{fig:tevprel2012} (bottom-left), is consistent at $68$\% CL with the expectation for 
the SM Higgs boson 
in the mass ranges $\MH=110{-}140\GeV$ and approximately $\MH=190{-}200 \GeV$.
Finally, the best signal-strength parameters in the different decay modes $\Htobb$, $\PH\to\PW\PW$, and $\Htogg$ were
determined for various fixed values of $\MH$. The result for $\MH=125\GeV$ is shown in \reffi{fig:tevprel2012} (bottom-right).

\section{Searches at the LHC}

The design parameters of the LHC and the two multipurpose experiments ATLAS and CMS
are chosen in such a way that the whole unexplored mass range, from the limit obtained 
at LEP up to the upper border of $\lsim1\TeV$
demanded by unitarity (see \refse{se:Hconstraints}), can 
be covered.  On 30 March 2010, the LHC took first proton--proton collision data at the
CM energy of $7\TeV$, an energy unprecedented at  accelerators at that time. 
Fully hadronic final states, arising from production in gluon fusion or weak-vector-boson 
fusion with subsequent decay via $\Htobb$ or $\PH\to\PW\PW (\PZ\PZ) \to 4\,$quarks 
yielding the largest production rates, are not considered in the search. 
The overwhelming background and the limited trigger capabilities for these final states do 
not allow for performing a sensitive search. 
The large instantaneous
luminosities of up to $6.8\times 10^{33}$ cm$^{-2}$s$^{-1}$ ($3.7\times 10^{33}$ cm$^{-2}$s$^{-1}$)
reached in 2012 up to June 
(2011) together with the large total inelastic cross section and the interaction rate of
$20\,\mathrm{MHz}$ lead to an average 
of 20 (10) 
overlayed proton--proton interactions in 2012 up to June (2011)  often referred to as ``pile-up''.
This challenge, which was not of relevance at Tevatron, imposes additional challenges on the event 
reconstruction especially on isolation criteria and a precise determination of the 
missing transverse momentum.
Only final states that contain at least one photon, electron, muon, or a hadronic tau-lepton decay
in association with large missing transverse momentum are considered in the searches.
The production rates for these topologies at the LHC and the ratio of cross sections at different CM energies
are shown in \reffi{fig:lhc7tevxsbr}. 
\begin{figure}
  \centering
  \begin{minipage}[b]{.47\textwidth}
    \includegraphics[width=.99\textwidth]{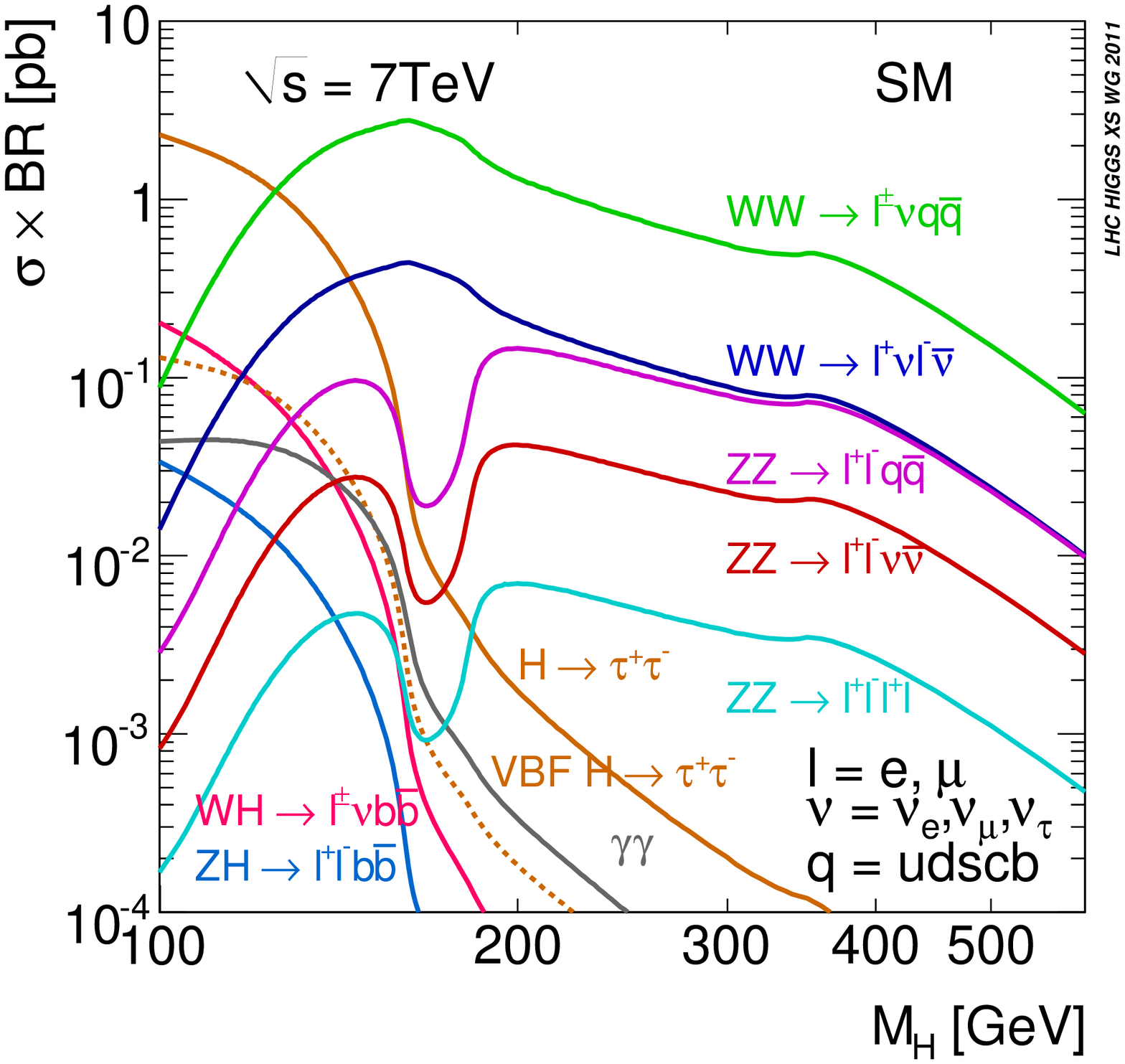}
  \end{minipage}
  \begin{minipage}[b]{.47\textwidth}
    \includegraphics[width=.99\textwidth,height=.95\textwidth]{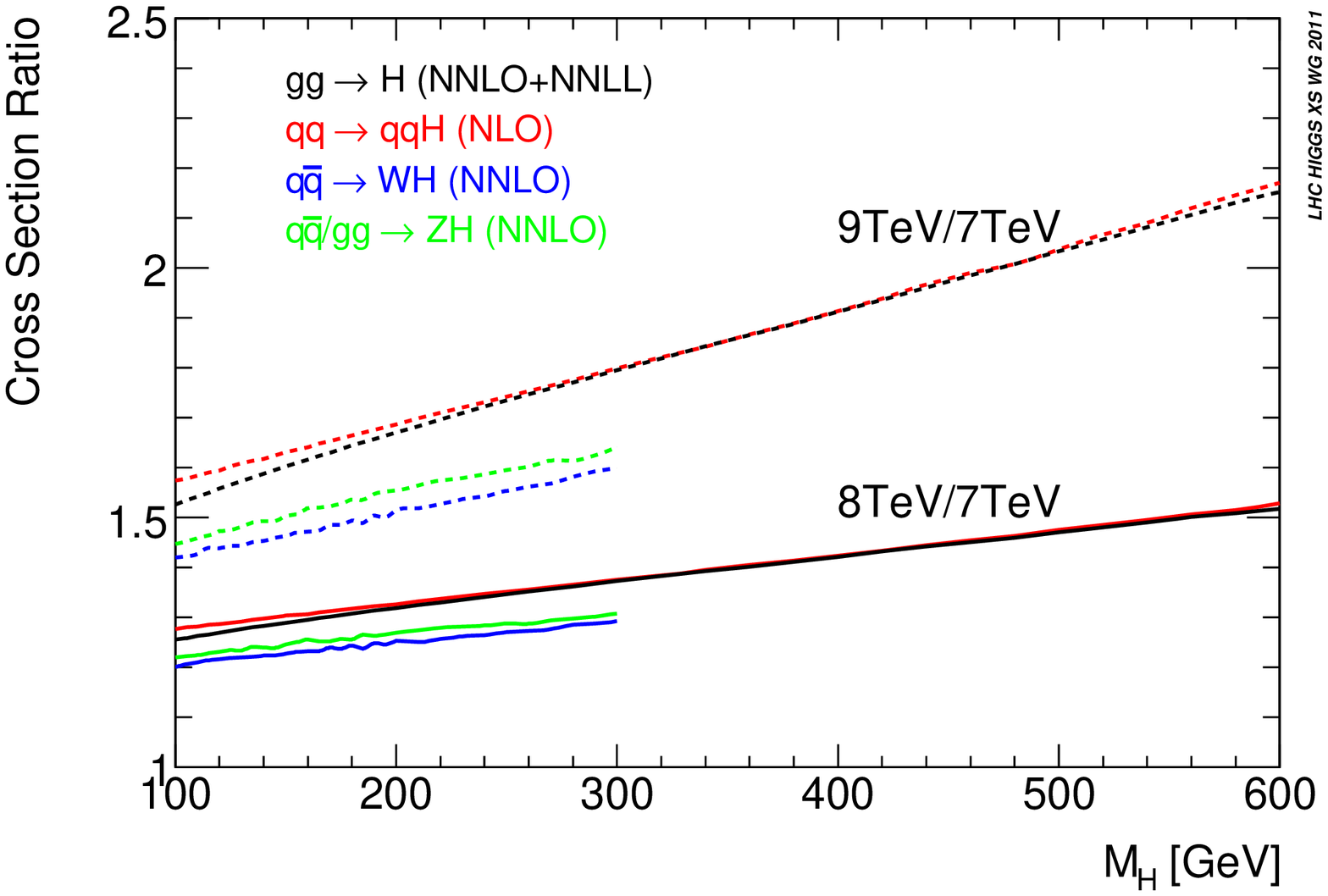}
  \end{minipage}
  \caption{Higgs-boson production rates at the LHC for CM energies of $7\TeV$ 
           in final states which are used to search for the Higgs boson (left)~\cite{Dittmaier:2011ti}
           and ratio of cross sections at different CM energies (right)~\cite{lhchxswg}.}
  \label{fig:lhc7tevxsbr}
\end{figure}
In contrast to the searches at Tevatron channels with low branching ratios, such as
$\Htogg$ and $\Htofl$ with branching ratios of $0.2\%$ and $0.013\%$ for a Higgs-boson mass
of $125\GeV$ are also 
accessible due to the larger production cross section at the LHC.  
In fact these two channels provide the highest sensitivity over a large mass range. 

The signal cross sections, the branching ratios, and their uncertainties are computed, as 
detailed in \refses{se:Hdecays} and \ref{sec:hadroncollcrosssec},
following the recommendations of the LHC Higgs Cross Section Working Group~\cite{Dittmaier:2011ti,Dittmaier:2012vm}.
Simulated events are generated using \Powheg\ interfaced to \Pythia\ for the dominant production mechanisms in gluon fusion~\cite{Bagnaschi:2011tu} 
and in weak-vector-boson fusion~\cite{Nason:2009ai}. The associated production with a weak gauge boson and a pair of top quarks 
is mainly simulated with \Pythia\ (CMS uses \Powheg\ for the decay channel $\Htobb$). 

The dominant background processes, in particular the irreducible ones apart from a very few exceptions, are estimated at least partially with 
data-driven techniques. Only in some analyses the normalization and shape of the final discriminating distribution are completely estimated
from data.
Quite often the normalization is estimated in a control region, defined, e.g., 
by inverting some of the selection
criteria, and the prediction for the ratio of the background expectations in the signal and control regions as well as the shape of the final discriminant 
are estimated from simulated events. The most advanced event generators are used to get as precise as possible predictions. Those include 
\Alpgen~\cite{Mangano:2002ea}, \Madgraph~\cite{Alwall:2011uj}, \MCatNLO~\cite{Frixione:2002ik,Frixione:2010wd}, and \Sherpa~\cite{Gleisberg:2008ta},
which perform a matching of matrix-element calculations and parton showers at NLO or LO in QCD, and dedicated programmes for gluon-induced di-boson 
production, such as \ggtoWW~\cite{Binoth:2005ua} and \ggtoZZ~\cite{Binoth:2008pr} at LO. For cross checks and simulation of the parton shower, of the underlying event,
and of pile-up the multipurpose event generators \Pythia~\cite{Sjostrand:2006za,Sjostrand:2007gs} and HERWIG~\cite{Corcella:2000bw,Bahr:2008pv} are used.

\subsection{Search with data collected in 2010}

\begin{sloppypar}
The highest cross sections with final states involving  at least one colour-neutral object are obtained 
in the mass range between $150\GeV$ and $170\GeV$ in the decay chains $\Htowwlnqq$  and 
$\Htowwll$ (see l.h.s.\ of \reffi{fig:lhc7tevxsbr}). The semi-leptonic decay of the pair of W bosons is not usable 
in this low-mass interval due to the large background from $\PW$+jets production. The first sensitivity 
at the LHC hence stemmed from the $\Htowwll$ final state. Both experiments performed a search in this 
final state with the data collected in 2010 at a CM energy of $7\TeV$ corresponding to integrated luminosities of
$36\,\ipba$~\cite{Aad:2011qi} and 
$35\,\ipba$~\cite{cms2010ww}. 
The expected sensitivity in the mass range $150{-}170\GeV$ was at the level of $2{-}3$ times the 
SM prediction (see Fig.\ref{fig:lhc2010results}, middle and right). 
\begin{figure}
  \centering
  \begin{minipage}[b]{.32\textwidth}
    \includegraphics[width=1.00\textwidth,height=5.5cm]{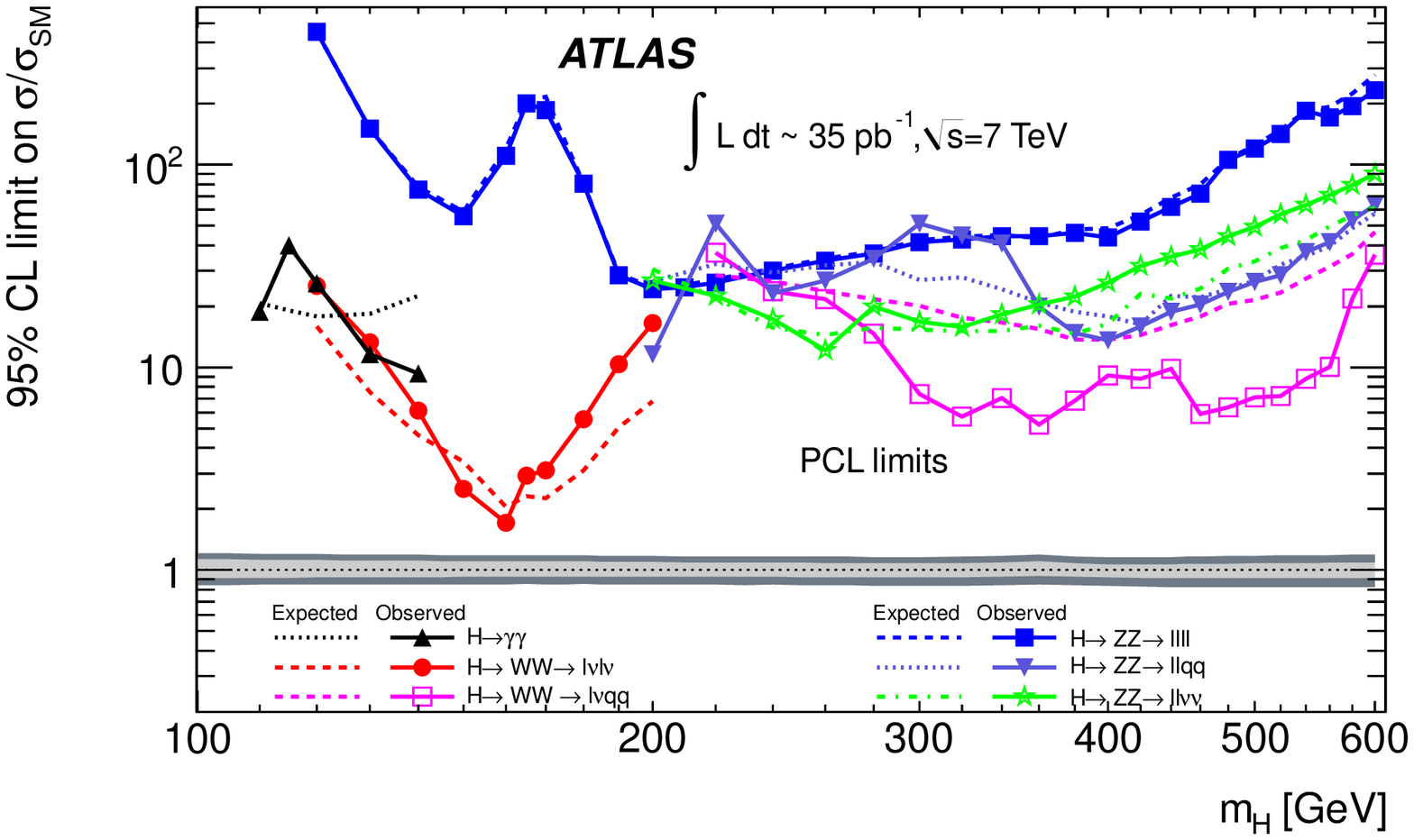}
  \end{minipage}
  \begin{minipage}[b]{.32\textwidth}
    \includegraphics[width=1.00\textwidth,height=5.5cm]{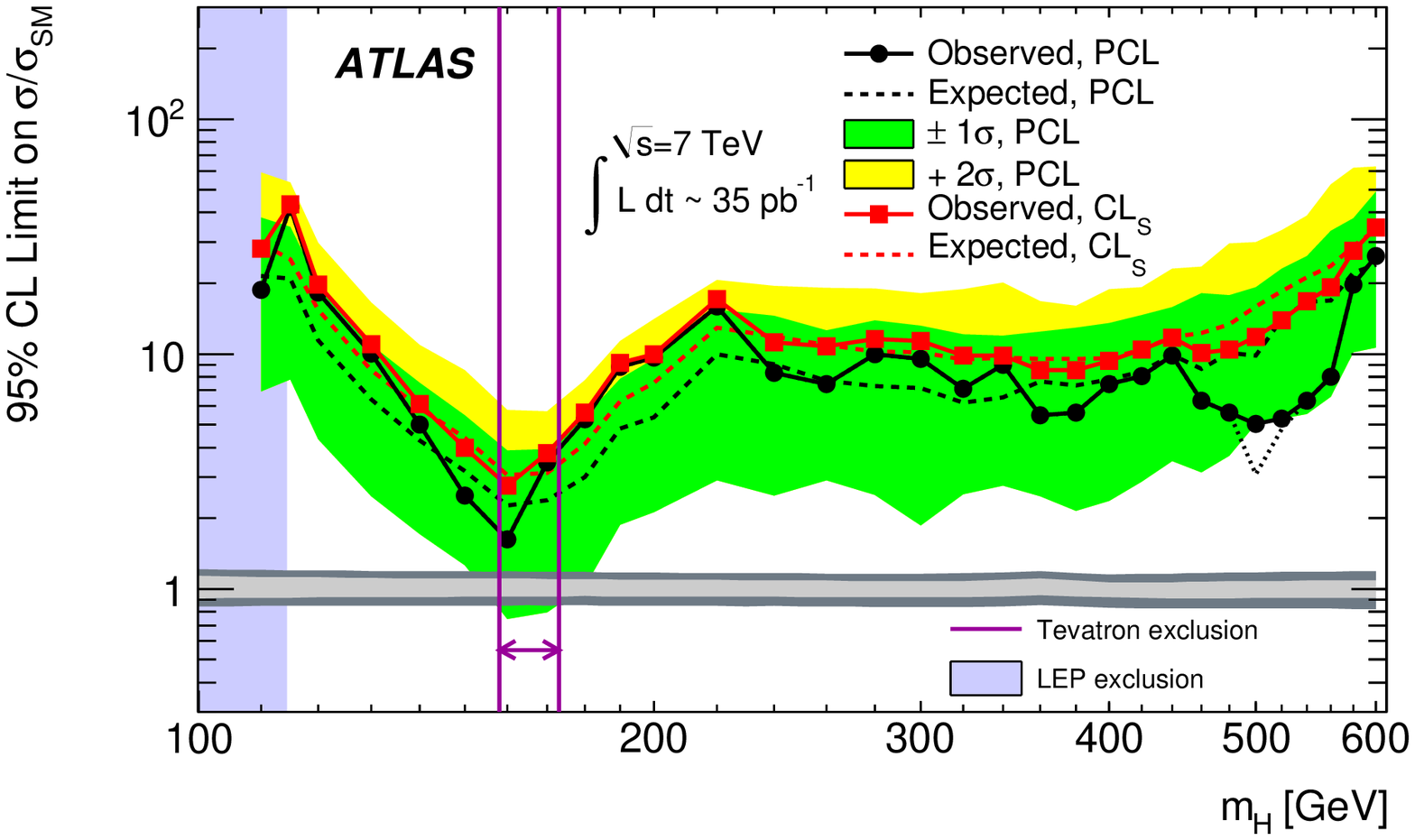}
  \end{minipage}
  \begin{minipage}[b]{.32\textwidth}
    \includegraphics[width=1.00\textwidth]{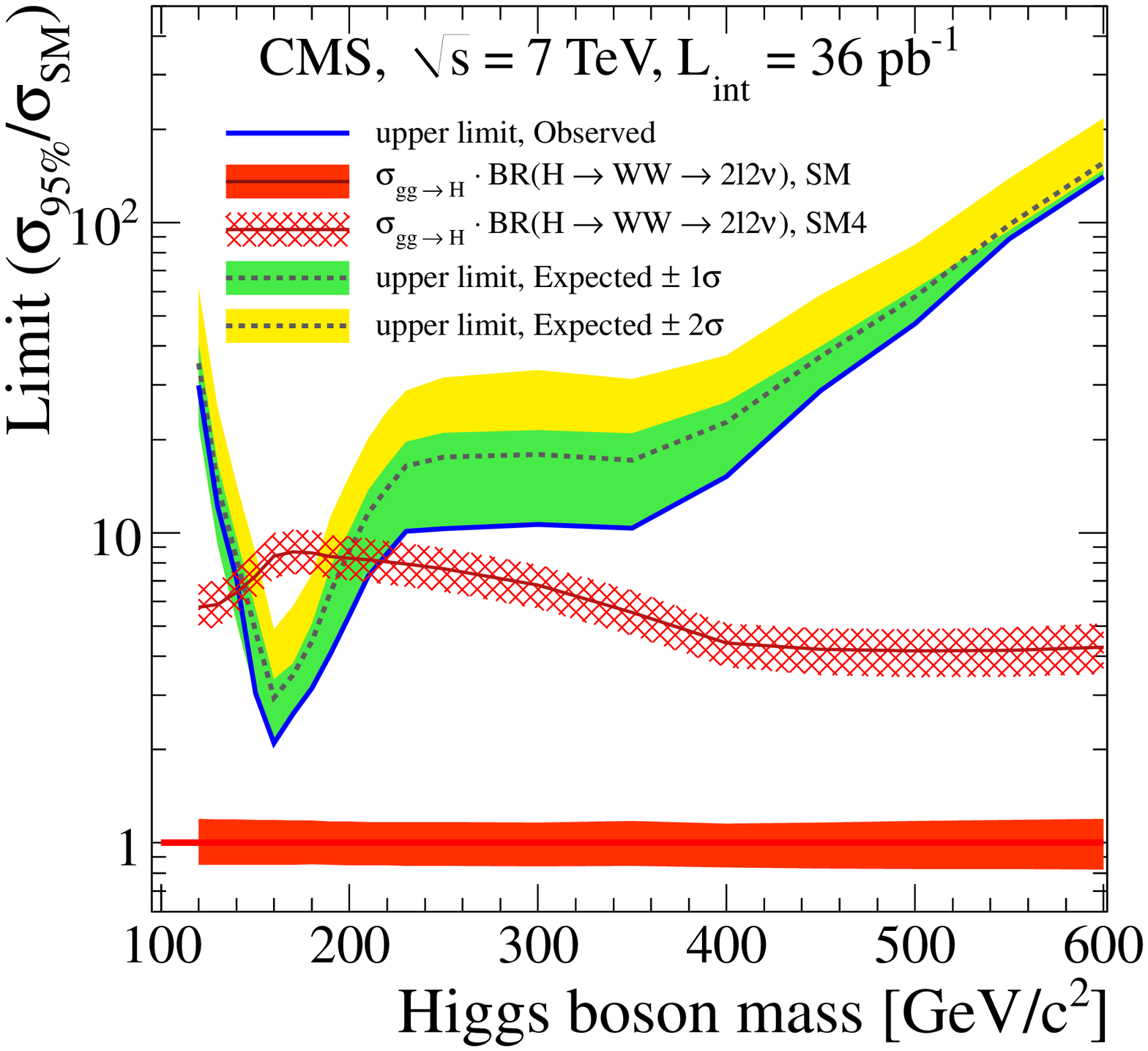}
  \end{minipage}
  \caption{Observed and expected exclusion limits on the signal-strength parameter $\mu$
    obtained from 2010 collision data: individual channels (left), their combination (middle) 
    in ATLAS~\cite{Aad:2011qi}, and the combination in CMS (right)~\cite{cms2010ww}.}
  \label{fig:lhc2010results}
\end{figure}
No significant deviation 
from the expected event yields from background processes was observed. ATLAS already investigated 
several other final states  based on 2010 data: $\Htogg$, $\Htofl$, $\Htoln$, $\Htolq$, and $\Htowwlnqq$. The individual 
sensitivities of each of these searches were at the best on the level of 15 times the SM prediction
(Fig. \ref{fig:lhc2010results}, left). 
\end{sloppypar}

\subsection{Searches with data collected in 2011 and up to June in 2012}

In 2011 each of the two experiments collected a data set corresponding to integrated luminosities of 
up to $5.1\,\ifba$ at a CM energy of $7\TeV$. In 2012 until June a data set corresponding to up to 5.9\,$\ifba$ 
per experiment was recorded at a CM energy of $8\TeV$. 
An overview of the search channels considered, the 
considered mass range in each channel, and the integrated luminosities analyzed is given in 
\refta{tab:lhcsearchchannels}. Three mass regions can be roughly distinguished:
(a) $110\GeV \le  \MH \le 150\GeV$, (b) $150\GeV\le \MH \le 200\GeV$, and (c) 
$200\GeV \le \MH \le 600\GeV$.
\begin{table}
	\caption{Search channels, mass ranges, integrated luminosities considered by ATLAS and CMS, and corresponding references.
The results used in the combination of the 2011 data are documented in \citere{:2012an} for ATLAS
and in \citere{Chatrchyan:2012tx} for CMS. An entry ``--'' indicates that this search channel
was not considered for this data set. In 2012 CMS restricted the mass range 
for all published searches to values below $160\GeV$. When two mass ranges are given, the search was 
performed in two disconnected ranges.}
\label{tab:lhcsearchchannels}
\begin{center}
\begin{tabular}{c||c|cc||c|cc}
\hline
           &\multicolumn{3}{c||}{ATLAS}&\multicolumn{3}{c}{CMS}\\ \hline 
channel    & mass  range & \multicolumn{2}{c||}{$\int L \,\rd t\,[\ifba$]}& mass range   &    \multicolumn{2}{c}{$\int L \,\rd t\,[\ifba$]} \\ 
           & $\MH$~[GeV] & 2011    & 2012                                              & $\MH$~[GeV]     & 2011  & 2012  \\ \hline 
$\Htogg$     & 110--150  & 4.8     \cite{ATLAS:2012ad}  & 5.9 \cite{:2012gk}       & 110--150         & 5.1 \cite{Chatrchyan:2012tw} & 5.3 \cite{:2012gu} \\ 
$\Htott$     & 110--150  & 4.7     \cite{Aad:2012mea}   & --                       & 110--145         & 4.9 \cite{Chatrchyan:2012vp} & 5.1  \cite{:2012gu}\\
$\Htobb$     & 110--130  & 4.7     \cite{:2012zf}       & --                       & 110--135         & 5.0 \cite{Chatrchyan:2012ww} & 5.1  \cite{:2012gu}\\ 
$\Htofl$     & 110--600  & 4.8     \cite{ATLAS:2012ac}  & 5.8 \cite{:2012gk}       & 110--600/160     & 5.1 \cite{Chatrchyan:2012dg} & 5.3  \cite{:2012gu}\\ 
$\Htowwll$   & 110--600  & 4.7     \cite{Aad:2012npa}   & 5.8 \cite{:2012gk}       & 110--600/160     & 4.9 \cite{Chatrchyan:2012ty} & 5.1  \cite{:2012gu}\\ 
$\Htolt$     & --       & --                           & --                       & 190--600         & 4.7 \cite{Chatrchyan:2012hr} & --   \\ 
$\Htolq$     & 200--600  & 4.7     \cite{:2012vw}       & --                       & 130--163,200--600 & 4.6 \cite{Chatrchyan:2012sn} & --   \\ 
$\Htoln$     & 200--600  & 4.7     \cite{ATLAS:2012va}  & --                       & 250--600         & 4.6 \cite{Chatrchyan:2012ft} & --   \\ 
$\Htowwlnqq$ & 300--600  & 4.7     \cite{Aad:2012me}    & --                       & --              & --    & --   \\ \hline
\end{tabular}
\end{center}
\end{table}     

The mass range above $600\GeV$, which is highly disfavoured by electroweak precision measurements, 
was not considered, because the theoretical tools or recipes described in
\citeres{Passarino:2010qk,Dittmaier:2011ti,Dittmaier:2012vm,Anastasiou:2012hx}
(see also references therein)
for a proper
description of the extremely broad Higgs-boson lineshape and their
interference with irreducible background for such large masses
were not available to the experiments until June 2012. 

Both experiments published their findings based on the complete 2011 data set~\cite{:2012an,Chatrchyan:2012tx}.
These data were partially reanalyzed with better reconstruction and identification tools, and optimized analysis techniques,
resulting in a significantly increased sensitivity. In combination with the analysis of the data set collected 
in 2012 up to June, the reanalyzed 2011 data are published in \citeres{:2012gk,:2012gu}.  CMS only considered the mass 
region below $160\GeV$ in the 2012 data analysis and reanalyzed the 2011 data in the channels $\Htogg$, $\Htobb$, 
$\Htott$, and $\Htofl$. ATLAS reanalyzed the channels $\Htogg$ and $\Htofl$ in the 2011 data, and so far has only investigated 
those two channels and the $\PH\to\PW^+\PW^-\to\Pe\mu\nu\bar\nu$ decay mode in 2012 data.

In the low-mass region $110{-}150\GeV$ the decays to photons, to a pair of tau leptons or bottom quarks,
and to a pair of weak gauge bosons ($\PW \PW$ and $\PZ\PZ)$, which decay to four 
leptons, were considered by both experiments.
The decays $\Htogg$ and $\Htofl$, with very low rate, provide an excellent resolution for the reconstruction of 
the invariant mass of the decay products of the Higgs-boson candidate with an accuracy of about $1{-}2 \%$. 
In the decay mode $\Htowwll$, which has a comparably high production rate, only the transverse mass can be reconstructed.
The decays $\Htott$ and $\Htobb$ allow for reconstructing the mass of the Higgs-boson candidate with limited resolution
in the range of $10{-}30\%$, due to the presence of several neutrinos in the final state or the limited energy resolution
for the reconstruction of jets compared to charged leptons and photons. CMS also considered the decay $\Htolq$ for 
mass values larger than $130\GeV$, where the mass can be reconstructed with an accuracy of roughly $10\%$. 

In the intermediate-mass region $150{-}200\GeV$ the sensitivity of the search is dominated by the decay mode $\Htowwll$, 
which has a large production rate, but allows only for reconstructing the transverse mass. 
The search for $\Htofl$ with low rate, but excellent mass reconstruction capabilities provides a sensitivity that is 
roughly one order of magnitude smaller.

\begin{sloppypar}
In the high-mass region $200{-}600\GeV$ various final states arising from the decay of the Higgs boson into
a pair of electroweak gauge bosons are considered. In addition to the final states that dominate the 
sensitivity in the intermediate-mass region, 
the decays $\Htolq$, $\Htoln$ are searched for by both experiments as well, while $\Htolt$ is analyzed only by CMS,
and $\Htowwlnqq$ only by ATLAS. 
The signal rates for $\Htofl$ and $\Htowwll$ get quite 
small for large $\MH$, and hence the sensitivity is increased by including decay modes of the 
weak gauge bosons with larger branching ratios. These final states are not considered for smaller masses as 
the backgrounds, dominated by $\PW$ and $\PZ$ production in association with jets, increase in importance 
in this region and cannot be suppressed sufficiently or cannot be estimated with sufficient precision from control regions in data 
to the required  level. Additionally the width of the reconstructed Higgs-boson mass distribution is no longer dominated 
by the experimental resolution, but by the natural width in the $\Htofl$ channels for masses above 
$350 \GeV$. Hence the relative sensitivity to the other 
decay modes, which have a worse mass resolution or only allow an approximate mass observable to be reconstructed, 
gets weaker. 
\end{sloppypar}

The channels $\Htowwll$ and $\Htofl$ contribute significantly to the overall sensitivity over the full mass range.
For small values of $\MH$ it is important to lower the threshold for the transverse momenta of the selected leptons
as much as possible, limited often by trigger rate requirements, and to retain simultaneously a low misidentification rate.
\begin{itemize} 
\item In the $\Htowwll$ analysis an excess of events with two leptons of opposite charge and large missing transverse 
momentum is searched for. Events are divided into separate categories according to jet multiplicity and lepton
flavours, where the two-jet category has selection  criteria designed to enhance sensitivity to the VBF production process.
ATLAS so far has only considered the $\Pe\mu$ final state for the 2012 data analysis due to the higher pile-up conditions with worse 
resolution for the missing transverse momentum and  the correspondingly
increased background from Drell--Yan production in the 
same-lepton-flavour final state. 
In events with no jets, the main background stems from non-resonant $\PW\PW$ production; in events with 
one jet, the dominant backgrounds are from $\PW\PW$ and top-quark production,
and in events with two jets from top-quark production. 
In CMS in the analysis of the 2011 data MVA classifiers are trained for a 
number of Higgs-boson masses 
to improve the separation of signal from backgrounds, 
and a search is made for an excess of events in the output distributions of the classifiers. 
In the 2012 analysis CMS uses sequential cuts, and the di-lepton invariant mass is used as final discriminant. ATLAS performs a 
cut-based analysis, 
optimized for different $\MH$ ranges, and the $\PW\PW$ transverse-mass distribution is used as the final discriminant. 
All background rates, except for very small contributions from $\PW\PZ$, $\PZ\PZ$, and $\PW\gamma^*$
are evaluated from data.
The distribution of the final discriminants in the analysis of the $8\TeV$ data show 
a clear broad excess in ATLAS and a less significant broad one in CMS (see \reffi{fig:masshwwllnn}). 
\end{itemize} 
\begin{figure}
  \centering
  \begin{minipage}[b]{.475\textwidth}
    \includegraphics[width=.9\textwidth,height=8cm]{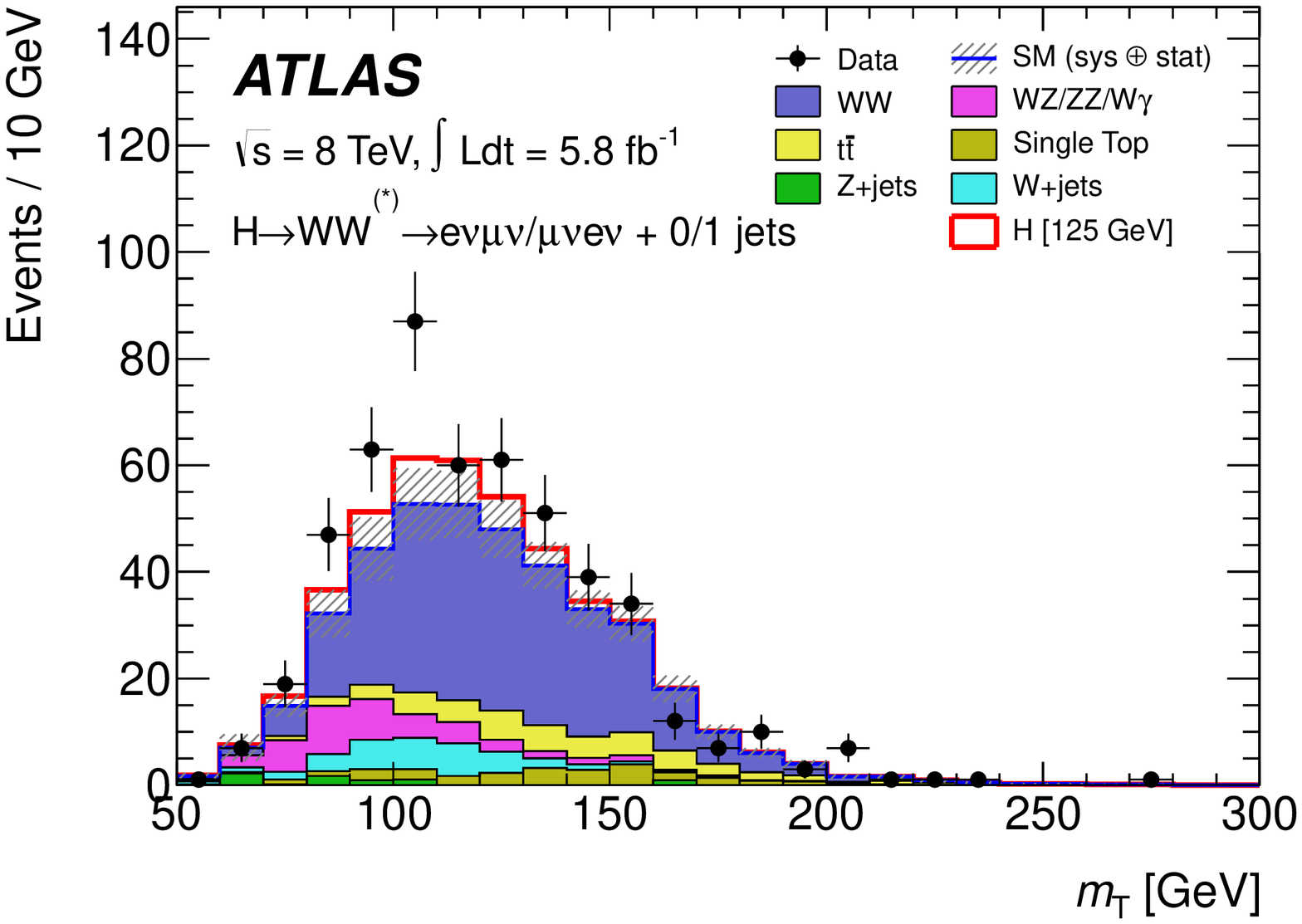}
  \end{minipage}
  \begin{minipage}[b]{.475\textwidth}
    \includegraphics[width=0.9\textwidth]{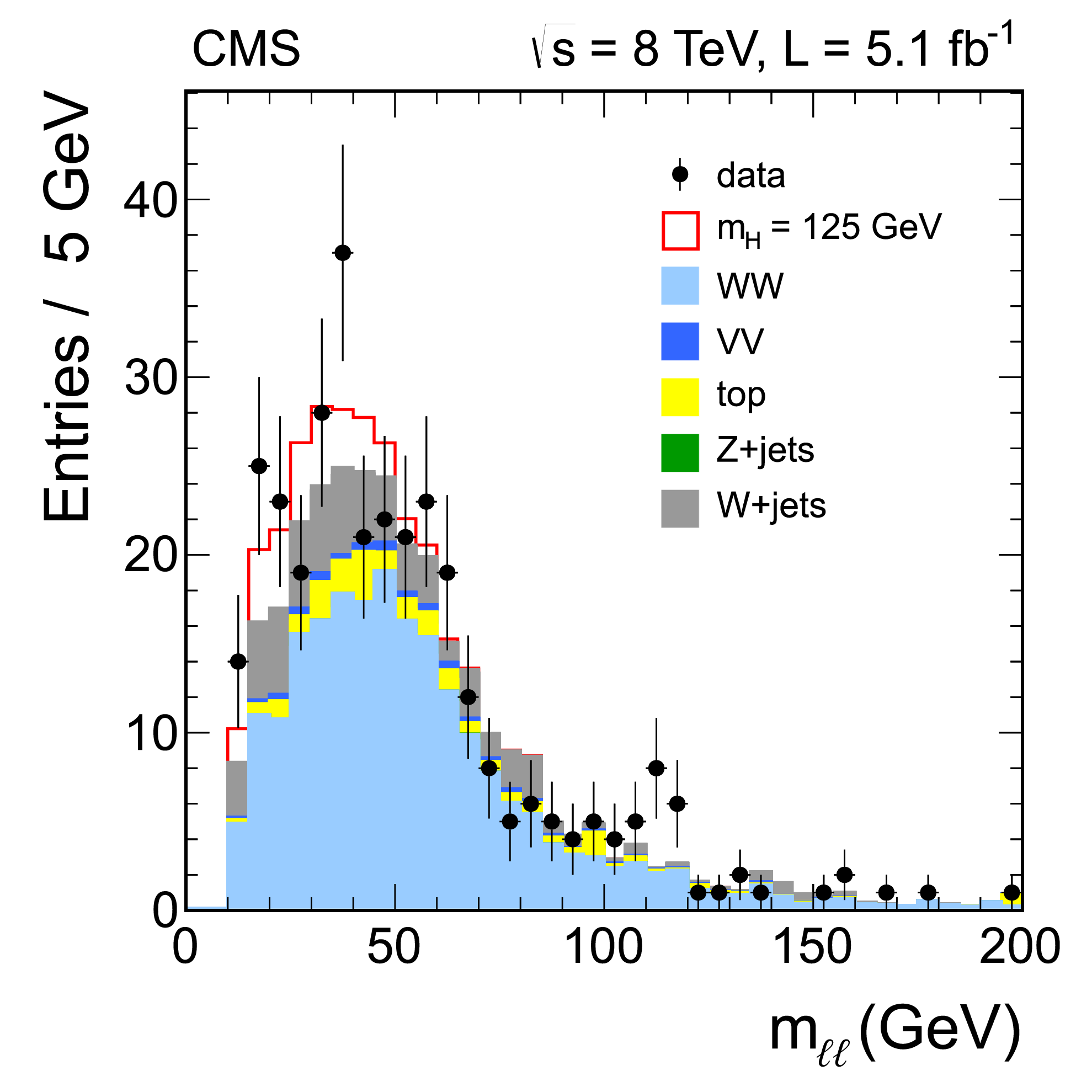}
  \end{minipage}
  \caption{Distributions of the final discriminant in the $\Htowwll$ search at $8\TeV$ in 2012 
           in ATLAS~\cite{:2012gk} (left) and CMS~\cite{:2012gu} (right).}
  \label{fig:masshwwllnn}
\end{figure}
\begin{itemize} 
\item In the $\Htofl$ channel an excess in the invariant-mass spectrum of the four isolated leptons over a small continuum background
is searched for. The 4$\Pe$, 4$\mu$, $2\Pe 2\mu$ final states are analyzed separately, since there are differences in the four-lepton 
mass resolutions and the background rates arising from jets misidentified as leptons. The main irreducible background
from non-resonant $\PZ\PZ$ production is mainly estimated from simulation. The smaller reducible backgrounds from $\PZ$+jets production, which mostly impacts the low four-lepton 
invariant-mass region, and  top-quark pair production are estimated from control regions in data, or at least the normalization is validated in dedicated 
control regions. The final discriminant in ATLAS is the four-lepton invariant-mass spectrum. CMS exploits the different kinematics and 
angular correlations to discriminate further the signal process from the irreducible $\PZ\PZ$ background. The information from the five angles,
which describe the production and decay kinematics in the Higgs-boson candidate rest frame, are combined in the so-called MELA observable\cite{Gao:2010qx}.
CMS uses a two-dimensional final discriminant consisting of the four-lepton invariant mass and the MELA output ($K_D$). A clear excess in the 
four-lepton invariant-mass spectra is observed in both experiments at masses of $125\GeV$ in ATLAS and $126\GeV$ in CMS, respectively (see \reffi{fig:hflmass}).
\end{itemize}
\begin{figure}
  \centering
  \begin{minipage}[b]{.475\textwidth}
    \includegraphics[width=.9\textwidth]{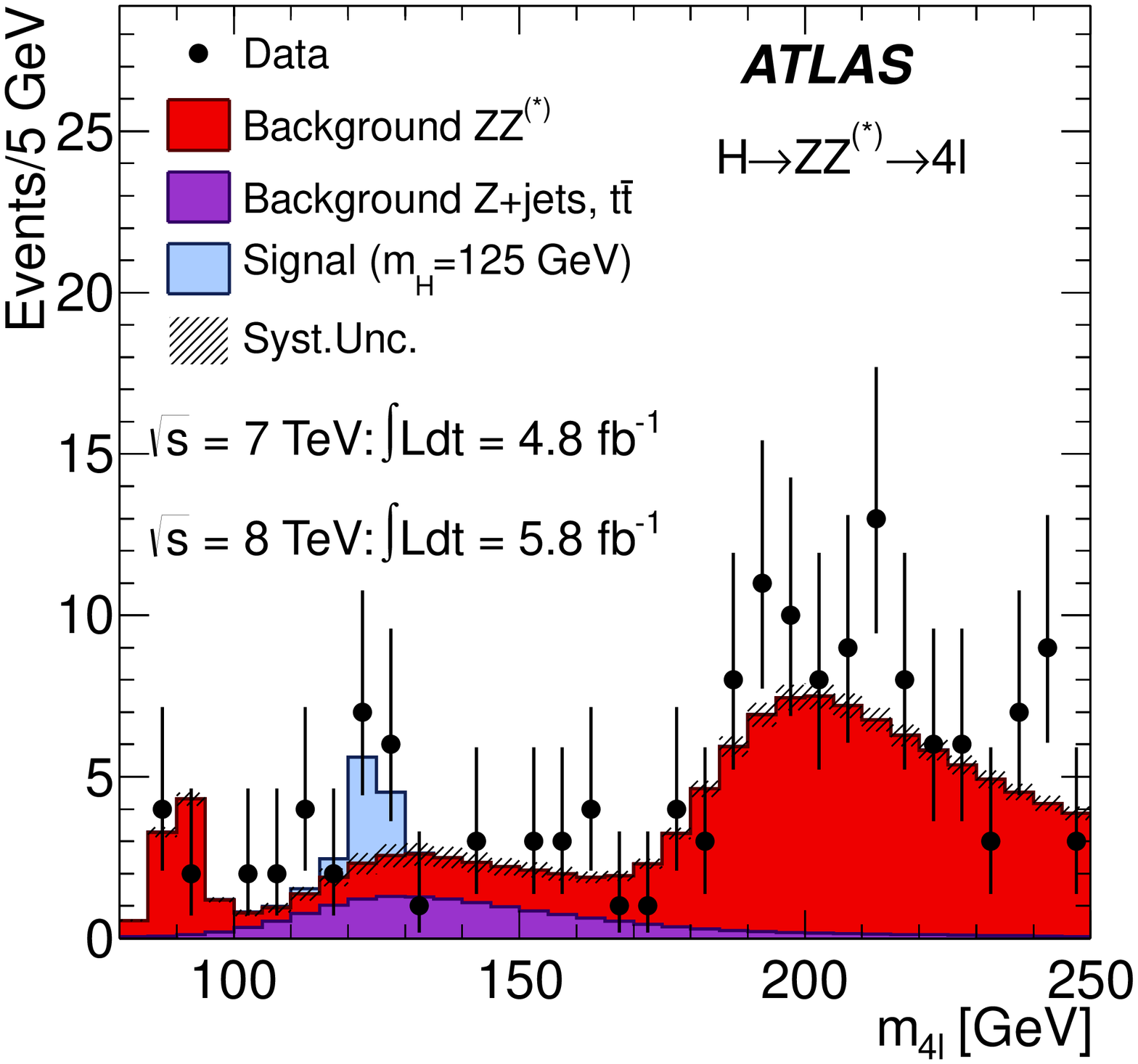}
  \end{minipage}
  \begin{minipage}[b]{.475\textwidth}
    \includegraphics[width=0.9\textwidth]{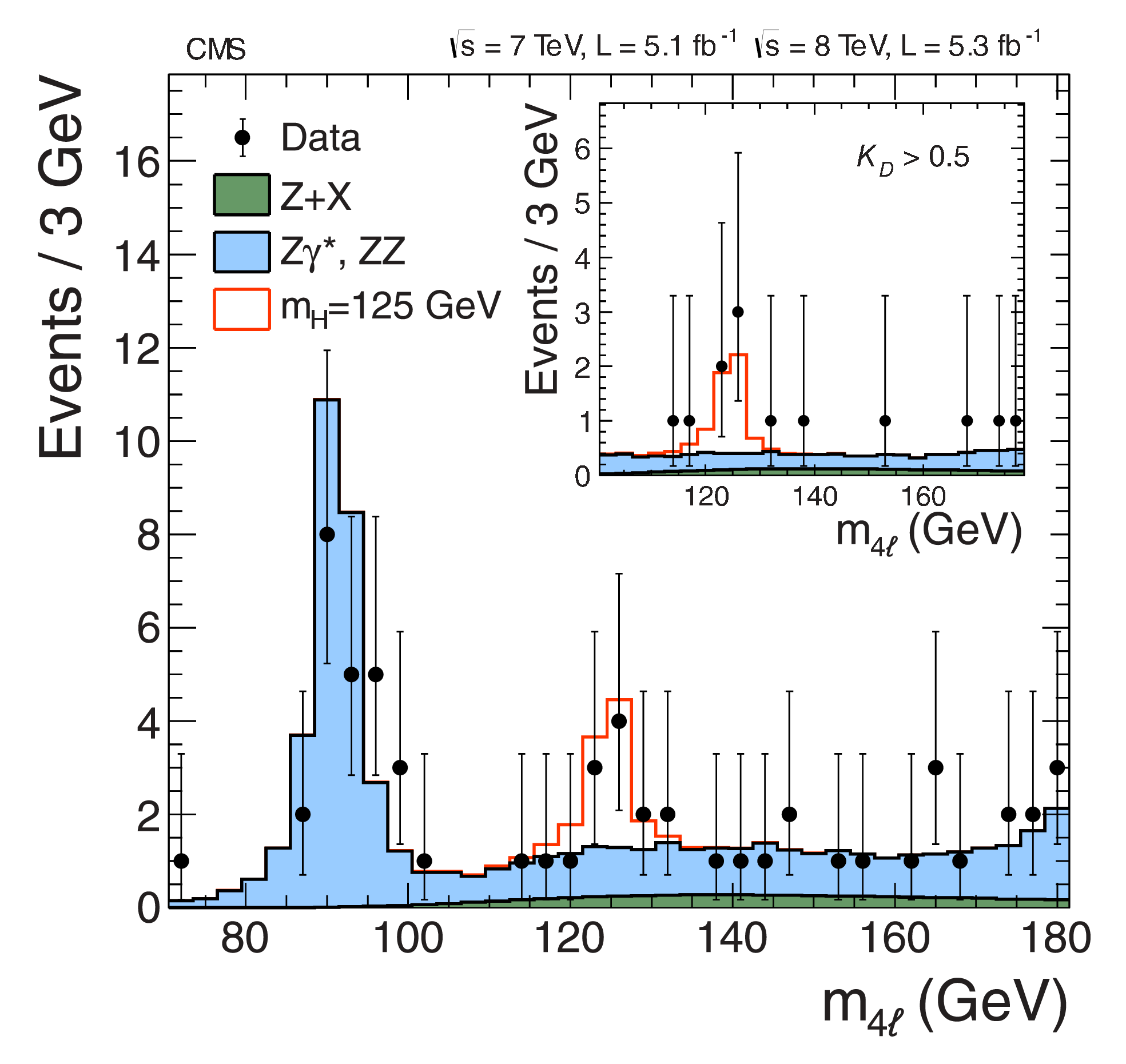}
  \end{minipage}
  \caption{Four-lepton invariant-mass spectra in ATLAS~\cite{:2012gk} (left) and CMS~\cite{:2012gu} (right) 
   for the combination of the 2011 and 2012 data.}
  \label{fig:hflmass}
\end{figure}
 
In addition to the two channels discussed above, where the full mass range is covered,
the channel $\Htoln$ provides the largest sensitivity in the high-mass range 
and dominates the sensitivity for $\MH$ values larger than $300\GeV$.
\begin{itemize} 
\item In the search for $\Htoln$ events with a pair of electrons or muons with opposite charge, consistent with the decay
of an on-shell $\PZ$, and large missing transverse momentum are selected.  After applying additional kinematical requirements, which 
depend on the Higgs-boson mass hypothesis, a transverse-mass observable is used as the final discriminant. The background 
predictions for weak-gauge-boson pair production are obtained from simulation, whereas the other backgrounds are determined from control samples 
in collision data. A significant fraction of the selected signal events stems from the decay $\Htowwll$ especially for lower masses.
In order to avoid double counting of signal yields, the selection is designed to have no overlap with the dedicated $\Htowwll$ analysis.
The normalization of most background processes is estimated from control regions in data. 
\item In the search for $\Htolq$ events with a pair of electron or muons and a pair of jets, which each have an invariant mass consistent with $\MZ$, are selected. 
The events  are categorized  according to the number of b-tagged jets in the final state (none or at least one in ATLAS; 0, 1, or 2 in CMS), in order to profit 
from the relatively large rate of b-jets from Z-boson decays present in the signal compared to the rate of b-jets found in the $\PZ$+jets background.
Additional kinematical requirements are applied depending on the Higgs-boson mass hypothesis.  CMS also exploits the different angular correlations between the 
final-state objects for the signal process compared to the background processes. The final discriminating variable is given by the invariant mass
of the two leptons and two jets. CMS also searches in the mass range below $2\MZ$, where the mass-window criteria on the di-lepton mass is replaced by an upper bound. 
The dominant background from $\PZ$+jet production is estimated from control samples in data, defined by sidebands in the di-jet invariant mass. 
\item The search for $\Htowwlnqq$ (only ATLAS) requires events with an isolated lepton and missing transverse momentum and 
two jets with an invariant mass compatible with 
a $\PW$ boson.  The mass of the Higgs-boson candidate can be constructed by solving a quadratic equation to solve for the component of the neutrino 
momentum along the beam axis exploiting the $\PW$ mass constraint for the lepton plus neutrino system. 
The analysis further classifies events by lepton flavour and by the number 
of additional jets (0, 1, or 2), where the two-jet channel is optimized for the VBF production process.
The final discriminant is given by invariant mass of the {\mathswitchr {\ell \nu \Pq\Pqbar}} system. The background is modeled with a smooth function. 
\item The search for $\Htolt$ (only CMS) selects events with electron or muon pairs, which are consistent with stemming from the decay of an on-shell $\PZ$ boson, 
and a pair of tau-lepton candidates, which either decay hadronically or into an electron or muon. Final states with only electrons or muons are not considered,
as they are already included in the dedicated $\Htofl$ search.  The final discriminant is given by the distribution of the 
di-lepton--di-tau invariant mass, constructed from the visible products of the tau-lepton decays, neglecting the effect of the accompanying neutrinos. 
The mass resolution for Higgs-boson candidates is $10{-}15$\%.
The dominant background from $\PZ\PZ$~ production is estimated from simulation. Other backgrounds are estimated from control regions in data.
\end{itemize} 
The final mass distributions (examples from ATLAS are shown in \reffi{fig:masdisthighmh}) agree with the expectations from background processes.
\begin{figure}
  \centering
  \begin{minipage}[b]{.475\textwidth}
    \includegraphics[width=.9\textwidth]{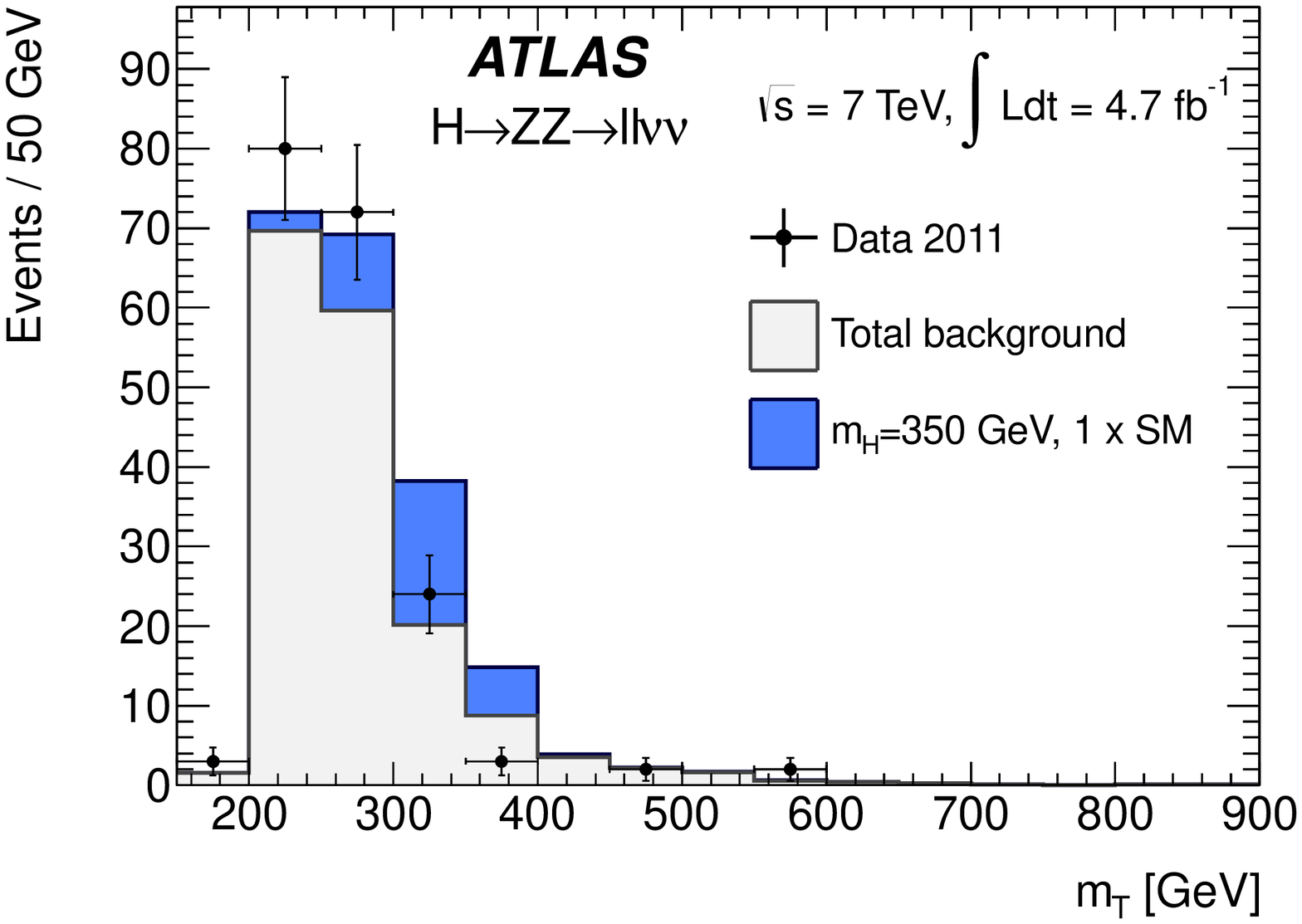}
  \end{minipage}
  \begin{minipage}[b]{.475\textwidth}
    \includegraphics[width=0.9\textwidth]{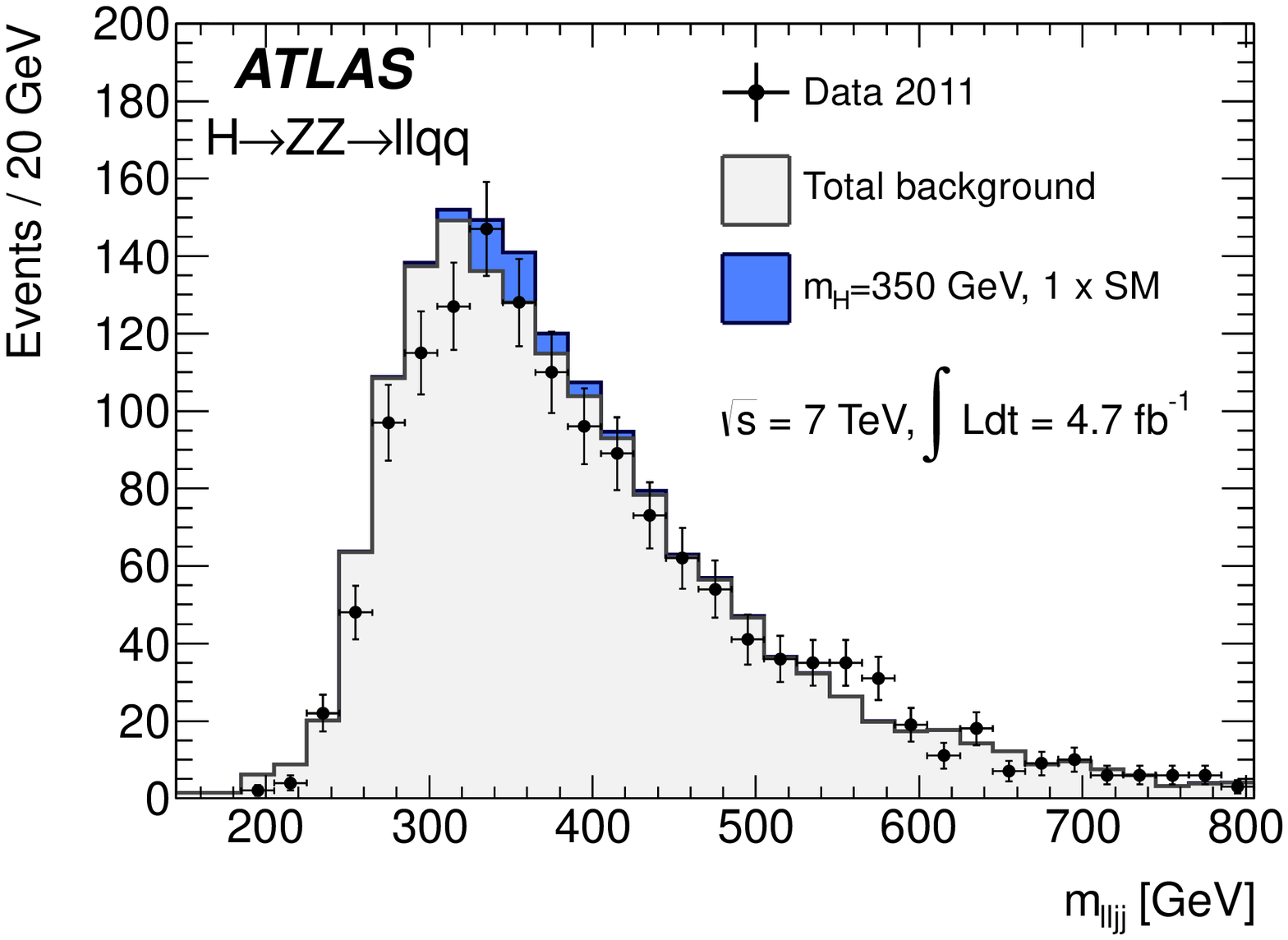}
  \end{minipage}
  \centering
  \begin{minipage}[b]{.475\textwidth}
    \includegraphics[width=.9\textwidth]{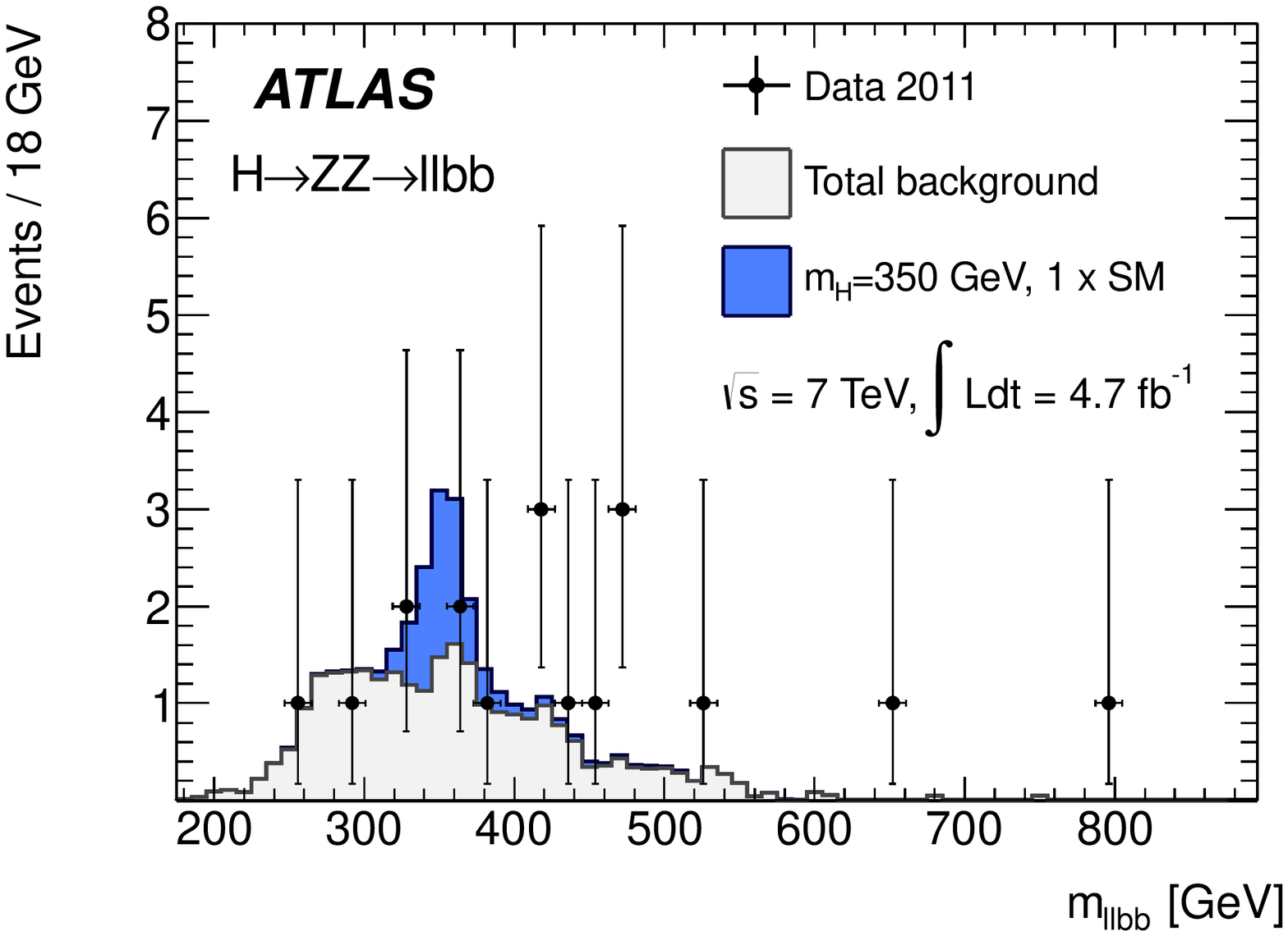}
  \end{minipage}
  \begin{minipage}[b]{.475\textwidth}
    \includegraphics[width=0.9\textwidth]{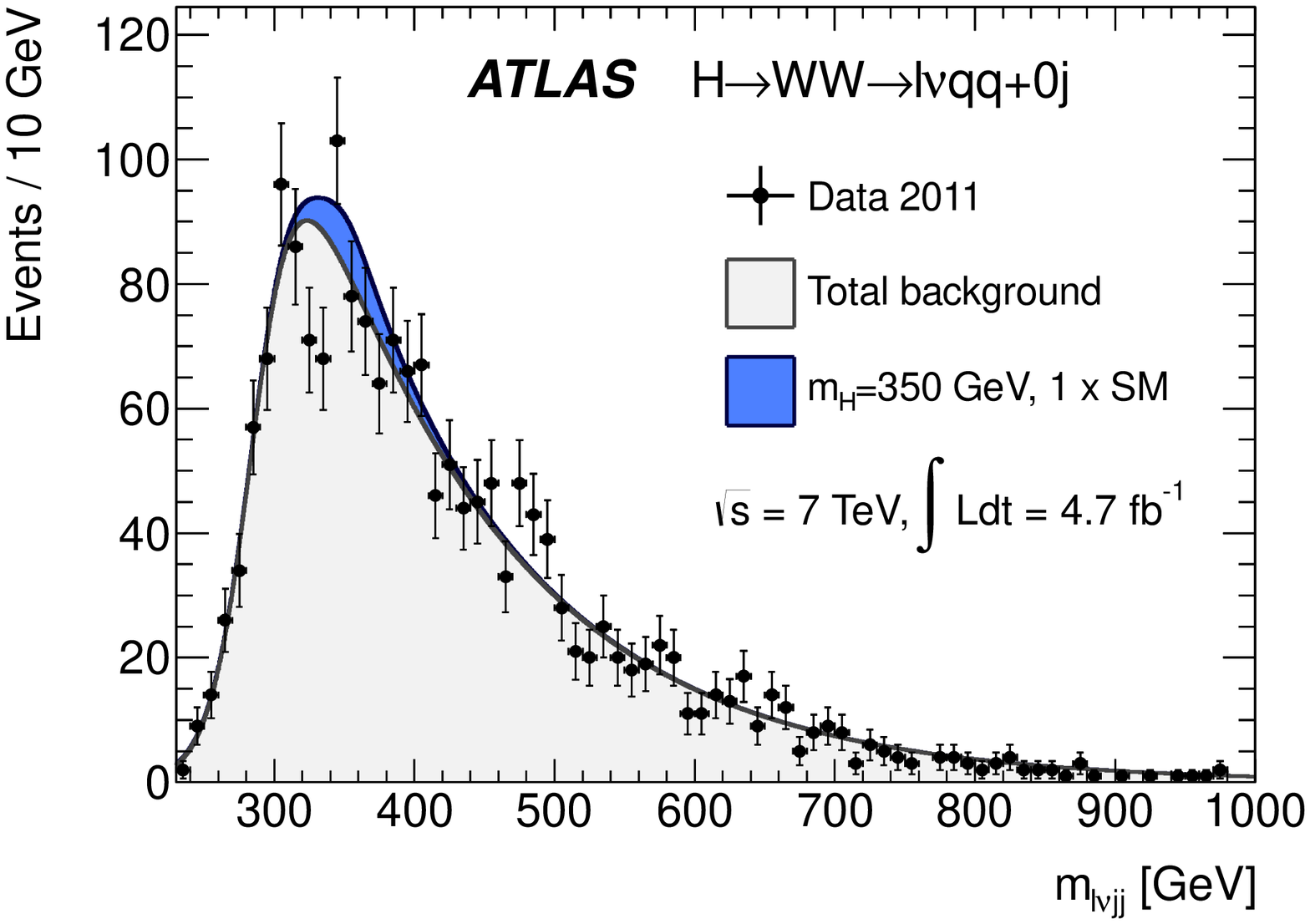}
  \end{minipage}
  \caption{Distributions of the final discriminants in the ATLAS searches~\cite{ATLAS:2012ae} in the high-$\MH$ region:
   $\Htoln$ (top-left), $\Htolq$ (top-right),  $\Htolb$ (bottom-left), and  $\Htowwlnqq$ (bottom-right).} 
  \label{fig:masdisthighmh}
\end{figure}

In the low-mass range in addition to $\Htowwll$ and $\Htofl$ the search is performed for $\Htogg$, $\Htott$, and $\Htobb$.
\begin{itemize} 
\item The $\Htogg$ analysis selects events with two isolated photons with large transverse momenta. The dominant reducible 
backgrounds are multi-jet production and single-photon production in association with jets with cross sections several orders 
of magnitude 
larger than the signal cross section. These backgrounds can be suppressed to a level of less than 25\% due to
the excellent discrimination capabilities of the LHC detectors between photon and jets. The irreducible background from di-photon production 
can only be discriminated from the signal process by an excellent reconstruction of the invariant mass of the di-photon system, which depends 
on the measurement of the energies of the photons and their opening angle. In both experiments the contribution from the resolution in the 
measurement of the opening angle was found to be negligible. The di-photon mass spectra after requiring two isolated photon candidates for 
the data from 2011 and 2012 are shown in \reffi{fig:hggmass}.  
\begin{figure}
  \centering
  \begin{minipage}[b]{.475\textwidth}
    \includegraphics[width=.9\textwidth,height=6.2cm]{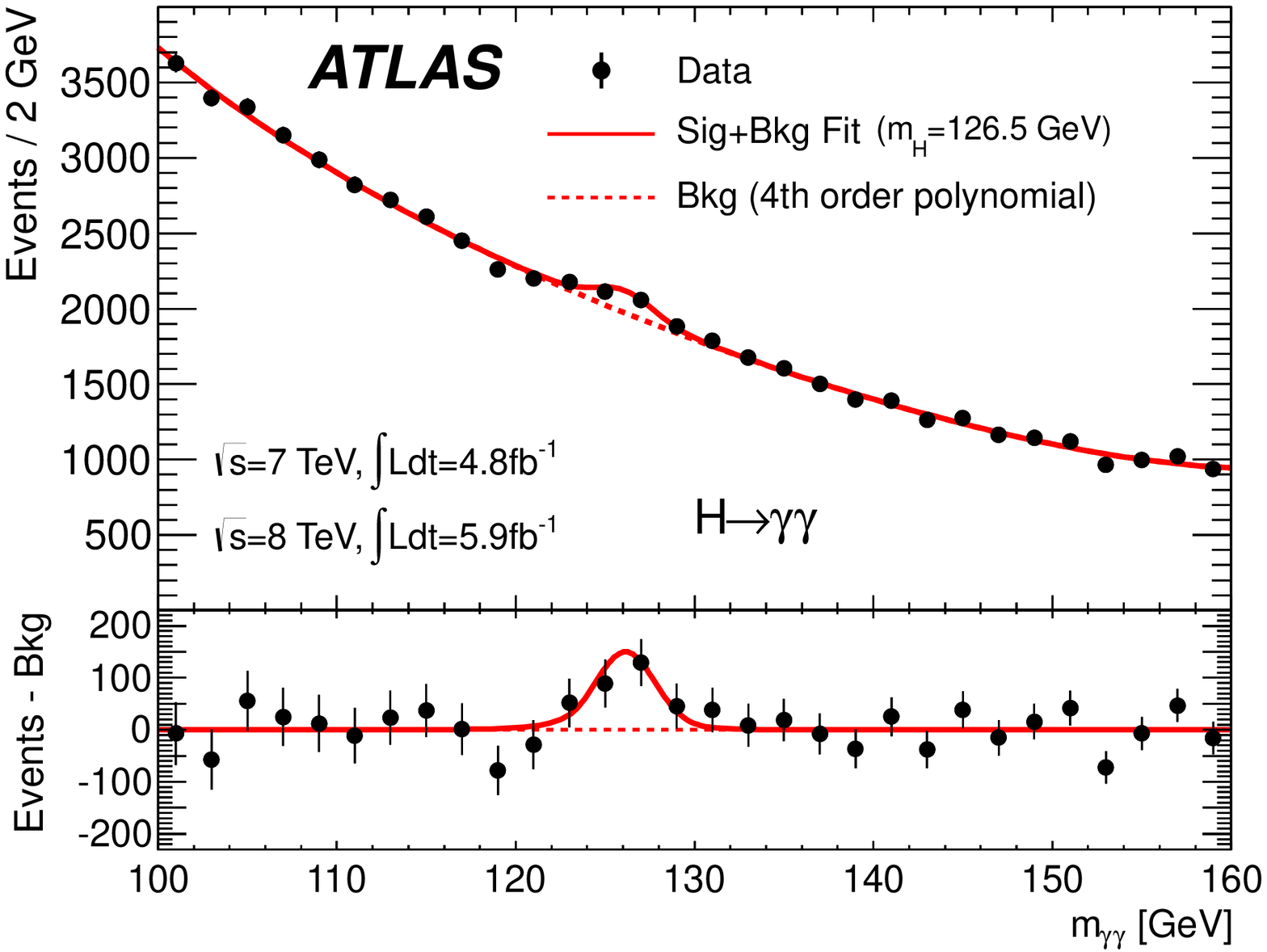}
  \end{minipage}
  \begin{minipage}[b]{.475\textwidth}
    \includegraphics[width=0.9\textwidth,height=6.5cm]{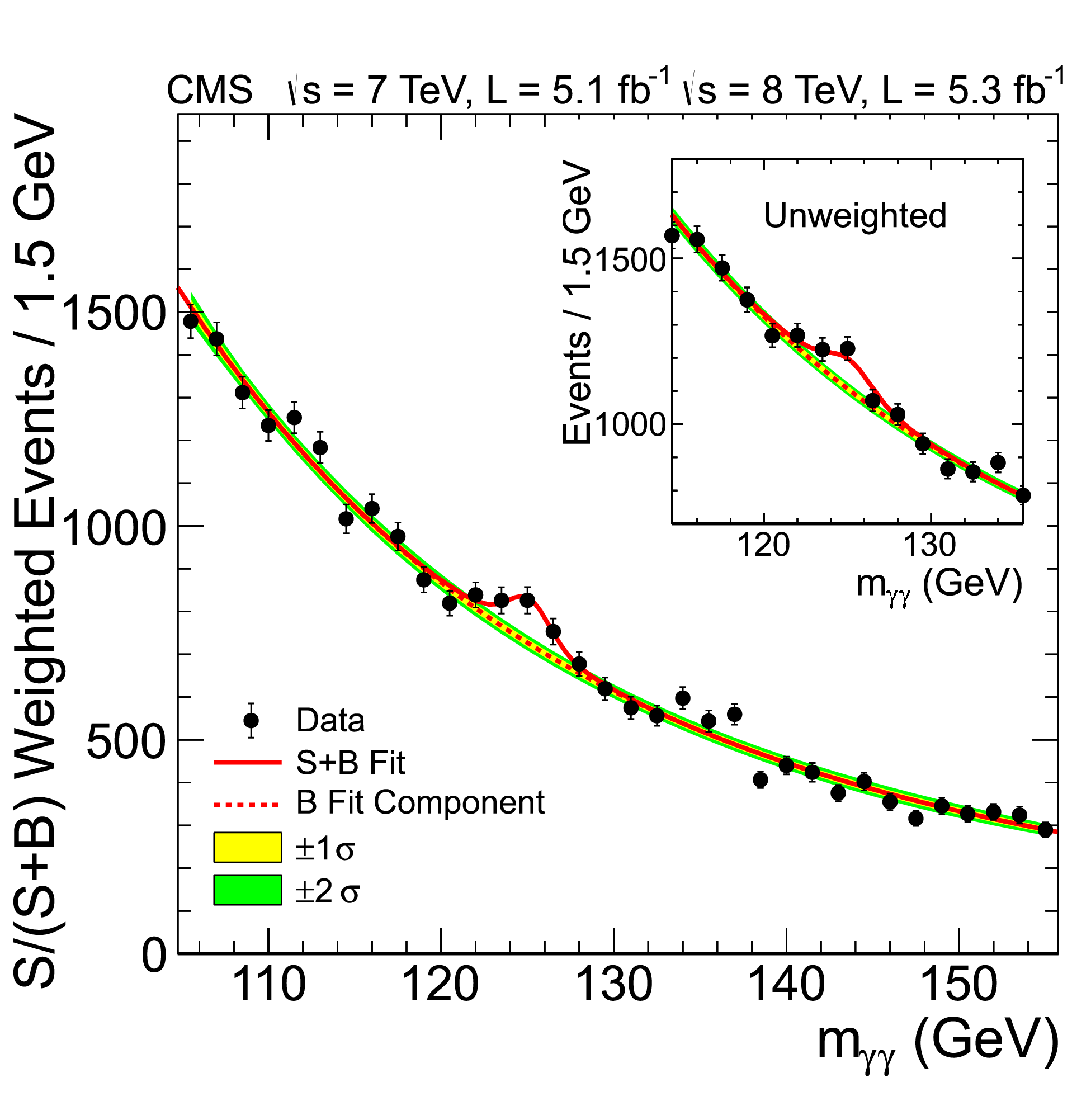}
  \end{minipage}
  \caption{Di-photon invariant-mass spectra after the inclusive selection
   in ATLAS~\cite{:2012gk} (left) and CMS~\cite{:2012gu} (right) for the combination of the 2011 and 2012 data.}
  \label{fig:hggmass}
\end{figure}
The sensitivity of the search can be enhanced by splitting the event sample 
in various categories with different ratios of expected signal over background event yields and different signal mass resolutions. 
Both experiments consider event classes exploiting the typical VBF signature of two additional jets with large separation in pseudorapidity 
and large di-jet invariant mass. ATLAS uses nine additional categories based on the pseudorapidity of each photon, whether it is reconstructed as a 
converted or unconverted photon, and the momentum component of the di-photon system transverse to the di-photon thrust axis.  CMS uses four additional 
categories depending on the score of a boosted decision tree.  The background in the signal region is estimated from a fit to the observed di-photon 
mass distribution in data. A clear excess in the invariant di-photon mass spectra is observed in both experiments at masses of $126\GeV$ in ATLAS and
$125\GeV$ in CMS, respectively. 
\item The $\Htott$ analysis searches for a broad excess in the reconstructed di-tau invariant-mass distribution, 
where the mass reconstruction is
mostly based on the ideas in \citeres{Ellis:1987xu,Elagin:2010aw} with a resolution between $10{-}30$\%. Events are classified according to the tau-lepton 
decay modes (into an electron, a muon, or hadrons plus neutrinos). ATLAS considers all decay-mode combinations, whereas CMS has not 
investigated yet the double electron and double hadronic final state. Depending on the final state the event samples are further divided into exclusive 
subcategories, which are optimized for the sensitivity to a particular production mode. The category optimized for the VBF signature provides the 
largest sensitivity. Events failing these selection requirements are separated further, e.g., in a category with two jets optimized for associated production $\PH\PW (\PH\PZ)$,  
with one jet with large transverse momentum for associated production and production in gluon fusion, and with either no jets or with one with a small transverse momentum
aiming at the gluon-fusion production process. The main irreducible background arises from $\PZ\to\tau\tau$ production, whose di-tau invariant-mass distribution is 
derived from data by selecting $\PZ\to\mu\mu$ events, in which the reconstructed muons are replaced with reconstructed particles 
from the decay of simulated $\tau$ leptons of the same momenta. The reducible backgrounds  from $\PW$+jets, Drell--Yan, and multi-jet production
are also evaluated from control samples in data. The distribution of the di-tau mass distribution for 2011 and 2012 data in CMS is shown in 
\reffi{fig:cmshtthbb} (left). No significant deviation from the background expectation is observed.
\begin{figure}
  \centering
  \begin{minipage}[b]{.49\textwidth}
    {\includegraphics[bb = 250 40 842 585,clip,width=.97\textwidth,height=.92\textwidth]{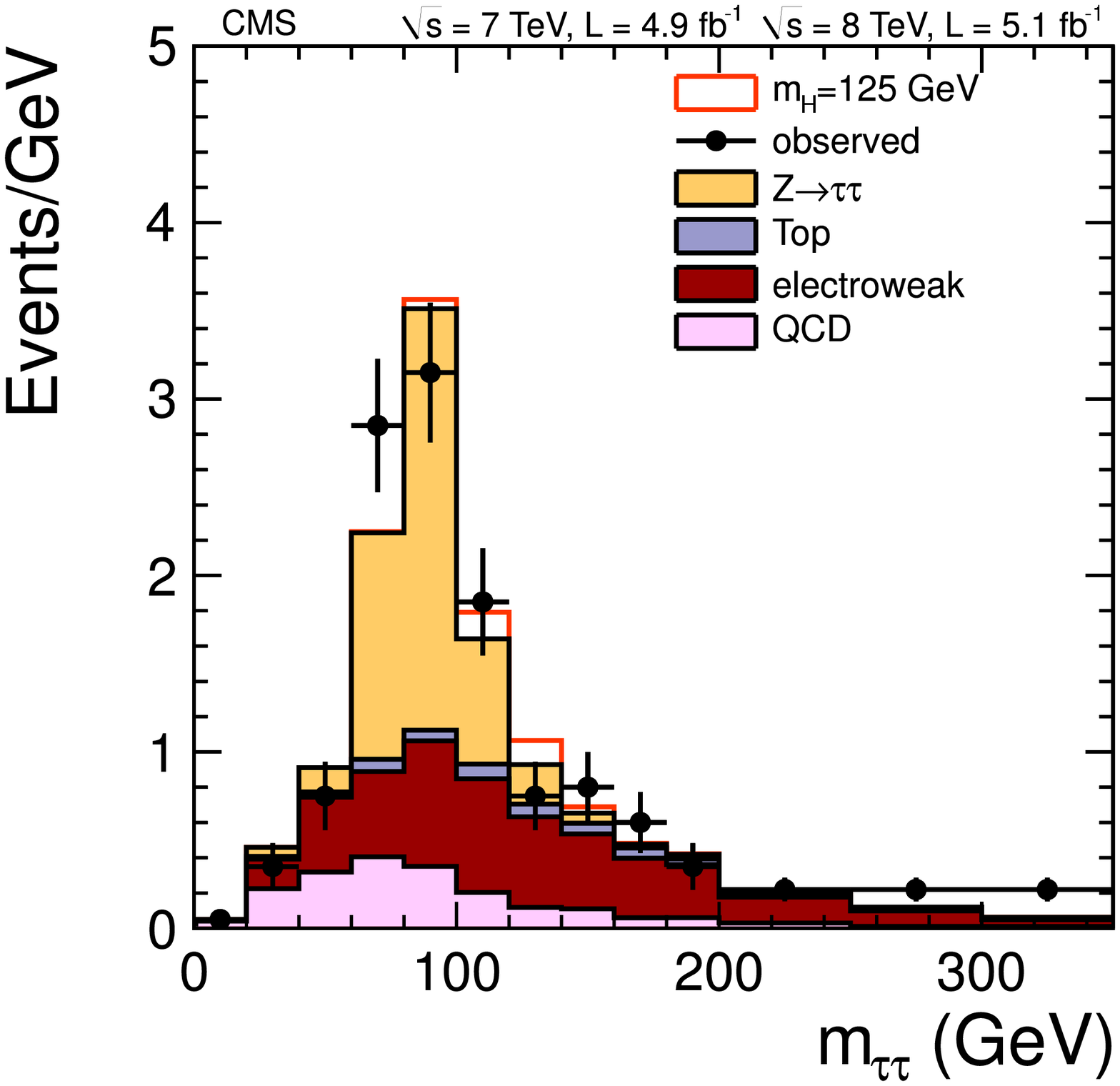}}
  \end{minipage}
  \begin{minipage}[b]{.49\textwidth}
    {\includegraphics[bb = 250 40 842 585,clip,width=0.97\textwidth]{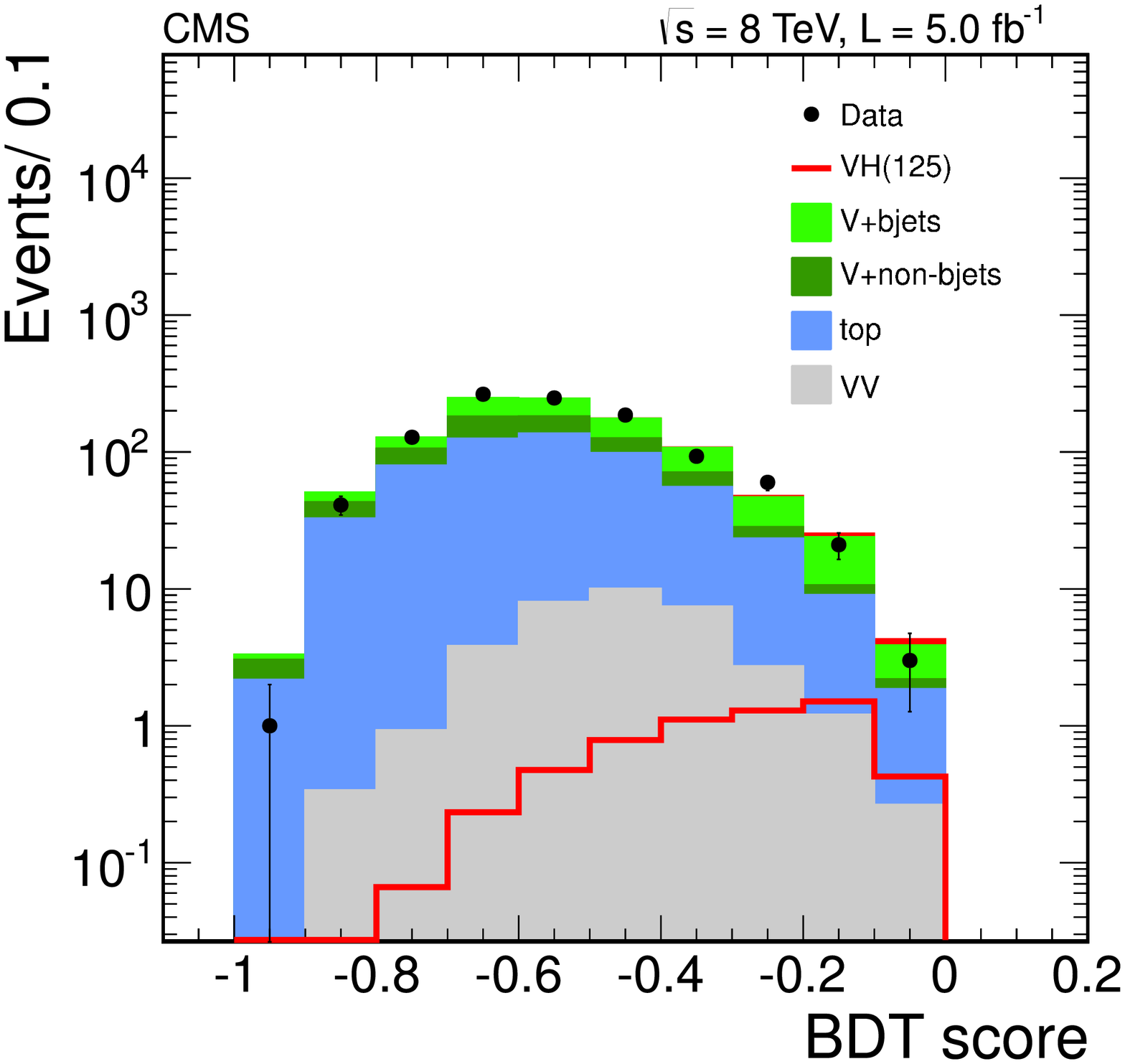}}
  \end{minipage}
  \caption{Distributions of the final discriminant for the search 
           in $\Htott$ combined for 2011 and 2012 data (left) and 
            $\Htobb$ for 2012 data (right) in CMS~\cite{:2012gu}.}
  \label{fig:cmshtthbb}
\end{figure}

\item The $\Htobb$ analysis selects events produced in association with a $\PW$ or $\PZ$ decaying via $\PZ\to\Pl^+\Pl^-$, $\PZ\to\nu\bar{\nu}$, or $\PW\to\Pl\nu$.  
The selection requires two b-tagged jets and either two leptons, one lepton and missing transverse momentum, or large missing transverse momentum.
CMS selects events with a high transverse momentum of the di-jet system in order to improve the mass resolution and to suppress background processes.
A multivariate classifier is trained for different $\MH$ values and its output is used as the final discriminant in CMS (see \reffi{fig:cmshtthbb}, right).
ATLAS subdivides the events in categories according to the transverse momentum of the weak gauge boson and uses the di-jet invariant mass of the 
b-tagged jets as the final discriminant. No significant deviation from the background expectation is observed.
\end{itemize} 

\subsection{Excluded mass ranges}
\label{se:lhcexclusion}

In most search channels and over almost the full investigated mass range from $110{-}600\GeV$
no significant excess with respect to the background-only hypothesis is observed.
This allows for the exclusion of a large part in the mass range of the SM Higgs boson.
In the combination of the different search 
channels the ratio of the cross section for the different production modes and the ratio of the decay branching ratios are
taken from the SM prediction. 
Upper limits on the signal strength $\mu$, which relates the excluded cross section to that for Higgs-boson production
in the SM, are derived based on the profile likelihood ratio \cite{Cowan:2010js} as test statistic using the CL$_{\mathrm{S}}$ 
technique \cite{Read:2002hq} to avoid exclusion of parameter space where no sensitivity is expected. Correlations in the 
theoretical and systematic  uncertainties, which are included via nuisance parameters in the likelihood function, 
among the various search channels and event categories are taken into account 
(see \citeres{comproc,:2012an,Chatrchyan:2012tx} for details).
The exclusion limits obtained in both experiments in the various search channels and their combination from the first 
analysis of the full 2011 data set are shown in \reffi{fig:limit2011channels}, for the combination in each experiment 
from the (re-)analysis of 2011 and 2012 data in \reffi{fig:limit2012}, and both are summarized in \refta{tab:lhclimits}.
\begin{figure}
  \centering
  \begin{minipage}[b]{.475\textwidth}
    \includegraphics[width=.9\textwidth,height=8cm]{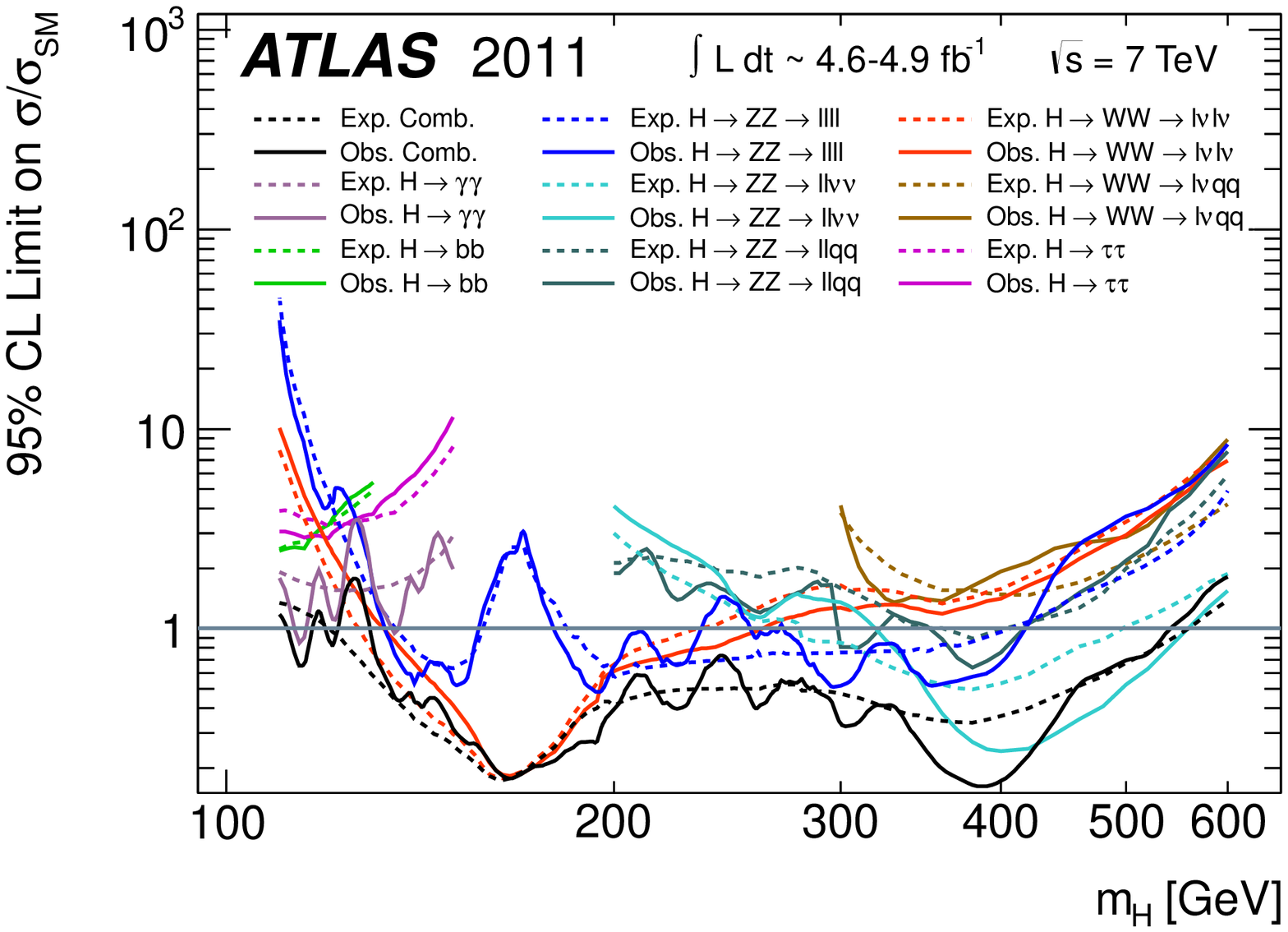}
  \end{minipage}
  \begin{minipage}[b]{.475\textwidth}
    \includegraphics[width=0.9\textwidth,height=8cm]{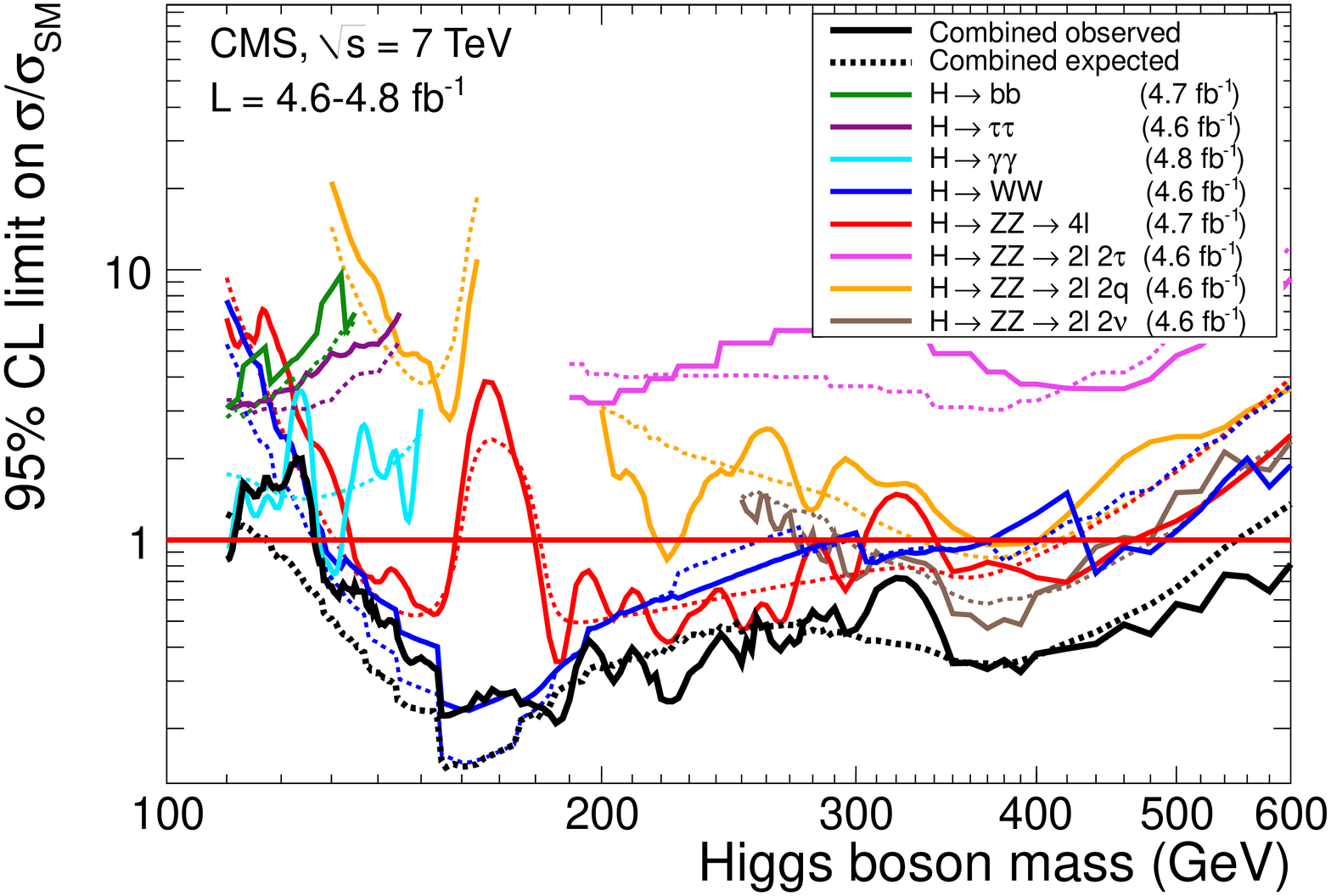}
  \end{minipage}
  \caption{Exclusion limits on the signal-strength parameter $\mu$
   as a function of the hypothetical Higgs-boson mass in the individual 
   channels and their combination obtained from the analysis of 2011 data 
   by  ATLAS~\cite{:2012an} (left) and CMS~\cite{Chatrchyan:2012tx} (right).}
  \label{fig:limit2011channels}
\end{figure}
\begin{figure}
  \centering
  \begin{minipage}[b]{.475\textwidth}
    {\includegraphics[width=.9\textwidth,height=0.66\textwidth]{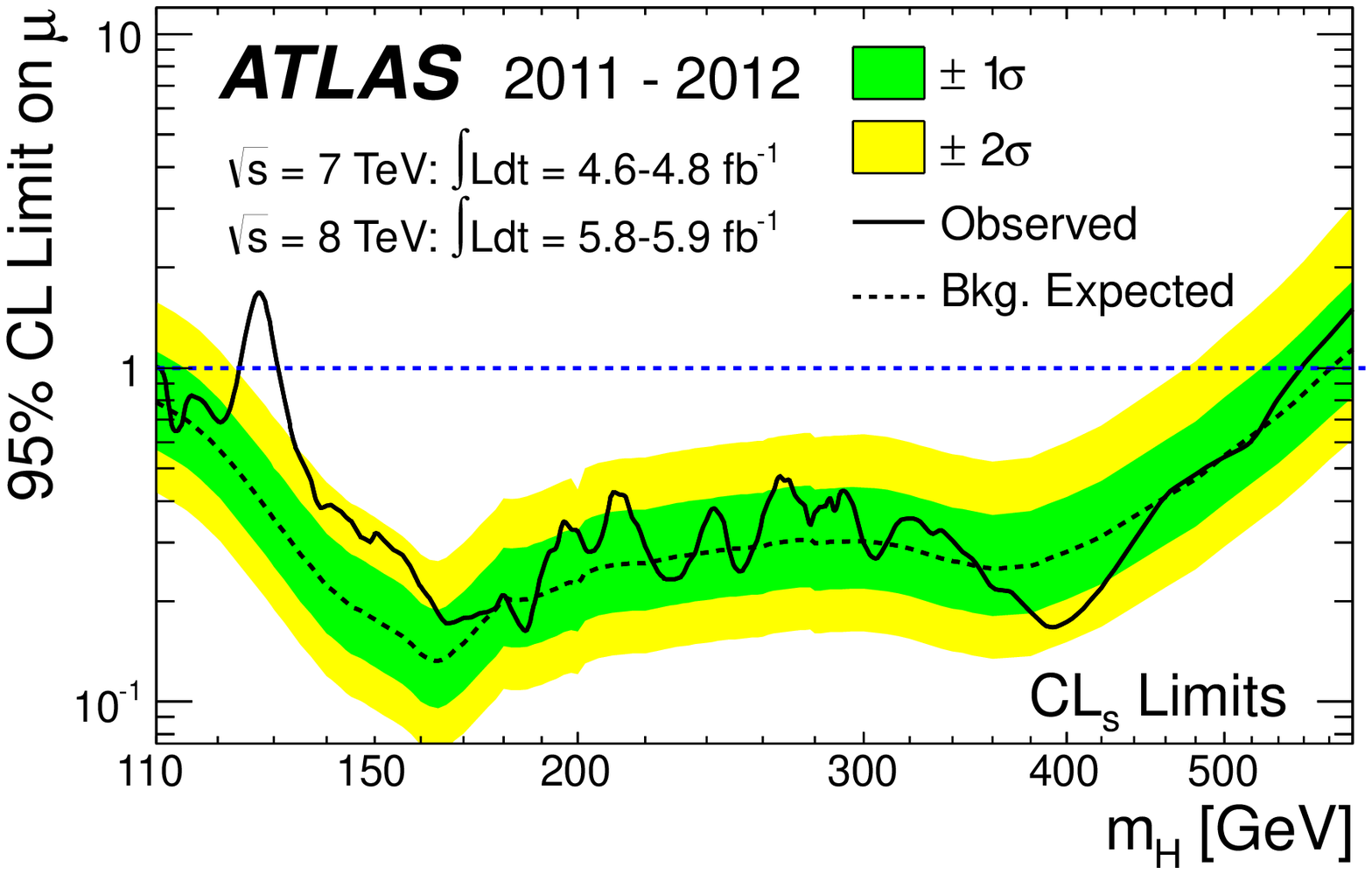}}
  \end{minipage}
  \begin{minipage}[b]{.475\textwidth}
    {\includegraphics[bb= 0 0 812 592,clip,width=0.9\textwidth,height=0.66\textwidth]{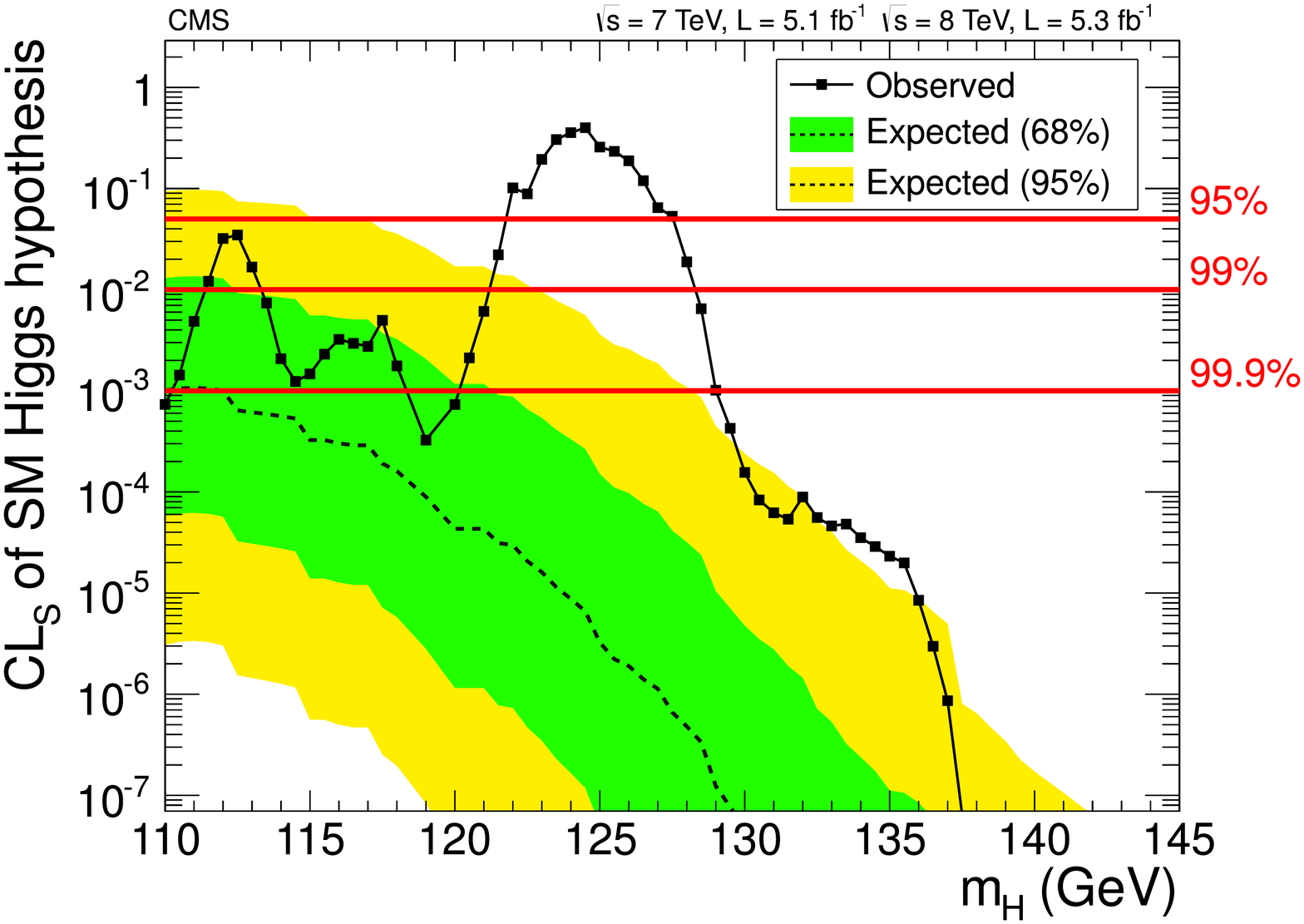}}
  \end{minipage}
  \caption{Exclusion limits on the signal-strength parameter $\mu$
   as a function of the hypothetical Higgs-boson mass
   obtained from the analysis of 2011 and 2012 data 
   by ATLAS~\cite{:2012gk} (left) and confidence level CL$_{\mathrm{S}}$ for rejecting the
   signal+background hypothesis as a function of the hypothetical 
   Higgs-boson mass by CMS~\cite{:2012gu} (right).}
  \label{fig:limit2012}
\end{figure}
\begin{table}
\caption{Mass ranges for the SM Higgs boson excluded by the combination of all searches of ATLAS and CMS:
         observed excluded range, expected excluded range for 2011 data alone \cite{:2012an,Chatrchyan:2012tx}, 
         and combination of reanalyzed 2011 data and 2012 data \cite{:2012gk,:2012gu}.}
\label{tab:lhclimits}
\begin{center}
{
\renewcommand{\tabcolsep}{5pt}
\begin{tabular}{c||c|c||c|c}
\hline
           & \multicolumn{2}{c||}{2011 data} & \multicolumn{2}{c}{2011+2012 data} \\ 
           & excl.~$\MH\,$[GeV]& exp.~excl.~$\MH\,$[GeV]& excl.~$\MH\,$[GeV] & exp.~excl.~$\MH\,$[GeV] \\ \hline 
ATLAS      & 111.4--116.6, 119.4--122.1 & 120--560 & 111--122,  131--559  & 110--582    \\  
           & 129.2--541               &         &                    &            \\ \hline 
CMS        & 127--600                 & 118--543 & 110--121.5, 127--600  & 110--543    \\ \hline 
\end{tabular}
}
\end{center}
\end{table}     
Already with the data set from 2011 a large mass range is excluded. However, in the low-mass range the observed 
limit is weaker than the expected one, which extends from $118{-}560\GeV$ (by applying a simple ``or'' of the 
two experiments). After adding the 2012 data set and re-analyzing the 2011 data the expected exclusion sensitivity
extends over the mass range of $110{-}582\GeV$. The mass range that
is excluded neither by ATLAS nor by CMS,
is the interval from $122{-}127\GeV$. The large deviations of the observed from  the expected limits are caused by the observed excesses 
of events discussed above. For a large mass range, reduced production cross sections of 20\% of the SM prediction are excluded as well, and the 
SM Higgs-boson hypothesis can be excluded with very high confidence level.

\subsection{Observation of a new particle}

When testing the compatibility with the background-only hypothesis, already in the 2011
data, tantalizing hints for the possible production of a new particle were observed,
which corresponded to local significance of 2.9 at a mass of $126\GeV$ in ATLAS~\cite{:2012an}
and a local significance of 3.1 at a mass of $124\GeV$ in CMS~\cite{Chatrchyan:2012tx}. These local significances
are already larger than the ones shown by the individual Tevatron experiments 
and quite similar to the combined significance published by CDF and D0.
When taking into account the ``look-elsewhere effect'' in the whole mass range 
considered from $110{-}600\GeV$ the global significances were found to be roughly 1
in ATLAS and 1.5 in CMS. After adding the 2012 data available up to June to the analysis 
the local significances ($p$-values) increased (decreased) to 5.9 
($1.7\times10^{-9}$) at $126.5\GeV$ in ATLAS and 5.0 ($2.8\times10^{-7}$) in CMS (see \reffi{fig:lhcp0}). 
\begin{figure}
  \centering
  \begin{minipage}[b]{.45\textwidth}
    \includegraphics[width=0.9\textwidth,height=0.76\textwidth]{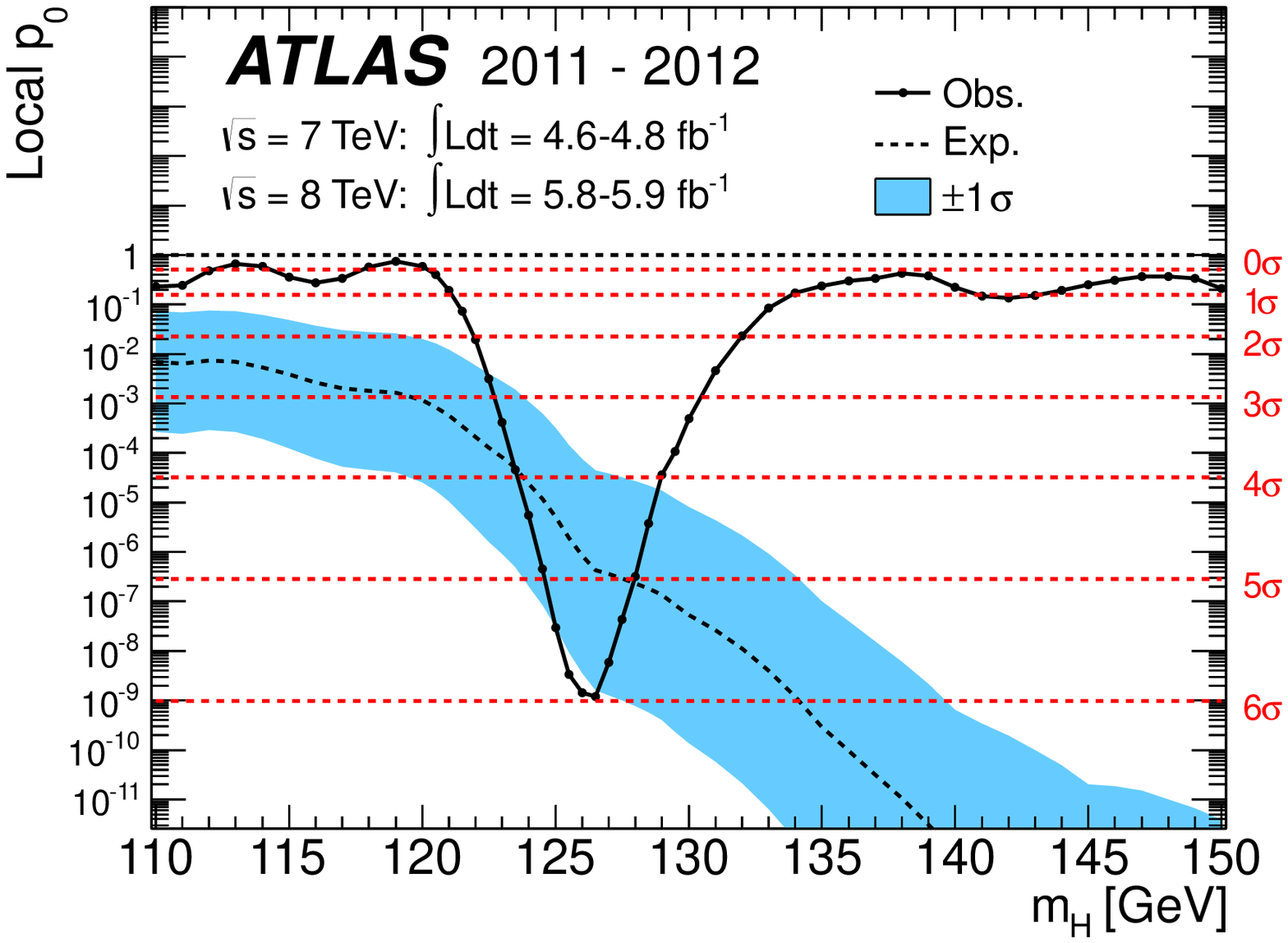}
  \end{minipage}
  \begin{minipage}[b]{.45\textwidth}
    \includegraphics[width=0.9\textwidth,height=0.76\textwidth]{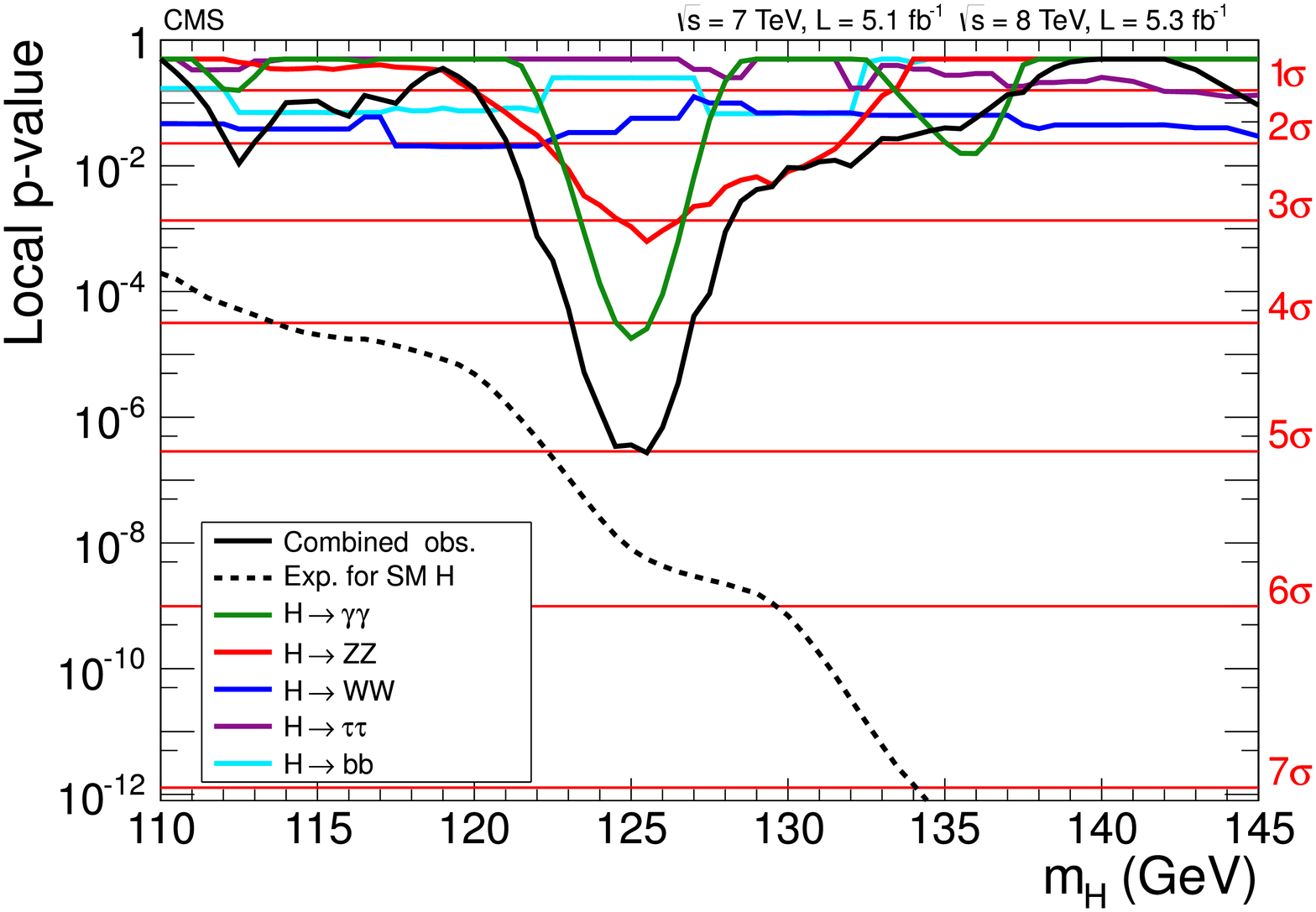}
  \end{minipage}
  \caption{Local $p$-values as function of the hypothetical  Higgs-boson mass in 
          ATLAS~\cite{:2012gk} (left) and CMS~\cite{:2012gu} (right).}
  \label{fig:lhcp0}
\end{figure}
An overview of the observed and expected significances in the various search channels
and their combination is given in \refta{tab:lhcsignificances}, 
ordered by the size of the excess. 
\begin{table}
\caption{Local significances in different search channels in ATLAS \cite{:2012gk} and CMS \cite{:2012gu}.
Z denotes the maximum significance observed at the mass value M$_{\mathrm{max.\,Z}}$ when scanning
the hypothetical Higgs-boson mass in the hypothesis test, and Z$_{\mathrm{exp.}}$ is the median 
expected significance for a SM Higgs boson at the same mass value. An entry ``--'' means that 
the information is not publicly available.
}
\label{tab:lhcsignificances}
\begin{center}
\begin{tabular}{c||c|c|c||c|c|c}
\hline
         &\multicolumn{3}{c||}{ATLAS}&\multicolumn{3}{c}{CMS}\\ \hline 
channel  & Z    &  Z$_{\mathrm{exp.}}$  & M$_{\mathrm{max.\,Z}}$ & Z   & Z$_{\mathrm{exp.}}$  & M$_{\mathrm{max.\,Z}}$\\ \hline
$\Htogg$   & 4.5  &  2.5         & 126.5 GeV    & 4.1 & 2.8         & 125   GeV    \\ 
$\Htofl$   & 3.6  &  2.7         & 125.0 GeV    & 3.2 & 3.8         & 125.6 GeV    \\ 
$\Htowwll$ & 2.8  &  2.3         & 125.0 GeV    & 1.6 & 2.4         &  --           \\
$\Htobb$+$\Htott$ & -- &  --       & --              & 0.4 & 2.4         &  --           \\ \hline 
combined & 5.9  &  4.9         & 126.5 GeV    & 5.0 & 5.8         & 125.5 GeV     \\ 
\hline
\end{tabular}
\begin{minipage}[t]{16.5 cm}
\vskip 0.5cm
\noindent
\end{minipage}
\end{center}
\end{table}     
The observed significance 
is due to the excesses observed in the channels with high mass resolution $\Htogg$,
$\Htofl$ and with large production rate $\Htowwll$.
No hints for the decay into a pair of fermions are observed yet, though the expected 
significance in the combination of $\Htott$ and $\Htobb$ is equal to the expected one 
in $\Htowwll$ in the CMS analysis.
Assuming that the observed signal is due to the SM Higgs boson, ATLAS observes an upwards fluctuation 
of roughly one standard deviation compared to the expectation at a mass of $126.5\GeV$ and 
CMS observes a downwards fluctuation of roughly one standard deviation at a mass of $125.5\GeV$.
The global significance is 5.1 (5.3) in ATLAS for the mass range $110{-}600\,(110{-}150)\GeV$
and 4.5 (4.6) in CMS for the mass range $110{-}145\,(110{-}130)\GeV$. These findings 
encouraged the two experiments ATLAS and CMS to proclaim the discovery of
a new particle and to entitle their publications~\cite{:2012gk,:2012gu} 
with ``Observation of a new particle / boson''.

\subsection{Anatomy of the new particle}

From the observation of the decay into a pair of particles with identical spin and with vanishing sum of the 
electric charges it can be concluded that the particle is electrically neutral, has integer 
spin, and hence is a neutral boson. The observation of the decay in two massless identical vector bosons, $\Htogg$, 
excludes the hypothesis of 
a spin-1 particle according to the Landau--Yang theorem~\cite{landauyang}.

A first crucial test of the compatibility of the observed event yields with the prediction for the SM Higgs boson
is provided by a fit of the signal-strength parameter $\mu$. The overall signal strength relative to the SM prediction
$\mu$, which assumes the ratio of production cross sections and ratio of branching ratios as predicted by the SM, 
is shown in \reffi{fig:lhcmu} as function of the hypothetical Higgs-boson mass. 
\begin{figure}
  \centering
  \begin{minipage}[b]{.475\textwidth}
    {\includegraphics[width=0.9\textwidth,height=0.7\textwidth]{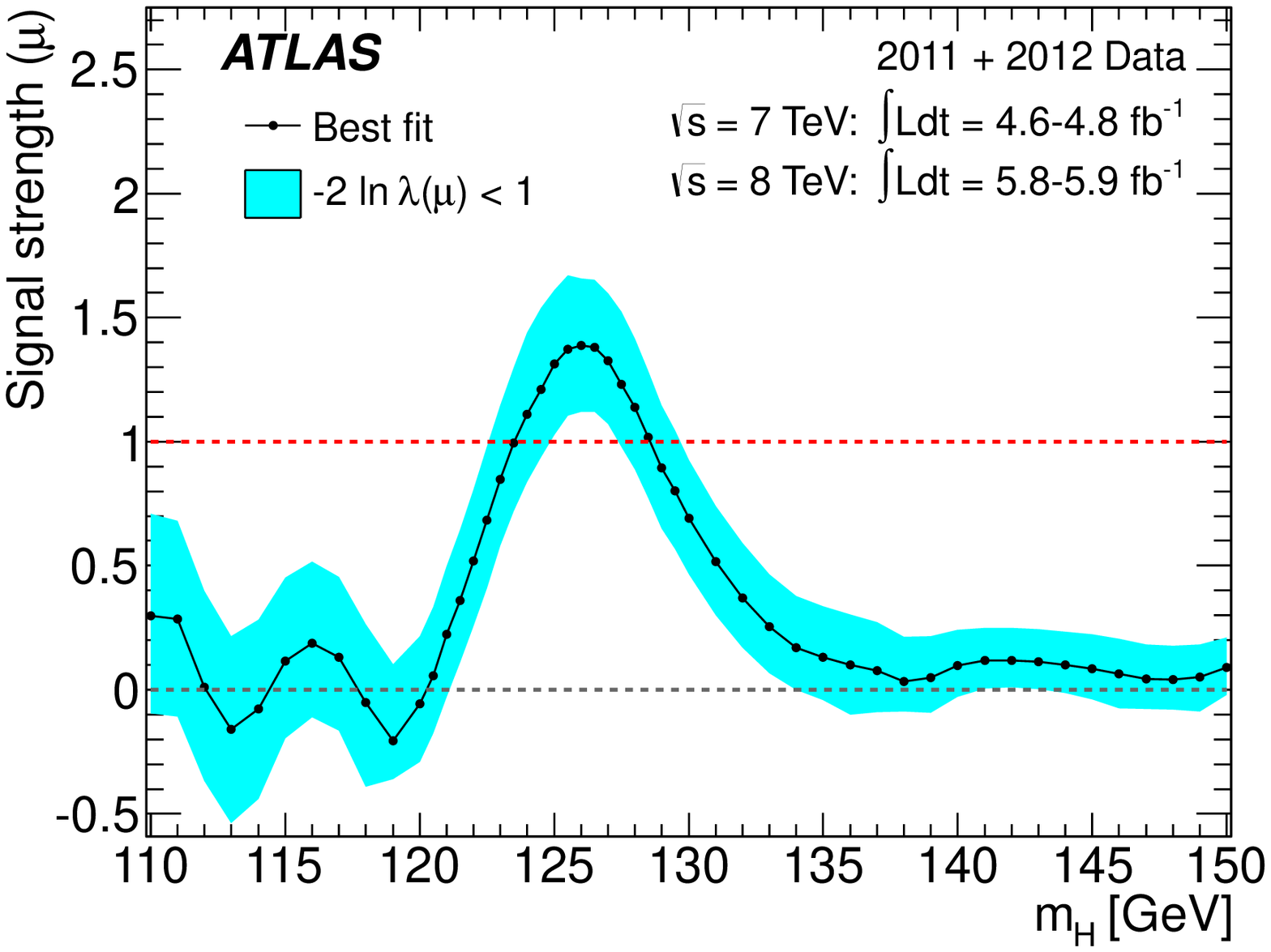}}
  \end{minipage}
  \begin{minipage}[b]{.475\textwidth}
    {\includegraphics[bb= 0 0 812 592,clip,width=0.9\textwidth,height=0.7\textwidth]{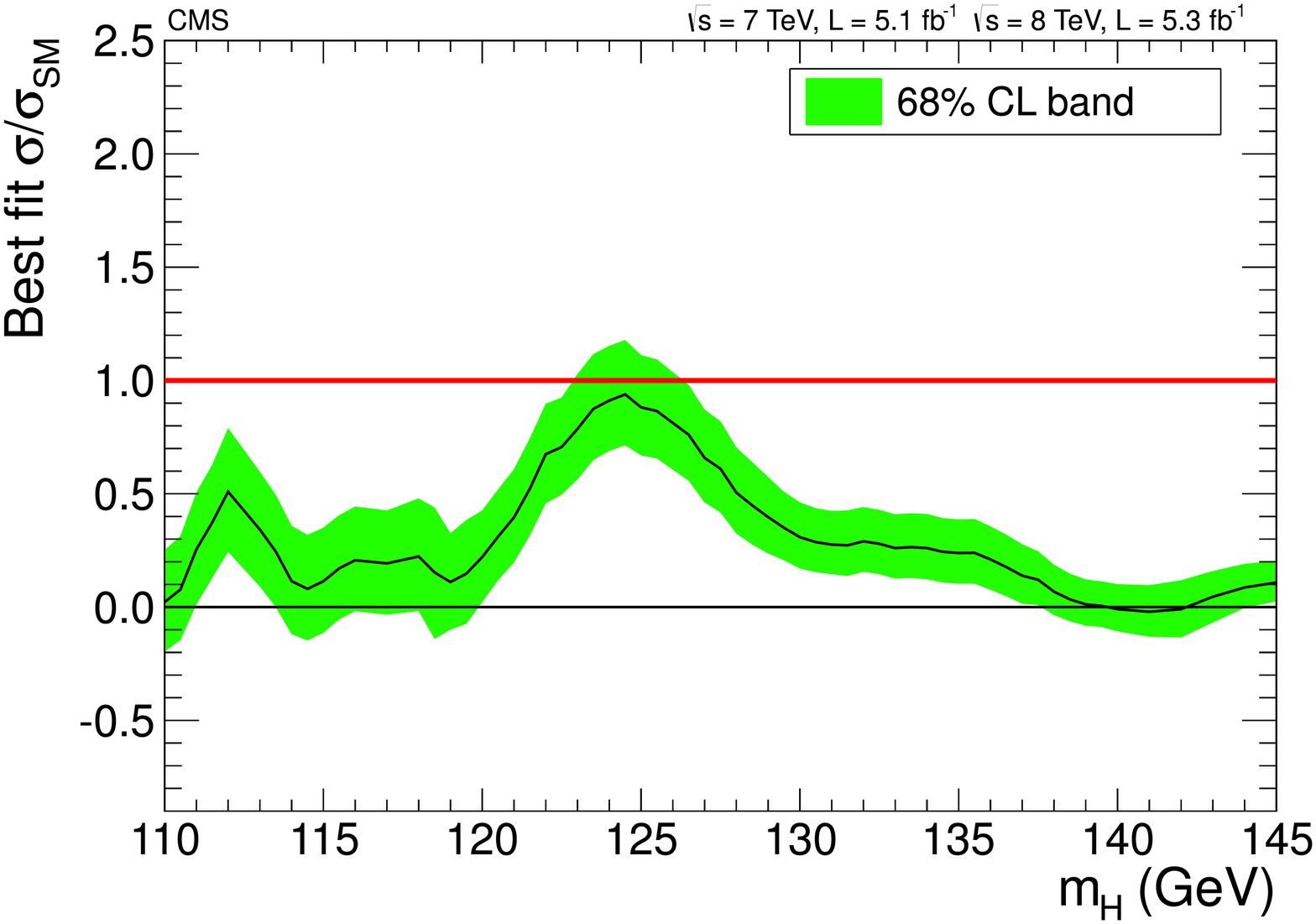}}
  \end{minipage}
  \caption{Best fit values for the signal strength $\mu$ 
           as function of the hypothetical Higgs-boson mass in 
           ATLAS~\cite{:2012gk} (left) and CMS~\cite{:2012gu} (right).}
  \label{fig:lhcmu}
\end{figure}
In both experiments the largest signal 
strengths of $1.4\pm0.3$ observed at a mass of $126.0\GeV$ (ATLAS) and of $0.87\pm0.23$ observed at a mass of $125.5\GeV$ (CMS) are compatible 
with unity. In a next step the best signal strength is determined for each decay mode of the new boson, which only 
assumes that the different production modes contribute as predicted by the SM. The fit results for the different 
decay modes, where the hypothetical Higgs-boson mass is fixed to the value that yields the largest overall signal strength, 
are also consistent with the SM prediction (\reffi{fig:lhcmudec}) within the still large uncertainties.
\begin{figure}
  \centering
  \begin{minipage}[b]{.45\textwidth}
    {\includegraphics[width=0.9\textwidth,height=0.7\textwidth]{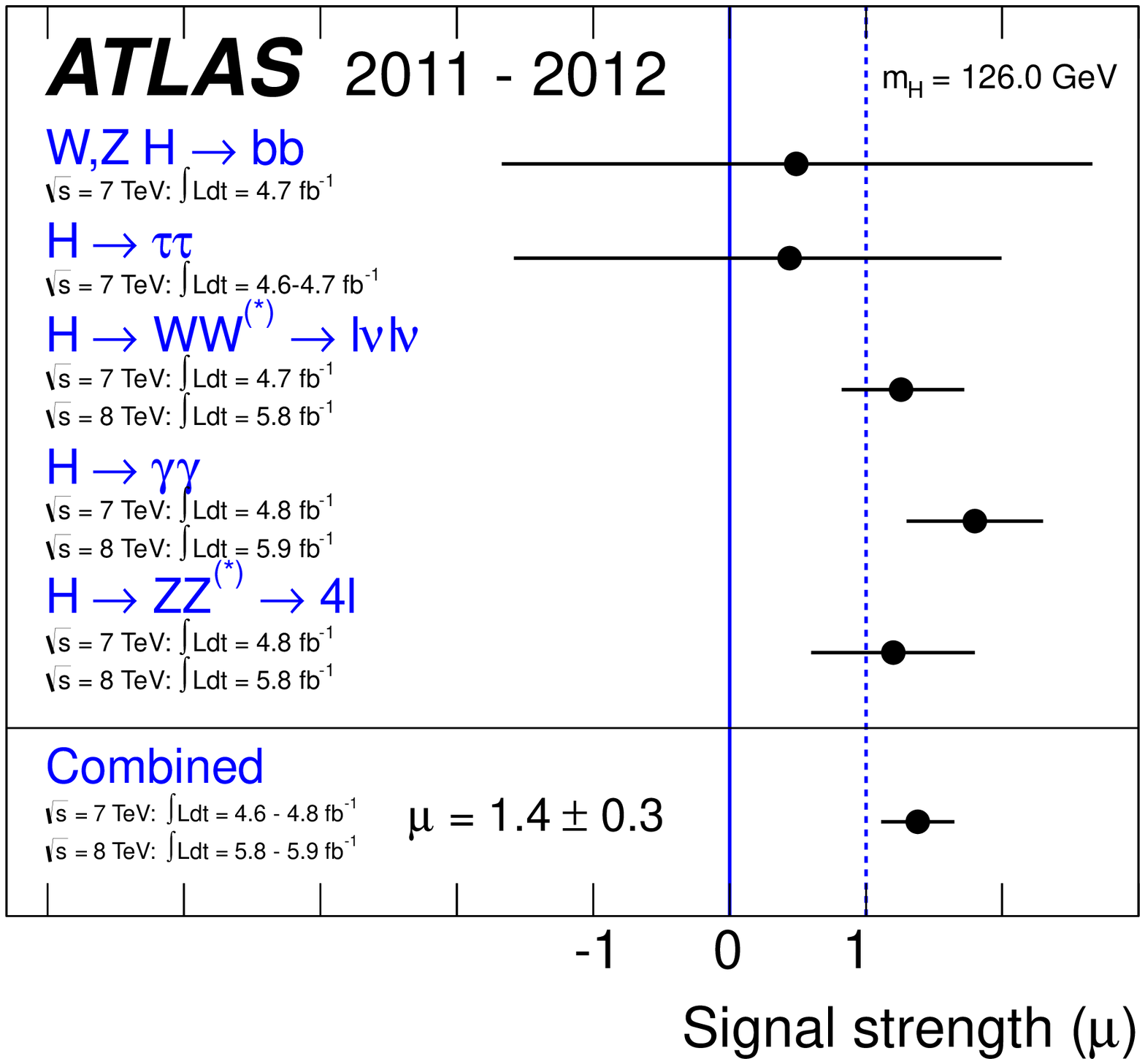}}
  \end{minipage}
  \begin{minipage}[b]{.45\textwidth}
    {\includegraphics[bb= 300 0 842 595,clip,width=0.9\textwidth,height=0.7\textwidth]{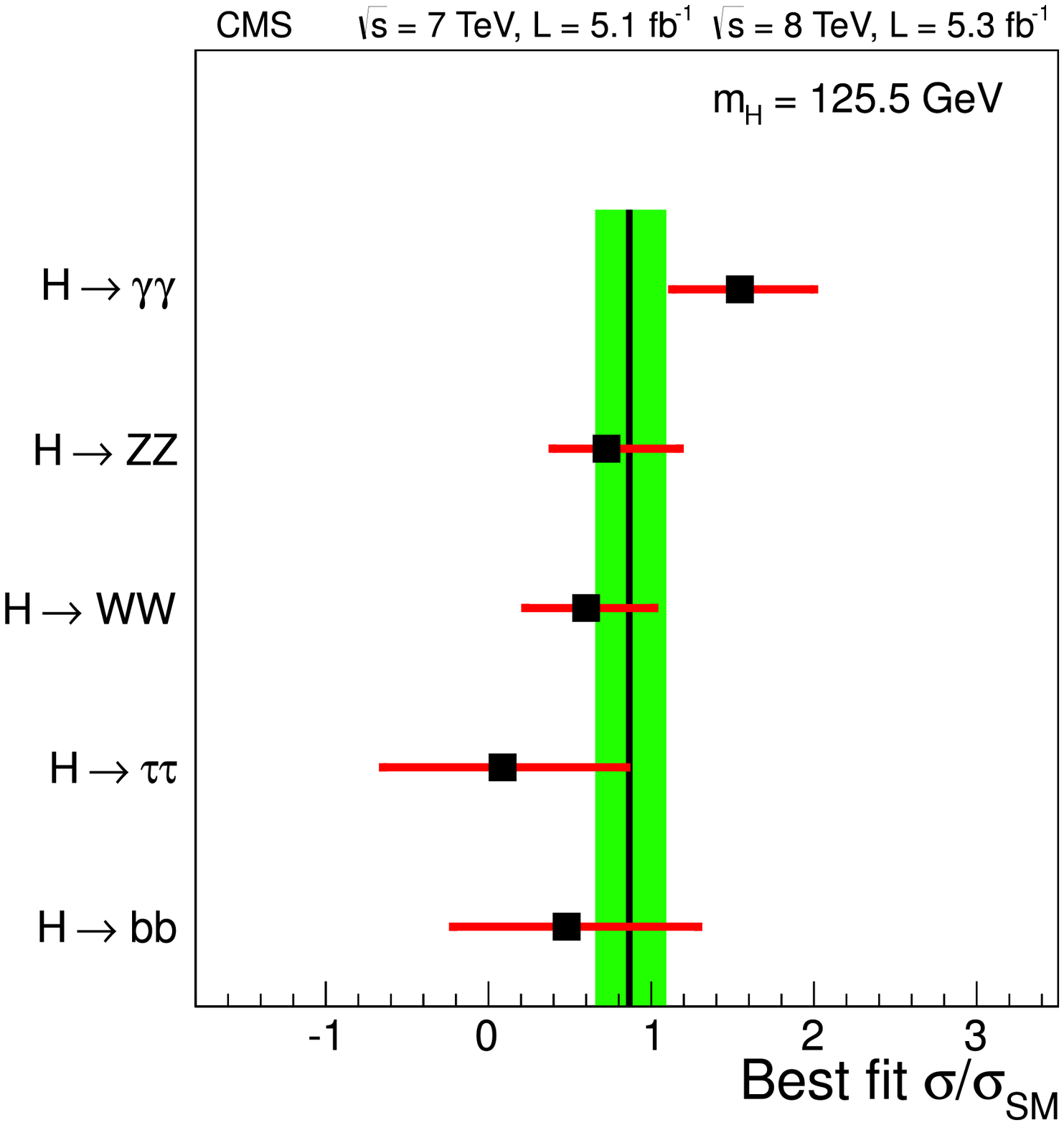}}
  \end{minipage}
  \caption{Best fit values for the signal strength $\mu$ in the different decay modes
           as function of the hypothetical Higgs-boson mass in 
           ATLAS~\cite{:2012gk} (left) and CMS~\cite{:2012gu} (right).}
  \label{fig:lhcmudec}
\end{figure}

The compatibility of a simultaneous determination of the best signal strength $\mu$ and the best mass value $\MH$ in several final 
states is demonstrated in \reffi{fig:lhcmumass}. The best mass value $\MH$ is determined in both experiments from the 
observed mass spectra in the $\Htogg$ and $\Htofl$ final states. The signal strengths in each final state 
(decay modes and classification categories) are treated as independent nuisance parameters.
The results for the mass of the new boson, as obtained by the two collaborations are 
$$
\mbox{ATLAS: } 126.0 \pm 0.4 (\mathrm{stat.}) \pm 0.4 (\mathrm{sys.}) \GeV \mbox{\cite{:2012gk}} \quad \mbox{and} \quad
   \mbox{CMS: } 125.3 \pm 0.4 (\mathrm{stat.}) \pm 0.5 (\mathrm{sys.}) \GeV \mbox{\cite{:2012gu}}.
$$
In both experiments the systematic uncertainty is dominated by the knowledge of the absolute energy scale of photons
and to a lesser extent of that of electrons. The mass values determined in the different final states within one experiment 
as well as the mass values determined in the two experiments are consistent with each other. Already now the precision of
the determination of the mass is better than one percent. Note that the measured mass value is perfectly
compatible with the $\MH$ range preferred by the fit of the SM to
electroweak precision measurements (see \refse{sec:blueband}).
The particle discovery, thus, is not only a triumph of the interplay of LHC measurements 
and corresponding theoretical predictions, but also a spectacular success of
electroweak precision physics---assuming the new particle is in fact the Higgs boson of the SM.
 
\begin{figure}
  \centering
  \begin{minipage}[b]{.475\textwidth}
    \includegraphics[width=0.9\textwidth,height=0.7\textwidth]{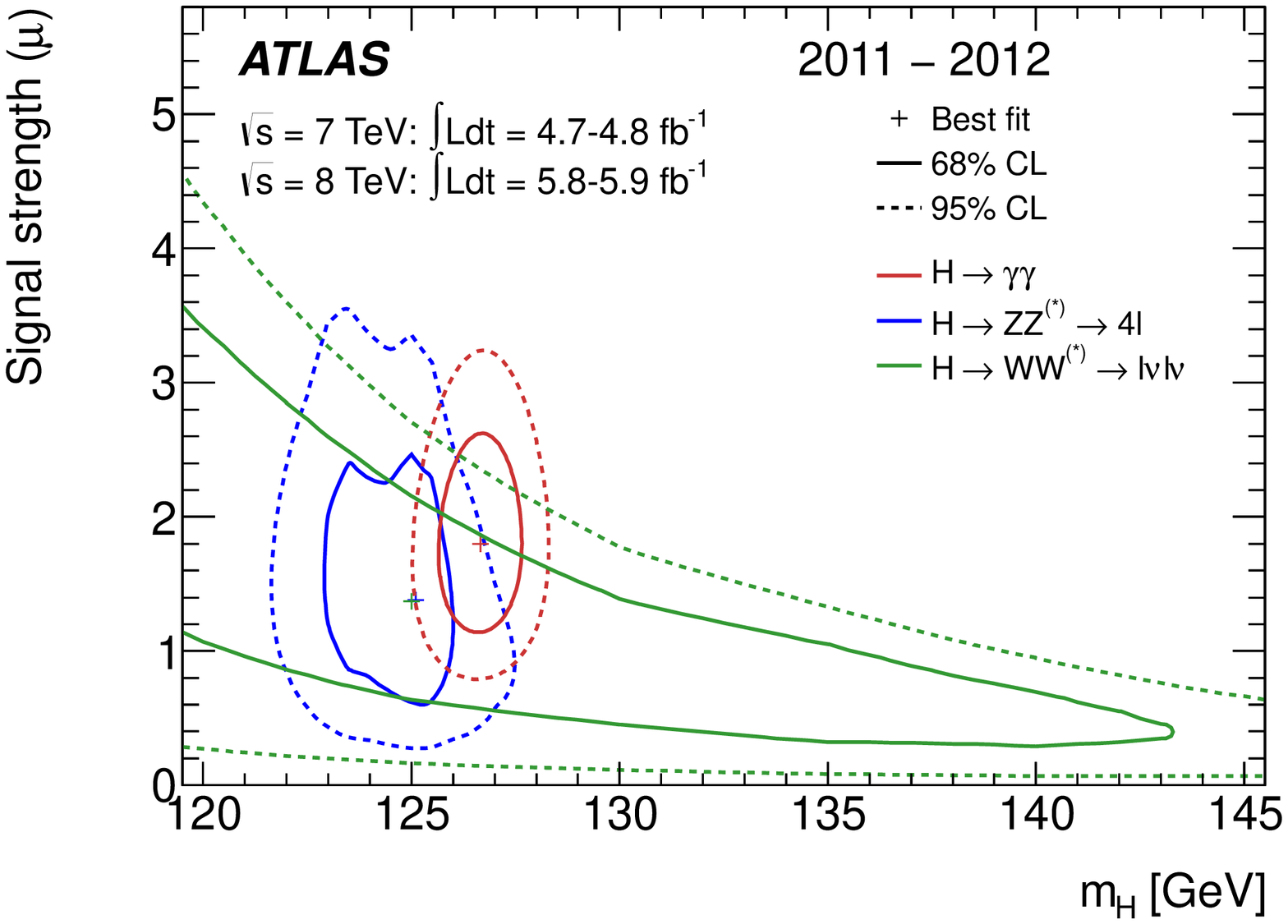}
  \end{minipage}
  \begin{minipage}[b]{.475\textwidth}
    \includegraphics[width=0.9\textwidth,height=0.7\textwidth]{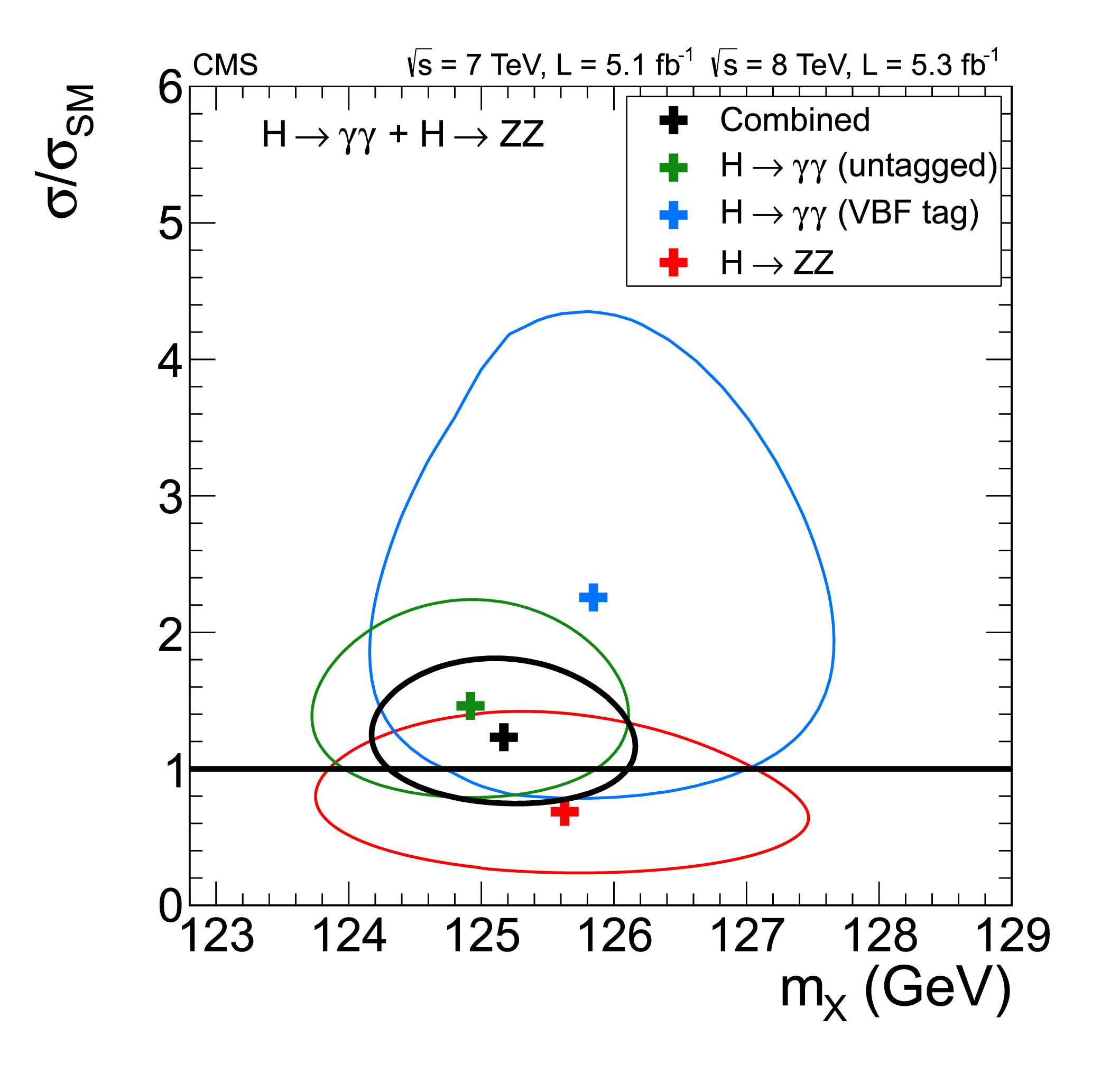}
  \end{minipage}
  \caption{Best fit values for the signal strength $\mu$ versus 
   hypothetical Higgs-boson mass $\MH$ in ATLAS~\cite{:2012gk} (left) and CMS~\cite{:2012gu} (right).}
  \label{fig:lhcmumass}
\end{figure}

\section{Conclusions and outlook}

Almost 50 years after the postulate of an elementary scalar boson,
which in the standard theory of electroweak interactions ---the Glashow--Salam--Weinberg model---
is the companion of the mechanism to describe the spontaneous breaking of the electroweak symmetry,
a new neutral boson with a mass of $126\GeV$ has been discovered in the search for the SM Higgs boson at the LHC.
So far all observed event yields in the different final states are consistent with the expectations 
for the SM Higgs boson, but also with predictions for a Higgs boson in several extended models.
Assuming that the excess observed at Tevatron in 
the $\Pb\Pbbar$ final state is not due to a statistical fluctuation, but 
due to the production of the particle discovered at the LHC, the decay of the new particle into 
a pair of b-quarks and probably also into a pair of tau leptons should be observed at LHC in the near future.

The allowed mass range for the Higgs boson could be confined unambiguously by the tremendous and challenging efforts 
in the searches at the colliders LEP, Tevatron, and LHC during the last decades: 
The combination of ALEPH, DELPHI, L3, and OPAL results exclude $\MH=0{-}114.4\GeV$,  
the preliminary combination of CDF and D0 results additionally exclude $\MH=147{-}180\GeV$ 
(and $\MH=100{-}103\GeV$), and finally at least one of ATLAS and CMS excludes the SM Higgs boson in
the ranges $\MH=110{-}122\GeV$ and $\MH=127{-}600\GeV$. 
In significant parts of the excluded mass ranges a Higgs boson with reduced couplings with respect to the SM 
prediction is also excluded.

The role of the Higgs boson in the zoo of elementary particles is rather special. It is the only elementary spin-0 boson 
and shows a very distinctive interaction pattern with all other particles. The scalar nature of the new boson can be affirmed 
by excluding the alternative integer-spin hypotheses from the measurement of angular correlations in production and decay
of the new boson. Assuming it is a spin-0 boson this will probably be possible with the data collected at LHC until the end of 2012.
In order to verify the consistency with being a Higgs boson, 
the Lorentz structure of the interactions with all particles
to which a coupling has been or will be observed, in particular to the massive gauge bosons, has to be determined. 
This includes the determination of its CP nature, which in the SM is predicted to be CP even. In models with extended
Higgs sectors CP-odd states or states of mixed CP nature are often 
predicted. 

The idea of the Higgs mechanism interprets particle masses as an interaction strength of each elementary massive particle 
to the vacuum expectation value of the Higgs field, whose particle excitation represents the Higgs boson.
As a result, all massive elementary particles couple to the Higgs boson proportional to their masses. 
In a first approach, the coupling strengths can be determined from 
the measurement of the products of production cross section times branching ratio, which can be extracted from the 
observed event yields in the different final states. 
Later, more sophisticated coupling analyses based on effective field theories, quantifying non-standard
effects in a model-independent way, may be carried out with higher precision. 
For a mass of $126\GeV$, for which the SM predicts the total Higgs-boson 
width to be $4\MeV$, the total width cannot be determined at the LHC with sufficient precision. As a consequence 
only ratios of couplings can be determined at the LHC in a completely model-independent way. However, with mild assumptions,
e.g.\ that the couplings to massive weak gauge bosons 
are not larger than predicted by the SM (valid in many extended Higgs sectors),
absolute coupling strengths can be measured as well. At a future $\Pep\Pem$ collider absolute coupling values can be determined
in a completely model-independent way
and probably with a significantly better precision. Finally, a determination of the
trilinear Higgs-boson self-coupling would be highly desirable in order to
at least partially reconstruct the Higgs potential. Owing to the low 
signal rates that are expected  for Higgs-boson pair production this will be very challenging at the LHC, both
in its run at a CM energy of $14\TeV$
and in its high-luminosity phase, but also future $\Pep\Pem$ colliders will have to be pushed
to their limits to this end.

Assuming the observed particle will be identified as a Higgs boson, the power to discriminate the SM Higgs boson from 
Higgs bosons in extended models depends on the achievable precision in the coupling determination. 
The possibly observed 
qualitative or quantitative differences between SM theory predictions and experiment in Higgs-boson observables 
could indicate the way,  how the SM may have to be extended, 
and tell which models are clearly disfavoured or can be ruled out.
In addition, the future research programme includes the search for additional Higgs bosons in non-standard scalar sectors,
the search for other new particles, and the investigation of processes like weak-gauge-boson scattering that
are most sensitive to the mechanism of electroweak symmetry breaking.

Any model for the unification of the fundamental strong and electroweak forces will predict new heavy particles, which in turn 
receive their mass most likely via spontaneous symmetry breaking. Accepting the Higgs mechanism as the correct modelling of
electroweak symmetry breaking, it is, thus, reasonable to expect that new Higgs bosons arise in SM
extensions. Conversely, precision measurements of {\it the} or more Higgs bosons serve as a window to non-standard theories
or even grand unification. Supersymmetric theories are typical examples sharing this feature.

Apart from these general considerations, there are at least two known phenomena where our established theoretical world, which consists
of the Standard Model of particle physics and general relativity, finds its limitations: The unknown constitution of {\it Dark Matter}
and the enigmatical role and size of {\it Dark Energy} in the universe. Clarifying the nature and profile of the new particle is of utmost
importance especially in the light of those conundrums, since extended Higgs models can accommodate Dark Matter candidates and Higgs fields
could in fact play a key role in the inflationary phase of the cosmological expansion that is driven by Dark Energy.

Particle physics has reached a crossroad, and Higgs precision physics will probably contribute a lot to steer 
the field in future directions.

\end{document}
